\documentclass{ws-ijmpe}
\usepackage[super,compress]{cite}
\usepackage[usenames,dvipsnames]{color}
\usepackage{nicefrac}
\pdfoutput=1
\usepackage{float,ulem}
\usepackage{graphicx}
\usepackage{epsfig,epstopdf,amsmath,amssymb}
\usepackage{xcolor}
\usepackage{subfig}
\usepackage{subcaption}
\usepackage{caption}

\newcommand{\be}{\begin{equation}}
	\newcommand{\ee}{\end{equation}}
\newcommand{\ba}{\begin{eqnarray}}
	\newcommand{\ea}{\end{eqnarray}}
\newcommand{\nn}{\nonumber\\}

\newcommand{\bas}{\begin{eqnarray*}}
	\newcommand{\eas}{\end{eqnarray*}}

\usepackage{scalerel}

\usepackage{soul}
\usepackage{hyperref}
\usepackage{wrapfig}

\usepackage{slashed}
\usepackage{comment}
\usepackage{float}
\usepackage{graphicx}
\begin{document}
\setcounter{tocdepth}{1} 

\markboth{P. Palni et al.}{Dynamics of Hot QCD Matter 2024 – Bulk Properties}

\catchline{}{}{}{}{}

 \title{Dynamics of Hot QCD Matter 2024 – Bulk Properties}

\author{
Prabhakar Palni$^1$\footnote{prabhakar@iitmandi.ac.in}~,
Amal Sarkar$^1$\footnote{amal@iitmandi.ac.in}~,
Santosh K. Das$^2$\footnote{santosh@iitgoa.ac.in}~,
Anuraag Rathore$^1$, Syed Shoaib$^1$, Arvind Khuntia$^3$, Amaresh Jaiswal$^4$, Victor Roy$^{4}$, Ankit Kumar Panda$^4$, Partha Bagchi$^4$, Hiranmaya Mishra$^4$, Deeptak Biswas$^4$, Peter Petreczky$^6$, Sayantan Sharma$^7$, Kshitish Kumar Pradhan$^8$, Ronald Scaria$^8$, Dushmanta Sahu$^8$, Raghunath Sahoo$^8$, Arpan Das$^9$, Ranjita K Mohapatra$^{10}$, Jajati K. Nayak$^{11}$, Rupa Chatterjee$^{11}$, Munshi G Mustafa$^{14}$, Aswathy Menon K.R.$^8$, Suraj Prasad$^8$, Neelkamal Mallick$^8$, Pushpa Panday$^{13}$, Binoy Krishna Patra$^{13}$, Paramita Deb$^{14}$, Raghava Varma$^{14}$, Ashutosh Dwibedi$^{17}$, Thandar Zaw Win$^{17}$, Subhalaxmi Nayak$^{17}$, Cho Win Aung$^{17}$, Sabyasachi Ghosh$^{17}$, Sesha Vempati$^{17}$, Sunny Kumar Singh$^{15}$, Manu Kurian$^{16}$, Vinod Chandra$^{15}$, Soham Banerjee$^4$, Sumit$^{13}$, Rohit Kumar$^{1}$, Rajkumar Mondal$^{5,11}$, Nilanjan Chaudhuri$^{5,11}$, Pradip Roy$^{5,12}$, Sourav Sarkar$^{5,11}$, Lokesh Kumar$^{18}$
(authors)\footnote{
		The contributors on this author list have contributed only to those sections of the report, which they cosign with their name. Only those have collaborated together, whose names appear together in the header of a given section.}}

  \address{
  $^{1}$ School of Physical Sciences, Indian Institute of Technology Mandi, Mandi, Himachal Pradesh, India \\
    $^{2}$ School of Physical Sciences, Indian Institute of Technology Goa, Ponda-403401, Goa, India\\
  $^{3}$ INFN Bologna, Italy\\
  $^{4}$School of Physical Sciences, National Institute of Science Education and Research, HBNI,
Jatni-752050, India\\
  $^{5}$Homi Bhabha National Institute, Training School Complex, Anushaktinagar, Mumbai -
400085, India\\
$^{6}$Physics Department, Brookhaven National Laboratory, Upton NY 11973, USA\\
$^7$The Institute of Mathematical Sciences, a CI of Homi Bhabha National Institute, Chennai, 600113, India\\
$^8$Department of Physics, Indian Institute of Technology Indore, Simrol, Indore 453552, India\\
$^9$ Department of Physics, Birla Institute of Technology and Science Pilani, Pilani Campus, Pilani, Rajasthan-333031\\
$^{10}$ Department of Physics, Rajdhani College, Bhubaneshwar, Odisha 751003, India \\
$^{11}$ Variable Energy Cyclotron Centre, 
1/AF, Bidhan Nagar, Kolkata-700064, India\\
$^{12}$ Saha Institute of Nuclear Physics, 1/AF Bidhannagar, Kolkata - 700064, India\\
$^{13}$ Department of Physics, Indian Institute of Technology Roorkee, Roorkee, Uttrakhand, 247667, India\\
$^{14}$ Department of Physics, Indian Institute of Technology Bombay, Powai, Mumbai-400076, India\\
$^{15}$Department of Physics, Indian Institute of Technology Gandhinagar, Gandhinagar, Gujarat 382355,
India\\
$^{16}$RIKEN, Brookhaven National Laboratory,
New York, New York 3000442, USA\\
$^{17}$ Department of Physics, Indian Institute of Technology Bhilai, Kutelabhata, Durg 491002, India\\
$^{18}$ Department of Physics, Panjab University Chandigarh 1610014, India\\
}

\maketitle

\begin{abstract}
The second Hot QCD Matter 2024 conference at IIT Mandi focused on various ongoing topics in high-energy heavy-ion collisions, encompassing theoretical and experimental perspectives. This proceedings volume includes 19 contributions that collectively explore diverse aspects of the bulk properties of hot QCD matter. The topics encompass the dynamics of electromagnetic fields, transport properties, hadronic matter, spin hydrodynamics, and the role of conserved charges in high-energy environments. These studies significantly enhance our understanding of the complex dynamics of hot QCD matter, the quark-gluon plasma (QGP) formed in high-energy nuclear collisions. Advances in theoretical frameworks, including hydrodynamics, spin dynamics, and fluctuation studies, aim to improve theoretical calculations and refine our knowledge of the thermodynamic properties of strongly interacting matter. Experimental efforts, such as those conducted by the ALICE and STAR collaborations, play a vital role in validating these theoretical predictions and deepening our insight into the QCD phase diagram, collectivity in small systems, and the early-stage behavior of strongly interacting matter. Combining theoretical models with experimental observations offers a comprehensive understanding of the extreme conditions encountered in relativistic heavy-ion and proton-proton collisions.

\end{abstract}


\ccode{PACS numbers:12.38.-t, 12.38.Aw}

\tableofcontents 

%
%
\section{Study of correlation between the relative transverse multiplicity activity in underlying event and transverse spherocity}


\author{Anuraag Rathore, Syed Shoaib, Arvind Khuntia, and Prabhakar Palni}

\bigskip

\begin{abstract}
This contribution studies the correlation between two global observables of event activity, namely the relative transverse multiplicity activity classifier $(R_{T})$ in the Underlying Event (UE) and transverse spherocity $(S_{0})$ in proton-proton collisions. This study aims to understand soft particle production using the differential study of $(R_{T})$ and $(S_{0})$. We have used the PYTHIA 8 Monte Carlo (MC) with different implementations of color reconnection and rope hadronization models to simulate proton-proton collisions at $\sqrt{s}$ = 13 TeV. The relative production of hadrons is also discussed in low and high transverse activity regions. Experimental confirmation of these results is feasible using ALICE Run 3 data, providing more insight into soft physics in the transverse region and enhancing our understanding of small system dynamics.
\end{abstract}




\subsection{Introduction}
High energy particle collisions are crucial for studying matter's building blocks and governing interactions. Differentiating between event topologies, like jetty events or hard scatterings with high transverse momentum and isotropic events with soft interactions, is essential. Event classifiers such as transverse spherocity $S_{0}$ and relative transverse event-averaged multiplicity classifier $R_{T}$ are vital. Spherocity measures an event's shape\cite{Acharya2019}, while $R_{T}$ is just the ratio of charged particle multiplicity versus event-averaged multiplicity for minimum bias events, offering insights into particle production variations. \\This study uses Pythia8, MC event generator with the Monash 2013\cite{Skands2014}. It also includes rope hadronization and color reconnection in Monash's framework. Rope hadronization involves color interactions forming 'ropes' among multiple strings in high-multiplicity collisions\cite{Acharya2023}, increasing hadron and specific particle production. Color reconnection optimizes parton color connections before hadronization\cite{Ortiz2019}, minimizing string length and energy, thus affecting particle distribution. Using these models or tunes, minimum-bias events are simulated for pp collisions at $\sqrt{s}$ = 13 TeV.

\begin{figure}[!htbp]
  \centering
  \includegraphics[width=6.1cm, height=5.5cm]{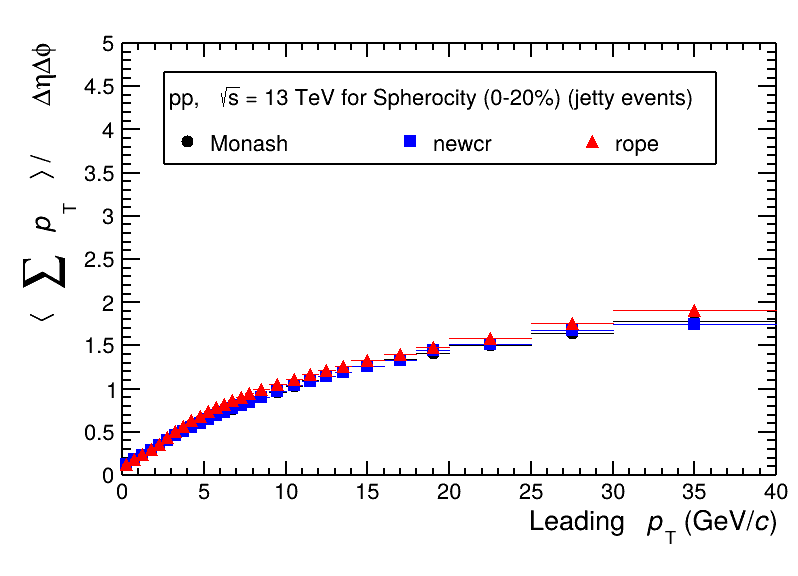}
  \includegraphics[width=6.1cm, height=5.5cm]{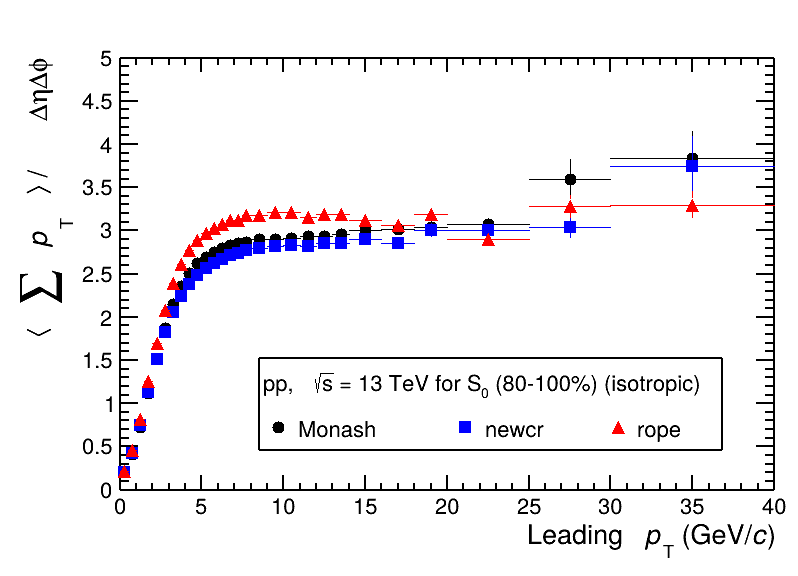}
  \caption{The energy density of charged particles as a function of leading $p_{T}$ in the transverse region for different spherocity classes: (left) for $S_{0}=0-20\%$, and (right) for $S_{0}=80-100\%$.}
  \label{ch_tunes}
\end{figure}

\begin{figure}[!htbp]
  \centering
  \includegraphics[width=6.2cm, height=6cm]{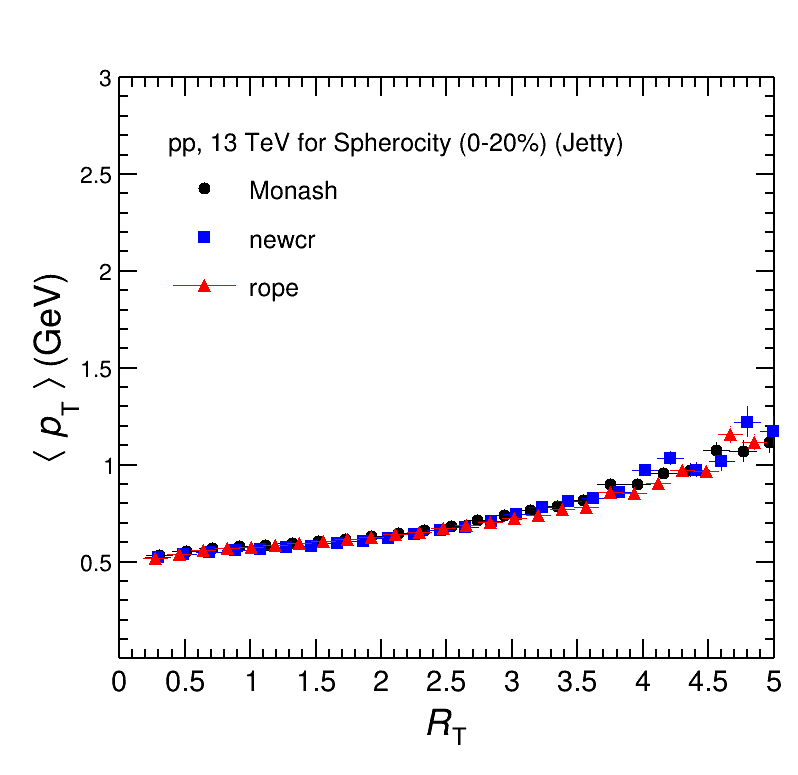}
  \includegraphics[width=6.2cm, height=6cm]{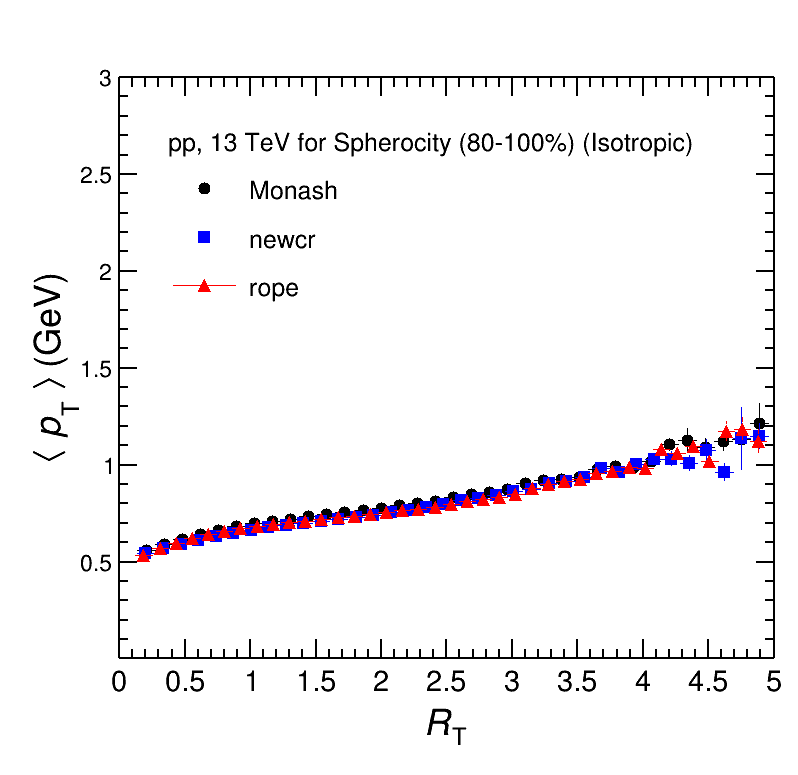}
  \caption{The average transverse momentum $\langle p_{T} \rangle$ as a function of $R_{T}$ in the transverse region for different spherocity classes: (left) for $S_{0}=0-20\%$, and (right) for $S_{0}=80-100\%$.}
  \label{meanpt_tunes}
\end{figure}

\begin{figure}[htbp!]
    \centering
    \includegraphics[width=0.45\textwidth]{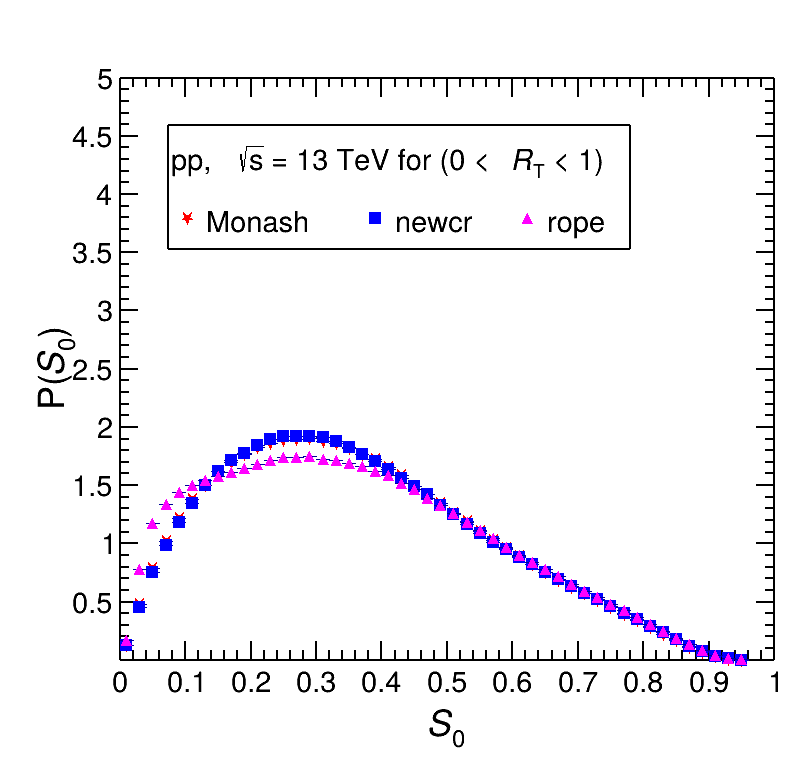}
    \hfill
    \includegraphics[width=0.45\textwidth]{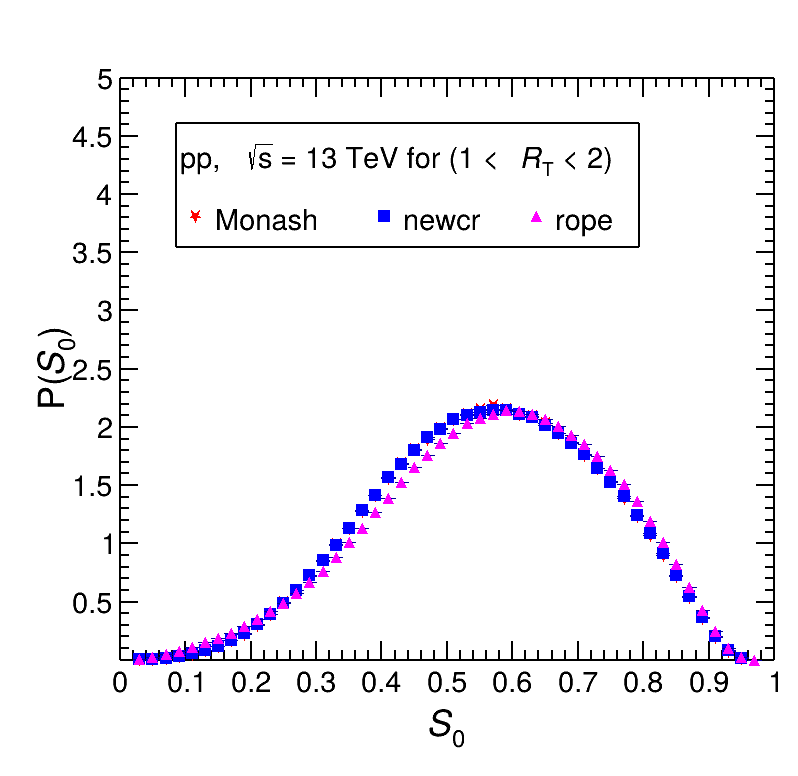}

    \vspace{0.5cm} 

    \includegraphics[width=0.45\textwidth]{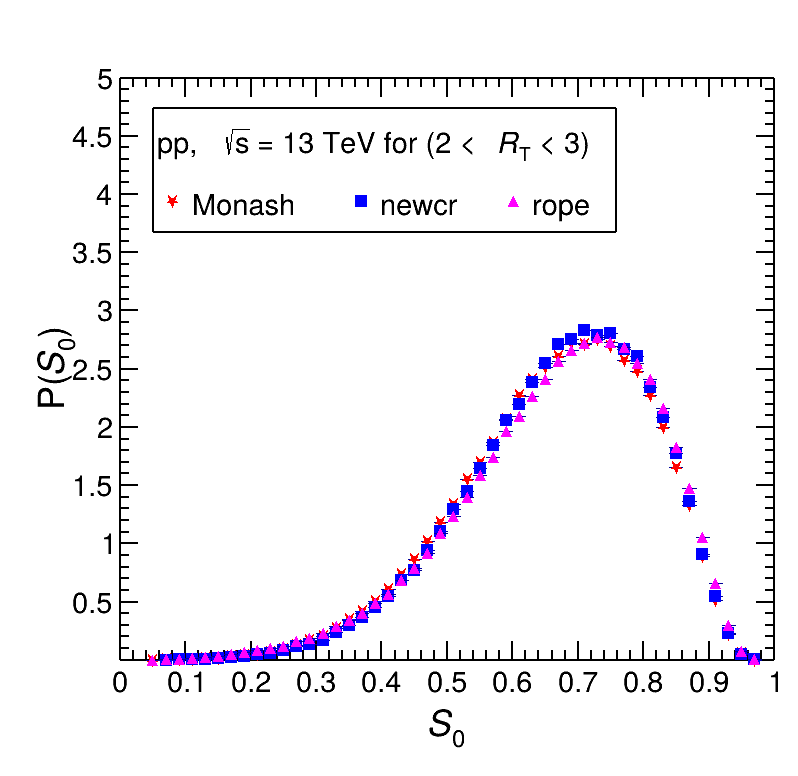}
    \hfill
    \includegraphics[width=0.45\textwidth]{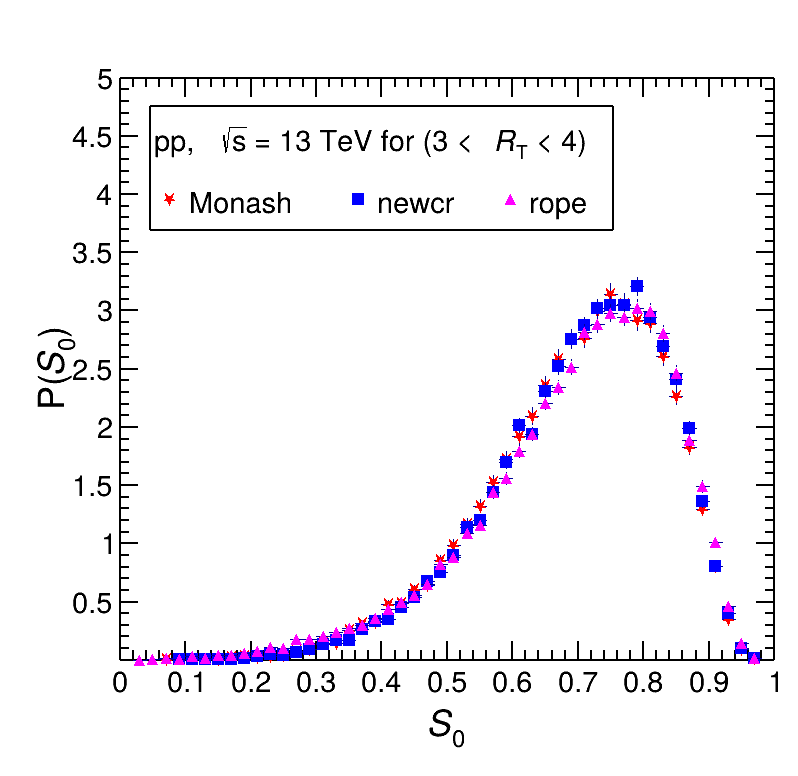}

    \caption{Spherocity plots for different $R_{T}$ values in the transverse region: 
    (top-left) $0 < R_{T} < 1$, (top-right) $1 < R_{T} < 2$, 
    (bottom-left) $2 < R_{T} < 3$, and (bottom-right) $3 < R_{T} < 4$.}
    \label{sph}
\end{figure}

\subsection{Event Generation and the Observables}
The UE characteristics depend on the leading charged particle's orientation\cite{Ortiz2019}, aligned with the highest transverse momentum parton. Events are divided into three regions: (i) toward, (ii) away, and (iii) transverse, based on the azimuthal angle difference with the leading particle's path. Particles in the toward region are within azimuthal angle $|\triangle\phi| < 60^\circ$, while those in the away region are influenced by hard scattering and characterized by $|\triangle\phi| > 120^\circ$. The transverse region is ideal for UE analysis. Transverse spherocity is introduced as an unique event shape observable, designed to differentiate events based on their geometric configuration\cite{Ortiz2024, Acharya2024} and is defined as 
\begin{equation}
    S_{0} = \frac{\pi^2}{4} \min_{\hat{n}} \left( \frac{\sum_i |\vec{p}_{T,i} \times \hat{n}|}{\sum_i \vec{p}_{T,i}} \right)^2
\end{equation}

Now $(R_{T})$ is defined as the ratio of multiplicity of the inclusive charged-particles to its event-averaged
multiplicity $(\langle N_{inc} \rangle)$ in the transverse region, serving as a tool to differentiate events based on transverse activity\cite{Palni2020}.

\begin{equation}
    R_{T} = \frac{N_{inc}}{\langle N_{inc} \rangle}
\end{equation}

\subsection{Results and Discussions}
In this work all the work has been done by taking Monash 2013 tune as reference to rope hadronisation and new color reconnection models. 
Here Fig. \ref{ch_tunes} shows energy density of charged particles implemented for different tunes for different spherocity classes i.e, for jetty events $(S_{0}=0-20\%)$ and the isotropic events $(S_{0}=80-100\%)$. The energy density first increases at around $p_{T}= 3-7$ Gev/c then there comes the plateau region. The energy density of the charged particles is coming out more for isotropic events as compared to the jetty events. Also more enhancement in the rope hadronisation is seen compared to the other tunes. And Fig. \ref{meanpt_tunes} shows average transverse momentum $\langle p_{T} \rangle$ as a function of $R_{T}$ for jetty and isotropic events with different tunes implemented. Initially, we see that $\langle p_{T} \rangle$ is more for isotropic events as compared to the jetty ones. Now Fig. \ref{sph} shows spherocity plots for different $R_{T}$ regions. Here as we are moving from low $R_{T}$ (low transverse activity) to high $R_{T}$ (high transverse activity), the peak of the spherocity shifts towards right, and also becomes narrower, indicating isotropic distribution of particles in the high transverse activity region.

\subsection{Summary and Conclusions}
We have conducted an extensive investigation into the soft particle production utilizing the relative transverse multiplicity activity event classifier $(R_{T})$ within the underlying event (UE) and transverse spherocity $(S_{0})$ pertaining to proton-proton (pp) collisions at $\sqrt{s}=13$ TeV. The energy density of charged particles shows a trend around $p_{T} \simeq 3-7$ GeV/c, then there comes the plateau region. Energy density of charged particles is coming out more for isotropic events as compared to the jetty events in transverse region. The differences in the tunes can be clearly seen for isotropic distribution. The $\langle p_{T} \rangle$ is notably higher for isotropic events compared to the jetty events at low $R_{T}$. However, at high $R_{T}$, the $\langle p_{T} \rangle$ for jetty events experiences a greater increase compared to the isotropic one. The peak of spherocity shifts towards right as well as it becomes narrower as we move from low $R_{T}$ to high $R_{T}$ region, indicating isotropic distribution of particles. Inclusion of rope hadronization with Monash tune enhances the modelling of high density partonic environments, leading to higher multiplicity. Differences in these models highlight the importance of accurately modeling hadronization, especially in dense partonic environments.

\section{Polarization and spin hydrodynamics in relativistic heavy ion collisions}

\author{Amaresh Jaiswal}

\bigskip

\begin{abstract}

This proceedings contribution discusses two challenging aspects encountered in the theoretical formulation and phenomenological application of relativistic spin hydrodynamics.

\end{abstract}


\ccode{PACS number(s): 12.38.--t, 12.38.Aw}



\subsection{Introduction}

In non-central heavy-ion collisions at relativistic collider facilities like the  Large Hadron Collider (LHC) and the Relativistic Heavy-Ion Collider (RHIC), intense magnetic fields and large angular momentum are produced during the early stages of evolution. These phenomena can interact with the intrinsic spin of constituent particles through mechanisms analogous to the Einstein-de Haas and Barnett effects. This interaction was theorized to induce spin polarization in the medium, which becomes observable in particles emitted during the freeze-out~\cite{Niida:2024ntm}. Subsequent experimental observations have confirmed these predictions, sparking considerable interest and advancing the study of spin polarization in such systems~\cite{Florkowski:2018fap}.

It is now well established that the evolution of strongly interacting matter formed in relativistic heavy ion collisions can be described via relativistic hydrodynamics. Observation of spin polarization of hadrons in relativistic heavy ion collisions necessitates the inclusion of angular momentum conservation in the formulation of relativistic hydrodynamics, leading the the framework of relativistic spin hydrodynamics. On the other hand, determination of the total number of transport coefficients in relativistic dissipative spin hydrodynamics encounters a significant challenge, largely due to the so-called pseudogauge freedom or pseudogauge symmetry~\cite{Dey:2023hft}.

In phenomenological application of relativistic spin hydrodynamics, models incorporating equilibrated spin degrees of freedom have successfully explained observations of global spin polarization. However, they have failed in explaining the correct sign of longitudinal spin polarization, leading to a puzzling scenario referred to as the ``polarization sign problem"~\cite{Becattini:2017gcx}. This discrepancy suggests the possibility of distinct origins for spin polarization. Further, it has led to the hypothesis that spin degrees of freedom in the transverse plane may not reach equilibration by the time of freeze-out~\cite{Banerjee:2024xnd}.

In this proceedings contribution, I discuss these two puzzles: (1) Pseudogauge freedom and transport in spin hydrodynamics, and (2) Longitudinal polarization sign problem.


\subsection{Pseudogauge freedom and transport in spin hydrodynamics}

In a given theory, the energy-momentum and spin tensors can be recast into alternative forms that satisfy the same conservation laws for energy, linear momentum, and angular momentum~\cite{Hehl:1976vr}
\begin{eqnarray}
T^{\prime \mu \nu} &=& T^{\mu\nu} + \frac{1}  {2} \partial_\lambda \left( 
\Phi^{\lambda, \mu \nu} 
+\Phi^{\nu, \mu \lambda}
+\Phi^{\mu, \nu \lambda} \right), \nonumber \\
S^{\prime \lambda, \mu \nu} &=&
S^{\lambda, \mu \nu} - \Phi^{\lambda, \mu \nu} + \partial_\rho Z^{\mu\nu, \lambda \rho}. \label{eq:PG}
\end{eqnarray}
Here $T^{\mu\nu}$ and $S^{\lambda, \mu \nu}$ are the energy-momentum tensor and spin tensor, respectively. The tensors $\Phi^{\lambda, \mu \nu}$ and $Z^{\mu\nu, \lambda \rho}$ are known as superpotentials which have the following symmetries with respect to the exchange of indices
\begin{eqnarray}
\Phi^{\lambda, \mu \nu} &=& -\Phi^{\lambda, \nu \mu}, \nonumber \\
Z^{\mu\nu, \lambda \rho} &=& - Z^{\nu\mu, \lambda \rho} \,\,\,=\,\,\, -Z^{\mu\nu, \rho \lambda}. \label{eq:PGsymmetries}
\end{eqnarray}
The forms of energy-momentum tensor and spin tensor vary for different choices of the superpotentials $\Phi^{\lambda, \mu \nu}$ and $Z^{\mu\nu, \lambda \rho}$. However, these forms differ in the local redistribution of the conserved quantities, while the global values of energy, linear momentum, and angular momentum remain conserved.

The flexibility in definition of conserved currents due to pseudogauge freedom leads to a puzzling situation when attempting to count the total number of transport coefficients in the theory. This can be seen by considering three pseudogauge choices most commonly used in literature: (1) Belinfante, (2) Canonical and (3) de Groot, van Leuween and van Weert (GLW). In the case of Belinfante pseudogauge choice, the superpotentials $\Phi^{\lambda, \mu \nu}$ and $Z^{\mu\nu, \lambda \rho}$ are fixed such that the spin tensor $S^{\lambda, \mu \nu}$ vanishes and the energy-momentum $T^{\mu\nu}$ is symmetric in the two indices. For the Canonical pseudogauge choice, the spin tensor $S^{\lambda, \mu \nu}$ is antisymmetric under exchange of all three indices and $T^{\mu\nu}$ is not symmetric in the two indices. In the GLW case, the tensor structure for $S^{\lambda, \mu \nu}$ is most general with antisymmetric in last two indices only and $T^{\mu\nu}$ is symmetric tensor. 

In these three cases, we see that the Belinfante pseudogauge does not retain information about evolution of spin polarization. The spin tensor in Canonical pseudogauge is not of the most general form due to antisymmetric in all three indices and therefore has less information regarding evolution of spin polarization. On the other hand, the anti-symmetric part of $T^{\mu\nu}$ compensates for this leading to appearance of transport phenomena for spin evolution in both $S^{\lambda, \mu \nu}$ and $T^{\mu\nu}$. The GLW has most general form of spin tensor and therefore all the transport phenomena for spin evolution is contained in $S^{\lambda, \mu \nu}$. This redistribution of contribution to spin evolution between $S^{\lambda, \mu \nu}$ and $T^{\mu\nu}$ leads to an unresolved issue in counting of transport coefficients for evolution of spin polarization. From theoretical perspective, it is important to resolve this issue in order to formulate a consistent framework for dissipative spin hydrodynamics.


\subsection{Longitudinal polarization sign problem}

\begin{figure}[t!]
\centerline{\includegraphics[width=0.57\textwidth]{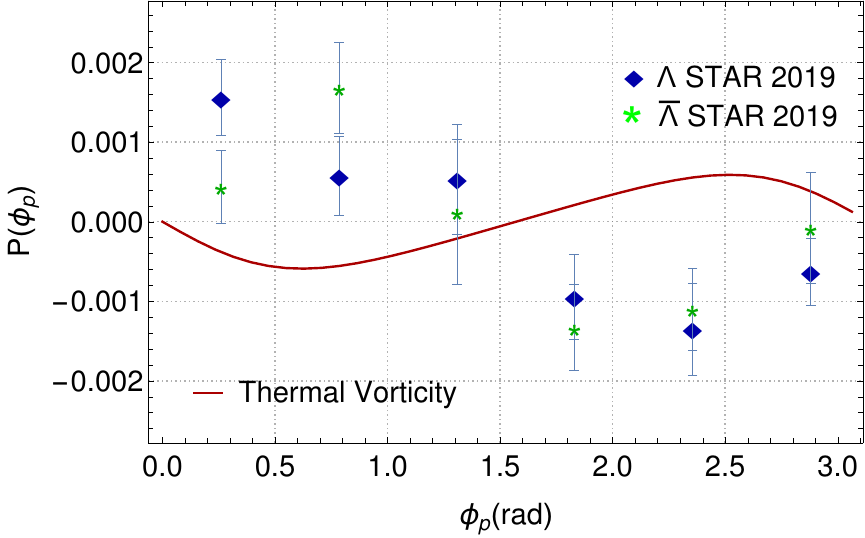}}
\caption{The longitudinal polarization shown as a function of the azimuthal angle $\phi_p$ for 30--60\% Au-Au collisions at $\sqrt{s_{\rm NN}} = 200$ GeV. Theoretical result using thermal vorticity is compared with the experimental data by STAR.}
\label{sign_prob}
\end{figure}

The longitudinal spin polarization of $\Lambda$ hyperons produced in Au-Au collisions at a beam energy of $\sqrt{s_{\rm NN}} = 200$~GeV has garnered significant interest~\cite{STAR:2019erd}. Experimental data reveal a quadrupole structure of longitudinal polarization in the transverse momentum plane, which notably disagrees in sign with most theoretical predictions—a discrepancy commonly referred to as the ``sign problem"~\cite{Becattini:2017gcx}; see Fig.~\ref{sign_prob}.

Previous studies have shown that incorporating thermal shear can help address the sign problem~\cite{Fu:2021pok, Becattini:2021suc, Becattini:2021iol}. However, additional assumptions are often required to fully reproduce the experimental data. These include, for instance, neglecting temperature gradients at freeze-out~\cite{Becattini:2021iol} or substituting the $\Lambda$ mass with the strange quark mass~\cite{Fu:2021pok}. Other approaches, such as employing projected thermal vorticity, have also proven effective in resolving the sign problem~\cite{Florkowski:2019voj}. However, the origin of this sign puzzle is the use of thermal vorticity in calculating longitudinal polarization at freezeout. Note that the thermal vorticity is the global equilibrium solution of the spin hydrodynamic equations~\cite{Florkowski:2018fap}. On the other hand, the global equilibrium condition may not be achievable due to the short lifetime of the fireball created
in relativistic heavy-ion collisions~\cite{Kapusta:2019sad, Ayala:2020ndx, Kumar:2023ghs, Hidaka:2023oze, Wagner:2024fhf, Banerjee:2024xnd}. Therefore, it is imperative to solve the hydrodynamic equations considering the evolution of spin polarization with appropriate initial conditions~\cite{Singh:2024cub}.


\subsection{Conclusion}

Two challenging aspects encountered in the theoretical formulation and phenomenological application of relativistic spin hydrodynamics were outlined: (1) Counting of transport phenomena in spin hydrodynamics due to pseudogauge freedom and, (2) Sign problem in longitudinal polarization. The resolution of these problems are crucial in further understanding of spin hydrodynamics and its application.


\section{Baryon stopping and EM fields in Heavy-Ion collisions}

\author{Victor Roy, Ankit Kumar Panda, Partha Bagchi, Hiranmaya Mishra}

\bigskip

\begin{abstract}
This study investigates the impact of baryon stopping on electromagnetic fields in low-energy ($\sqrt{s}_{NN} \sim 4- 20 $ GeV) heavy-ion collisions. Using a Monte-Carlo Glauber model and incorporating a novel deceleration ansatz, we demonstrate significant alterations in the magnitude and time evolution of electromagnetic fields when baryon stopping is considered. Our findings suggest that the interplay between baryon stopping and finite conductivity could lead to longer-lasting fields, potentially influencing various observables in heavy-ion collisions at low energies.
\end{abstract}

\subsection{Introduction}
Heavy-ion collisions at relativistic energies produce strong electromagnetic (EM) fields, which play a crucial role in various phenomena observed in these collisions. At lower collision energies, baryon stopping becomes increasingly important \cite{BRAHMS:2009wlg,BRAHMS:2003wwg,NA49:1998gaz}, potentially affecting the evolution of these EM fields. This work explores the interplay between baryon stopping and EM fields in low-energy heavy-ion collisions, with a focus on Au+Au collisions at low energies $\sqrt{s_{NN}} = 4 - 20$ GeV.

\subsection{Methodology}
We employ a Monte-Carlo Glauber (MCG) model to simulate the vacuum evolution of the electromagnetic fields produced in high-energy heavy-ion collisions. The usual MCG model incorporates the nuclear charge density distribution
and the energy-dependent nucleon-nucleon scattering cross-subsection to generate event-by-event distribution of nucleons inside the colliding nucleus and hence could be used to calculate the number of binary collisions and participants for a given centre of mass energy and impact parameters of a collision.
To account for baryon stopping, we introduce a novel deceleration ansatz:
\begin{equation}
    \beta(\tau) = \mathcal{A} \left[1 - \tanh \left(\frac{\tau - \tau_h}{\Delta\tau}\right)\right].
\end{equation}

\begin{figure}[ht] 
    \centering
    \includegraphics[scale=0.25]{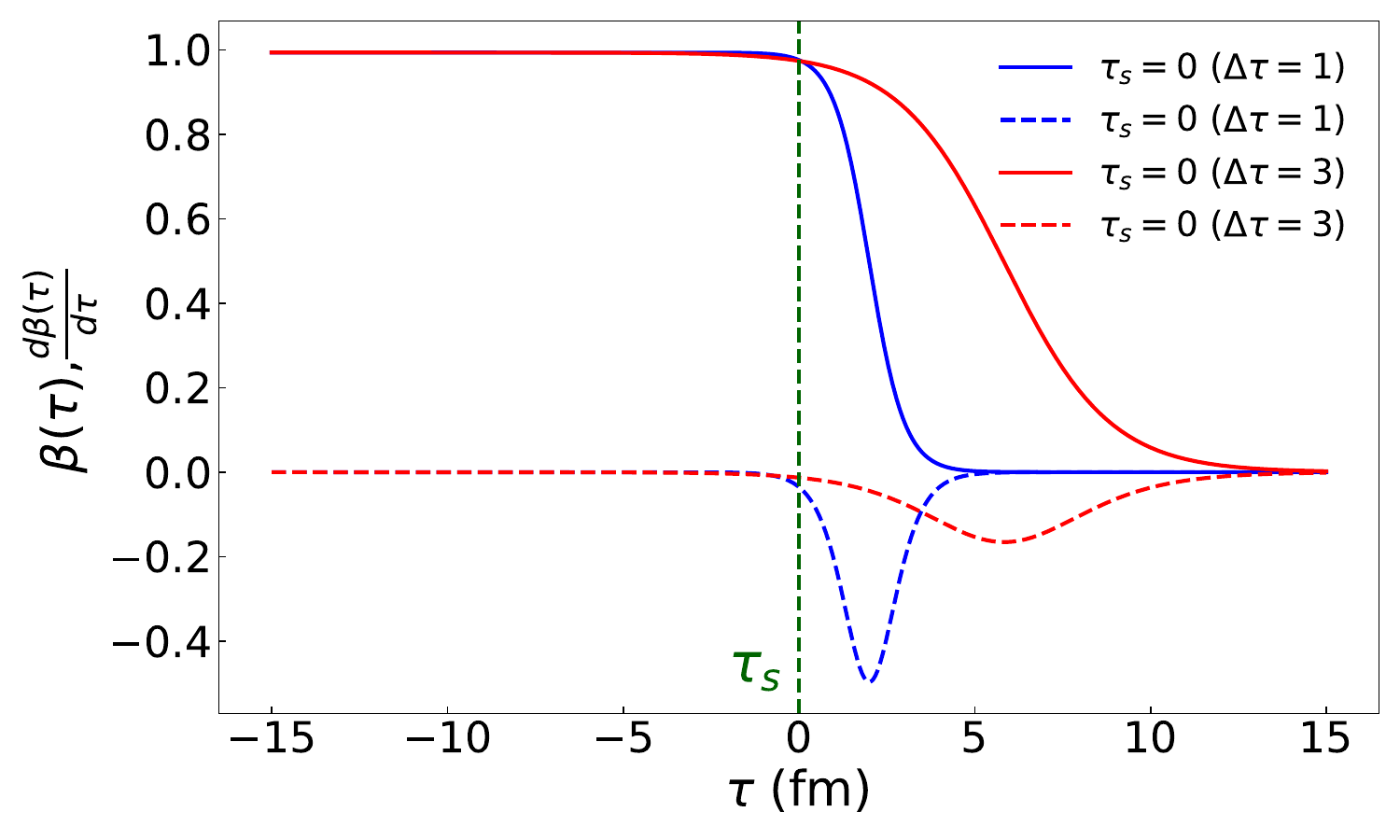} 
    \caption{Parametrized velocity (solid lines) and the corresponding acceleration (dashed lines) for $\Delta\tau=1$ (blue lines) and 3 fm (red lines).}
    \label{fig:vel_parm} 
\end{figure}

This ansatz offers following advantages:
\begin{itemize}
    \item It is free from kinks, providing a smooth deceleration profile.
    \item It can be applied to collisions at any energy.
    \item The deceleration can be controlled by adjusting parameters.
\end{itemize}
The initial velocity $\beta_{NN}$ is related to the collision energy $\sqrt{s_{NN}}$ by:
\begin{equation}
    \beta_{NN} = \left(1.0 - \frac{4m^2}{s_{NN}}\right)^{1/2} = 2\mathcal{A}.
\end{equation}
Further we define a starting time $\tau_s$ for individual nucleon-nucleon
collision as the time when $\beta(\tau_s)/\beta_{NN} \sim 0.98$, ensuring a realistic initial deceleration. The model parameters are tuned to reproduce known characteristics of baryon stopping:

\begin{itemize}
    \item At high energies, each binary collision typically results in 1 unit of rapidity loss
    \item For $\sqrt{s_{NN}} = 4$ GeV, we observe:
    \begin{itemize}
        \item Approximately 1.2 units of rapidity loss.
        \item An 80\% change in velocity.
    \end{itemize}
\end{itemize}

Figure \ref{fig:vel_parm} illustrates the velocity profile $\beta(\tau)$ and its derivative for different values of $\Delta\tau$ , demonstrating the flexibility of our ansatz in modeling various deceleration scenarios.

EM fields are calculated at the point of observation ${\bf{r}}_{\text{obs}}$ at time $t_{\text{obs}}$ using relativistic field equations with retarded time \cite{Tong:lecturenotes}:
\begin{equation}
e{\bf{B}}({\bf{r}}_{\text{obs}},t_{\text{obs}}) = -\mathcal{C}  Z \alpha_{EM} \left[ \frac{\hat{\bf{R}} \times {{\beta}(t')}}{\gamma^2 k^3 R^2} + \frac{(\hat{{\bf R}} \cdot \dot{{\beta}}(t')) (\hat{\mathbf{R}} \times {\beta}(t')) + k \hat{{\bf R}} \times \dot{{\beta}}(t')}{k^3 R}\right]_{t'} ,
\end{equation}
\begin{equation}
e{\bf{E}}({\bf{r}}_{obs},t_{obs}) = \mathcal{C} Z \alpha_{EM}  \left[ \frac{\hat{\bf{R}}-{\beta}(t^{\prime})}{\gamma^2 k^3 R^2} 
 + \frac{\hat{R} \times [(\hat{\bf{R}}-{\beta}(t^{\prime}))\times \dot{{{\beta}}}(t')]}{k^3 R } \right]_{t'},
\end{equation}
where $t^{\prime}= t_{obs}- \frac{R}{c}$ is the retarted time. Here $\mathcal{C}(=fm^{-2}/m_{\pi}^{2} \sim 2$) is a numerical factor, $\alpha_{EM}= \frac{1}{137}$ is the fine structure constant, and $Z$ is the atomic number of each nucleus (we consider symmetric collisions). The relative position $\bf{R}(t^{\prime}) = \bf{r}_{\text{obs}}-\bf{r'}(t')$, and the unit vector along it is defined as $\hat{\bf{R}} = \frac{\vec{R}}{R}$, the factor $k = 1- \hat{\bf{R}} \cdot \bm{\beta}(t')$, and $\gamma= \frac{1}{\sqrt{1-\beta^2}}$ is the Lorentz factor.

\subsection{Results and Discussion}

Our simulations reveal several important insights into the interplay between baryon stopping and electromagnetic fields in low-energy heavy-ion collisions:

\subsubsection{Impact of Baryon Stopping on EM Fields}
\begin{figure}[h]
   \centering
    \includegraphics[scale=0.3]{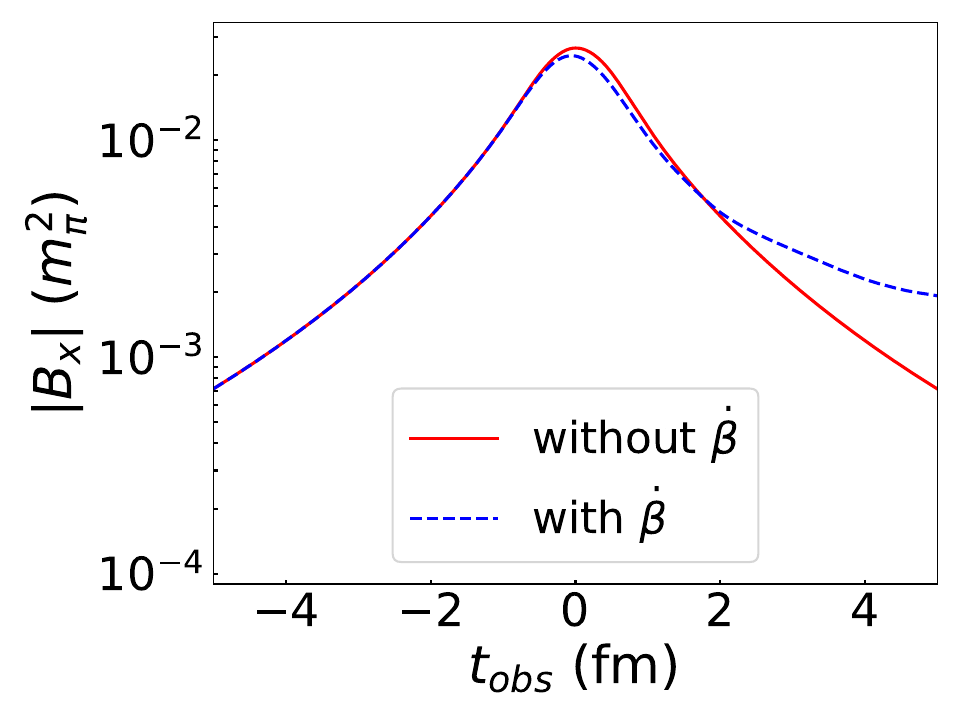}
    \includegraphics[scale=0.3]{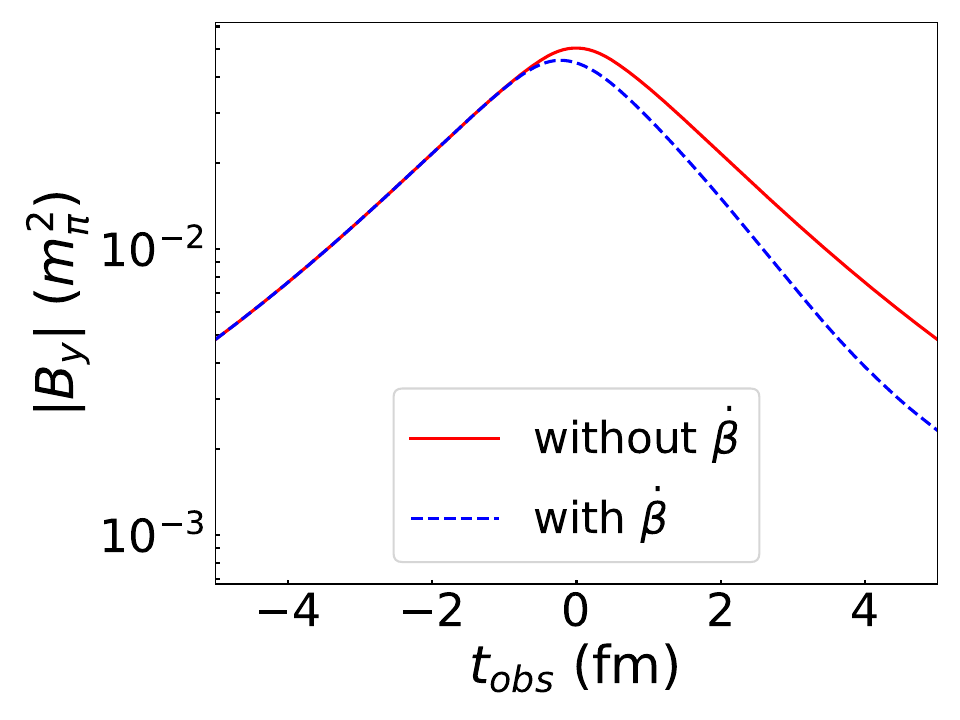}
    \includegraphics[scale=0.3]{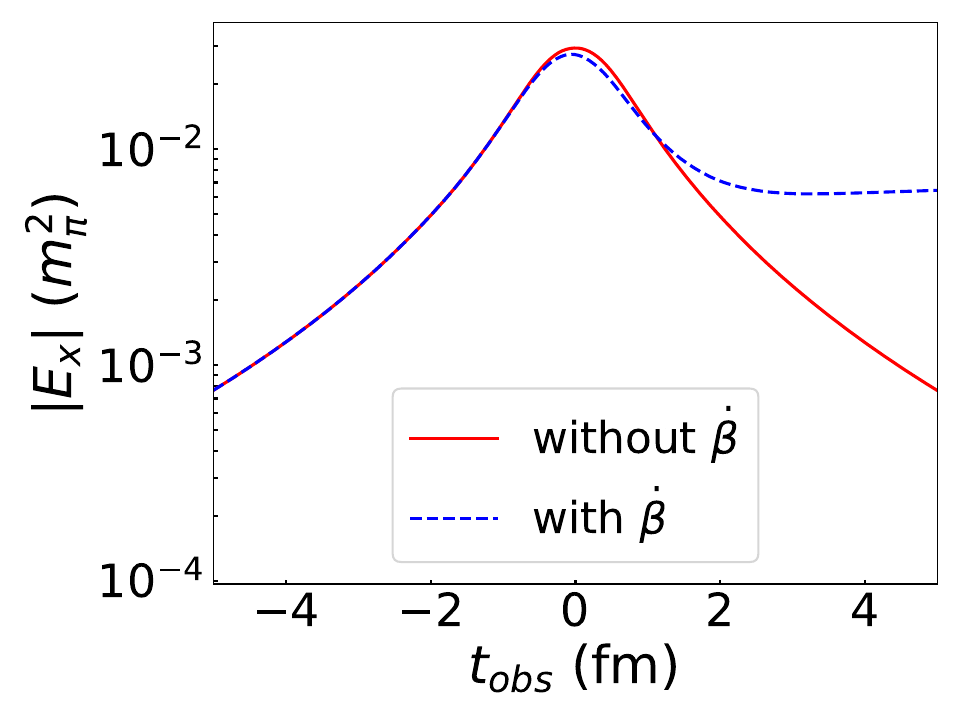}
    \includegraphics[scale=0.3]{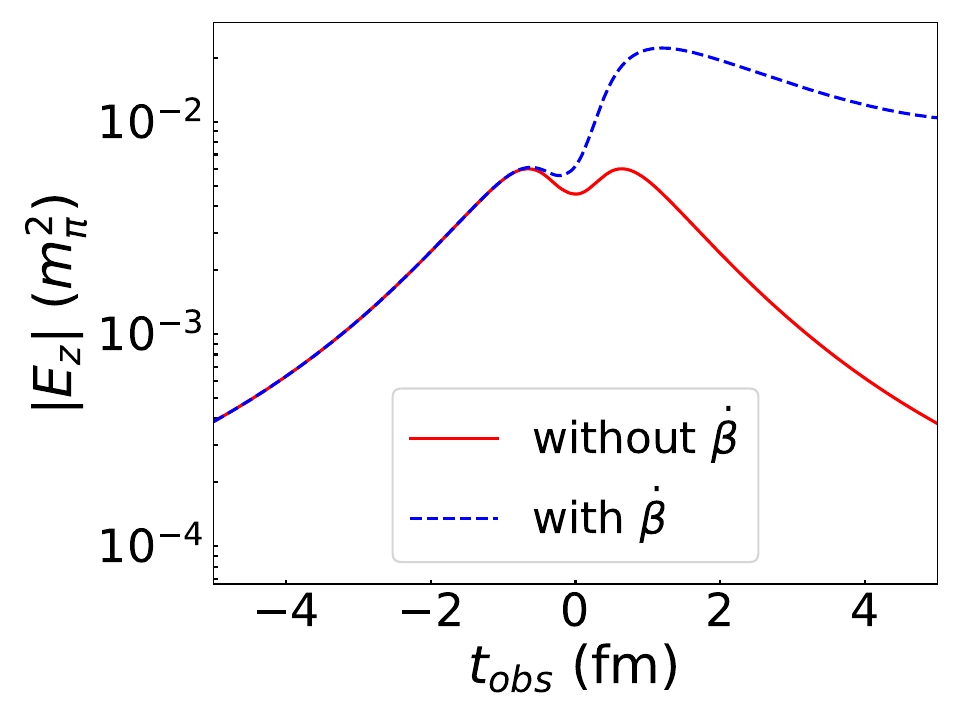}
  \caption{(Color online) Comparison of $|B_x|, |B_y|$ (top two panels), and  $|E_x|, |E_z|$ (bottom two panels) with (blue dashed lines) and without deceleration (red solid lines) at $\sqrt{s_{NN}} = 4$ GeV with $b$ = 3 fm at $\textbf{r}_{\text{obs}}= (0,0,0)$.}\label{fig:compwithwithout}
\label{fig:2}
\end{figure}

The incorporation of baryon stopping significantly alters both the magnitude and time evolution of electromagnetic fields. In particular referring to Fig.(\ref{fig:2}) we observe: 

\begin{itemize}
    \item For central collisions (b = 3 fm), we observe that the magnetic field component $B_x$ experiences a slower decay when baryon stopping is considered. This prolonged field duration could have significant implications for various observables in heavy-ion collisions.
    
    \item In peripheral collisions (b = 12 fm), the y-component of the magnetic field ($B_y$) dominates over other components (not shown here). The effect of baryon stopping is least in the peripheral collisional as expected since the number of participant is very small in those collisions. 
    
    \item The electric field components show similar modifications, with the $E_z$ component exhibiting the most pronounced changes due to baryon stopping.
\end{itemize}

\subsubsection{Dependence on Collision Parameters}

We investigated the sensitivity of our results to various collision parameters. The deceleration parameter $\Delta\tau$ plays a crucial role in determining the field evolution. We observe a non trivial dependence of field time evolution on $\Delta\tau$; some of the field components sustain for a longer period for larger $\Delta\tau$ but rest of the components show opposite trend.
 The observation point $\mathbf{r}_\text{obs}$ significantly influences the observed field components. We found that fields measured at points away from the collision axis show distinct behavior compared to those on the axis. The collision energy $\sqrt{s_{NN}}$ affects both the initial field strengths and the degree of baryon stopping. Our model successfully captures the energy dependence of these effects.
 
\subsubsection{Implications for Observables}

The modified electromagnetic fields due to baryon stopping could have far-reaching consequences for various observables in heavy-ion collisions:

\begin{itemize}
    \item The prolonged duration of strong fields could enhance the production of dileptons and photons, particularly in the low transverse momentum region.
    
    \item Directed flow of charged particles may be significantly affected, especially in peripheral collisions where the field asymmetry is most pronounced.
    
    \item The interplay between baryon stopping and finite conductivity of the medium suggests the possibility of even longer-lasting fields, which could influence the evolution of the quark-gluon plasma.
\end{itemize}

Our results underscore the importance of considering baryon stopping in electromagnetic field calculations, especially for low-energy heavy-ion collisions where this effect is most prominent. The enhanced fields and their prolonged durations could provide new insights into the properties of strongly interacting matter under extreme conditions.

\subsection{Conclusion}
This study demonstrates the significant impact of baryon stopping on EM field evolution in low-energy heavy-ion collisions. Our improved velocity ansatz and consideration of multiple collisions would possibly provide a foundation for explaining net baryon rapidity distribution in future work. The interplay between baryon stopping and finite conductivity suggests the possibility of longer-lasting EM fields, which could have important implications for observable phenomena in heavy-ion collisions at low energies.

\section{Chiral pseudo-critical line in a hadron resonance gas model} 
\author{Deeptak Biswas, Peter Petreczky, Sayantan Sharma }

\bigskip

\begin{abstract}
We investigate the chiral transition in a hadron resonance gas (HRG) model at moderate and high baryon density by including a mean-field repulsive interaction among baryons. We have fixed the strength of the repulsion by comparing the HRG model's estimations of the higher-order baryon susceptibilities with those from the lattice QCD. We have identified the pseudo-critical line by analyzing the temperature variation of the renormalized chiral condensate and calculated the curvature coefficients $\kappa_2$ and $\kappa_4$. Our findings have revealed a non-zero value of $\kappa_4$ for the first time. Additionally, we discuss potential implications for heavy-ion collisions while considering strangeness neutrality.
\end{abstract}


\ccode{PACS numbers:}
\subsection{Introduction}

 The chiral symmetry, which is broken in the QCD vacuum, is effectively
 restored at a pseudo-critical temperature $T_{pc}=156.5(1.5)$ MeV \cite{HotQCD:2018pds}.
 We address different aspects of this chiral transition from a hadron resonance gas model (HRG) perspective. 
 In Ref~.\cite{Biswas:2022vat}, we determined a pseudo-critical temperature $T_{pc} = 161.2(1.7)$ MeV at 
 zero-baryon density within an ideal HRG model. The estimated curvature coefficient of the pseudo-critical line 
 was in excellent agreement with the lattice QCD results~\cite{HotQCD:2018pds, Borsanyi:2020fev}. At finite baryon 
 densities, the non-resonant interactions among the (anti-)baryons are necessary to explain various bulk observables 
and the conserved charge fluctuations~\cite{Vovchenko:2016rkn}. In Ref.\cite{Huovinen:2017ogf} repulsive interaction 
among (anti-) baryons was introduced in the mean-field approximation to address the deviation of ideal HRG model 
results from lattice QCD  for the difference between second and fourth-order fluctuations. In Ref.~\cite{Biswas:2024xxh}, 
we extended the study at higher baryon densities and examined the limit of the applicability of this model in the context 
of chiral transition. We constrained the strength of the repulsive interaction by comparing with the state-of-the-art 
lattice QCD results of net-baryon number fluctuations, which allowed us to calculate the pseudo-critical line up to $
\mu_B=750$ MeV.

\subsection{Mean-field repulsion in the HRG model}\label{secII}
With the inclusion of repulsive interactions at the mean-field level, the pressure of the interacting (anti-)baryon ensemble at temperature $T$ and baryon chemical potential $\mu_B$ is~\cite{Huovinen:2017ogf}
\begin{equation}
P^{B\{\bar{B}\}}_{int}=T\sum \limits_{i\in B \{\bar B \}} \int 
g_i \frac{d^3 p}{(2 \pi)^3} \text{ln}\bigg[1+ e^{-\beta(E_i-\mu_i^{eff}})\bigg] 
+\frac{K}{2} n_{b\{\bar{b}\}}^2.
\label{Eq.Pint}
\end{equation}
Here $n_b$ and $n_{\bar{b}}$ denote the densities of baryons and anti-baryons, respectively. The effective 
chemical potential for the $i$th hadron is $\mu_i^{eff}=B_i \mu_B - K n_{b\{\bar{b}\}}$, where $K$ is the 
mean-field coefficient.  The number densities can be solved using the following pair of transcendental equations:
\begin{equation}
n_{{b}}=\sum_{i\in B}\int g_i \frac{d^3 p}{(2 \pi)^3}\:\frac{1}{e^{\beta{(E_{i}-\mu_B+K n_b)}}+1}
~~~{\text{and}}~~~
n_{{\bar{b}}}=\sum_{i\in \bar{B}}\int g_i \frac{d^3 p}{(2 \pi)^3}\:\frac{1}{e^{\beta{(E_{i}+\mu_B+K n_{\bar{b}})}}+1}.
\label{Eq.nb}
\end{equation}
The total pressure in this HRG model is partial sum of the non-interacting mesons and interacting baryons and anti-baryons. 
We have used an extended list of hadrons in our HRG model, which consists of the quark model predicted states along with 
the experimentally confirmed states with mass up to $3$ GeV, thus calling as a QMHRG model~\cite{Alba:2017mqu}. 
Interactions among the baryons and anti-baryons are imprinted in the baryon number susceptibilities, and we have used the 
continuum estimates of these quantities measured using lattice QCD to better calibrate the phenomenological interaction coefficient $K$.
The baryon number susceptibilities are defined as
\begin{equation}
\chi^{B}_{n}=\frac{\partial^n \left[P(\mu_B/T)/T^4\right]}{\partial (\mu_B/T)^n },~~~n=2,4,6,8
\end{equation}
\begin{figure}
\includegraphics[width=6.2cm]{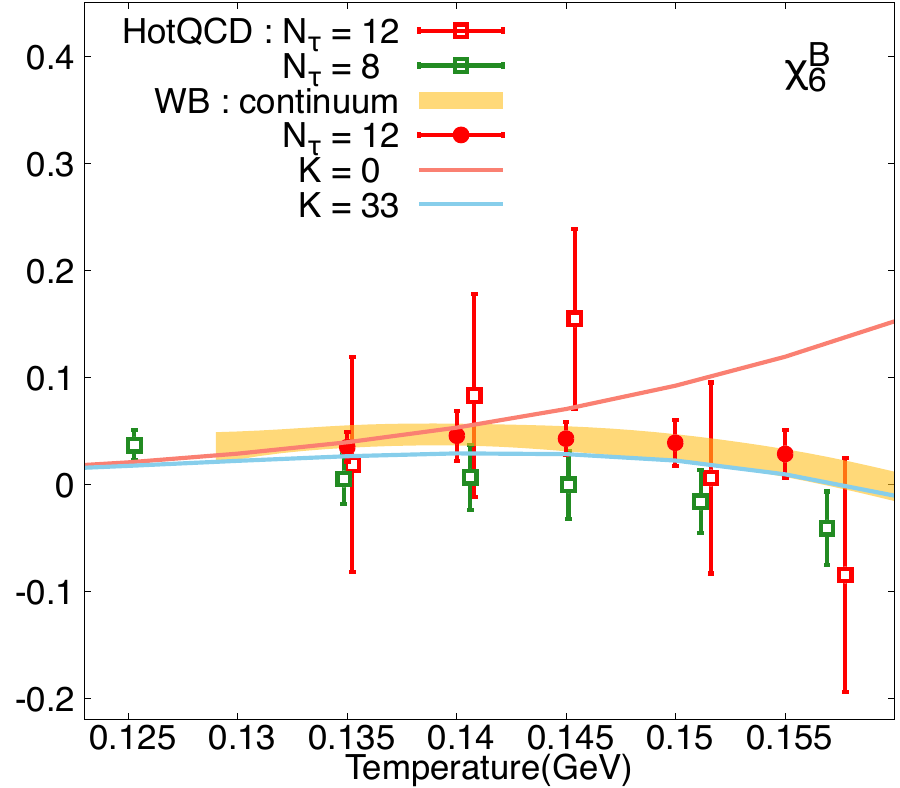}
\includegraphics[width=6.2cm]{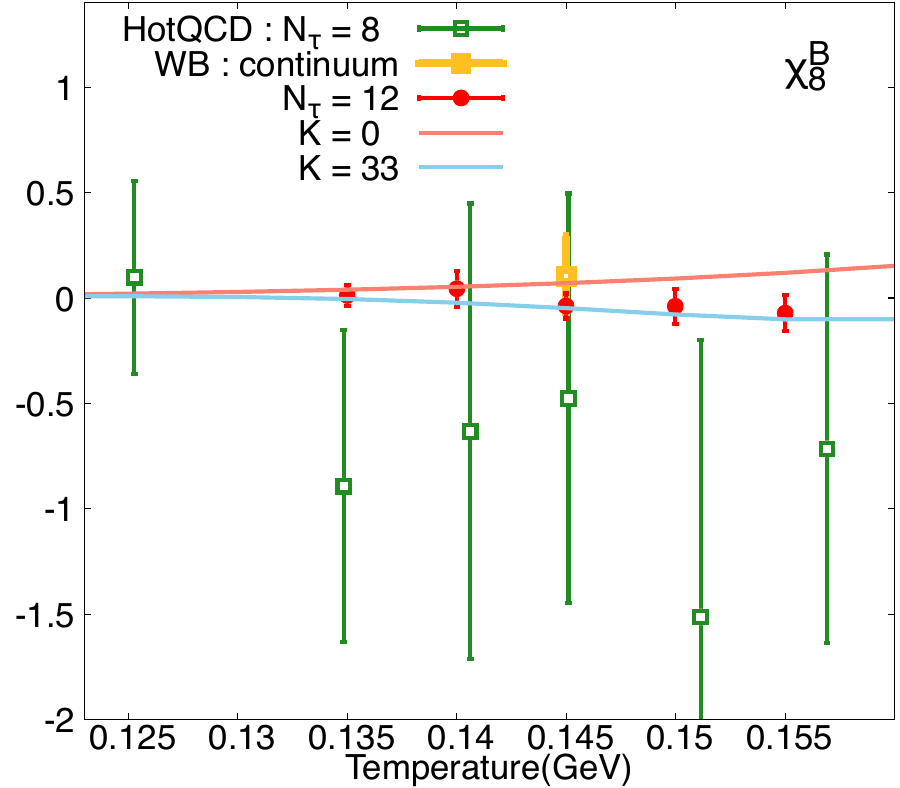}
\caption{The sixth (left) and eighth (right) order baryon number fluctuation results compared between ideal QMHRG (red line), QMHRG with repulsive mean-field interactions (blue line) models, and lattice QCD results with physical quark masses. HotQCD results for $N_\tau = 8$ and $12$ are from Ref~\cite{Bazavov:2020bjn}. Wuppertal-Budapest (WB) results for $N_\tau = 12$ (red circle), and continuum limit (yellow band) are from Ref.~\cite{Borsanyi:2018grb} and \cite{Borsanyi:2023wno} respectively. The plots have been taken from Ref.\cite{Biswas:2024xxh}.}
\label{fig:chi68B}
\end{figure}
In previous studies, a typical value of $K=56.25~\text{GeV}^{-2}$ was considered to describe the hadron spectra at the 
chemical freeze-out~\cite{Sollfrank:1996hd} and for measuring baryon number susceptibilities~\cite{Huovinen:2017ogf}. 
We have instead constrained the value of $K$ by suitably reproducing the lattice results of $\chi^{B}
_{n}$ for $n=2,4,6~\text{and}~8$ since it is known that the lattice results deviate from QMHRG model estimates 
for $T > 150$ MeV. We found out that a mean-field repulsion among all (anti-)baryons with a strength $K=33~\text{GeV}
^{-2}$ reproduces the temperature dependence of the lattice data of these susceptibilities. In Fig.~\ref{fig:chi68B} 
we show a comparison of higher-order fluctuations $\chi_6^B$ and $\chi_8^B$, between an ideal HRG model and repulsive 
QMHRG model with $K=33~\text{GeV}^{-2}$. 

\subsection{Chiral transition in the repulsive mean-field QMHRG model}\label{secIII}
Within this model, we have calculated the renormalized chiral condensate~\cite{Bazavov:2011nk} defined as 
\begin{eqnarray}
\Delta^R_l=d+ m_s r_1^4 \left[\langle\bar{\psi}\psi\rangle_{l,T}-\langle\bar{\psi}\psi\rangle_{l,0}\right], ~\text{where}~ \langle\bar{\psi}\psi\rangle_{l,T} - \langle\bar{\psi}\psi\rangle_{l,0}=  
\frac{\partial P}{\partial m_l}
\label{Eq.relation1}
\end{eqnarray}
Here $m_l$ and $m_s$ are the light and strange quark mass respectively, and $P$ is the total pressure. We consider 
$2+1$ flavor scenario with $m_u=m_d=m_l$. $d=r_1^4 m_s (\lim_{m_l \rightarrow 0} \langle \bar \psi \psi \rangle_{l,0})^R$, 
where the parameter $r_1=0.3106$ fm~\cite{MILC:2010hzw} is derived from the static quark ant-quark potential. Using 
the FLAG 2022 values for the vacuum chiral condensates for 2+1 flavor case~\cite{Aoki:2021kgd}, we have estimated 
$d=0.022791$~\cite{Biswas:2022vat}. Calculating the mass derivative of pressure with respect to $m_l$ necessitates 
including the mass derivatives of hadrons and resonances, details of which are in Ref.~\cite{Biswas:2022vat}.
\begin{figure}
\subfloat{\includegraphics[width=6.2cm]{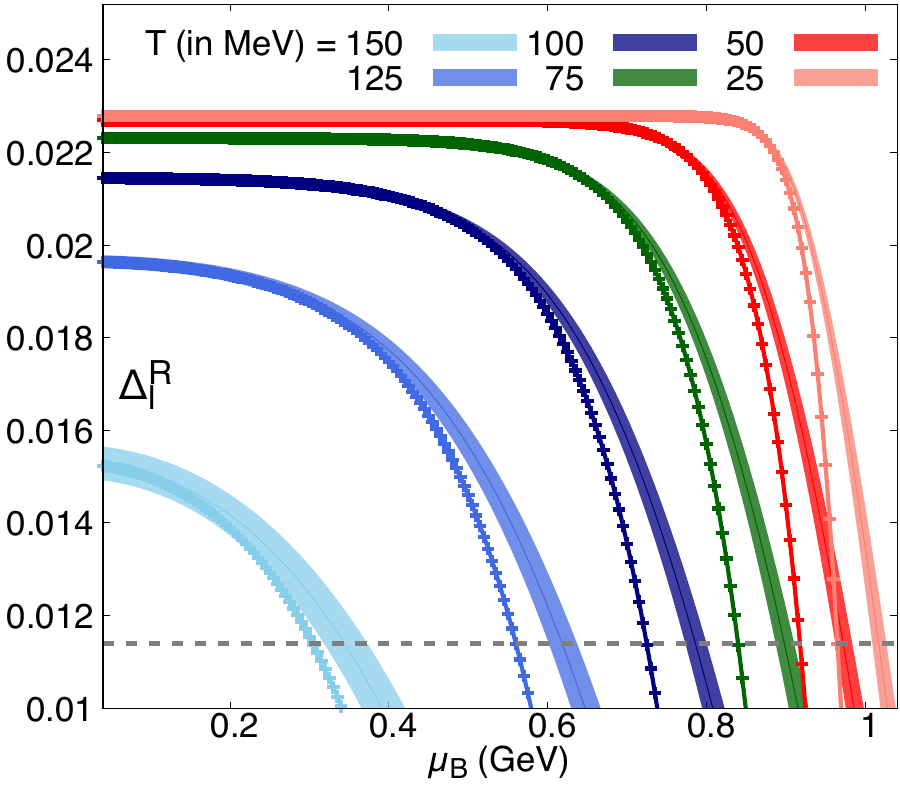}}
\subfloat{\includegraphics[width=6.2cm]{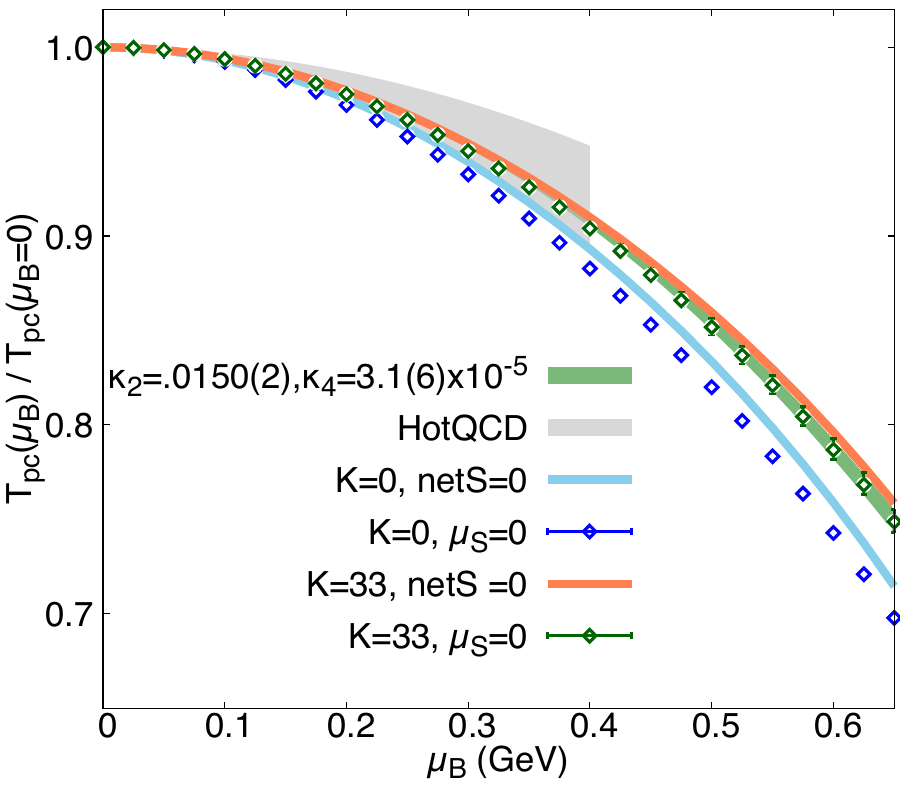}}
\caption{Left: The $\mu_B$ dependence of the renormalized chiral condensate for different temperature values $T=25$-$150$ MeV. Right: Pseudo-critical lines from ideal and repulsive QMHRG model, both for the $\mu_S=0$ and $\text{net S}=0$ case (see the text for details). 
}
\label{fig:chiral_1}
\end{figure}
The variation of the renormalized chiral condensate with baryon chemical potential $\mu_B$ for different values temperature from $25-150$ MeV are shown in the left panel of Fig.\ref{fig:chiral_1}. The bands represent the QMHRG 
model data including the mean-field repulsion, and the width of these bands arises due to the uncertainties in the 
quark mass derivatives. The ideal QMHRG model results are shown as points connected by solid lines. There is no 
variation with $\mu_B$ at lower $\mu_B$, and the extent of this flat portion of the curve is larger for temperatures 
$T<100~$ MeV. As the $\mu_B$ increases, $\Delta^R_l$ falls faster in an ideal QMHRG compared to when interactions 
are included since presence of repulsive interactions saturates the number density of the baryons.

We have next estimated the pseudo-critical line from the variation of $\Delta^R_l$ with $\mu_B$. For a given value of $
\mu_B$, the pseudo-critical temperature $T_{pc}$ is determined at the point where $\Delta^R_l$ drops to half of the zero 
temperature value. In our earlier work~\cite{Biswas:2022vat}, this condition provided a $T_{pc}(\mu_B=0) = 161.2(1.7)$ MeV, which is in good agreement with the state-of-the-art lattice QCD results. We have estimated $T_{pc}(\mu_B)$ for a 
large range of $\mu_B$ using this criterion. We show the pseudo-critical lines for the ideal (blue line with points) and mean-field QMHRG model (green line with points) in Fig.~\ref{fig:chiral_1}, from Ref.~\cite{Biswas:2024xxh}. Since 
the $T_{pc}(\mu_B=0)$ estimates vary between ideal and mean-field QMHRG models and in lattice QCD, we have used 
this value to normalize the $T_{pc}$. Parameterizing the pseudo-critical line $T_{pc}(\mu_B)$ with the following ansatz~\cite{HotQCD:2018pds}, $\frac{T_{pc}(\mu_B)}{T_{pc}(0)}=1-\kappa_2 \left(\frac{\mu_B}
{T_{pc} (0)}\right)^2 - \kappa_4 \left(\frac{\mu_B}{T_{pc} (0)}\right)^4 $ we obtain the curvature coefficients 
$\kappa_2=0.0150(2)$, $\kappa_4=3.1(6)\times 10^{-5}$. The results obtained after performing the fit (green band) 
is consistent with continuum estimated pseudo-critical line from lattice QCD (gray band)~\cite{HotQCD:2018pds, 
Borsanyi:2020fev}. We observe a finite value of $\kappa_4$, which was not reported earlier, since current 
lattice QCD data for this quantity is consistent with zero with large uncertainties.

\subsection{Implication of strangeness neutrality}
The strangeness neutrality condition is usually considered to mimic the absence of net strangeness in the initial 
colliding nuclei in heavy-ion collisions. Employing this criterion requires an explicit calculation of $\mu_S$ 
that satisfies the condition $n_S=0$. Constraining the charge chemical potential $\mu_Q=0$, to realize isospin 
symmetric condition and choosing $n_Q/n_B = 0.5$, we calculate the pseudo-critical line shown in Fig.~\ref{fig:chiral_1}, 
for the ideal(blue band) and mean-field QMHRG(orange band) models. For a given $\mu_B$, imposing the strangeness 
neutrality restricts the phase space density of the strange hadrons, thus the resulting $T_{pc}(\mu_B)$ values are 
higher than those in the $\mu_S=0$ case. This also corroborates with the recent findings within the chiral mean-field 
model (NJL) at finite strangeness~\cite{Ali:2024nrz}. 

\begin{figure}
\subfloat{\includegraphics[width=6.2cm]{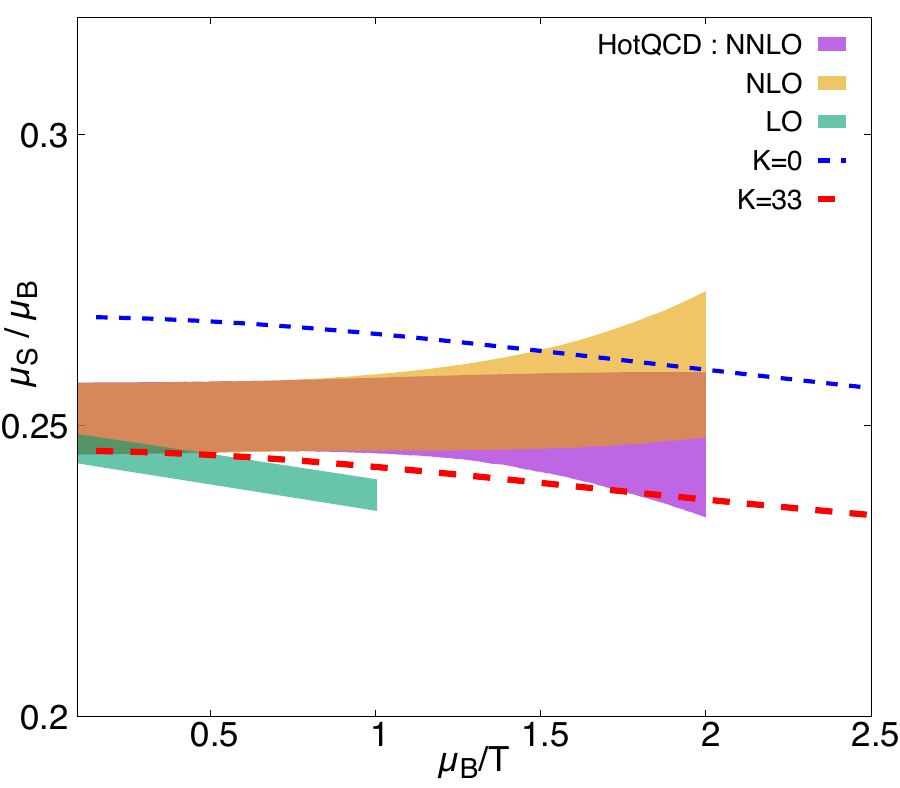}}
\subfloat{\includegraphics[width=6.2cm]{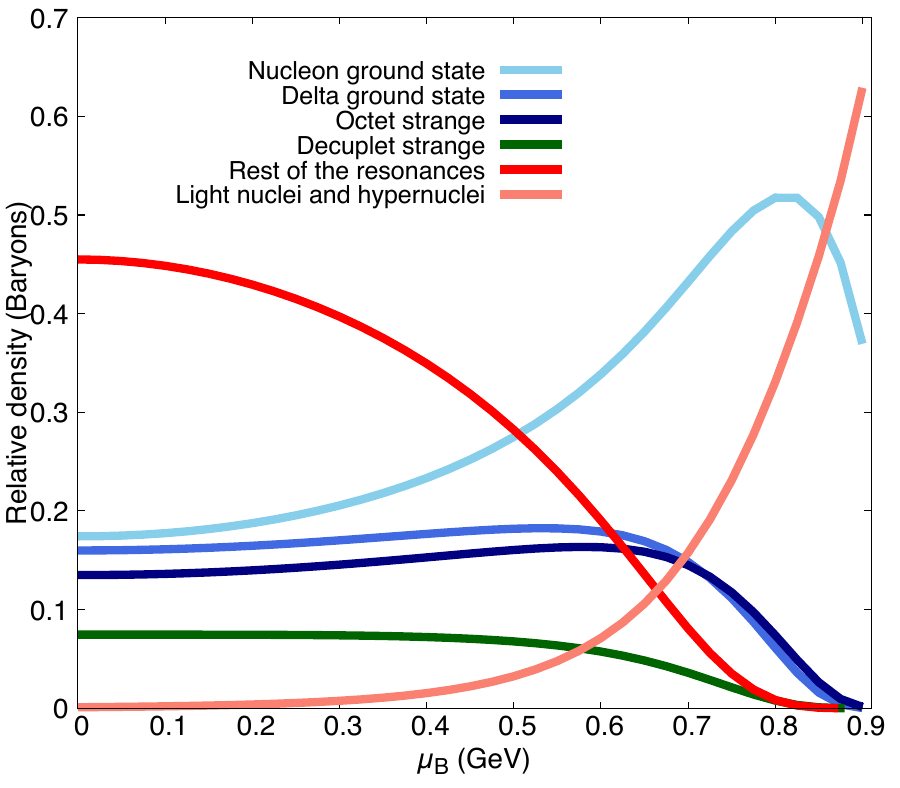}}
\caption{Left: The $\mu_B/T$ dependence of the $\mu_S/\mu_B$ (see the text for details). Right: Relative abundances of baryons and nuclei, hyper-nuclei states along the pseudo-critical lines.  
}
\label{fig:fig_3}
\end{figure} 
We have also compared the $\mu_S/\mu_B$ calculated within QMHRG model with recent continuum extrapolated lattice 
results~\cite{Bollweg:2024epj} in the left panel of Fig.\ref{fig:fig_3}. The lattice calculation uses a Taylor series 
expansion for second-order cumulants up to NNLO in $\mu_B$ along the pseudo-critical line, whereas we have explicitly 
evaluated the $\mu_S$ imposing $\text{netS}=0$ for a given $T$ and $\mu_B$. We observe a good agreement between the 
lattice result and mean-field QMHRG estimates (red dashed line), whereas the ideal QMHRG model result (blue dashed) 
deviates from the lattice. 
We have considered the same values of $K$ for repulsive interactions among strange and non-strange baryons, whereas 
these strengths have to be different in a more realistic model. Future lattice QCD results of higher-order strange 
fluctuations might provide insights in this direction.

\subsection{Light nuclei as attractive interaction channel}
To develop a more realistic model of hadrons, one should also consider attractive interaction among nucleons at large
baryon densities near the nuclear saturation densities. This feature is necessary to understand the nuclear liquid-gas 
transition and neutron star equation of state. In a more simplistic approach, we have included the light and hyper-nuclei 
states within our mean-field QMHRG model to mimic the attractive interactions among nucleons and hyperons. It is 
important to note that we have not included repulsive interactions among the light nuclei and hypernuclei. The 
relative contribution of the light and hypernuclei states to the net-baryon density along the pseudo-critical line is shown in the right panel of Fig.~\ref{fig:fig_3}. The contribution of these nuclei states increases significantly 
around $\mu_B=700$ MeV and beyond. This indicates the necessity of their inclusion to extend the present framework 
beyond $\mu_B=700$ MeV.

\subsection{Summary}
We have extended the QMHRG model with repulsive mean-field interactions among baryons by better constraining the 
interaction strength using lattice data for baryon number fluctuations. This enabled us to extend the pseudo-critical 
line upto $\mu_B\sim 700$ MeV and extracting a non-zero value of $\kappa_4=3.1(6)\times 10^{-5}$ for the first time. 
We also discuss how this model could be extended to understand QCD phase diagram at even larger $\mu_B$.



\section{Understanding proton number cumulants with a modification to van der Waals hadron resonance gas model}

\author{Kshitish Kumar Pradhan, Ronald Scaria, Dushmanta Sahu, and Raghunath Sahoo}

\bigskip

\begin{abstract}
The fluctuations of proton number cumulants are studied within an interacting hadron resonance gas model that includes van der Waals-type repulsive as well as attractive interactions among the hadrons.  In this study, we have explored the possible temperature ($T$) and baryochemical potential ($\mu_B$) dependency of the van der Waals (VDW) attractive and repulsive parameters $a$ and $b$, respectively. We find that both the VDW parameters can be parameterized as a function of $T$ and $\mu_B$. Hence, the conventional VDW hadron resonance gas (VDWHRG) model with constant $a$ and $b$ parameters is now replaced with a modified van der Waals hadron resonance gas (MVDWHRG) model that incorporates the $T$ and $\mu_B$ dependence of these VDW parameters. This leads to a significant change in the thermodynamics of the hadron gas as well as the higher-order fluctuations while going towards the high $\mu_B$ and low $T$ region. We use a simple parameterization to study the higher-order net proton fluctuations as a function of the centre of mass energy, $\sqrt{s_{NN}}$, and a reasonable agreement with experimental results is observed.
\end{abstract}


\ccode{PACS numbers:}


\subsection{Introduction}

In recent years, one of the primary goals of both theoretical and experimental high-energy physics communities is to study the quantum chromodynamics (QCD) phase diagram characterized by temperature ($T$) and baryochemical potential ($\mu_B$). At high $T$ and vanishing $\mu_B$, the lattice QCD (lQCD) predicts a smooth crossover transition from hadronic matter to a deconfined quark-gluon plasma state. Theoretical models, however, predict a first-order phase transition at high $\mu_B$ and low $T$
which ends at a possible critical endpoint (CEP). It has been suggested that the energy dependence of higher-order fluctuations of conserved charges like net-baryon, net-charge, net-strangeness, etc., can show a non-monotonic behaviour near the critical point. Therefore, fluctuations of these conserved charges have become one of the important experimental observables in an attempt to locate the CEP. The fluctuation of conserved charges in experiments can be related to the thermodynamic susceptibilities calculated in different theoretical models. Here, we use hadron resonance gas (HRG) models to study net proton fluctuations, which are studied as a proxy for net baryon fluctuations in experiments. The ideal HRG model assumes a system of non-interacting point-like hadrons and resonances. This model remains quite successful in explaining the thermodynamic properties from lQCD calculations up to a temperature, $T\sim$ 150 MeV and also the particle ratios from experimental results. However, it fails to explain the higher-order fluctuations measured in experiments. An extension of the IHRG model is done by including van der Waals' attractive and repulsive interaction, known as the VDWHRG model, which improves the fluctuation calculations. However, it is still far from explaining the higher-order fluctuations as observed in experiments. We attempt to modify the VDWHRG model by considering the $T$ and $\mu_B$ dependence of VDW attractive and repulsive parameters $a$ and $b$, respectively. This modified VDWHRG (MVDWHRG) shows considerable agreement with experimental results on net proton fluctuations from the latest results of the STAR experiment.

\subsection{Formalism}

We use a $\chi^2$ minimization technique to fit the pressure and energy density obtained in the VDWHRG model to that of lQCD results \cite{Bazavov:2017dus} as a function of temperature for different values of $\mu_B/T$. This results in obtaining the VDW parameters $a$ and $b$ separately for each case of $\mu_B/T$ \cite{Pradhan:2023etz}. The obtained VDW parameters for each value of $\mu_B/T$ and corresponding $\chi^2$ values are shown in Table~\ref{table}. The details are given elsewhere \cite{Pradhan:2023etz}. It is observed that both the VDW parameters are decreasing with $\mu_B/T$. Therefore, the strength of VDW interaction reduces as one goes to the low $T$ or high $\mu_B$ region. We fit the obtained $a$ and $b$ parameters with a negative exponential function, and hence, the parameters can now be quantified as functions of $\mu_B/T$ as
\begin{equation}
    \begin{split}
    \label{eqn1}
    a = p_{1}\exp(p_{2}\frac{\mu_{B}}{T}),\\
    b = p_{3} \exp(p_{4}\frac{\mu_{B}}{T}).
    \end{split}
\end{equation}
The values of constant parameters $p_i$ in above equation are given by, $p_{1}$ = 1.66 $\pm$ 0.05 GeV fm$^{3}$, $p_{2}$ = -0.88 $\pm$ 0.04, $p_{3}$ = 541.93 $\pm$ 15.98 GeV$^{-3}$, and $p_{4}$ = -0.61 $\pm$ 0.03. The new approach is termed as modified VDWHRG (MVDWHRG), where the VDW parameters are no longer constants but vary as a function of $T$ and $\mu_B$. We then estimate higher-order net proton number cumulants in this model using $n^{th}$ order susceptibilities. These thermodynamic susceptibilities are obtained as the $n^{th}$ order derivative of pressure ($P$) with respect to $\mu_B$ as
\begin{equation}
    \label{eqn2}
    \kappa_{n} = \frac{\partial^n}{\partial (\mu_{B}/T)^n}\Big(\frac{P}{T^4}\Big).
\end{equation}
The first-order derivative gives the proton number density, whereas the second, third, and fourth-order derivatives give variance ($\sigma$), Skewness ($S$), and kurtosis ($\kappa$), respectively. Then, we can define the cumulant ratios as
\begin{equation}
    \label{eqn3}
    S\sigma = \frac{\kappa_3}{\kappa_2}, \hspace{0.5cm} \frac{S\sigma^3}{M} = \frac{\kappa_3}{\kappa_1}, \hspace{0.5cm}   \kappa\sigma^2 = \frac{\kappa_4}{\kappa_2}.
\end{equation}

\begin{table}[pt]
\tbl{\label{table}VDW parameters obtained for different values of $\mu_B/T$ using a $\chi^2$ minimization technique}
{\begin{tabular}{@{}cccc@{}} \toprule
$\mu_{B}/T$ & $a$ (GeV fm$^3$) & $r_{B}$ (fm) &
$\chi^2$ \\
\colrule
 0.0 & 1.650 $\pm$ 0.05 & 0.635 $\pm$ 0.05 & 1.06/20 \\ 
 1.0 & 0.786 $\pm$ 0.064 & 0.515 $\pm$ 0.05 & 0.91/20 \\ 
 2.0 & 0.275 $\pm$ 0.025 & 0.425 $\pm$ 0.05 & 1.88/20 \\ 
 2.5 & 0.150 $\pm$ 0.05 & 0.385 $\pm$ 0.15 & 3.5/20 \\ 
 \botrule
\end{tabular}}
\end{table}

\subsection{Results and Discussion}
In order to calculate the net proton number fluctuations and compare them with the experimental results, the freezeout parameters, $\mu_B$ and $T$, need to be obtained as a function of beam energy. The $T$ and $\mu_B$ parameters at different energies are obtained separately for all three models, Ideal HRG, VDWHRG, and MVDWHRG, by fitting the experimental results on particle multiplicities. Then, we fit a polynomial function to parameterize $T$, and $\mu_B$ as a function of $\sqrt{s_{NN}}$. The details are given elsewhere~\cite{Pradhan:2023etz}. The kinematic acceptance cuts, $0.4<p_T<2.0$ GeV/c and $|y|<0.5$, are also considered in this study. The heavier resonance decay effects can affect the proton number fluctuation to a greater extent. Therefore, we include the fluctuation due to protons produced from resonances along with those due to primordial protons. While comparing theoretical results with experimental observations, it is also important to include the correction for the global baryon number conservation. We use the latest net proton number fluctuations data from the STAR experiment at Relativistic Heavy Ion Collider (RHIC) \cite{STAR:2021iop, STAR:2022vlo} for the comparison. 
\begin{figure*}[ht!]
\begin{center}
\includegraphics[scale = 0.2]{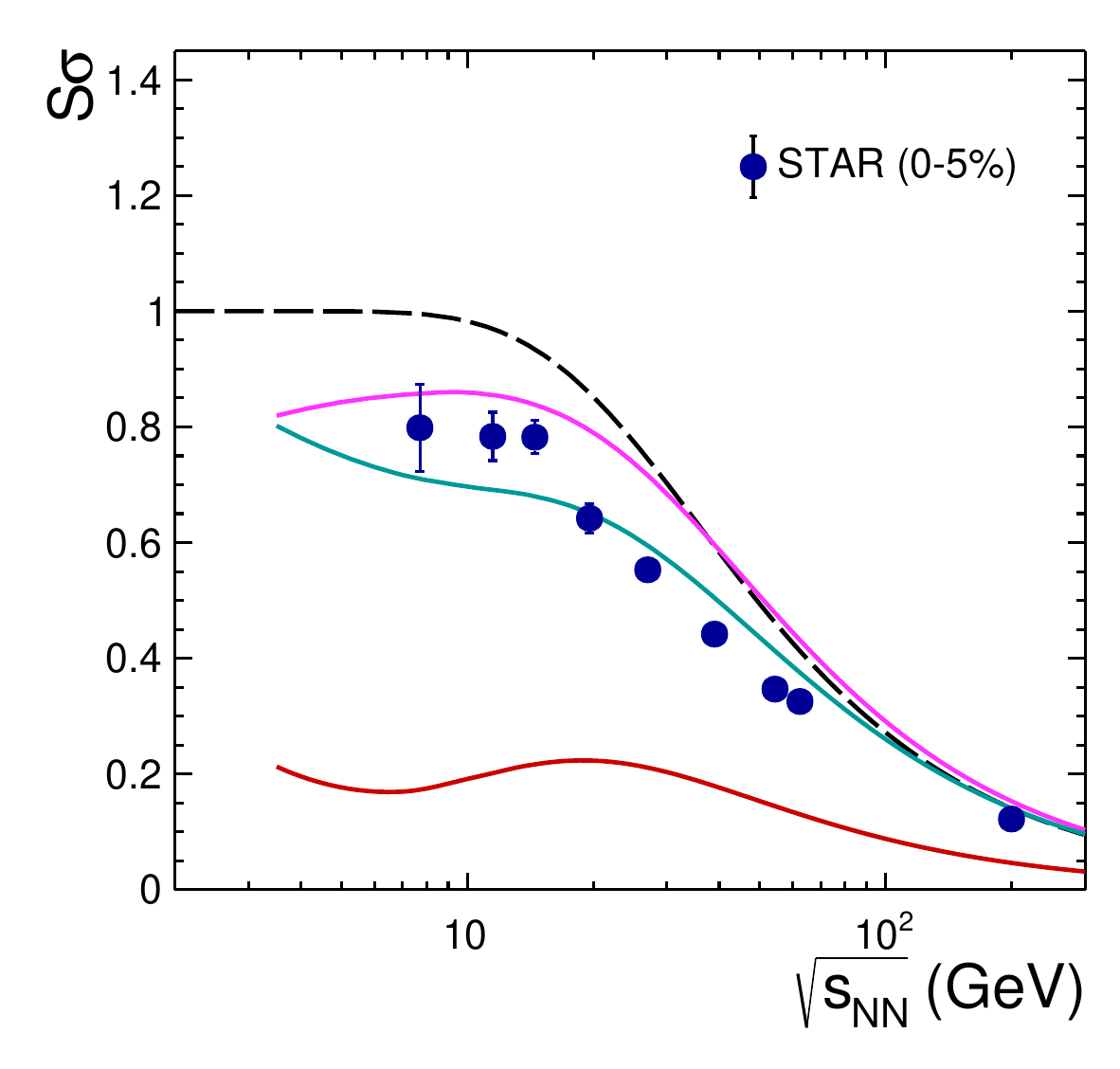}
\includegraphics[scale = 0.2]{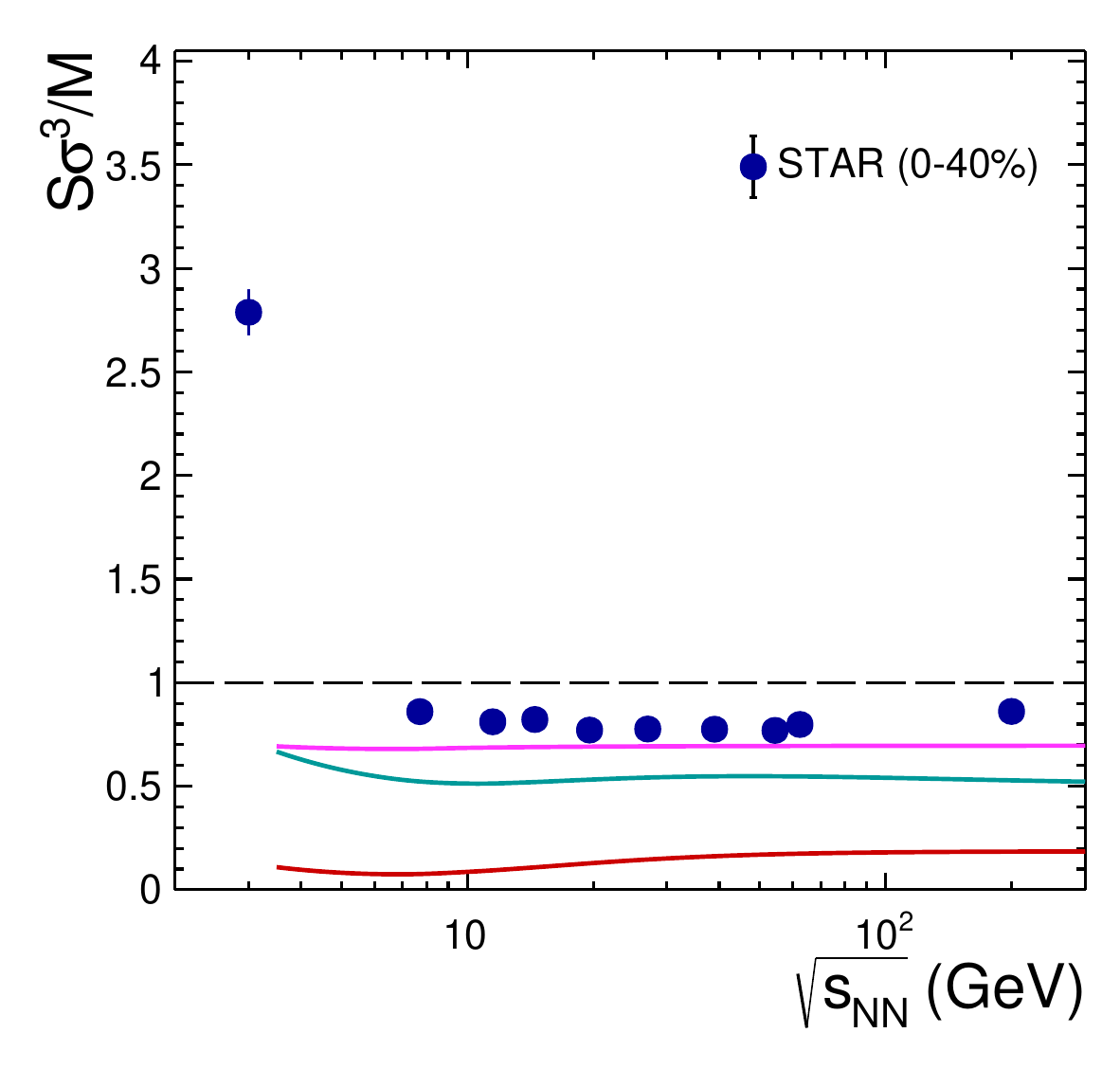}
\includegraphics[scale = 0.2]{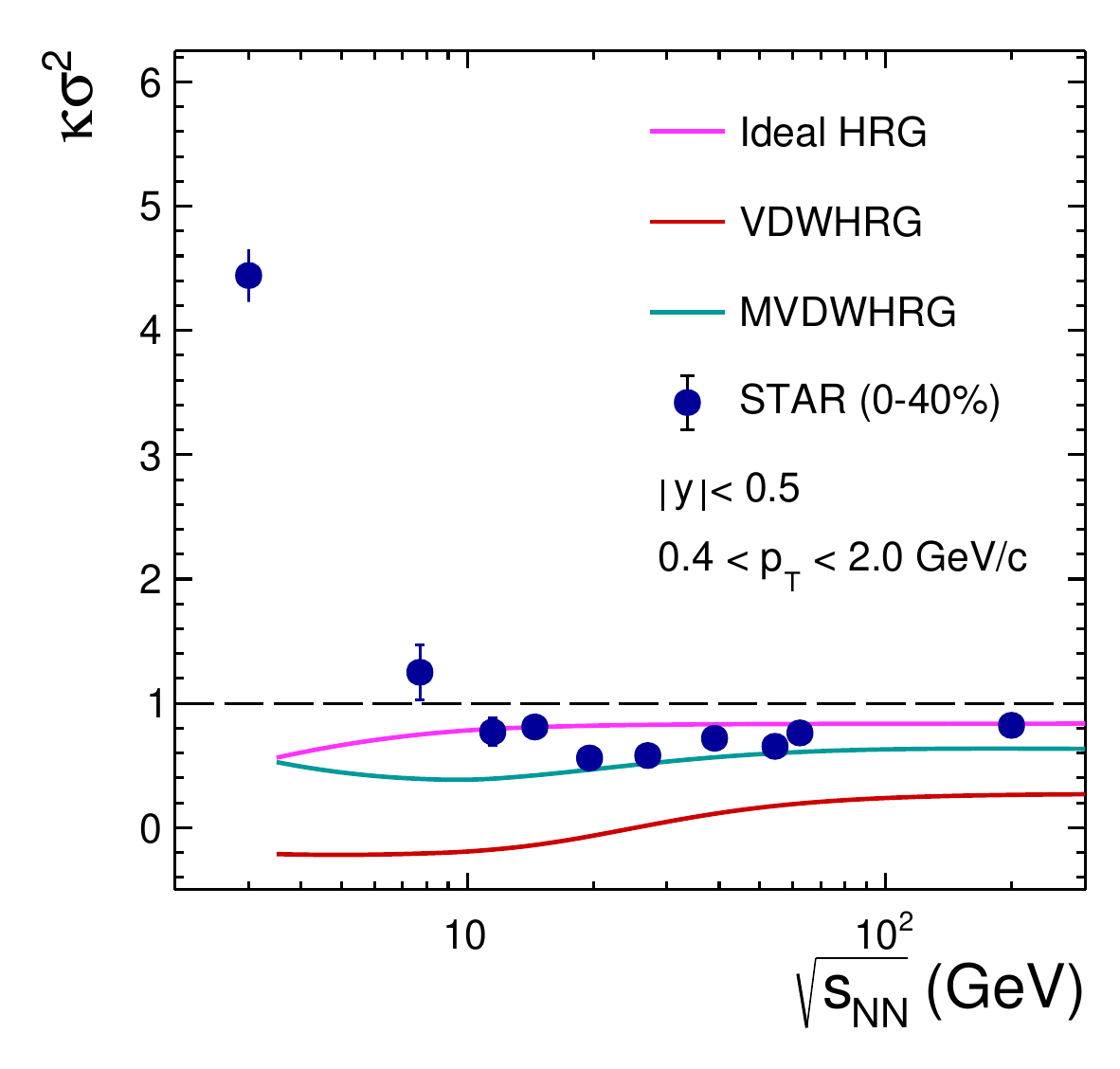}
\caption{(Colour Online) The net proton cumulant ratios as functions of centre-of-mass energies. The cyan solid line represents the results from MVDWHRG, compared with the ideal HRG (magenta), VDWHRG (orange) along with those obtained in RHIC BES-I experiment~\cite{STAR:2021iop, STAR:2022vlo} (blue markers).}
\label{fig1}
\end{center}
\end{figure*}

In Fig.~\ref{fig1}, the energy dependence of the cumulant ratios defined in Eq.~(\ref{eqn3}) is shown. The solid blue markers represent the net proton fluctuation as measured in the RHIC Beam Energy Scan I (BES-I)  experiment. The dashed black is from the Skellam predictions, whereas the solid magenta line is for the ideal HRG calculations. The solid orange curve is for the VDWHRG model, which uses constant VDW parameters, $a=$ 329 MeV fm$^3$ and $b=$ 3.42 fm$^3$ \cite{Vovchenko:2015pya}. The results obtained in the MVDWHRG model are shown in a solid cyan curve and are compared with other models as well as with experimental data. For $S\sigma$ in the left panel of Fig.~\ref{fig1}, one can observe that the ideal HRG model fairly explains the experimental data at low energy. However, it fails at high energy and goes in line with what is obtained from Skellam predictions. The VDWHRG model underestimates the experimental data at all energies. The MVDWHRG model, however, shows better agreement with experimental data than the other models. In the middle panel, the $S\sigma/M^3$ results show that all the models deviate from the experimental results, though the deviation for the VDWHRG model is maximum. The inclusion of the resonance decay effect results in more variation of results from the Skellam distribution. In the right panel, for the $\kappa\sigma^2$, the MVDWHRG is in considerable agreement with experimental data at high energy. However, at low energy, all the models deviate from it. It can observed from Fig.~\ref{fig1} that while going towards low energy, the MVDWHRG results approach the ideal HRG model calculation. This can be explained on the basis of $T$ and $\mu_B$ dependence of VDW parameters in the MVDWHRG model. At low energy, which corresponds to high $\mu_B$, the VDW parameters decrease; hence, the strength of VDW interaction decreases, and the model tends towards ideal HRG. A detailed study of the higher-order fluctuations and the effect of acceptance cuts, resonance decay contribution, and global baryon conservation are also explicitly studied~\cite{Pradhan:2023etz}.

\subsection{Summary}
In this study, we attempt to parameterize the van der Waals (VDW) parameters as a function of temperature ($T$) and baryochemical potential ($\mu_B$) by fitting our model to the available lQCD data of thermodynamic variables at different $\mu_B/T$. A qualitative agreement is observed between experimental data and our model estimates. We thus provide stringent limits to the non-critical fluctuations, which are essential for the QCD critical point search at RHIC.

\section{Diffusion of multiple conserved charges in hot and dense hadronic matter}

\author{Hiranmaya Mishra, Arpan Das, and Ranjita K. Mohapatra}

\bigskip

\begin{abstract}
Strongly interacting matter produced in heavy ion collision experiments can have multiple conserved charges, e.g., baryon number, strangeness, and electric charge. Since hadrons can carry different conserved charges, the spatial inhomogeneity of one conserved charge will also result in the diffusion process of other conserved charges. In such a situation, a diffusion matrix can describe the diffusion process associated with these conserved charges. We estimate this diffusion coefficient matrix for the hadronic phase using the Boltzmann kinetic theory description, considering the relaxation time approximation at finite temperature and density. We consider the well-celebrated hadron resonance gas (HRG) model with and without excluded volume corrections to model the hadronic matter. In our calculation, we also incorporate the Landau-Lifshitz frame condition. This frame condition makes the diagonal diffusion coefficients positive definite. Our calculation indicates that the off-diagonal components of the diffusion matrix can be significant, which can affect the charge diffusion in a fluid with multiple conserved charges. We also find that the excluded volume corrections in the diffusion matrix estimation are insignificant. 
\end{abstract}




\subsection{Introduction}
\label{sec:intro}
In the search for QCD critical point in heavy-ion collision (HIC) experiments, the fluctuation and correlation studies of QCD conserved charges can play an important role \cite{Stephanov:1999zu,Pal:2020ucy}. It has been suggested that fluctuations in conserved quantities, such as net baryon number (B), net electric charge (Q), and net strangeness (S), on an event-by-event basis may serve as indicators of QGP formation and the quark-hadron phase transition \cite{Asakawa:2015ybt}. Because symmetries protect the conserved charges and the associated currents, the diffusion process can give rise to the time evolution of these conserved quantities. At ultra-relativistic collision energies, due to almost vanishing net baryon density at the mid-rapidity region, the diffusion dynamics is not expected to be significant. However, in the low-energy nuclear collisions at the Facility for Antiproton and Ion Research (FAIR) at Darmstadt and in Nuclotron-based Ion Collider fAcility (NICA) at Dubna a baryon-rich medium can be produced \cite{Friman:2011zz}. Moreover, in the beam energy scan (BES) program at RHIC, the study of a nuclear medium with finite net baryon density is in progress, where baryon diffusion becomes relevant \cite{Mohanty:2011nm}. 

The diffusion process can be described by the Fick's law, which relates the diffusion current $(\Delta\vec{J}_q)$ to spatial inhomogeneity of the related charge density $n_q(t,\vec{x})$, i.e., $\Delta J^i_q=\kappa _{qq}D^i(\beta \mu_q)$. Here $D^i$ is the spatial component of $D^{\mu}=\nabla^{\mu\alpha}\partial_{\alpha}$, $\nabla^{\mu\alpha}=g^{\mu\alpha}-u^{\mu}u^{\alpha}$, $u^{\mu}$ is the fluid flow vector normalized as i.e. $u^{\mu}u_{\mu}=1$, $\beta=1/T$, $\mu_q$ is the chemical potential corresponding to the conserved charge $q$.
 
Due to the presence of multiple conserved charges in the QCD medium, i.e. the simple Fick's law as above, now gets modified. Since hadrons and quarks can carry multiple conserved charges, i.e., B, Q, S, the diffusion current of each charge will depend on the gradient of other charges. Since the gradients of every single charge density can generate a diffusion current of any other charges, the generalized Fick's law can be written involving crossed-diffusion as 
\begin{align}
    \begin{pmatrix}
\Delta J^{i}_B \\
\Delta J^{i}_Q \\
\Delta J^{i}_S
\end{pmatrix}=\begin{pmatrix}
\kappa_{BB} & \kappa_{BQ} & \kappa_{BS} \\
\kappa_{QB} & \kappa_{QQ} & \kappa_{QS}\\
\kappa_{SB} & \kappa_{SQ} & \kappa_{SS}
\end{pmatrix}
    \begin{pmatrix}
D^{i}(\beta\mu_B)\\
D^{i}(\beta\mu_Q) \\
D^{i}(\beta\mu_S)
\end{pmatrix}.\nonumber
\end{align}
$\kappa_{qq^{\prime}}$ denotes the multicomponent diffusion matrix. The coefficients $\kappa_{qq}$ are the diagonal diffusion coefficients. On the other hand, $\kappa_{qq^{\prime}}$ with $q\neq q^{\prime}$ are the cross-diffusion coefficients which are particularly important for low energy HICs.

We estimate here the diffusion matrix elements for the hot and dense hadronic medium within the hadron resonance gas (HRG) model using the Boltzmann kinetic theory within relaxation time approximation as discussed in Refs. \cite{Fotakis:2021diq} to find the explicit expressions for the same. We explicitly take into account the Landau-Lifshitz matching condition in the local rest frame and show that the diagonal components of the diffusion matrix are manifestly positive definite. This is the novel aspect of our calculation. 

\subsection{Formalism}
\label{formalism} 
In the Boltzmann kinetic theory approach, the phase-space evolution of the single-particle distribution function $f_a(x,p_a)$ ($a=1, N$, being the species index) in a mixture of multicomponent species can be expressed as,
\begin{align}
p_a^{\mu}\partial_{\mu} f_a(x,p_a)=(u_{\mu} p^{\mu}_a)\sum_b{\cal C}_{ab}(x,p) \equiv \mathcal{C}_a.
\label{boltzmann}
\end{align}
Here, we do not consider the effect of any external force. $\mathcal{C}_{ab}$ is the collision operator having both gain and loss terms,
\begin{align}
& \mathcal{C}_{ab}[f]=\frac{1}{2}\sum_{c,d}
\int dP_b dP_c
{ ^\prime}dP_d{^\prime}\times\left[f_cf_d{\tilde f}_a{\tilde f}_b
-f_af_b\tilde f_c\tilde f_d\right]
W(ab|cd).
\label{cab}
\end{align}
Here $W(ab|cd)$ represents the scattering rate corresponding to the binary scattering $a+b\rightarrow c+d$. $f_a=f_a(x,p_a)$, $\tilde f_a=(1-\kappa f_a/g_a)$ with $\kappa=\pm 1$ for fermions and bosonic particles while $\kappa=0$ for classical particles respectively and $dP_a=\frac{d^3 p_a}{(2\pi)^3}$ is the integration measure. $g_a$ is the spin degeneracy factor. 

Using the single-particle distribution function, the energy-momentum tensor can be expressed as,
\begin{align}
T^{\mu\nu}=\sum_a\int \frac{d^3p_a}{(2\pi)^3}\frac{p^\mu_a p^\nu_a}{E_a}f_a =-Pg^{\mu\nu}+\omega u^\mu u^\nu+\Delta T^{\mu\nu},
\label{equ9ver1}
\end{align}
and the current corresponding to a conserved charge $q$ given as
\begin{align}
J_q^\mu=\sum_a q_a \int \frac{d^3p_a}{(2\pi)^3} \frac{p^{\mu}_a}{E_a}f_a=n_qu^\mu+\Delta J_q^\mu.
\label{equ10ver1}
\end{align}
$P$, $\varepsilon$ is the pressure and energy density, respectively
and  $\omega=\varepsilon+P$ is the enthalpy.  $\Delta T^{\mu\nu}$, and $\Delta J_q^\mu$ are the dissipative corrections to the energy-momentum tensor and the conserved currents. These dissipative corrections are associated with the out-of-equilibrium contribution of the distribution functions. This out-of-equillibrium contribution can be obtained by solving the Boltzmann equation. 
For this, we write the distribution function in powers of the Knudsen number and truncate such an expansion in the lowest order  as \cite{Fotakis:2021diq}, 
\begin{align}
f_a(x,p)=f_a^{(0)}(p)\left(1+\phi^a(x,p)\right), ~~f_a^{(0)}& = g_a \exp(-\beta u\cdot p_a+\beta\sum_q q_a\mu_q).
\label{equ3ver1}
\end{align}
Once we know $\phi^a(x,p)$ we can obtain $\Delta T^{\mu\nu}$ and $\Delta J_q^\mu$.  
In general, $\phi^a(x,p)$ can contain different dissipative parts, but since we are interested in the diffusion processes, we write $\phi_a\equiv \phi_a(p^{\mu}_a)\simeq -\sum_q B_a^q p_a^\mu D_\mu\alpha_q$, here, $\alpha_q =\mu_q/T$, $\mu_q$ corresponding to different chemical potentials (e.g. $q=B,S,Q$). This leads to, 
\begin{align}
    \Delta J^{i}_q = & \sum_{a} q_a\int \frac{d^3p_a}{(2\pi)^3}\frac{p_a^{i}}{E_a}f_a^{(0)}\phi_a =\sum_{q^{\prime}}\kappa_{qq^{\prime}}D^{i}\alpha_{q^{\prime}}
    \label{equ12ver3}
\end{align}
here the diffusion matrix $\kappa_{qq^{\prime}}$ can be identified as, 
\begin{align}
    \kappa_{qq^{\prime}} =  \sum_a q_a \int \frac{d^3p_a}{(2\pi)^3}\frac{p_a^2}{3E_a}f_a^{(0)}B^{q^{\prime}
    }_a.
\label{equ14ver1}
\end{align}
The function $B_a^q$ can be obtained using the Boltzmann equation. However, the solution for $B_a^q$ so obtained is not unique. A proper hydrodynamic frame choice can fix this arbitrariness. Here, we consider the Landau Lifshitz choice of the flow velocity where in the local rest frame \cite{Albright:2015fpa} $\Delta T^{0i}=0,\Delta J^0_q=0$. This eventually allows us to find a closed form expression of the diffusion coefficients as given in Eq. \eqref{equ14ver1} (for a detailed derivation, see Ref. \cite{Das:2021bkz}), 
\begin{align}
    & \kappa_{qq^{\prime}}
     = \sum_a\int \frac{d^3p_a}{(2\pi)^3}\frac{\tau_a p_a^2}{3E_a^2}\left(q_a-\frac{n_qE_a}{\omega}\right) \left(q^{\prime}_a-\frac{n_q^{\prime}E_a}{\omega}\right) f_a^{(0)}.
     \label{equ45ver1}
\end{align}
This makes the expression of $\kappa_{qq}$ positive definite. Also note that $\kappa_{qq^{\prime}}$ is symmetric with respect to the change $q\leftrightarrow q^{\prime}$ \cite{PhysRev.38.2265}. 
To obtain different $\kappa_{qq^{\prime}}$, we use the ideal HRG (IHRG) model to estimate thermodynamic quantities, i.e., energy density, pressure, and enthalpy as needed. We also consider the excluded volume HRG (EVHRG) model, which takes into account the repulsive interaction among finite size hadrons explicitly. Moreover, the relaxation times of various hadrons can be calculated using the hard sphere scattering approximations. For the binary scattering process, the inverse of the thermal averaged relaxation time can be expressed as $\tau_a^{-1}\equiv \sum_bn_b\langle\sigma_{ab}v_{ab}\rangle$, where $n_b$ denotes the number density of scatterers and $\langle \sigma_{ab}v_{ab}\rangle$ represents thermal averaged cross section \cite{PhysRevC.92.035203,Das:2021qii,Gondolo:1990dk}. 
 \begin{align}
    \langle \sigma_{ab}v_{ab}\rangle =\frac{\sigma}{8Tm_a^2m_b^2K_2(m_a/T)K_2(m_b/T)} & \times\int_{(m_a+m_b)^2}^{\infty}ds\times \frac{[s-(m_a-m_b)^2]}{\sqrt{s}}\nonumber\\
    &\times [s-(m_a+m_b)^2]K_1(\sqrt{s}/T). 
\end{align}
$\sigma=4\pi R^2$ is the hard sphere scattering cross section of hardons with radius $R = 0.5$ fm. For the numerical estimation of $\kappa_{qq^{\prime}}$ we consider all the hadrons and their resonances up to a mass cutoff $\Lambda=2.6$ GeV, as is listed in Ref.~\cite{ParticleDataGroup:2008zun}.

\subsection{Results and discussions}
\begin{figure*}
    \centering
    \begin{minipage}[t]{0.45\textwidth}
        \centering
        \includegraphics[scale=0.3]{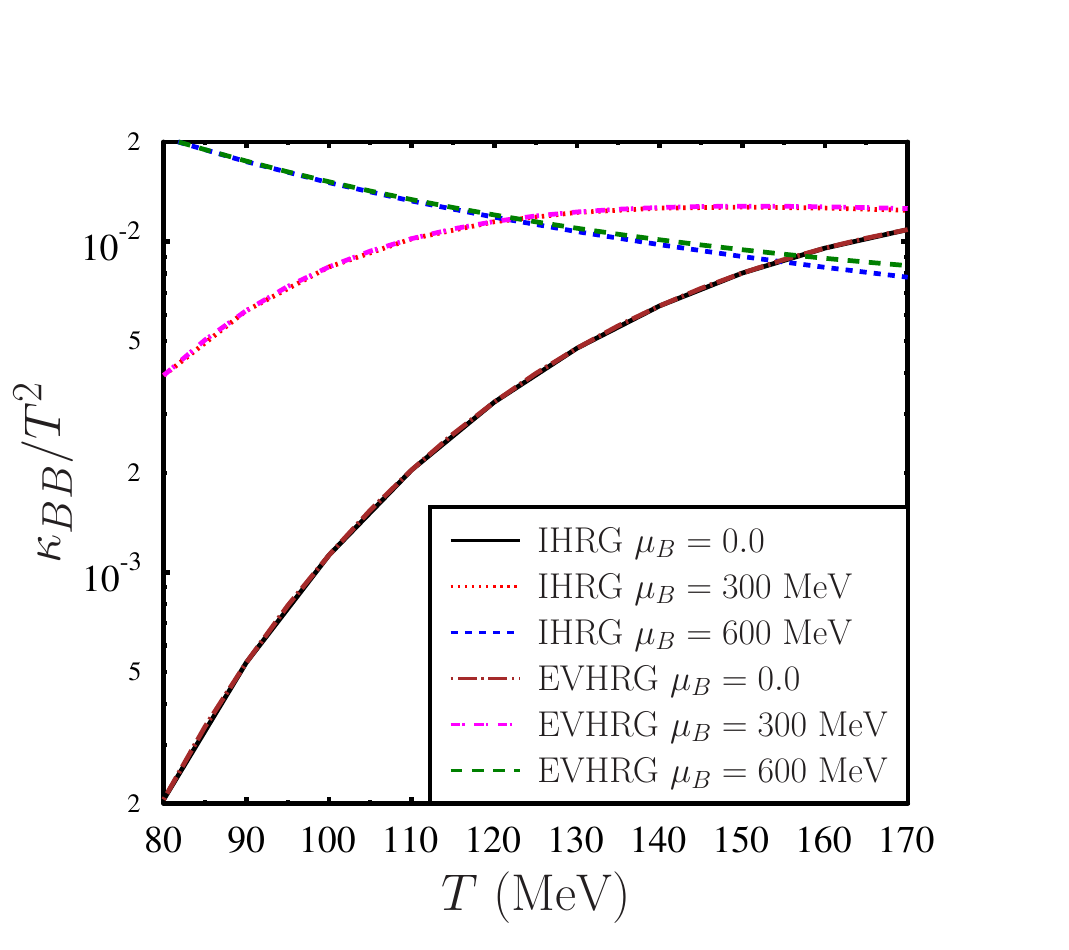}
    \end{minipage}%
    \hfill
    \begin{minipage}[t]{0.45\textwidth}
        \centering
        \includegraphics[scale=0.3]{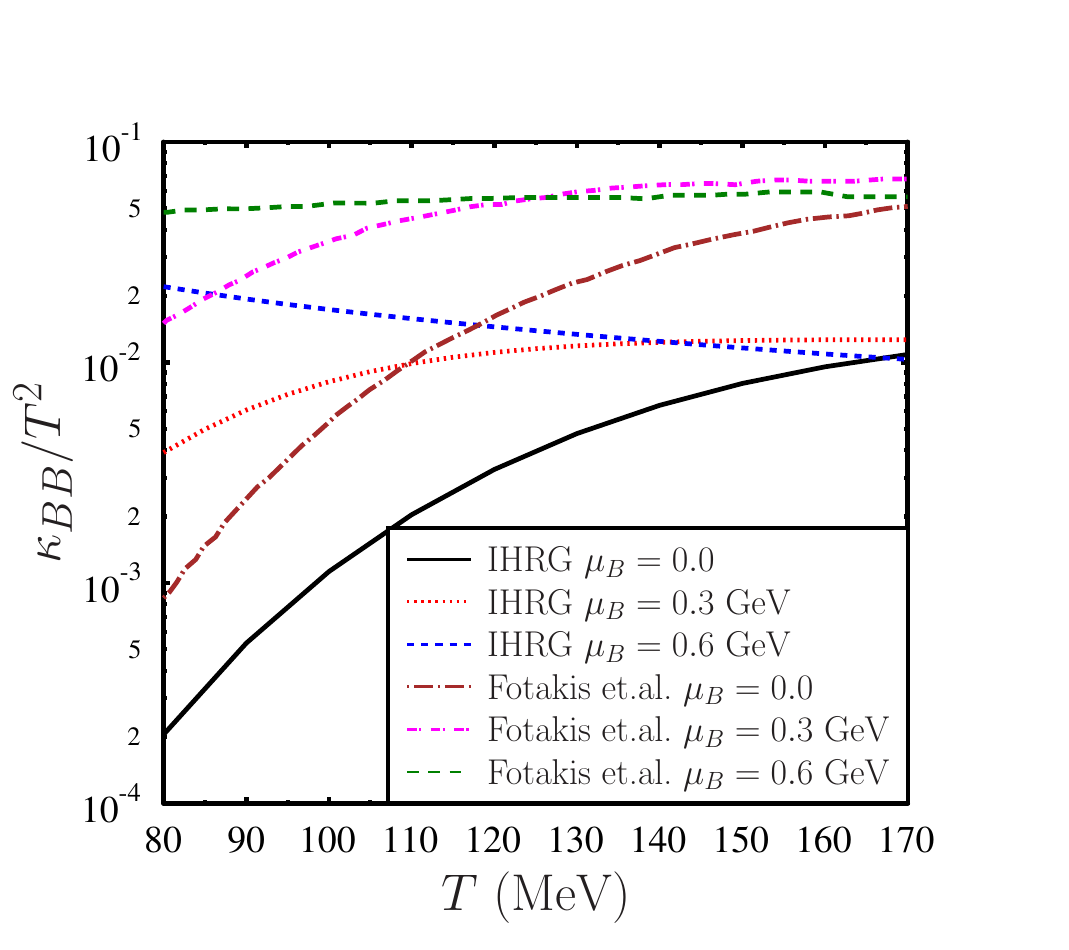}
    \end{minipage}
    \caption{Left: variation of  $\kappa_{BB}/T^2$ with $T$ and $\mu_B$ for the IHRG and EVHRG models. Here we also assumed that $\mu_Q=0=\mu_S$. Right: variation of  $\kappa_{BB}/T^2$ with $T$ and $\mu_B$ for the IHRG model. But here we consider $n_S=0$ and $\mu_Q=0$. We compare our results with the results obtained in Ref. \cite{Fotakis:2019nbq}. For a detailed discussion, check Ref. \cite{Das:2021bkz}.}
    \label{kappabb}
\end{figure*}

In Fig. \ref{kappabb}(left) we show the variation of $\kappa_{BB}/T^2$ with $T$ and $\mu_B$ for $\mu_Q=0$ and $\mu_S=0$. Estimated values of $\kappa_{BB}/T^2$ are almost similar in IHRG and EVHRG models. In Fig. \ref{kappabb}(right) we compare our results with the results obtained in Ref. \cite{Fotakis:2019nbq}. Since IHRG and EVHRG results are very similar, we show the results for IHRG model in this case. We note that in Fig. \ref{kappabb}(right) the results are shown for the conditions $n_S = 0$ and $\mu_Q=0$ which is qualitatively different from $\mu_Q=0$ and $\mu_S=0$ condition of Fig. \ref{kappabb}(left). 

\begin{figure*}
    \centering
    \begin{minipage}[t]{0.45\textwidth}
        \centering
        \includegraphics[scale=0.3]{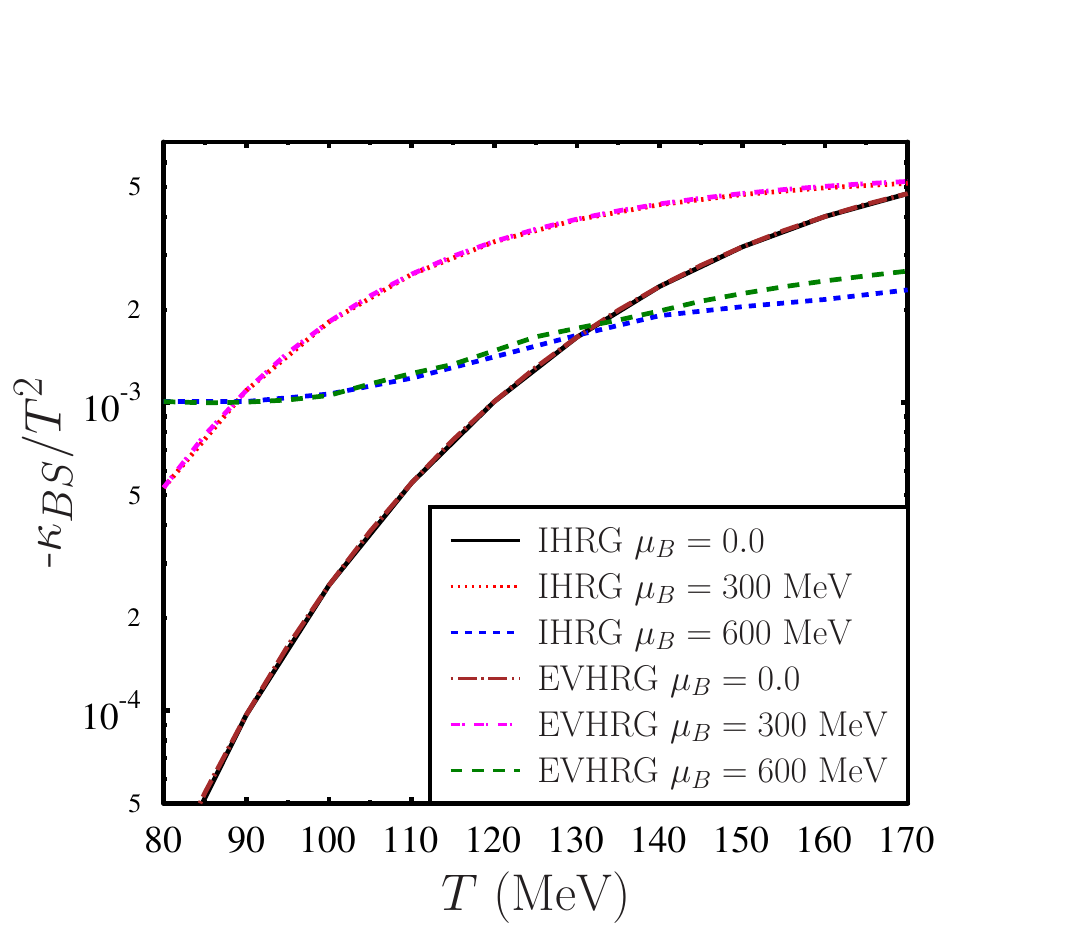}
    \end{minipage}%
    \hfill
    \begin{minipage}[t]{0.45\textwidth}
        \centering
        \includegraphics[scale=0.3]{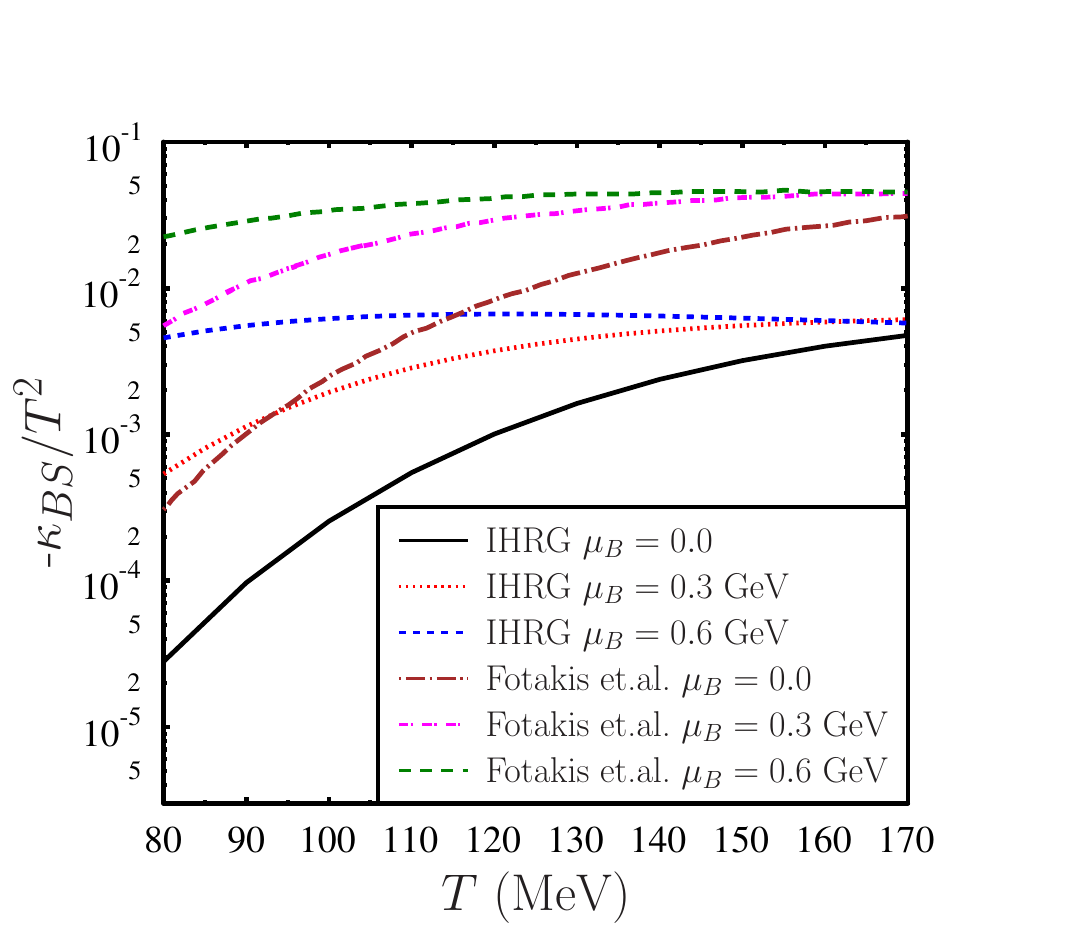}
    \end{minipage}
    \caption{Left: variation of the cross diffusion coefficient $(-\kappa_{BS}/T^2)$ with $T$ and $\mu_B$ for $\mu_Q=0=\mu_S$. We show results for IHRG and EVHRG models. Right: we compare our result as obtained in the IHRG model considering $n_S=0$ and $\mu_Q=0$, with the result obtained in Ref. \cite{Fotakis:2019nbq}. $n_S$ is the net strangeness number density. For a detailed discussion, check Ref.~\cite{Das:2021bkz}.}
    \label{kappabs}
\end{figure*}

In of Fig. \ref{kappabs}(left) we show the variation of $\kappa_{BS}/T^2$ with $T$ and $\mu_B$ for $\mu_Q=0$ and $\mu_S=0$ while in Fig. \ref{kappabs}(right) we plot the same for $\mu_Q=0$ and $n_S=0$. A nonvanishing value of $\kappa_{BS}$ indicates the generation of baryon current due to the gradient in number density of hadrons containing strangeness. It should be mentioned that $\kappa_{BS}/T^2$ results for $n_S=0$, $\mu_Q=0$ and $\mu_S=0$, $\mu_Q=0$ are different, particularly for a large value of baryon chemical potentials. We emphasize that $\kappa_{BS}$ is small but not negligible as compared to $\kappa_{BB}$.

\subsection{Summary}
In this article, we discussed the framework for calculating the diffusion coefficients within the framework of kinetic theory. We found the mathematical expression of the diffusion coefficients and estimated these transport coefficients for hot and dense hadronic matter. We modeled the hadronic matter using the ideal hadron resonance gas model and its excluded volume extensions. In the derivation of diffusion coefficients, we explicitly take into account the Landau frame choice. Due to such a frame choice, the diagonal components of the diffusion matrix turn out to be manifestly positive definite. However, the sign of the cross-diffusion coefficients can be either positive or negative. We explicitly presented results for the $\kappa_{BB}$ and $\kappa_{BS}$, which represent baryon diffusion current due to the inhomogeneity in the baryon number density and strangeness number density, respectively. $\kappa_{BB}$ is larger than the $\kappa_{BS}$, but $\kappa_{BS}$ is not negligible. The non-negligible cross-diffusion coefficient indicates that the diffusion process of conserved charges is different at finite baryon density as compared to baryon free matter.

\section{Mean free path of photons in relativistic heavy ion collisions}

\author{Jajati K. Nayak, Rupa Chatterjee}

\bigskip

\begin{abstract}
Electromagnetic probes, such as photons and dileptons, play a key role in diagnosing the initial temperature of the hot and dense quark-gluon plasma (QGP) matter created in relativistic nuclear collisions at very high energies. This is due to their large mean free path $\lambda$, which allows them to escape the medium without significant interactions. Unlike hadronic particles, which experience multiple scatterings and are affected by the evolving medium, electromagnetic probes carry undistorted  information from the initial stages of the expanding system. In this work an attempt has been made to revisit the estimation of mean free paths of photons in QGP phase for a temperature range predicted by hydrodynamics for heavy ion collisions at $\sqrt{s_{NN}}=200$ GeV at RHIC and   $\sqrt{s_{NN}}=2.76$ TeV at the LHC. The mean free paths have been estimated for a plasma expanding via (1+1)D and (2+1)D hydrodynamical expansions. For the  (1+1)D case, photons with low energy ($E_{\gamma}<  0.2$ GeV) coming from a high temperature ($>250$ MeV) source are found to have shorter mean free path compared to the expansion scale of the system. While the high energy photons have always larger mean free paths. A similar qualitative nature of the mean free path has also been observed for a more realistic (2+1)D hydrodynamic model calculations although the $\lambda$ values are found to be larger on a quantitative scale compared to the (1+1)D case. 
\end{abstract}


\ccode{PACS numbers:}


\subsection{Introduction}
The mean free path of photons in relativistic heavy ion collisions is expected to be large compared to the size of the produced medium, enabling them to escape the hot and dense matter with minimal interaction and thus serve as a valuable probe of the initial temperature and other thermodynamic properties~\cite{phot1j, gabor, phot3j, phot4j, Das:2022lqh}. The matter produced at RHIC and LHC encounter both quark gluon plasma  and hadronic phases. It is thus important to estimate the mean free paths of photons in both QGP and hadronic media.

It is widely known that in such dense medium, photons have a high mean free path due to their weak interactions. They do not to undergo multiple scatterings, which allow them to carry information directly from their point of origin. The mean free path depends on factors like the photon’s energy, the temperature, and the density of the medium. High energy photons tend to have longer mean free paths due to decreased scattering probabilities, making them even more direct probes of early-stage collision dynamics.

Kapusta {\it et. al.}~\cite{kapustaprd} in 1991 have estimated the mean free paths of photons at various energies from both quark gluon plasma and hadron gas at different temperatures and found to be large. Based on the calculation they proposed that photons having energy of about one-half to several GeV can be a good signal of the formation of a quark-gluon plasma in high-energy collisions. That estimation gave an quantitative idea of photon mean free paths which justified the use of photon and lepton pair measurements as good probe of heavy ion collision to infer the initial temperature of the  produced system. The initial temperature in other words can justify or disprove the formation of quark gluon plasma. During that time heavy ion experiments were being carried out at CERN's Super Proton Synchrotron (SPS) and BNL's Alternating Gradient Synchrotron(AGS) facilities.  Later, heavy ion experiments at RHIC and at the  LHC provided significant evidences of the formation of quark gluon plasma in those collisions.

In the mean time significant advancement has been made  in hydrodynamic model calculations which provides a reasonably well  explanation of the bulk properties of the produced matter. No other calculations of the photon mean free path have been made since that time. The calculation of rate of the photon production which is an input to the estimation of mean free path has been modified significantly in last couple of decades. 
The state-of-the-art complete leading order~\cite{arnold2001} as well as NLO rates 
of thermal photons production from QGP~\cite{nlo} are available for quite some time now. There has been advancement in the photon production from the hadronic matter~\cite{trg} as well  which  includes the meson-meson and meson-baryon bremsstrahlung [see Ref.~\cite{gabor} and references therein for detail]. It is to be noted that Kapusta {\it et. al.}~\cite{kapustaprd} calculated the mean free path of photons ignoring the expansion of the matter. 

Here in this work an attempt has been made to revisit the calculation with expansion of the system and considering the rate of photon production upto leading order in $\alpha_s$. The  formalism for the calculation of mean free path has been discussed in the next section. Then in Sec. \ref{sec:rateandhydro} rate of photon production and hydrodynamics model calculations have been discussed. Then results are shown in Sec.~\ref{sec:results} and Sec.~\ref{sec:conclusions}  summaries the work.  
\subsection{ Mean free path of photon in heavy ion collisions \label{sec:meanfreepath}}
The relaxation time for a static system can be calculated from the rate equation as shown in \cite{kapustaprd}. The solution to the rate equation with initial zero photon density, at $t=0$, can be written as 
\begin{equation}
 \frac{dn}{d^3p}=\frac{dn^{eq}}{d^3p}\left(1-exp(-t/\tau) \right)
\end{equation}
The equilibration time is 
\begin{equation}
 \tau=\frac{dn^{eq}/{d^3p}}{dR/d^3p} 
\end{equation}
For an expanding QGP we can write 
\begin{eqnarray}
 \tau &=& \frac{dn/{d^3p}}{dR/d^3p} \nonumber\\
 &=& E\frac{dn}{d^3p}/E\frac{dR}{d^3p}
\end{eqnarray}
Where, $n$ is the number density of photons with momentum $p$ at any temperature $T$ and $EdR/d^3p$ is the rate of production and described in the next section. The mean free path $\lambda =c \times \tau$ and $c=1$ here.   

\subsection{Photon production and evolution of the system \label{sec:rateandhydro}}
In QGP, photon emissions are considered from Compton processes, $q(\bar{q}) g \rightarrow q(\bar{q}) \gamma$, annihilation processes $q \bar{q} \rightarrow g \gamma$ and bremsstrahlung processes $g q \rightarrow g q \gamma$, $q q \rightarrow q q \gamma$, $q q \bar{q}\rightarrow q q \gamma$ and $g q \bar{q} \rightarrow g \gamma$. The rate is given in \cite{arnold2001} 
\begin{equation}
 \frac{dN}{d^4x d^3p}=(\frac{1}{2\pi})^3 A(p)( ln[T/m_q(T) ] +\frac{1}{2} ln(2E/T)+C_{tot}(E/T))  
\end{equation}
Where, $E=p$ for mass less photons.

Thermal mass $m_q^2=4\pi \alpha_s T^2/3$,  $A(p)=2\alpha N_c\sum_s q_s^2\frac{m_q^2(T)}{E}f_D(E)$, $s$ is active quark flavors, $N_c$ =3, $q_s$ is the fractional quark charges and $f$ is the Fermi-Dirac distribution function. 
$C_{tot}(E/T)$ is given by the following expression, 
$$C_{tot}(E/T)=C_{2\rightarrow 2}(E/T)+C_{brem}(E/T)+C_{aws}(E/T) $$
Since all $C_i(E/T)$ contains non-trivial integrations, it is parametrised after numerical integration in \cite{renk2003} 
The parametrisation of rate upto order $\alpha_s$ \cite{renk2003} is used here. 
Au+Au collisions at 200A GeV and Pb+Pb collisions at 2.76A TeV are considered for both (1+1)D and (2+1)D ideal longitudinally boost invariant  hydrodynamical model evolution~\cite{rc_pramana}. The initial parameters for the hydrodynamic framework are been tuned to reproduce the experimental data for charged particle multiplicity and  hadronic spectra both at RHIC and LHC~\cite{pramana}. A smooth initial density distribution has been considered for the initial state in central collisions and net baryon density is considered to be negligible at both energies. A lattice based EOS is taken for the (2+1)D model calculation with a transition temperature from QGP to hadronic phase of about 170 MeV~\cite{jfcc}. 
\subsection{Results}\label{sec:results}

\begin{figure}
\begin{center}
\includegraphics[scale=0.4]{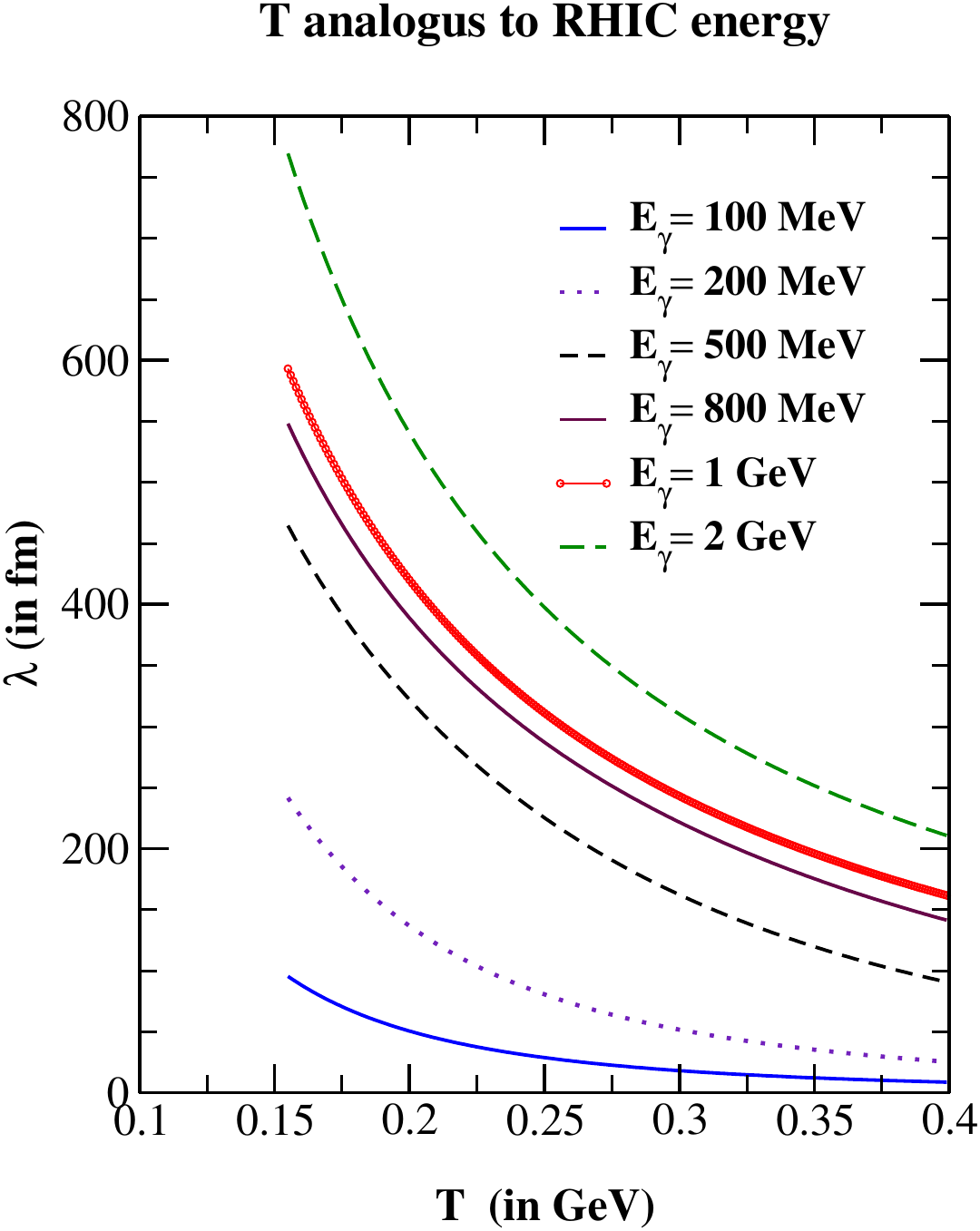}
\caption{ Mean free path of photons versus temperature coming with different energies, from (1+1) dimensional hydrodynamic calculation. }  
\label{fig1j}
\end{center}
\end{figure} 
The photon mean path $\lambda$ from (1+1)D hydrodynamic calculation has been shown in Fig.~\ref{fig1j}. The $\lambda$ values are plotted as a function of temperature T at different photon energies. The mean free path has been found to be extremely large around the QGP hadronic transition temperature and it decreases towards the hot and dense initial stage. A photon with larger energy results in  significant enhancement in the value of  $\lambda$ for a fixed temperature. 
The typical system size in Au+Au collisions is expected to be of the order of 10--20 fm which is way too smaller than the $\lambda$ of  the photons emitted with energy $\ge$ 0.2 GeV. 
\begin{figure}
\begin{center}
\includegraphics[scale=0.4]{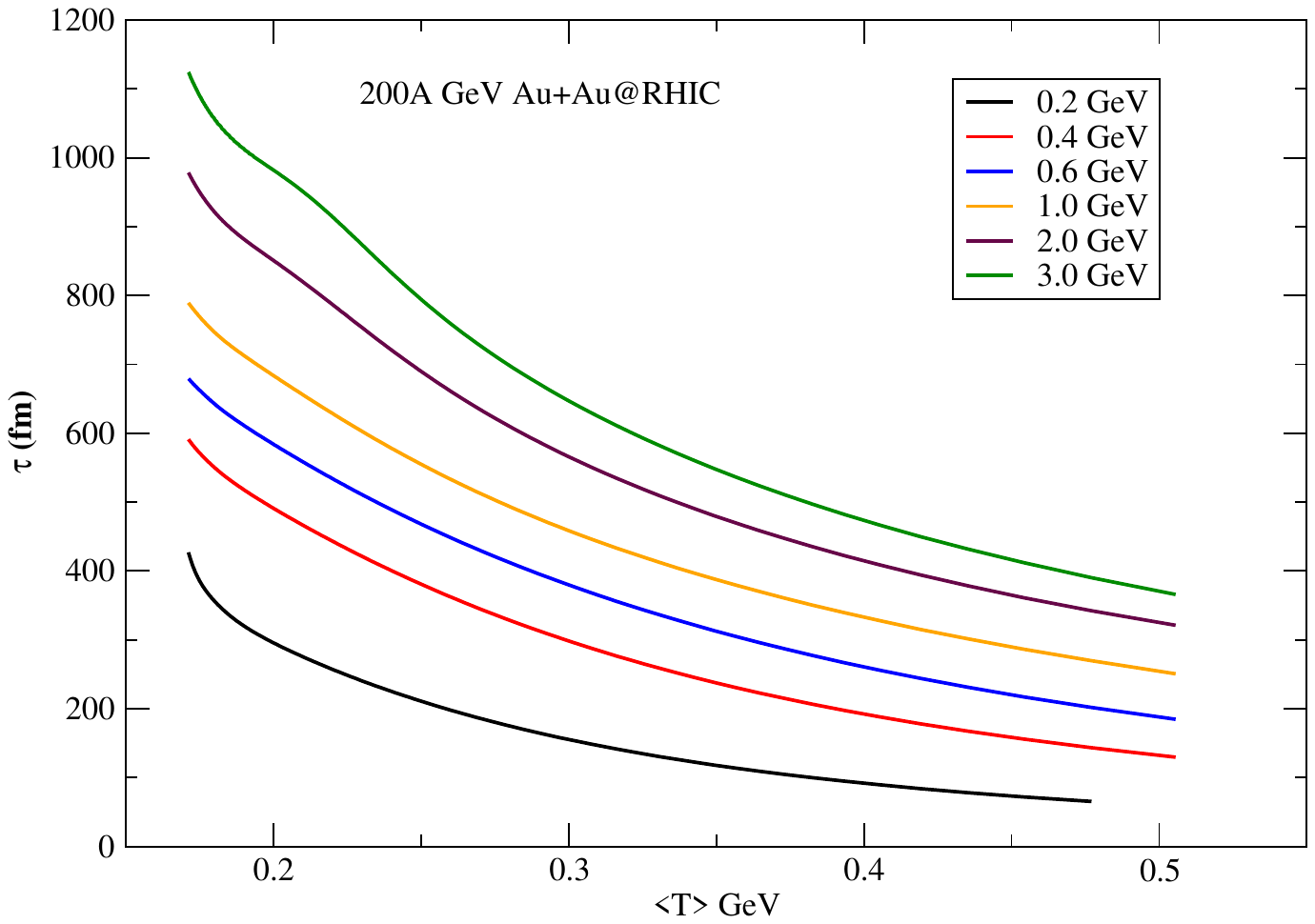}
\caption{ Mean free path of photons of different energies from expanding plasma from a (2+1)D hydrodynamical model calculation using rates from~\cite{arnold2001,trg}. }  
\label{fig2}
\end{center}
\end{figure} 

The results from a more realistic (2+1)D hydrodynamic model calculation at 200A GeV at RHIC has been shown in Fig.~\ref{fig2}. The temperature dependent mean free path values of the photons with energy rage 0.2 to 3 GeV in this case also shows similar qualitative nature as observed for the earlier case od (1+1)D calculation. However, the lambda values are estimated to be larger (about 20\%) for the (2+1)D case. The generic nature of the $\lambda$ values are expected to be same even when initial state fluctuation and viscosity are included in the hydrodynamic model calculation. 

\subsection{Summary and Conclusions}
\label{sec:conclusions}
The photon mean free path $\lambda$ has been estimated using state of the art QGP photon rates and ideal hydrodynamical model framework in collisions of heavy ions at relativistic energies. 
These calculations reveal that $\lambda$ values are highly sensitive to both the temperature of the medium and the energy of the emitted photons. The results show that the photon mean free path is considerably larger than the size of the system, allowing them to escape with minimal interaction. These results show the importance of photons as reliable probes of the early thermal stages in heavy-ion collisions, as they provide a direct signal from the QGP with little distortion. However it is observed that photons from a high temperature phase coming with very low energy, of the order of a few hundred MeV, may have shorter mean free path compared to the expansion of the system.

\section{Investigation of initial state in relativistic nuclear collisions through photon probes}

\author{Rupa Chatterjee}

\bigskip

\begin{abstract}
Photons emitted in relativistic nuclear collisions serve as a powerful probe for investigating the produced initial state  and spatial geometry in those collisions. Recent studies have highlighted the potential of direct photon anisotropic flow  to explore the alpha-clustered structures in light nuclei, such as $^{12}$C and $^{16}$O. The unique  nuclear structures of clustered nuclei can give rise to  distinct spatial anisotropies when collided at relativistic energies and subsequently large  anisotropic flow of emitted photons, providing valuable insights into the initial state and clustered structures.
 
\end{abstract}



\subsection{Introduction}

Significant progress has been made in both theoretical model calculations and experimental analysis of direct photon spectra and anisotropic flow parameters in relativistic heavy ion collisions at top RHIC and LHC energies~\cite{rhic,lhc,phe2, ratio, rc_tau0, v2_v3, p1,v2_1, v2_2, viscometer,gabor,phot6}. The prompt photon spectra from $p+p$ collisions have been well-described by next-to-leading order (NLO) perturbative QCD calculations, utilizing appropriate parton distribution functions over a wide range of transverse momentum. The high $p_T \ (>3 \ {\rm GeV})$ part of the direct photon spectrum from heavy ion collisions is  dominated by prompt contributions and the  NLO pQCD calculations  provide a reasonable explanation in that $p_T$ range. The low $p_T$ region of the direct photon spectrum is dominated by thermal radiations  which can be explained  using state-of-the-art thermal rates and relativistic hydrodynamic framework. However, a simultaneous explanation of both spectra and the anisotropic flow parameters ($v_2$ and $v_3$) of direct photons continues to pose a significant challenge~\cite{rc_pramana}.

Collisions of small systems  at relativistic energies can provide valuable information about the initial state produced in these collisions. The high multiplicity proton-lead collisions at LHC, as well as the deuteron-gold collisions at RHIC have shown significant evidence of collective behaviour in hadronic observables.  These systems can achieve temperatures comparable to those found in central heavy ion collisions due to their compact size and large energy density.

\begin{figure}
    \centering
        \includegraphics[width=0.44\textwidth,clip=true]{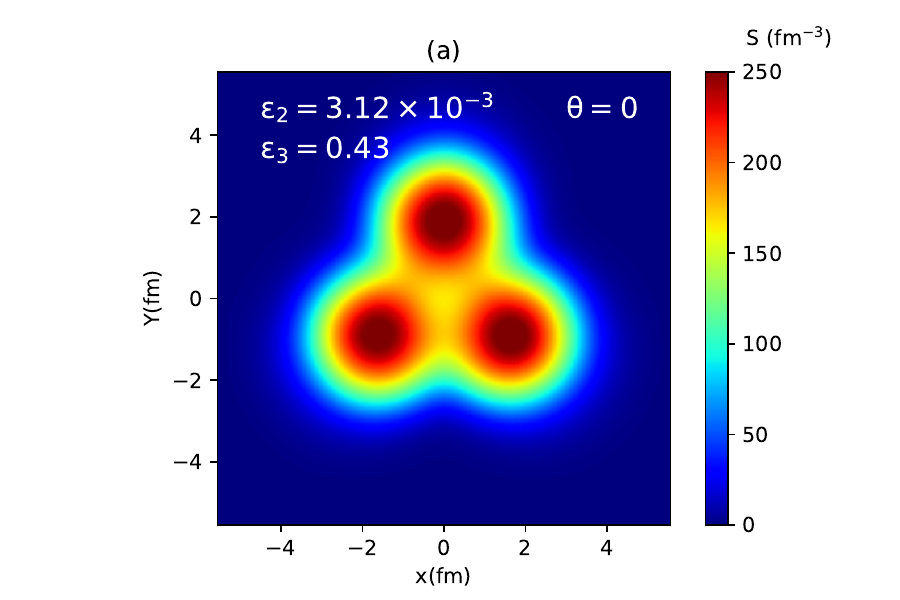}
    \hspace{0.05\textwidth}
        \includegraphics[width=0.44\textwidth,clip=true]{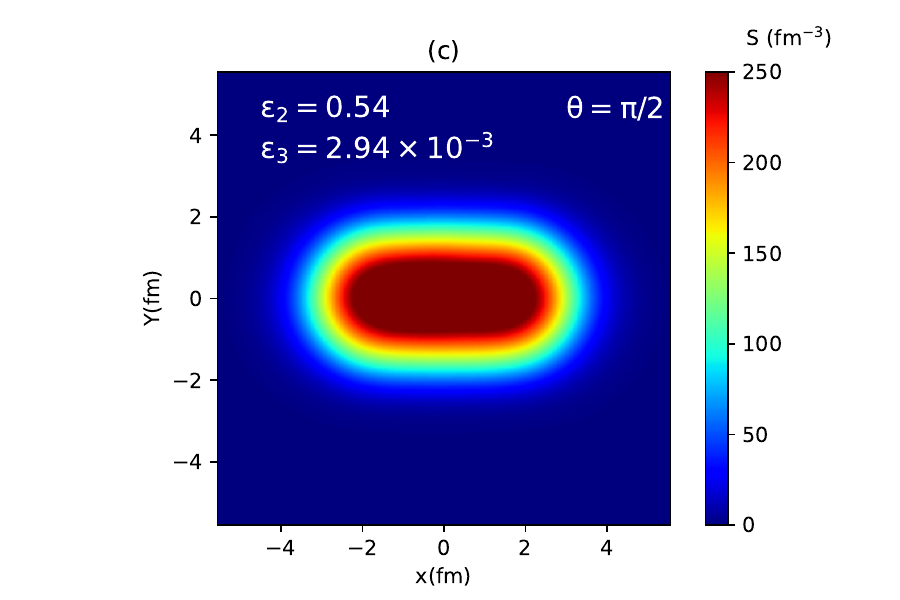}
\caption{(Color online) Distribution of entropy density at the formation time $\tau_0$ on transverse (x-y) plane for central (b $\sim$ 0 fm ) C+Au collisions at 200A GeV for (a) 0 and (b) $\pi/$2 orientation angles (see Ref.~\cite{rc_cluster} for detail).}
\label{Fig1}
\end{figure}
It has also been shown that  thermal photons produced in collision of small systems can outshine the prompt contribution in the low $p_T$ region of direct photon spectrum~\cite{chun_prl}.  This enhancement of thermal radiation can thus serve as a clear signature of the presence of a hot and dense QGP phase  in small collision systems and also provide validation for hydrodynamic behavior in these collisions.


\subsection{Photons from collision of alpha clustered light nuclei}

The presence of alpha-clustered structures is particularly significant in light nuclei, as it plays a crucial role in understanding certain nuclear reactions and properties such as binding energy and specific nuclear states. The carbon and oxygen nuclei can be thought of as being composed of 3 and 4 alpha particles, respectively, arranged in specific geometric configurations. For the carbon nucleus, the 3 alpha clusters are arranged at the vertices of an equilateral triangle, while in the case of oxygen, the 4 alpha clusters form a tetrahedral structure.
\begin{figure}
\centerline{\includegraphics*[width=8.4 cm]{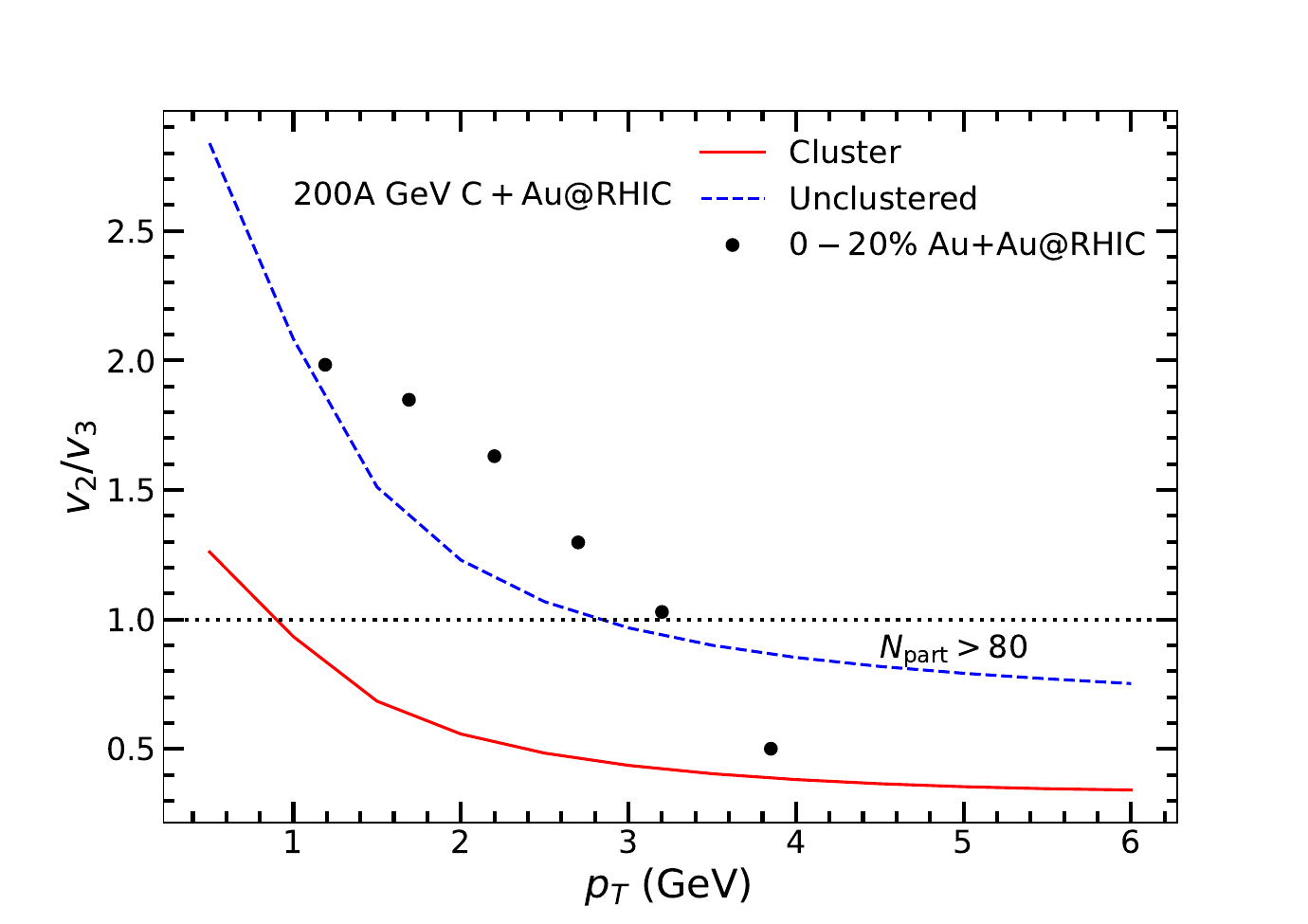}}
\caption{(Color online) The ratio of thermal photon $v_2$ and $v_3$ from clustered and unclustered C+Au collisions at 200A GeV at RHIC as a function of $p_T$~\cite{cluster_ebye}.}
\label{fig2r}
\end{figure}

Recent studies have highlighted the potential of using relativistic nuclear collisions  to investigate the ground state structure of clustered light nuclei~\cite{c1,c2,c3}. The presence of alpha-clustered structures  leads to distinct collision geometries when these nuclei collide with heavy nuclei at relativistic energies. Most central collisions are particularly valuable, as they provide an opportunity to study nuclear structures more effectively and  a clearer understanding of the underlying geometric configurations.

A recent proposal for dedicated $^{16}$O+$^{16}$O collision runs at 7 TeV at the LHC offers the opportunity for experimental verification of cluster structures at such energies. Additionally, the system size of $^{16}$O+$^{16}$O  collisions is comparable to high-multiplicity proton-proton and peripheral lead-lead collisions, providing a unique opportunity to investigate the origin  of collective behavior in small collision systems.

The collision of a triangular $\alpha$ clustered  carbon ($^{12}$C) with Au nuclei at RHIC energy gives rise to different collisions geometries depending on the orientation of the incoming light nuclei. It has been shown that particular orientation of the carbon nucleus can results in large initial ellipticity ($\epsilon_2$), whereas, some other orientation gives rise to large triangular eccentricity 
($\epsilon_3$) (see Fig.\ref{Fig1})~\cite{rc_cluster}. As a result, different orientations of the light clustered nuclei would  gives rise to qualitatively different anisotropic flow parameters of photons even in most central collisions with a heavy nuclei. 

It is to be noted that the most central collision of unclustered carbon with gold gives rise to marginal elliptic or triangular flow parameters of photon due to the initial isotropic energy  density distribution in the transverse plane. The photon spectra from  the unclustered and clustered collision cases are found to be close to each other and can not differentiate between the two cases~\cite{rc_cluster}.
\begin{figure}[ht]
    \centering
        \includegraphics[width=0.45\textwidth]{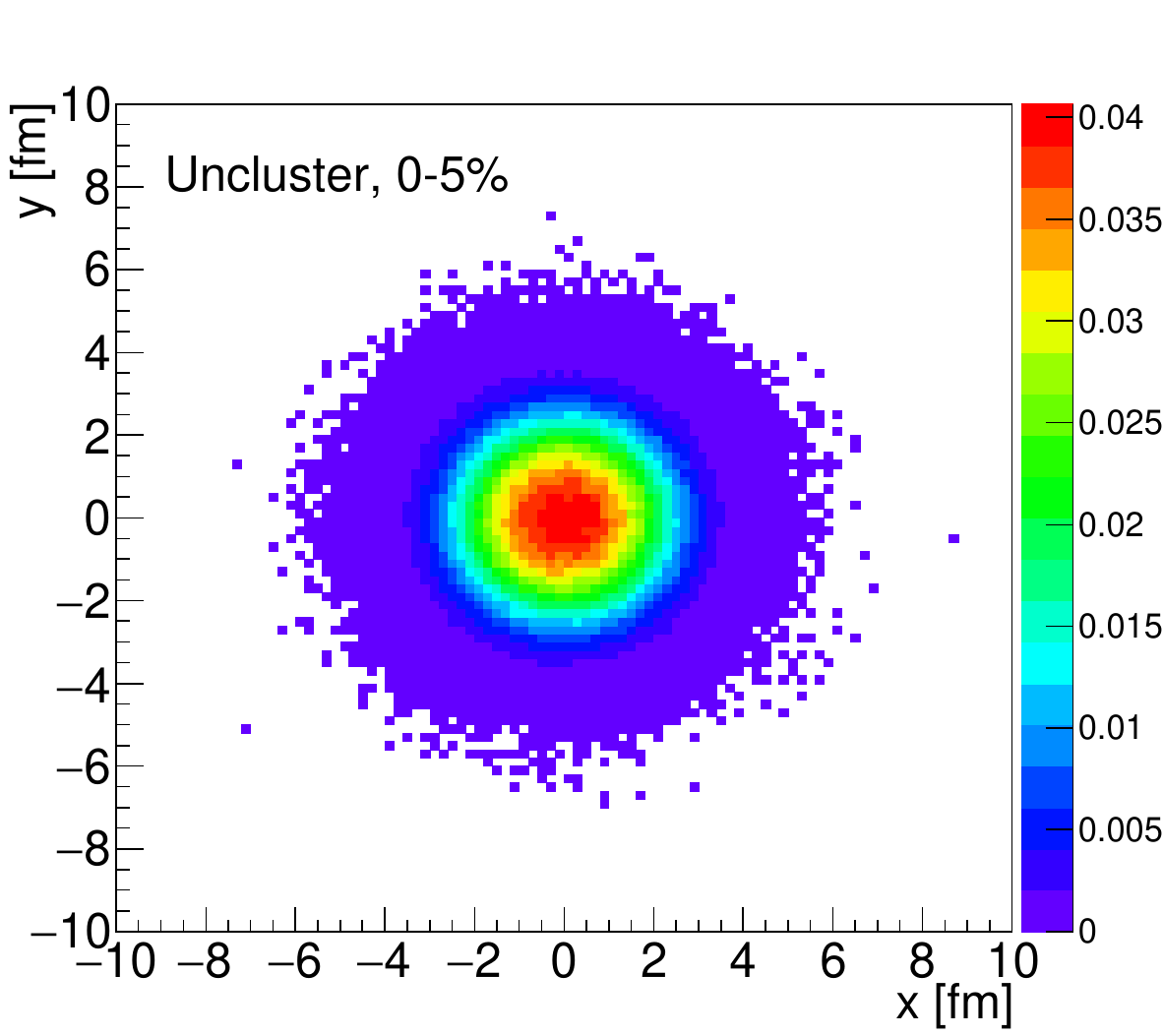}
    \hspace{0.05\textwidth}
        \includegraphics[width=0.45\textwidth]{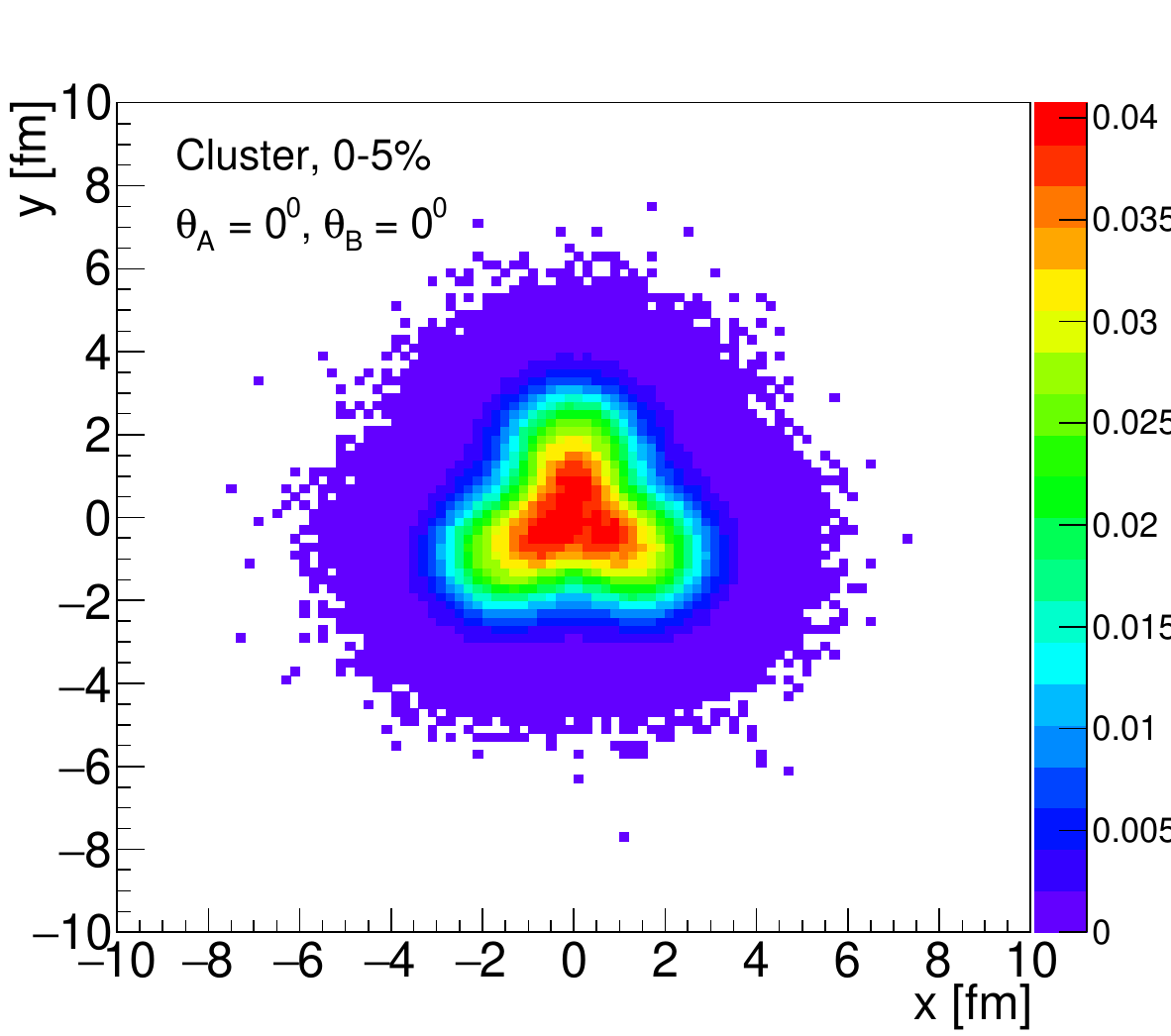}
    \caption{Nucleon density distribution on the transverse plane for 0--5\% O+O collisions at LHC for (a) unclustered and (b) clustered configurations~\cite{sanchari}.}
    \label{fig3}
\end{figure}

A more realistic calculation using a (2+1) dimensional hydrodynamical model framework has shown that the event averaged initial $\epsilon_3$ is much larger than $\epsilon_2$ for most central collision of clustered carbon with a heavy nuclei~\cite{cluster_ebye}. As a result, the final thermal (as well as direct) photon triangular flow is estimated to be significantly larger than the elliptic flow parameter in  presence of clustered structure in the initial state. Thus, experimental determination of direct photon anisotropic flow in  such collisions can be useful to know more about the initial state clustered structures.

The experimental data for individual  anisotropic flow of direct photons are still not in good agreement with the results from  theoretical model calculations. It has been shown that ratio of photon anisotropic flow ($v_2/v_3$) as a function of $p_T$ explains the data better compared to the individual anisotropic flow parameters for  heavy ion collisions~\cite{ratio}. 

The ratio of triangular and elliptic flow of photons from collision of clustered and unclustered C+Au collisions at RHIC using hydrodynamical model calculations has been shown in Fig. \ref{fig2r}.  The photon $v_2/v_3$  from model calculations show different nature  as a function of $p_T$. The  photon $v_3(p_T)$ from clustered case is significantly larger than the $v_2(p_T)$ which results in much larger ($ >1$) value of the ratio in the region $p_T <4$ GeV. On the other hand, for the unclustered case, the ratio is estimated to be less than one (as $v_2 \ > \ v_3$) for $p_T \ >$ 1 GeV.

Direct photon observables from O+O collisions at LHC energy will be complementary to these results from C+Au collisions at RHIC. The GLISSANDO initial condition has been used to study the initial state produced in clustered O+O collision at different values of impact parameter. The initial nucleon density distributions in the transverse plane for central O+O collisions are shown in Fig. \ref{fig3}.  In case of clustered O+O collisions, the initial distribution shows significant initial $\epsilon_3$ for most central collisions, whereas, for the unclustered case the distribution is somewhat isotropic. It has also been observed that the tetrahedron clustered  structure in oxygen gives rise to substantially large initial $\epsilon_4$ for the most central collisions which would result in large $v_4$ for hadrons and photons. The qualitative nature of the initial anisotropies are found to be different for peripheral collisions than for central O+O collisions~\cite{sanchari}.

\subsection{Summary and conclusions}
Relativistic nuclear collisions can  be valuable to study the clustered structures in light nuclei and the initial nucleon-level geometry can be efficiently studied using the  electromagnetic probes. The triangular or tetrahedral clustered structures in  carbon and oxygen nuclei respectively lead to initial spatial anisotropies when collided at relativistic energies, which subsequently result in large final-state momentum space anisotropies of the produced particles. 

The photon anisotropic flow parameters are expected to display significant qualitative and quantitative differences in $p_T$ dependent nature depending on the initial clustered configurations. In addition,  the ratio of photon $v_2$ and $v_3$ can serve as a valuable parameter for probing the initial clustered structure by minimizing the  non-thermal contributions to the direct photon anisotropic flow measurements.



\section{Studying the medium anisotropy in p--O and p--C collisions at the LHC with exotic \texorpdfstring{$\alpha$}{alpha}-cluster density profiles}

\author{Aswathy Menon K.R., Suraj Prasad, Neelkamal Mallick and Raghunath Sahoo}

\bigskip

\begin{abstract}
Small system collectivity and the effects of exotic $\alpha$-clustered nuclear structure on the final state anisotropic flow coefficients are studied by performing p--O and p--C collisions at $\sqrt{s_{\rm NN}}$~=~9.9~TeV using a multi phase transport model (AMPT). In addition, a model-independent sum of Gaussians (SOG) density profile is also implemented in $^{16}\rm O$ and $^{12}\rm C$ nuclei for comparison. We observe a clear dependence of the production yield, initial eccentricities, and the final-state anisotropic flow coefficients on the nuclear density profiles, collision centrality, and the colliding system. This work thus serves as a transport model-based prediction for the O--O and p--O collisions planned at the LHC energies in the upcoming years.
\end{abstract}


\ccode{PACS numbers:}


\subsection{Introduction}

Searching for the signatures of Quark-Gluon Plasma (QGP) and comprehending its properties, is a major goal of relativistic nuclear collisions at the Large Hadron Collider at CERN and the Relativistic Heavy-Ion Collider at BNL. However, the lifetime of QGP is extremely small that it cannot be probed directly. Thus, the collider physicists resort to indirect signatures of QGP, one being the collectivity of the medium formed in relativistic heavy-ion collisions. Small collision systems like p--Pb or pp conventionally serve as the baseline, as a QCD medium formation is least expected here. However, the recent experimental observations suggest that the heavy-ion-like features are seen even in such small collision systems~\cite{CMS:2016fnw} and this motivates the experimentalists to study small system collectivity in detail. The transverse collective flow is usually studied via momentum-space azimuthal anisotropy which is understood as the medium response to the initial spatial anisotropy in non-central nuclear collisions. The azimuthal anisotropy is quantified by the coefficients of Fourier expansion  ($v_{\rm n}$) of the azimuthal distribution of the final state particles. It is known that the anisotropic flow coefficients are sensitive to the nuclear shape, deformation, charge density distribution and energy (or entropy) density fluctuations in the collision overlap area~\cite{Behera:2023nwj}. Therefore, to investigate medium collectivity in small collisions, and to explore the possible effects of an $\alpha$-clustered initial nuclear density profile on the final state flow coefficients, we focus our study on p--O and p--C collisions at $\sqrt{s_{\rm NN}}$~=~9.9~TeV within a multi-phase transport (AMPT) model, with SOG and $\alpha$-cluster nuclear density profiles implemented for $^{16}\rm O$ and $^{12}\rm C$ nuclei. 

Currently, the LHC is looking forward to O--O and p--O collisions in its ongoing Run~3 while such collisions have already been performed at RHIC in its Geometry Scan Program. These collisions are of immense experimental interest because they bridge the multiplicity gap between pp, p--Pb and Pb--Pb collisions. In addition, the p--O collisions at LHC and RHIC energies resemble cosmic rays interacting with the atmospheric oxygen, thus helping us constrain various theoretical models dealing with cosmic air showers. Therefore, our study can timely serve as a transport model prediction for these upcoming experiments.

\subsection{Methodology}
 We use the string-melting version of a multi phase transport model (AMPT) with the settings reported in Ref.~\cite{R:2024eni} to simulate p--O and p--C collisions. For estimating flow coefficients, a two-particle Q-cumulant method~\cite{R:2024eni} is employed where the substantial non-flow effects are reduced by introducing a suitable pseudorapidity gap. The colliding $^{16}\rm O$ and $^{12}\rm C$ nuclei are configured for two different nuclear density profiles viz. the $\alpha$-clustered profile and the Sum of Gaussians (SOG) density profile, the details of which are briefed below.

\begin{itemlist}
 \item \textbf{$\alpha$-cluster} : In light nuclei having 4$n$ number of protons and neutrons, nucleons can cluster into $n$ groups of $\alpha$-particles (i.e. $^{4}\rm He$ nucleus) forming a stable geometrical shape. This is referred to as $\alpha$-clustering. In $^{16}\rm O$ nucleus, the nucleons can arrange themselves into four $\alpha$-clusters, at the four corners of a regular tetrahedron. Likewise, in $^{12}\rm C$, the three $\alpha$-clusters are positioned at the three corners of an equilateral triangle. The nucleons inside each $\alpha$-cluster are sampled according to the three-parameter Fermi (3pF) distribution~\cite{R:2024eni}. The values of various parameters used, corresponding to each nucleus can be found in Ref.~\cite{R:2024eni}~. The spatial orientation of the tetrahedron and the triangle is randomized before each collision for both the target and the projectile nuclei. 
 
 \item \textbf{Sum of Gaussians (SOG)} : Representing the nuclear charge density, $\rho(r)$ in a model-independent fashion to extract the charge density parameters, is a method much preferred by nuclear physics experimentalists. Here, $\rho(r)$ is fitted with a sum of two Gaussian functions, 
\begin{equation}
       \rho (r) = C_{1}e^{-a_{1}r^{2}} + C_{2}e^{-a_{2}r^{2}}.
\end{equation}
The parameters of SOG used to simulate the $^{16}\rm O$ and $^{12}\rm C$ nuclei are tabulated in Ref.~\cite{R:2024eni}~. 
\end{itemlist}

\subsection{Results and Discussions}

\begin{figure*}[ht!]
\centering \includegraphics[scale=0.29]{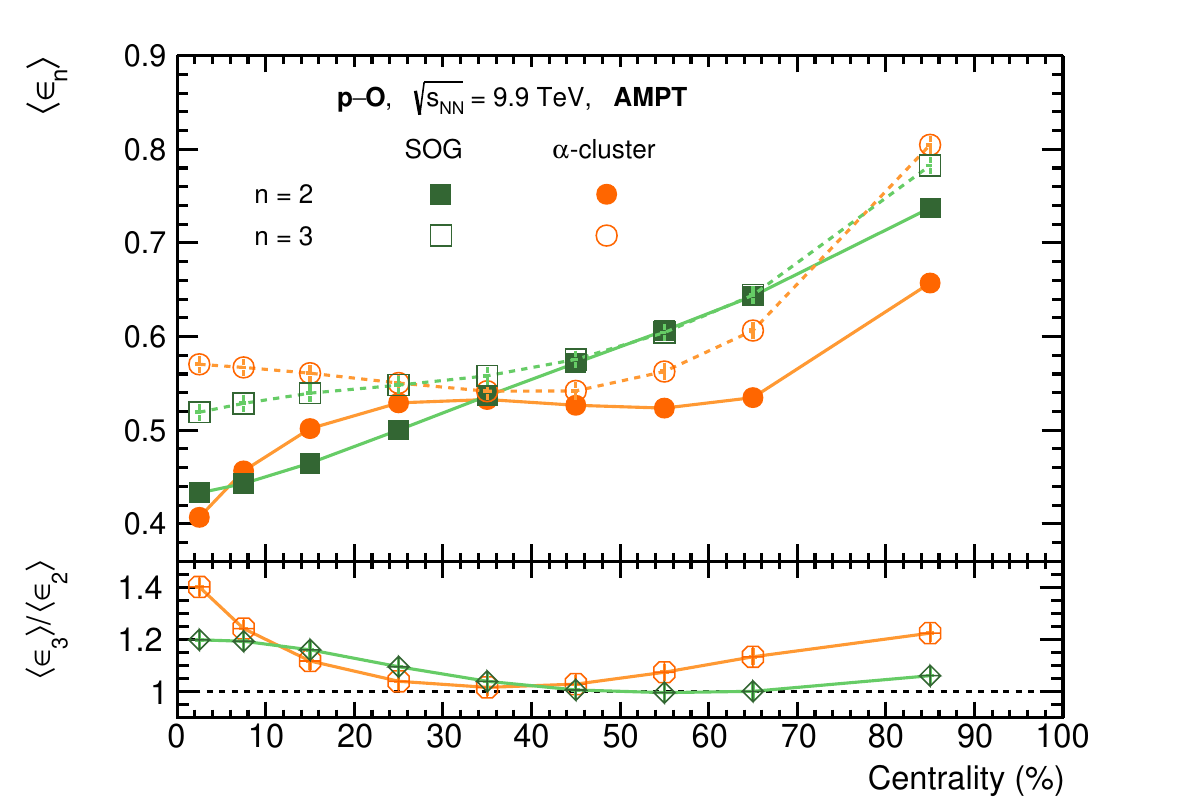}
\centering\includegraphics[scale=0.29]{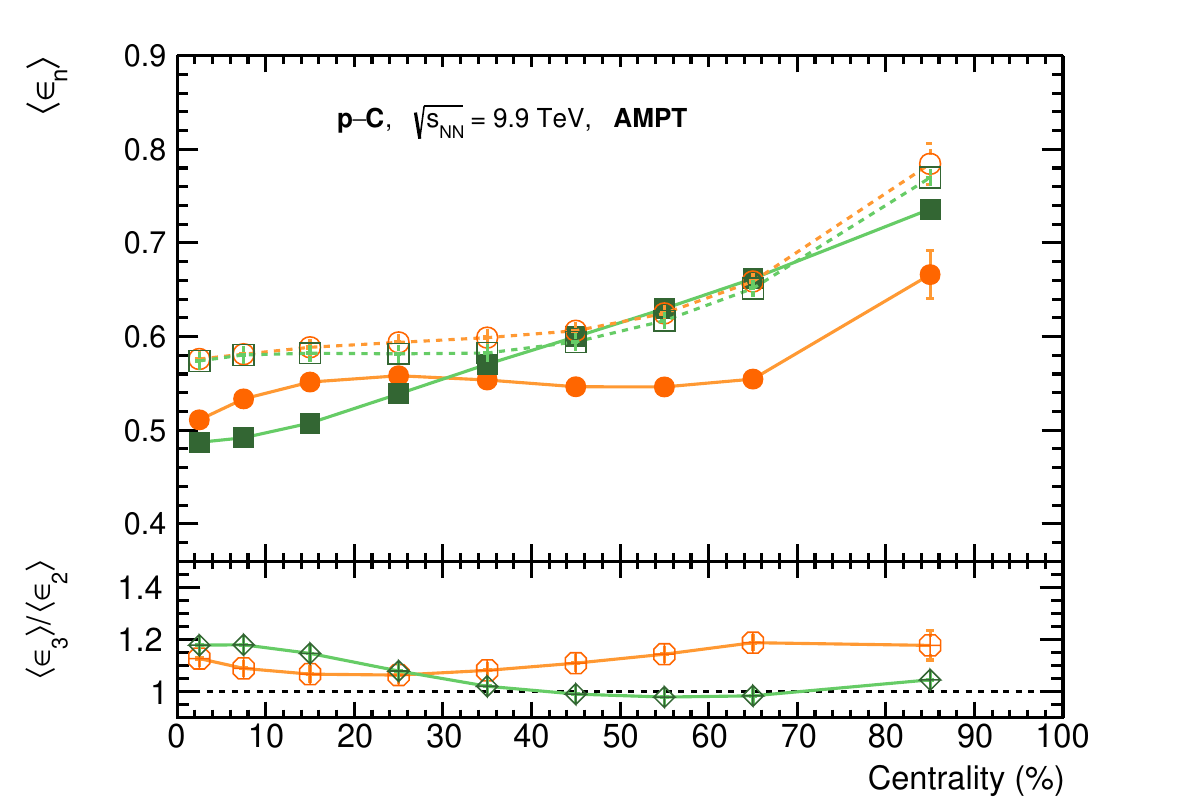}
\caption{ Centrality dependence of $\langle\epsilon_{2}\rangle$ and $\langle\epsilon_{3}\rangle$ for p--O (left) and p--C (right) collisions at $\sqrt{s_{\rm NN}}$~=~9.9~TeV using AMPT for SOG and $\alpha$--cluster nuclear density profiles. Ratio $\langle\epsilon_{3}\rangle$/$\langle\epsilon_{2}\rangle$ as a function of centrality is plotted in the lower panel~\cite{R:2024eni}.}
\label{fig:ecc}
\end{figure*}

Eccentricity ($\langle \epsilon_{2} \rangle$) and triangularity ($\langle \epsilon_{3} \rangle$) can quantify the initial spatial anisotropy of the collision overlap region and hence are studied in Fig.~\ref{fig:ecc} for p--O and p--C collision systems as a function of collision centrality and nuclear density profiles. 

We observe that the $\langle \epsilon_{2} \rangle$ rises steadily with centrality for the SOG density profile case in both p--O and p--C collisions. On the contrary, for the $\alpha$-clustered case, the centrality dependence of $\langle\epsilon_{2}\rangle$ has a wavy nature which is uniquely preserved for both the collision systems. This is in line with the results of $\alpha$-clustered O--O collisions~\cite{Behera:2023nwj}. Such contrasting patterns in the SOG and $\alpha$-cluster profiles can be owed to the difference in the geometrical distribution of the nucleons inside the nuclei~\cite{R:2024eni}. The centrality dependence of $\langle \epsilon_{3} \rangle$ shows qualitative and quantitative similarity for both the density profiles in the p--C system. In contrast, for p--O collisions, $\langle \epsilon_{3} \rangle$ due to the $\alpha$-clustered nuclear structure, drops to a minimum for mid-central collisions, followed by a strong rise towards the peripheral collisions. The bottom panel of Fig.~\ref{fig:ecc} depicts the ratio $\langle \epsilon_{3} \rangle$/$\langle \epsilon_{2} \rangle$ where we see a hike in the most central p--O collision for the $\alpha$-cluster case. Surprisingly, this central hike is not developed in p--C collisions, and this could be attributed to the extra $\alpha$-cluster that is present in $^{16}\rm O$ nucleus. Thus, despite the colliding species being the same, a difference in the nuclear density profile is seen to lead to significant differences in the initial spatial anisotropies.

\begin{figure*}[ht!]
\includegraphics[scale=0.15]{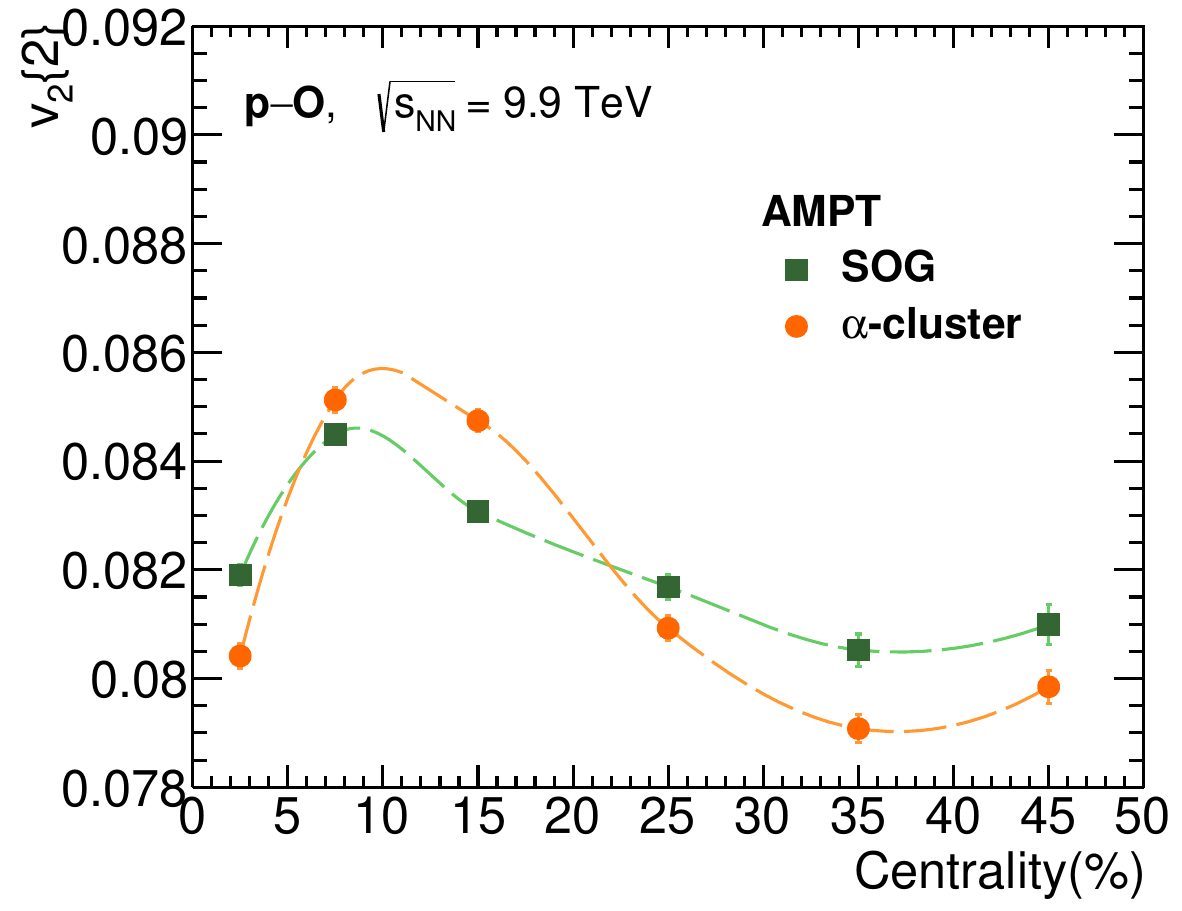}
\includegraphics[scale=0.15]{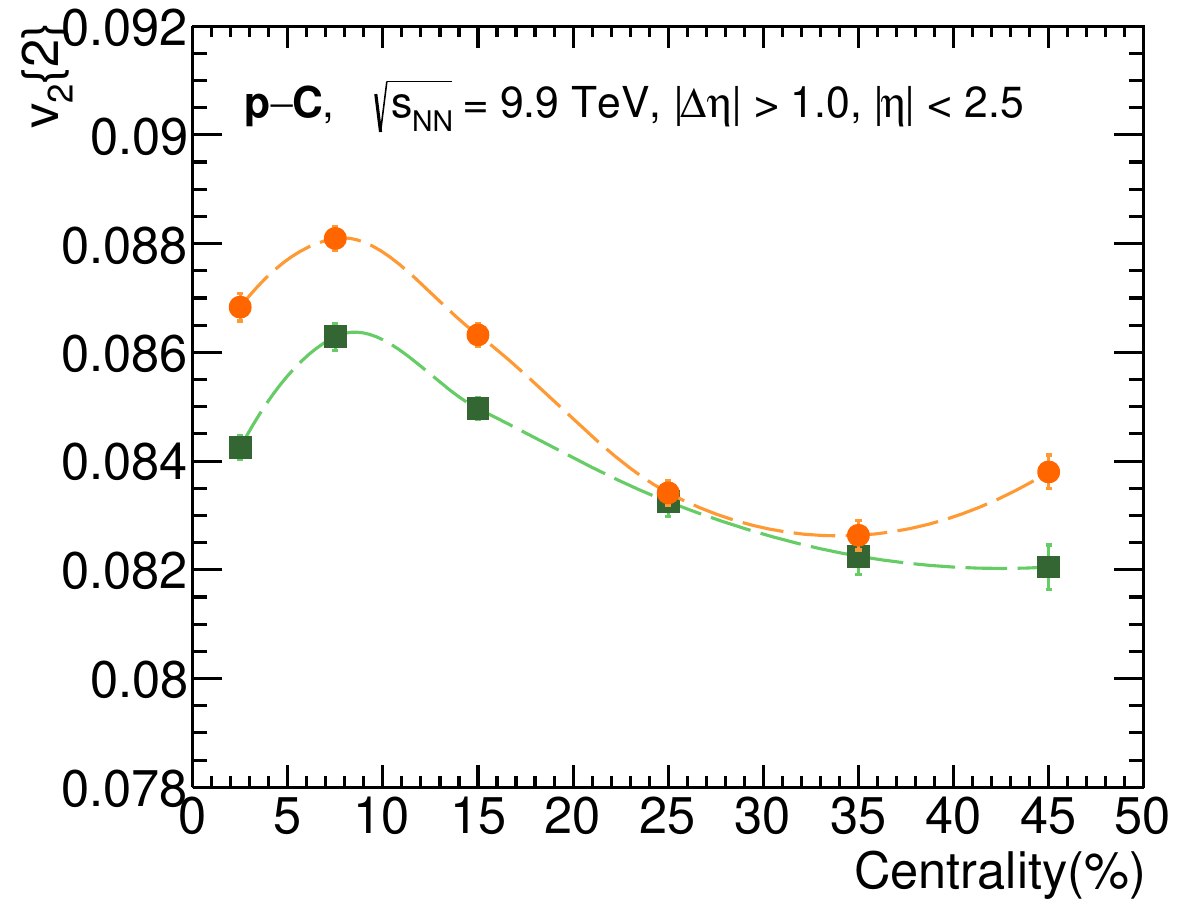}
\includegraphics[scale=0.15]{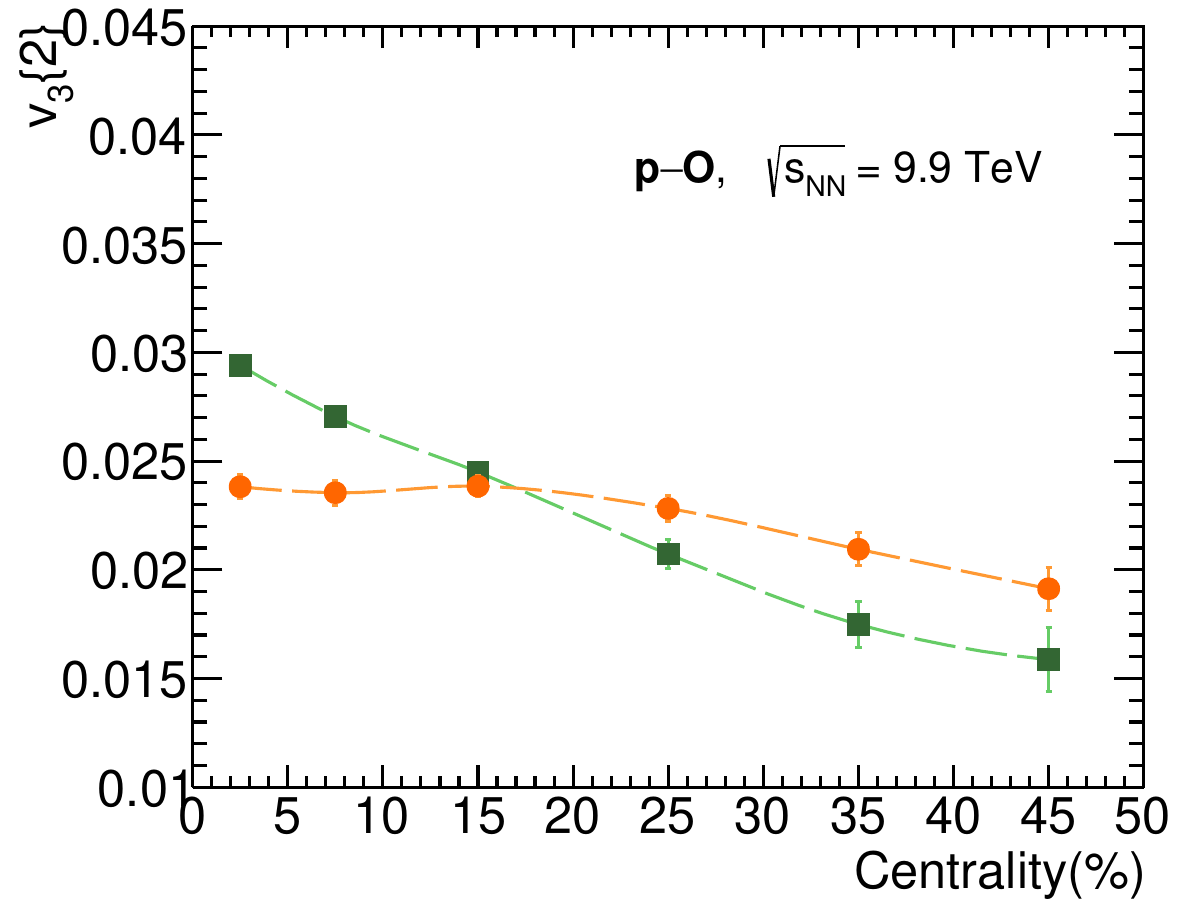}
\includegraphics[scale=0.15]{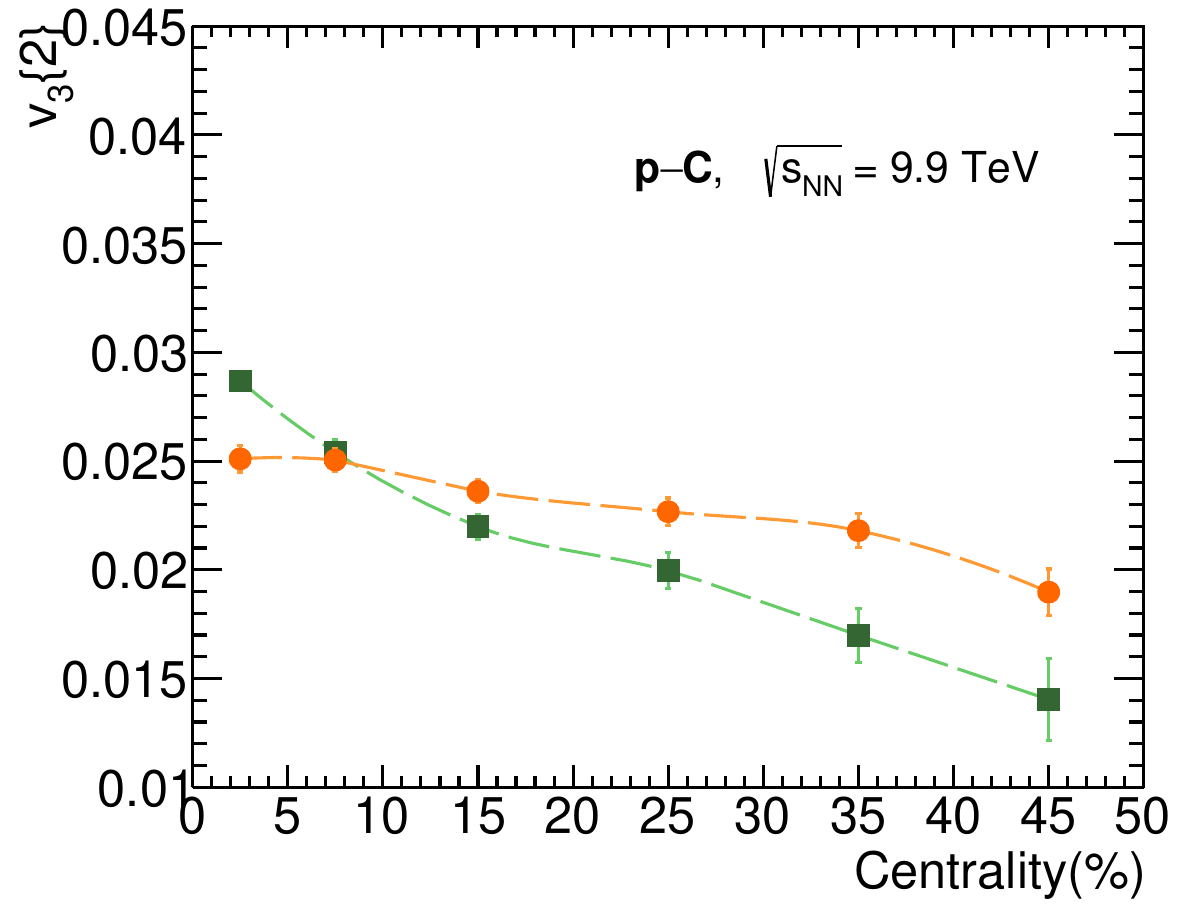}
\caption{~ $v_{2}\{2\}$ and $v_{3}\{2\}$ as a function of centrality for SOG and $\alpha$-cluster density profiles in p--O and p--C collisions at $\sqrt{s_{\rm NN}}$~=~9.9~TeV~\cite{R:2024eni}.}
\label{fig:v2v3}
\end{figure*}

In Fig.~\ref{fig:v2v3}, we study the centrality dependence of the second and third-order azimuthal anisotropic flow coefficients, viz. elliptic flow ($v_{2}$) and triangular flow ($v_{3}$). 
As expected for small collision systems, the initial anisotropies are not very well carried forward to the final state beyond central collisions, possibly due to a very small number of participants. For both p--O and p--C collisions, $v_{2}$ is the largest in the (5-10)\% centrality class, irrespective of the kind of the nuclear density profile. While $v_{2}$ from $\alpha$-cluster case maintains its wavy trend in both p--O and p--C collisions, $v_{2}$ in the SOG case is seen to continuously drop down after the most central collisions. The magnitude of $v_{2}$ is higher in p--C collisions than in p--O, similar to the magnitudes of $\langle \epsilon_{2} \rangle$, owing to a smaller system with large eccentricity fluctuations. The trend for $v_{3}$ is such that it decreases with centrality for both p--O and p--C collision systems with the magnitude of slope for the SOG case being larger than the corresponding $\alpha$-clustered density profile case. On comparing $v_{2}$ and $v_{3}$, we recognize that the overall centrality dependence is much less for $v_{2}$ than for $v_{3}$. Apart from these, to understand the response of the formed medium, if any, to the evolution of anisotropic flow from the initial spatial anisotropies, we also study the ratios, $v_{2}$/$\langle \epsilon_{2} \rangle$, $v_{3}$/$\langle \epsilon_{3}\rangle$ and $v_{3}$/$v_{2}$ for both clustered and unclustered nuclear density profiles in the two collision systems, p--O and p--C~\cite{R:2024eni}.

\subsection{Summary}
This work studies the initial spatial anisotropies ($\langle \epsilon_{2} \rangle$ and $\langle \epsilon_{3} \rangle$) and the final-state anisotropic flow coefficients ($v_{2}$ and $v_{3}$) for their dependency on the nuclear density profile, collision centrality, and the collision system selections by simulating p--O and p--C collisions at $\sqrt{s_{\rm NN}}$~=~9.9~TeV using AMPT. Here, the flow coefficients are estimated using the two-particle Q-cumulant method. We find that for a given collision species, a difference in the nuclear density profile naturally leads to differences in the initial spatial anisotropies. Due to the small number of participants in these collisions, the centrality dependence of $\langle \epsilon_{2} \rangle$ is not seen to reflect well in the final state $v_{2}$ beyond central collisions. The system formed in p--C collisions is subjected to larger density fluctuations, owing to a small system size, and hence, the magnitudes of $\langle \epsilon_{2} \rangle$ and subsequently $v_{2}$ are larger in p--C than in p--O collisions. In a nutshell, using ultra-relativistic nuclear collisions as a tool to probe the effects of $\alpha$-clustered nuclear structure, our study aims to provide a good prediction for the upcoming p--O collisions planned for the LHC Run~3.

\section{Causal third-order viscous hydrodynamics from kinetic theory }

\author{Pushpa Panday, Amaresh Jaiswal, and Binoy Krishna Patra}

\bigskip

\begin{abstract}
In the present work, we derive a linearly stable and causal theory of relativistic third-order viscous
hydrodynamics from the Boltzmann equation with relaxation-time approximation. We employ a Chapman-Enskog-like iterative solution
of the Boltzmann equation to obtain the viscous correction to the distribution function. Our derivation highlights the necessity of incorporating a new dynamical degree of freedom, specifically an irreducible tensor of rank three. This
differs from the recent formulation of causal third-order theory from the method of moments which
requires two dynamical degrees of freedom: an irreducible third-rank and a fourth-rank tensor. We
verify the linear stability and causality of the proposed formulation by examining perturbations
around a global equilibrium state.
\end{abstract}




\subsection{Introduction}
 Relativistic dissipative hydrodynamics has been successfully applied to study
the collective behaviour of QGP \cite{gale,Heinz:2013th}. Relativistic dissipative hydrodynamics is formulated through an order-by-order expansion in powers of the space-time gradients of hydrodynamical variables, with ideal hydrodynamics corresponding to the zeroth order. The first-order theory, often called the Navier-Stokes theory, is recognized as ill-defined because of acausality and numerical instability. Causality was restored in the second-order Israel-Stewart (IS) theory, though stability remains uncertain. However, IS theory introduces some undesirable effects, such as reheating of the expanding medium and the appearance of negative longitudinal pressure. Additionally, the scaling solutions of IS theory diverge from transport results for large viscosities, suggesting the limitations of the second-order approach. It has been proposed that empirically including higher-order terms greatly enhances agreement with transport results, emphasizing the need to develop relativistic dissipative hydrodynamics beyond the second-order IS theory.\par 
Several researchers have investigated the formulation of relativistic third-order dissipative fluid dynamics using different frameworks. Recently, the linear stability and causality of a third-order theory formulated in [\refcite{Jaiswal:2013vta}] were analyzed, showing acausal and unstable behaviour [\refcite{Brito:2021iqr}]. To address this, a heuristic modification was proposed, introducing a new dynamical degree of freedom. In this article, we derive a linearly stable and causal third-order theory of relativistic viscous hydrodynamics directly from the Boltzmann equation under the relaxation-time approximation. For this purpose, we incorporate viscous corrections to the distribution function derived from a Chapman-Enskog-like iterative solution of the Boltzmann equation. This derivation highlights the critical inclusion of a new dynamical degree of freedom, specifically an irreducible rank-3 tensor, within this framework. To validate our formulation, we examine its linear stability and causality by analyzing perturbations around a global equilibrium state. 
\subsection{Third-order viscous evolution equation}

The hydrodynamic evolution of a system, without net conserved charges, is determined by
the conservation equations of energy and momentum. The conserved energy-momentum
tensor ($T^{\mu\nu}$) can be represented in terms of the single-particle phase-space distribution function
and decomposed into hydrodynamical tensor degrees of freedom. For a system of massless particles, bulk viscosity vanishes, and $T^{\mu\nu}$ can be written as

\begin{equation}\label{energy}
	T^{\mu\nu}=\int dP ~p^{\mu}p^{\nu}f(x,p)=\epsilon u^{\mu}u^{\nu}-P\Delta^{\mu\nu}+\pi^{\mu\nu},
\end{equation}

where, $dP \equiv g\,d^3{{\boldsymbol p}}/\left[(2\pi)^3|{\boldsymbol p}|\right]$ is Lorentz invariant momentum integral measure with $g$ being the degeneracy factor. Here, $p^{\mu}$ is the particle four-momentum and $f(x,p)$ is the single-particle phase-space distribution function with $x^\mu$ representing the position four-vector. In the tensor decomposition $\epsilon$ and $P$ are energy density and thermodynamic pressure, respectively, $\pi^{\mu\nu}$ is the shear-stress tensor and $u^{\mu}$ is the fluid four-velocity defined in the Landau frame, $u_{\mu}T^{\mu\nu}=\epsilon u^{\nu}$. Moreover, we have the orthogonality condition $u_\mu\pi^{\mu\nu}=0$ and introduced the notation $\Delta^{\mu\nu} \equiv g^{\mu\nu} - u^{\mu}u^{\nu}$ as the projection operator orthogonal to $u^{\mu}$. \par
The first-order expression for $\pi^{\mu\nu}$ can be expressed in terms of small deviation ($\delta f$) from equilibrium ($f=f_{eq}+\delta f, \delta f\ll f_{eq}$) as

\begin{equation}
    \pi^{\mu\nu} = \Delta^{\mu\nu}_{\alpha\beta}\int dP \, p^{\alpha} \, p^{\beta} \, \delta f,
\end{equation}

where, $\Delta^{\mu\nu}_{\alpha\beta}\equiv \frac{1}{2}\!\left( \Delta^{\mu}_{\alpha}\Delta^{\nu}_{\beta} + \Delta^{\mu}_{\beta}\Delta^{\nu}_{\alpha} \right) - \frac{1}{3}\Delta^{\mu\nu}\Delta_{\alpha\beta}$ is a traceless and doubly symmetric projection operator, which is orthogonal to $u_{\mu}$ as well as $\Delta_{\mu\nu}$.\par
The non-equilibrium phase-space distribution function can be obtained by solving the
kinetic equation such as Boltzmann equation. The relativistic Boltzmann transport equation (RBTE) under relaxation-time approximation (RTA) for the collision term is given by \cite{anderson}

\begin{equation}\label{boltz}
    p^{\mu}\partial_{\mu}f = - \frac{u\cdot p}{\tau_R}\,\delta f,
\end{equation}

where $\tau_R$ is the relaxation time. We employed the Chapman-Enskog like iterative solution to the RBTE and obtained the first-order expression for $\pi^{\mu\nu}$ as

\begin{equation}\label{pimunu1}
    \pi^{\mu\nu} = 2 \tau_R\beta_{\pi}\sigma^{\mu\nu}, \quad \beta_{\pi}=\frac{4}{5}P.
\end{equation}

Following the methodology discussed in [\refcite{Denicol:2010xn}], the second and third-order viscous evolution equation is obtained as

\begin{align}\label{pidot-2nd}
\dot{\pi}^{\langle\mu\nu\rangle}+\frac{\pi^{\mu\nu}}{\tau_\pi}=  2 \beta_\pi \sigma^{\mu\nu}+2 \pi_\gamma^{\langle\mu} \omega^{\nu\rangle \gamma} 
-\frac{10}{7} \pi_\gamma^{\langle\mu} \sigma^{\nu\rangle \gamma}-\frac{4}{3} \pi^{\mu\nu} \theta,
\end{align}
and
\begin{align}\label{pidot-3rd}
\dot{\pi}^{\langle\mu\nu\rangle}= & -\frac{\pi^{\mu\nu}}{\tau_\pi}+2 \beta_\pi \sigma^{\mu\nu}+2 \pi_\gamma^{\langle\mu} \omega^{\nu\rangle \gamma}-\frac{10}{7} \pi_\gamma^{\langle\mu} \sigma^{\nu\rangle \gamma} 
-\frac{4}{3} \pi^{\mu\nu} \theta+\frac{25}{7 \beta_\pi} \pi^{\rho\langle\mu} \omega^{\nu\rangle \gamma} \pi_{\rho \gamma}-\frac{1}{3 \beta_\pi} \pi_\gamma^{\langle\mu} \pi^{\nu\rangle \gamma} \theta \nonumber\\
& -\frac{38}{245 \beta_\pi} \pi^{\mu\nu} \pi^{\rho \gamma} \sigma_{\rho \gamma}-\frac{22}{49 \beta_\pi} \pi^{\rho\langle\mu} \pi^{\nu\rangle \gamma} \sigma_{\rho \gamma} 
 -\frac{24}{35} \nabla^{\langle\mu}\left(\pi^{\nu\rangle \gamma} \dot{u}_\gamma \tau_\pi\right)+\frac{4}{35} \nabla^{\langle\mu}\left(\tau_\pi \nabla_\gamma \pi^{\nu\rangle \gamma}\right) \nonumber\\
& -\frac{2}{7} \nabla_\gamma\left(\tau_\pi \nabla^{\langle\mu} \pi^{\nu\rangle \gamma}\right)+\frac{12}{7} \nabla_\gamma\left(\tau_\pi \dot{u}^{\langle\mu} \pi^{\nu\rangle \gamma}\right) 
 -\frac{1}{7} \nabla_\gamma\left(\tau_\pi \nabla^\gamma \pi^{\langle\mu\nu\rangle}\right)+\frac{6}{7} \nabla_\gamma\left(\tau_\pi \dot{u}^\gamma \pi^{\langle\mu\nu\rangle}\right) \nonumber\\
& -\frac{2}{7} \tau_\pi \omega^{\rho\langle\mu} \omega^{\nu\rangle \gamma} \pi_{\rho \gamma}-\frac{2}{7} \tau_\pi \pi^{\rho\langle\mu} \omega^{\nu\rangle \gamma} \omega_{\rho \gamma} 
 -\frac{10}{63} \tau_\pi \pi^{\mu\nu} \theta^2+\frac{26}{21} \tau_\pi \pi_\gamma^{\langle\mu} \omega^{\nu\rangle \gamma} \theta,
\end{align}

respectively.
 $\omega^{\mu\nu}\equiv \left(\nabla^{\mu}u^{\nu}-\nabla^{\nu}u^{\mu}\right)/2$ is the vorticity tensor. The second-order space-like derivatives of shear-stress tensor in the third-order shear viscous evolution equation causes instability and acausality. In the following, we restore causality in the formulation of relativistic third-order viscous hydrodynamics.

\subsection{Restoring causality at third-order}

In [\refcite{Brito:2021iqr}], authors proposed a mechanism to restore causality in third-order hydrodynamic theories by promoting the space-like gradients of the shear-stress tensor to a new hydrodynamical variable which is a third rank tensor, i.e, 

\begin{equation}\label{eq:csl_mec}
    \nabla^{\langle\mu}\pi^{\nu\lambda\rangle} \to \rho^{\mu\nu\lambda}.
\end{equation}

The complete third-order formulation required the evolution equation for $\pi^{\mu\nu}$ and $\rho^{\mu\nu\lambda}$. These equations are obtained as

\begin{align}
&\dot{\pi}^{\langle\mu\nu\rangle}=  -\frac{\pi^{\mu\nu}}{\tau_\pi}+2 \beta_\pi \sigma^{\mu\nu}+2 \pi_\gamma^{\langle\mu} \omega^{v\rangle \gamma}-\frac{10}{7} \pi_\gamma^{\langle\mu} \sigma^{v\rangle \gamma} 
 -\frac{4}{3} \pi^{\mu\nu} \theta+\frac{24}{7 \beta_\pi} \pi^{\rho\langle\mu} \omega^{v\rangle \gamma} \pi_{\rho \gamma}-\frac{5}{21 \beta_\pi} \pi_\gamma^{\langle\mu} \pi^{v\rangle \gamma} \theta \nonumber \\
& -\frac{52}{245 \beta_\pi} \pi^{\mu\nu} \pi^{\rho \gamma} \sigma_{\rho \gamma}-\frac{15}{49 \beta_\pi} \pi^{\rho\langle\mu} \pi^{v\rangle \gamma} \sigma_{\rho \gamma}-\frac{2}{7} \tau_\pi \omega^{\rho\langle\mu} \omega^{v\rangle \gamma} \pi_{\rho \gamma}-\frac{4}{7} \tau_\pi \pi^{\rho\langle\mu} \omega^{v\rangle \gamma} \omega_{\rho \gamma} 
 -\frac{8}{63} \tau_\pi \pi^{\mu\nu} \theta^2 \nonumber\\
 &+\frac{26}{21} \tau_\pi \pi_\gamma^{\langle\mu} \omega^{v\rangle \gamma} \theta-\nabla_{\gamma}\rho^{\gamma<\mu\nu>}+\frac{1}{7\beta_{\pi}}\pi^{\gamma<\mu}\pi^{\nu>\beta}\omega_{\gamma\beta},
\end{align}
 and

 \begin{align}
	&\dot{\rho}^{<\mu\nu\lambda>}+\frac{1}{\tau_\rho}\rho^{\mu\nu\lambda}=~
	\frac{3}{7}\nabla^{<\mu}\pi^{\nu\lambda>}-~\frac{18}{7}\dot{u}^{<\mu}\pi^{\nu\lambda>}-\frac{187}{81}\rho^{\mu\nu\lambda}\theta-\frac{10}{7}\tau_{\pi}\dot{u}^{<\mu}\pi^{\nu\lambda>}\theta-\frac{36}{7}\tau_{\pi}\omega^{\gamma<\mu}\pi^{\nu}_{\gamma}\dot{u}^{\lambda>}\nonumber \\
 &-\frac{389}{441\beta_{\pi}}\pi^{\gamma<\mu}\pi^{\nu}_{\gamma}\dot{u}^{\lambda>}-\frac{343}{105\beta_{\pi}}\dot{u}_{\gamma}\pi^{\gamma<\mu}\pi^{\nu\lambda>}+\frac{18}{7}\tau_{\pi}\dot{u}_{\gamma}\omega^{\gamma<\mu}\pi^{\nu\lambda>}-\frac{6}{7}\tau_{\pi}\omega^{\gamma<\mu}\nabla_{\gamma}\pi^{\nu\lambda>} \nonumber\\
&- \frac{6}{7}\tau_{\pi}\omega^{\gamma<\mu}\nabla^{\nu}\pi^{\lambda>}_{\gamma}-\frac{6}{7}\tau_{\pi}\pi_{\gamma}^{<\mu}\nabla^{\nu}\omega^{\lambda>\gamma}-\frac{47}{63\beta_{\pi}}\pi^{<\mu\nu}\nabla_{\gamma}\pi^{\lambda>\gamma}-\frac{11}{21\beta_{\pi}}\pi^{\gamma<\mu}\nabla_{\gamma}\pi^{\nu\lambda>} \nonumber\\
&-\frac{665}{441\beta_{\pi}}\pi^{\gamma<\mu}\nabla^{\nu}\pi^{\lambda>}_{\gamma}. 
\end{align}

The linear contributing terms to the evolution equation of $\pi^{\mu\nu}$ and $\rho^{\mu\nu\lambda}$ are $-\frac{\pi^{\mu\nu}}{\tau_{\pi}}+2\beta_{\pi}\
    \sigma^{\mu\nu}-\nabla_{\gamma}\rho^{\gamma<\mu\nu>}$ and $\frac{1}{\tau_\rho}\rho^{\mu\nu\lambda}+\frac{3}{7}\nabla^{<\mu}\pi^{\nu\lambda>}$, respectively. The causality and stability analysis of these linear equations leads to the constraints on the transport coefficients. Our results for the transport coefficient, $\eta_{\rho}=\tau_{\rho}=\tau_{\pi}=5\eta/(\epsilon+P)$, is found to be consistent with these constraints. Therefore, we conclude that the present formulation of relativistic third-order viscous hydrodynamics is linearly causal and stable.\par

    We demonstrated that a causal relativistic third-order theory requires the
inclusion of a new dynamical degree of freedom,i.e., an irreducible tensor of rank 3. Additionally, by considering perturbations around
a global equilibrium state, we showed that transport coefficients are in compliance
with stability and causality constraints.

\section{Fluctuations of conserved charges Finite size PNJL model with MRE approximation}

\author{Paramita Deb, Raghava Varma, and Amal Sarkar}

\bigskip

\begin{abstract}
Baryon, charge and strangeness fluctuations have been investigated in finite size Polyakov loop enhanced Nambu--Jona-Lasinio model. Multiple reflection expansion method has been incorporated to include the surface and curvature energy along with the system volume. The results of different fluctuations are then compared with the available experimental data from the heavy ion collision. Results show both qualitative and quantitative similarities with the experimental data.
\end{abstract}


\ccode{PACS numbers: {12.38.AW, 12.38.Mh, 12.39.-x}
}


\subsection{Introduction}
The primary objectives of Relativistic Heavy Ion Collider (RHIC)
at BNL and Super Proton Synchrotron (SPS) at CERN 
experiments are to study the strongly interacting matter at finite temperature and density and to
map the QCD phase diagram and estimate the critical end point (CEP). In order to conduct experiment
near the critical region, suitable experimental observables such as conserved quantities are needed.
Specifically,the fluctuations of these conserved quantities show non-monotonous behavior near the
critical region. Generally, the fluctuations of  experimental observables are defined as the variance and
higher non-Gaussian moments of the event-by-event distribution of the observable at each event in
ensemble of many events \cite{STAR}. 
Further more, the magnitude of these fluctuations of the conserved quantities such as net-baryon $(\Delta{B})$, net-strangeness $(\Delta{S})$ and net-charge $(\Delta{Q})$ as well as the correlation length ($\xi$) diverge at the critical point \cite{rajagopal2000}. 
The QCD inspired models \cite{fukushima,ratti, deb2009, deb2010a, deb2010b, deb2011, deb2013} further indicate that the net conserved quantum numbers ($B$, $Q$ and $S$) are related to the conserved number susceptibilities ($\chi_x= \langle(\delta N_x)^2\rangle/VT $ where $x$ can be either $B$, $S$ or $Q$ 
and $V$ is the volume), which generally agree with the lattice QCD models.
In contrast to the experiments, most of the earlier QCD inspired theoretical approaches do not consider the finite volume effect such as the Nambu--Jona-Lasinio (NJL) model {\cite {nambu1961}} and it's extension the Polyakov loop Nambu--Jona-Lasinio (PNJL) model {\cite{fukushima,ratti,pisarski2000,fukushima1,hansen2007}}.
A way to improve the finite volume study by incorporating the surface and curvature effects 
that bound the finite volume of PNJL system would be through the multiple reflection expansion (MRE) method {\cite {bloch1970,grunfeld2018}}. 
MRE describes a sphere with proper boundary conditions rather than a cube, which is more natural to study the problem of QGP fireball in heavy ion collision. 
In the light of above discussions, the current work will emphasis on the 3 flavor finite size finite density PNJL model with MRE formalism as well as infinite volume PNJL model to study the susceptibilities of different conserved charges and compare them to both the recent experimental findings.

\vskip -1.6in
\subsection{The PNJL Model}
The PNJL model consists of
both four quark and six quark interaction terms and a Polyakov loop which represents the gluonic degrees of freedom. The model nicely captures the confinement and chiral symmetry restoration properties of strongly interacting matter. We shall consider the 2+1 flavor PNJL model with six quark interactions as given below,
\begin{align}
   {\cal L} &= {\sum_{f=u,d,s}}{\bar\psi_f}\gamma_\mu iD^\mu
             {\psi_f}-\sum_f m_{f}{\bar\psi_f}{\psi_f}
              +\sum_f \mu_f \gamma_0{\bar \psi_f}{\psi_f}\nonumber \\
      &+{\frac{g_S}{2}} {\sum_{a=0,\ldots,8}}[({\bar\psi} \lambda^a
        {\psi})^2+
            ({\bar\psi} i\gamma_5\lambda^a {\psi})^2]
       -{g_D} [det{\bar\psi_f}{P_L}{\psi_{f^\prime}}+det{\bar\psi_f}
            {P_R}{\psi_{f^\prime}}\nonumber \\
                 &-{\cal {U^\prime}}(\Phi[A],\bar \Phi[A],T)
\end{align}
where $f$ denotes the flavors $u$, $d$ or $s$, respectively.
The matrices $P_{L,R}=(1\pm \gamma_5)/2$ are  the
left-handed and right-handed chiral projectors, and the other terms
have their usual meaning, described in details in
Refs.~\cite{deb2009, deb2010a, deb2010b, deb2011, deb2013,deb2013}.
The gluon dynamics is described by the chiral point couplings between quarks (present in the NJL part) and a background gauge field representing the Polyakov Loop dynamics. 
The Polyakov loop potential can be expressed as,
\begin{equation}
\frac {{\cal {U^\prime}}(\Phi[A],\bar \Phi[A],T)} {T^4}= 
\frac  {{\cal U}(\Phi[A],\bar \Phi[A],T)}{ {T^4}}-
                                     \kappa \ln(J[\Phi,{\bar \Phi}])
\label {uprime}
\end{equation}
where $\cal {U(\phi)}$ is a Landau-Ginzburg type potential commensurate
with the Z(3) global symmetry. Here we choose a form given in
\cite{ratti},
\begin{equation}
\frac  {{\cal U}(\Phi, \bar \Phi, T)}{  {T^4}}=-\frac {{b_2}(T)}{ 2}
                 {\bar \Phi}\Phi-\frac {b_3}{ 6}(\Phi^3 + \bar \Phi^3)
                 +\frac {b_4}{  4}{(\bar\Phi \Phi)}^2,
\end{equation}
where ${b_2}(T)=a_0+{a_1}exp(-a2{\frac {T}{T_0}}){\frac {T_0}{T}}$,
$b_3$ and $b_4$ being constants. 
The parameters of the polyakov loop potential are typically determined by fitting a few
thermodynamic quantities such as pressure as a function of temperature which have been obtained a-priori from Lattice QCD computations. The set of values chosen have been represented below \cite{deb2009}
$ T_0 =175 MeV$, $ a_0= 6.75 $ , $ a_1= -9.0 $, $ a_2 = 0.25 $, $ b_3 = 0.805 $, $ b_4 = 7.555$, $ \kappa = 0.1$. The parameter values for the quark interaction part can be obtained by fitting the physical observables like pion mass, kaon mass, pion decay constant obtained from the experimental observation $ m_u=5.5 MeV $, $ m_s = 134.76 MeV$ , $ \Lambda = 631 MeV $, $ g_S \Lambda^2 = 3.67 $ , $ g_D \Lambda^5 = 9.33 $.


Multiple reflection expansion technique was first formulated by \cite{bloch1970} to determine the distribution of eigenvalues of the equation $\bigtriangleup \phi + E \phi=0$ for a finite volume $V$ but for sufficiently smooth surface with boundary condition 
$\partial \phi/ \partial n = \kappa \phi$. 
 For a spherical system $R_1 = R_2 = R$ and the density of states for the quarks in the fireball can be written as
\begin{equation}
\rho_{i,MRE}(p,m_i,R)= 1 + {\frac { 6\pi^2} {pR}} f_{i,S} + {\frac {12\pi^2}
                         {({pR})^2}} f_{i,C} 
\end{equation}  
where the surface contribution to the density of states is
\begin{equation}
f_{i,S} = - {\frac {1} {8\pi}} (1 - {\frac {2} {\pi}}arctan {\frac {p} {m_i}}) 
\end{equation}
and the curvature contribution is given by 
\begin{equation}
f_{i,C} = {\frac {1} {12\pi^2}} [1 - {\frac {3p} {2m_i}} ({\frac {\pi} {2}} -
           arctan {\frac {p}{m_i}})]
\end{equation}
The freeze-out curve $T(\mu_B)$ in the $T-\mu_B$ plane and the dependence of 
the baryon chemical potential on the center of mass energy in nucleus-nucleus 
collisions can be parametrized as demonstrated by 
\begin {equation}
T(\mu_B) = a - b\mu_B^2 - c\mu_B^4
\end {equation}
where $a = (0.166 \pm 0.002) $ $GeV$, $b = (0.139 \pm 0.016) $ ${ GeV^{-1}}$,  
$c = (0.053 \pm 0.021) $ $GeV^{-3} $ and
\begin {equation}
\mu_B (\sqrt s_{NN}) = d/{(1+ e\sqrt s_{NN})}
\end {equation}
with $d$, $e$ as given by Karsch et al. \cite{karsch-strange2011}.

The ratio of baryon ($\mu_B$) to strangeness chemical potential ($\mu_S$) on the freeze-out  curve shows a weak dependence on the collision energy
\begin{equation}
{\frac{\mu_S}{\mu_B}} \sim 0.164 + 0.018 \sqrt s_{NN}
\end{equation}

{\subsection{Results}}
The experimental results for volume independent cumulant ratios of net-proton,
net-kaon and net-charge number distribution have been obtained for all BES energies 
${\sqrt s_{NN}}= 7.7, 11.5, 14.5, 19.6, 27, 39, 62.4$ and $200 {GeV}$ for top central and peripheral collisions {\cite{STAR-proton2014, STAR-charge2014, STAR-kaon2018}}. We have presented our results for finite size PNJL model including the MRE formalism for six-quark interactions and compared them against the experimental data.

\begin{figure}[htb]
\centering
\includegraphics[scale=0.7]{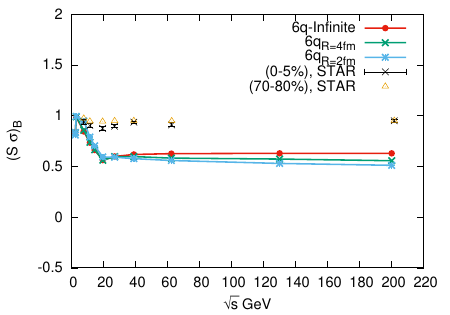}
\includegraphics[scale=0.7]{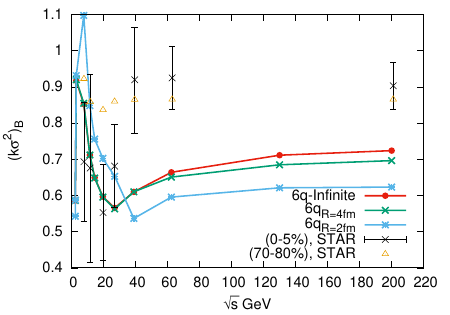}
\caption{(Color online) $S\sigma$ (left panel) and $k{\sigma}^2$ (right panel) of baryon fluctuation have been plotted with respect to collision centrality.} 
\label{c3c2B}
\end{figure}

 In this Figure (\ref{c3c2B}), we have plotted $S\sigma$ (which is the ratio of the third order to second order moment) and $k{\sigma}^2$ (ratio of fourth order to second order moment) of the baryon fluctuation with respect to collision energy for the PNJL model with infinite volume and also for the
finite size with $R=2fm$ and $R=4fm$. Here $S$ and $k$ represent the skewness and 
the kurtosis of higher order fluctuations, respectively. 
 
\begin{figure}[htb]
	\centering
	\includegraphics[scale=0.7]{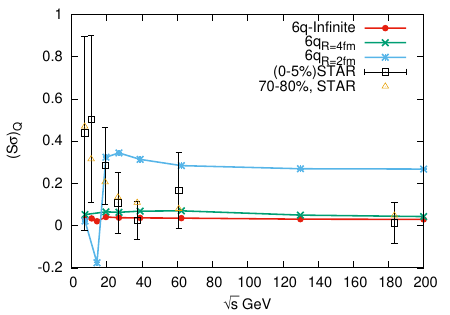}
	\includegraphics[scale=0.7]{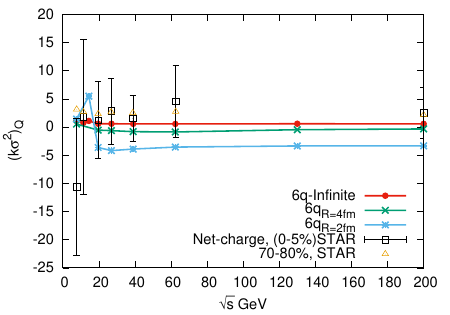}
	\caption{(Color online) $S\sigma$ (top panel) and $k{\sigma}^2$ (bottom panel)
		of charge fluctuation have been plotted with respect to collision centrality.}
	\label{c3c2Q}
\end{figure}

\begin{figure}[htb]
	\centering
	\includegraphics[scale=0.7]{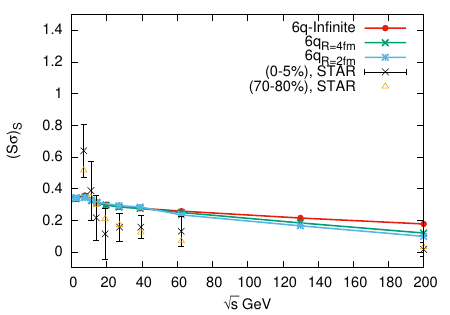}
	\includegraphics[scale=0.7]{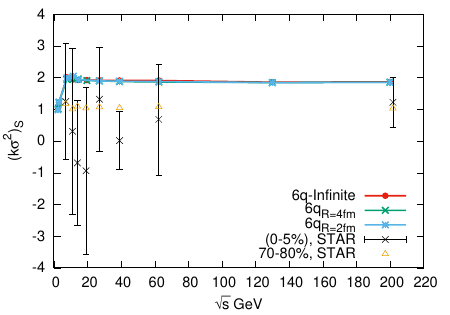}
	\caption{(Color online) $S\sigma$ (top panel) and $k{\sigma}^2$ (bottom panel)
		of strangeness fluctuation have been plotted with respect to collision
		centrality.}
	\label{c3c2S}
\end{figure}

Interestingly, the plot of the $S\sigma$ with collision energy (Figure \ref{c3c2B}) shows higher value of fluctuation around $\sqrt{s}= 11.5 GeV$. After $\sqrt{s} = 11.5 GeV$ the fluctuations decrease sharply; more so for the model data compared to the experimental data. Beyond $\sqrt{s} = 20 GeV$ the fluctuation becomes constant similar to the experimental value. Necessarily, the nature of fluctuation may indicate that the critical region lies below $\sqrt{s} = 20 GeV$. 
Similar nature has been also observed for the net-proton fluctuation in experiment. A comparison of the model (PNJL with MRE) data with the experimental data for $(0\%-5 \%)$ centrality 
and $(70\%-80 \%)$ centrality is also presented.
It may be remarked that the available 
experimental data is for net-proton fluctuations while the baryon fluctuation are calculated here. Meanwhile, the $k{\sigma}^2$ shows a 
sharp rise in fluctuation around $\sqrt{s} = 11.5 GeV$ and a minimum around 
$20-40 GeV$ for collision centrality. Interestingly, the initial values match quite well with the experimental value though for large collision energies (after 
$\sqrt{s}= 40 GeV$) they differ.
Figure \ref{c3c2Q} shows the $S\sigma$ and $k{\sigma}^2$ for charge 
fluctuation with respect to the collision centrality. As evident from the plot, $S\sigma$ attains a maxima around collision energy of $20-40 GeV$.
The value of fluctuation for infinite volume is significantly large for $R=2 fm$ than for $R=4 fm$.
In contrast, the experimental result shows almost a steady value of $k{\sigma}^2$ after initial turbulance at lower beam energy \cite{STAR-charge2014}. 
Figure ({\ref {c3c2S}}) shows the variation of strangeness fluctuations
( $C_3 / C_{2_S}$ and $C_4 / C_{2_S}$ ) with respect to
different collision energies for the PNJL model. In recent experimental data no 
significant deviation of the strangeness fluctuations has been found with respect to the Poisson expectation value within statistical and systematic
uncertainties for both the moments {\cite {STAR-kaon2018}}. $S\sigma$ for PNJL 
model with infinite volume and finite volume show similar behavior as 
the experimental results.

\begin{figure}[htb]
\centering
\includegraphics[scale=0.9]{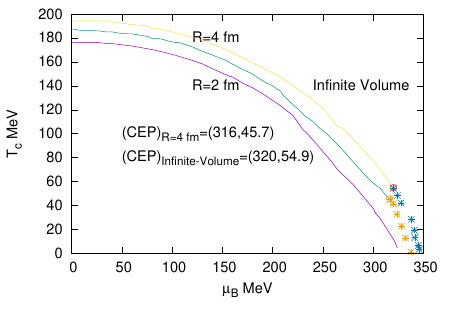}
\caption{(Color online)Phase diagram of the PNJL model for infinite and finite
volume system with $R = 4 fm$ and $R=2fm$. 
CEP for infinite
volume system is ($\mu_B, T_C = (320, 54.9)  MeV$) and for finite volume 
system with $R=4 fm$ is ($\mu_B, T_C = (316,45.7) MeV$).}
\label{tcmub}
\end{figure}

The phase diagram is one of the 
most interesting phenomenon pertaining to the strongly interacting matter. In the region where the temperature is less than $T_C$ and the chemical potential is greater than 
$\mu_B$, the chiral and deconfinement transitions are first order. The critical end point (CEP) separates the crossover transition from the first order phase transition. In the figure ({\ref{tcmub}}) the phase diagram for the PNJL model with infinite volume and 
finite volume with MRE approximation are shown. The position of CEP shifts to the  increasingly lower temperatures as the volume decreases and completely vanishes. 

\subsection{summary}
In summary, we have discussed the properties of net baryon, net charge and net strangeness fluctuations in nuclear matter within the PNJL model with MRE formalism. Model results suggest that as the temperature is increased at zero baryon chemical potential, there is a smooth croosover of the  order parameters for both chiral and deconfinement transitions in accordance with lattice QCD results.
An attempt has been also made to compare the model predicted values with experimental results. A good qualitative and quantitative 
similarity  between the reformulated experimental data and the model data has been observed. The phase diagram for the 6 quark PNJL model 
shows as the system size decreases, the position of the critical points shift towards the lower temperature and chemical potential until it vanishes completely at $R = 2 fm$. The study of various equilibrium thermodynamic measurements of the fluctuations using PNJL model would be helpful in determining the finite temperature finite density behavior of the hadronic sector.
\newcommand{\beas}{\begin{eqnarray}}
\newcommand{\eeas}{\end{eqnarray}}
\newcommand{\nno}{\nonumber}

\newcommand{\ep}{\epsilon}
\newcommand{\vk}{\vec k}
\newcommand{\vq}{\vec q}
\newcommand{\vp}{\vec p}
\newcommand{\bp}{\boldsymbol{p}}
\newcommand{\al}{\alpha}
\newcommand{\la}{\lambda}
\newcommand{\na}{\nabla}
\newcommand{\D}{\Delta}
\newcommand{\Ep}{\mathcal{E}}

\section{Fluid aspect and Wiedemann-Franz law violation in Nuclear and graphene systems}
\author{Ashutosh Dwibedi, Thandar Zaw Win, Subhalaxmi Nayak, Cho Win Aung, Sabyasachi Ghosh and Sesha Vempati }

	\bigskip
	
	\begin{abstract}
		The Wiedemann-Franz (WF) law, a hallmark of nonrelativistic electron transport, relates a Lorenz ratio between the thermal and electrical conductivities, which remains fixed in conventional metals, owing to the Fermi gas and Fermi liquid theory. Interestingly, quarks or hadrons in the matter produced in RHIC or LHC experiments do not follow this law. This high-energy nuclear physics-based many-body system (QCD phase diagram) is expected in the temperature and (baryon) chemical potential within the order of MeV-GeV scales. While a condensed matter physics-based many-body system is expected within meV-eV scales of temperature and Fermi energy (chemical potential). Though metals with Fermi energy from $2$ eV to $10$ eV, follow the WF law whereas graphene can reach the WF law violation domain by lowering its Fermi energy via doping. The interesting point is that towards a smaller chemical potential, quarks in RHIC/LHC matter and electrons in graphene both show fluid aspects with violation of WF law. The present study will broadly highlight and attempt to dig into the reason.
	\end{abstract}

\ccode{PACS numbers:}

	\subsection{Introduction}
	 	Hydrodynamics is the collective and coherent motion of the constituent particles of a system, which has been well practiced in mechanical engineering and other relevant branches for a long time. Extension of this traditional Non-Relativistic Hydrodynamics(NRHD)\cite{FD1} towards Relativistic Hydrodynamics(RHD)\cite{RHD} is eventually developed for the description of relativistic matter like quark-gluon plasma(QGP), produced in RHIC and LHC experiments. The matter produced in RHIC or LHC is expected to be a strongly coupled quark-gluon plasma(sQGP), which can be described by dissipative hydrodynamics\cite{PhysRevC.86.014902} with a shear viscosity to entropy density ratio close to its quantum lower bound or KSS bound($\frac{\eta}{s}\approx \frac{1}{4\pi}$)\cite{kovtun2005viscosity}. The shear viscosity expression\cite{gavin1985transport} in the framework of relaxation time approximation(RTA) based kinetic theory indicates that lower viscosity means shorter relaxation time and strongly coupled system. In the hydrodynamic regime, there will be violation of the Wiedemann-Franz(WF) law as recently observed in hot QCD matter \cite{sahoo2019wiedemann} and hadronic matter \cite{pradhan2023conductivity, singh2023effect}.\\
	 In the context of hydrodynamics, condensed matter physics never required the application of hydrodynamical theory until the recent observation of electron hydrodynamics(eHD) in graphene system\cite{sulpizio2019visualizing}. Graphene is a single layer of graphite in which the carbon atoms are arranged in a honeycomb lattice\cite{RevModPhys.83.837}.Hydrodynamic behavior is observed within the quasiparticles such as electrons\cite{sulpizio2019visualizing} and holes instead of diffusive behavior that is typically observed in metals. This theoretical framework of (eHD)\cite{sulpizio2019visualizing} in graphene system may be called graphene hydrodynamics(GHD) since it is neither NRHD nor RHD, as discussed in recent Refs\cite{TZW, Cho} and references therein. Interestingly, by lowering the values of Fermi energy or  chemical potential $\mu$ via controlling the doping, the graphene system will go from Fermi liquid (FL) $(\frac{\mu}{k_BT}>1)$ to Dirac fluid (DF) $(\frac{\mu}{k_BT}<1)$ domain and show WF law violation. The present draft has attempted to explore this interesting similarity between QGP and graphene systems in the context of fluid aspect and WF law violation. After addressing a brief formalism part in Sec.~(\ref{sec:sec2}), we have gone through the discussion and conclusion on the present topic in Sec.~(\ref{sec:sec3}).

	 \subsection{Formalism}
	 \label{sec:sec2}
			For any fluid description of charged particles, we start with the energy-momentum tensor and charge current respectively, as,
	\beas
	T^{\mu\nu} &=& T_0^{\mu\nu} + T_D^{\mu\nu},~J^{\mu}=J^{\mu}_{0}+J^{\mu}_{D}~,
	\eeas
	where $T_0^{\mu\nu}$ and $J_{0}^{\mu}$ represent the ideal part and $T_D^{\mu\nu}$ and $J^{\mu}_{D}$ account for the dissipative part. One usually decompose the ideal part of $T^{\mu\nu}$ and $J^{\mu}$  in terms of the building blocks - energy density $\ep$, pressure $P$, number density $n$, fluid four-velocity $u^\mu$ and metric tensor $g^{\mu \nu}$. Keeping in mind the dispersion relation of graphene electrons $E=pv_{F}$ that differs from the ultra-relativistic dispersion $E=pc$, one can form the ideal part of $T^{\mu\nu}$ and $J^{\mu}$ in the DF domain of graphene as:
	\beas
	T_0^{\mu\nu} &=&  \ep \frac{u^\mu u^\nu}{v_F^2} - P\bigg(g^{\mu\nu} -\frac{u^\mu u^\nu}{v_F^2}\bigg)~,
	\nno\\
	J_0^\mu &=& -ne \frac{u^\mu}{v_F}~,
	\eeas
	where $v_{F}$ is the Fermi velocity of electrons in graphene, which may be regarded as the ultimate speed in the graphene world. In GHD, one can write the fluid four-velocity as $u^{\mu}=\gamma_F (v_F, \vec{u})$ with $\gamma_F= 1/\sqrt{1-u^2/v_F^2}$~\cite{TZW,Cho}. Now let us come back to the the disspative part of the $T^{\mu\nu}$ and $J^{\mu}$. One part of $T_D^{\mu\nu}$ will be proportional to the velocity gradients, whose respective proportional constants will be shear and bulk viscosities. Its remaining part will be connected with heat current $q^\mu$, which is proportional to $\nabla^{\mu}(\frac{\mu}{T})$ with thermal conductivity $\kappa$ being the proportionality constant. Similarly, $J^\mu_D$ will also be proportional to the electric field $\tilde{E}^{\mu}$ with proportional constant - electrical conductivity $\sigma$. They can be written in mathematical form as:
	
	\beas
	&&  q^{\mu}= \kappa\nabla^{\mu}T,\label{A3}\\
	&& J^{\mu}_{D}=\sigma \tilde{E}^{\mu}. \label{A2}
	\eeas
 Following the arguments that we have used for $T^{\mu\nu}_{0}$ and $J^{\mu}_{0}$, Eq.~(\ref{A3}) and (\ref{A2}) can be written for graphene in the DF domain by replacing speed of light $c$ with Fermi velocity $v_{F}$ of electrons. For a graphene sheet in the DF regime, one can calculate the thermodynamic parameters- $n$, $\epsilon$, $P$ from the equilibrium Fermi-Dirac distribution function $f_{0}=1/[e^{(u^{\mu}p_{\mu}-\mu)/k_BT}]$ and the transport coefficients - $\sigma$ and $\kappa$ from the off-equilibrium part of the distribution $\delta f$ \cite{Cho}. These novel hydrodynamic transport coefficients and thermodynamic parameters differ from the usual parameters of the metallic transport regime. 
	We know that the WF law is valid in metals\cite{ashcroft2022solid}, i.e.,
	\beas
	&&\frac{\kappa}{\sigma T}=\frac{\pi^{2}}{3}\frac{k_{B}^{2}}{e^{2}}.\label{A4}
	\eeas
	In contrast, using the hydrodynamic aspects of graphene in the DF domain, one obtains,
	\beas
	&&\frac{\kappa}{\sigma T}= 9\left(\frac{f_{3}(A)+f_{3}(A^{-1})}{f_{2}(A)-f_{2}(A^{-1})}\right)^{2} \frac{k_{B}^{2}}{e^{2}}\label{A5},
	\eeas
	where $f_{j}$ is the usual Fermi integrals of order $j$, $f_{j}(A)=\frac{1}{\Gamma(j)}\int \frac{dx~ x^{j-1}}{e^{x}A^{-1}+1}$ with $A=e^{\mu/T}$. The Eq.~(\ref{A5}) suggests a violation of WF law in graphene in the DF domain\cite{Sci_16_Dirac,win2024wied}. A similar kind of expression for the ratio of thermal to electrical conductivity can also be expected for QGP\cite{jaiswal2015relativistic}. 
	 	
\subsection{Discussions and Conclusion}
\label{sec:sec3}
\begin{figure}
	\centering 
	\includegraphics[scale=0.30]{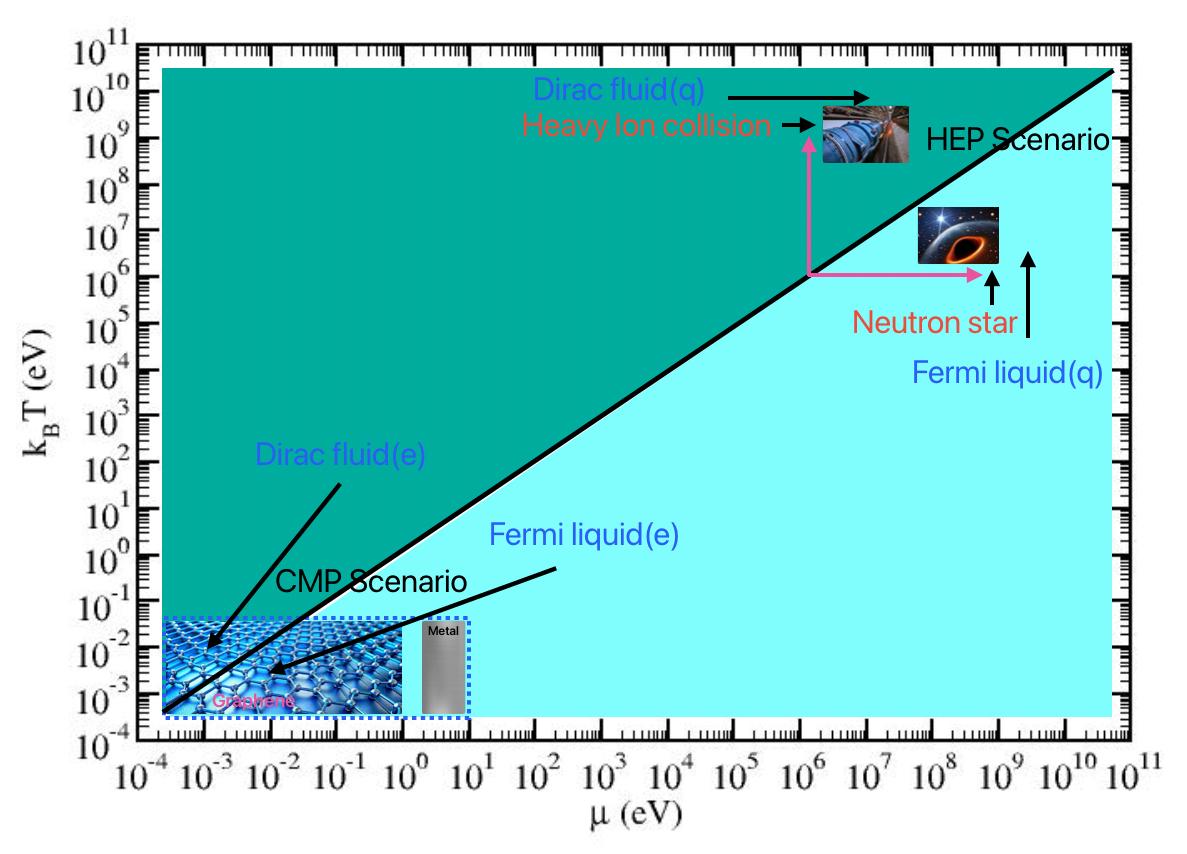}
	\caption{The Condensed Matter Physics (CMP) and High Energy Physics (HEP) locations and their Dirac Fluid (DF) and Fermi Liquid (FL) domains in $k_BT$-$\mu$ plane.} 
	\label{fig:mu_T}
\end{figure}

			Here we will briefly discuss the different regions of the $\mu$-$T$ plane, where hydrodynamics is necessary to describe the dynamics of the many-body system ranging from the Condensed Matter Physics (CMP) to the High Energy Physics (HEP) domains. Let us start from the low temperature ($k_{B}T\sim$ meVs) and low chemical potential ($\mu\sim$ meVs) region. We can observe from Fig.~(\ref{fig:mu_T}) that this region corresponds to graphene in the DF domain. In this DF region, the free electrons of graphene may be described by the usual relativistic hydrodynamics with $c$ replaced by $v_{F}$ as described in Sec.(\ref{sec:sec2}). As one moves further along the $\mu$ axis fixing $T$ to the region where $\mu \gg k_{B}T$, the hydrodynamic behavior of electrons in graphene is lost, and the electron dynamics are governed by the FL theory. The usual metals at room temperature ($k_{B}T\sim$ 25 meV) lie further right in the $\mu$ axis ($\mu\sim$ 2-10 eV) where the hydrodynamic behaviors are suppressed, and electrons obey FL theory. 
Now, we will shift our attention to the HEP systems where one has temperature and chemical potential in the order of MeVs. Interestingly, here also, one can observe the fluid aspect in the region $\frac{\mu}{k_{B}T}\ll1$ (similar to the DF domain in the graphene system). When we go to $\frac{\mu}{k_{B}T}\gg1$ domain, one can use the degenerate gas formalism of quarks and hadrons, expected in the neutron star environment (shown in the inset at the top right corner of Fig.~(\ref{fig:mu_T})). 

For both QGP and graphene cases, fluid properties are observed in $\frac{\mu}{k_{B}T}\ll1$ domain, along with the WF law violation. One may link the fluid aspect and WF law violation as a concluding remark, but certainly, more future research works in this direction are required.

\section{Revisiting shear stress evolution: Non-resistive Magnetohydrodynamics with momentum-dependent relaxation time}

\author{SUNNY KUMAR SINGH, MANU KURIAN, VINOD CHANDRA}

\bigskip

\begin{abstract}
Strong magnetic fields are expected to exist in the early stages of heavy ion
collisions and there is also an increasing evidence that the energy dependence of the
cross-sections can strongly affect the dynamics of a system even at a qualitative level.
This led us to the current study where we developed second-order non-resistive
relativistic viscous magnetohydrodynamics (MHD) derived from kinetic theory using an
extended relaxation time approximation (ERTA), where the relaxation time of the usual
relaxation time approximation (RTA) is modified to depend on the momentum of the
colliding particles. A Chapman-Enskog-like gradient expansion of the Boltzmann
equation is employed for a charge-conserved, conformal system, incorporating a
momentum-dependent relaxation time. The resulting evolution equation for the shear
stress tensor highlights significant modifications in the coupling with the dissipative
charge current and magnetic field. The results were compared with the recently
published exact analytical solutions for the scalar theory where the relaxation time
is directly proportional to momentum. Our approximated results show a very close
agreement with the exact results. This can lead us to transport coefficient calculations
for various theories for which exact results are not yet known.
\end{abstract}


\ccode{PACS numbers:}


\subsection{Introduction}

High-energy heavy-ion collisions at RHIC and LHC produce a hot, dense quark-gluon plasma (QGP), whose evolution is effectively modeled by relativistic hydrodynamics, including dissipative effects \cite{1s}. Early collisions generate strong magnetic fields that decay but can still influence QGP dynamics and electromagnetic fields \cite{2s}. Recent experiments suggest magnetic fields significantly affect the QGP's transport and thermodynamic properties, including phenomena like the chiral magnetic effect and magnetic catalysis \cite{3s}. This work investigates the magnetic field’s impact by deriving second-order viscous magnetohydrodynamics (MHD) equations using an extended relaxation time approximation, where relaxation time varies with momentum. This framework enables a detailed study of shear stress evolution, highlighting modifications from momentum-dependent relaxation time and magnetic fields, crucial for accurately modeling QGP dynamics.

\subsection{Second-order shear stress evolution}

\begin{figure}
\centerline{\includegraphics[width=7cm]{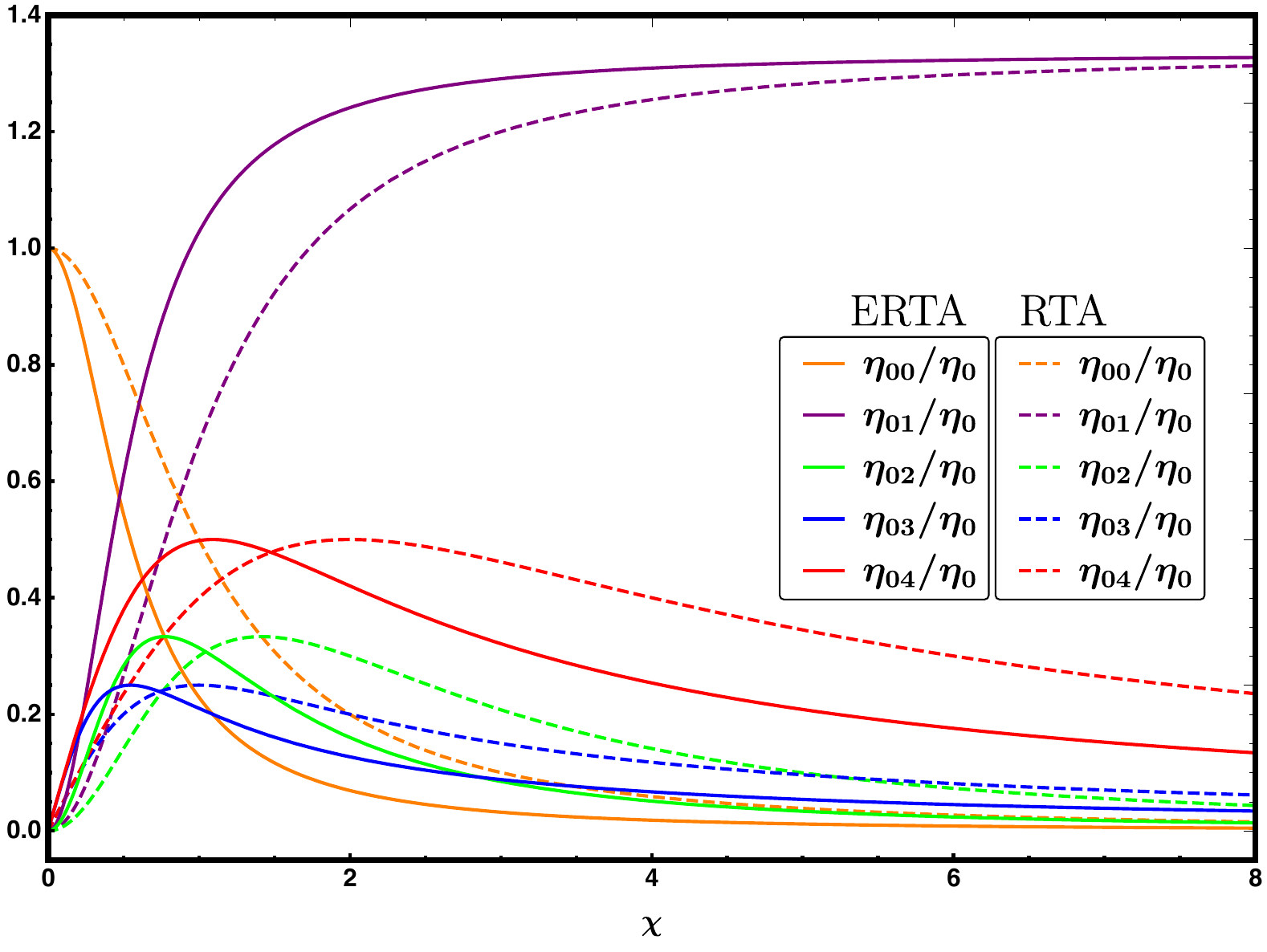}}
\caption{The dashed lines are the ERTA results for $\ell=0.5$ whereas the solid lines are the results from RTA. The x axis is given by $\chi= \frac{q B \tau_0(X)}{T}$}
\label{fig_sunny}
\end{figure}

The energy-momentum tensor of a conformal quark-gluon plasma (QGP) can be written as,
\begin{align}
&T^{\mu\nu}=\int{dP\,p^{\mu}p^{\nu}(f+\bar{f})}=\varepsilon u^{\mu} u^{\nu} - P \Delta^{\mu\nu} + \pi^{\mu\nu},\label{1.2}
\end{align}
where $\epsilon$ and $P$ are the energy density and pressure of the fluid. Near-equilibrium distribution function $f$ of the quarks and gluons are obtained from the underlying microscopic theory-kinetic theory. In order-by-order expansion, we have 
\begin{align}
     f=f_0+ \delta f_{(1)}+ \delta f_{(2)} + \cdots , 
\end{align}
where $\delta f_{(i)}$ with $(i=1, 2, 3, ...)$ is the  gradient correction of the distribution function to the $i^{\text{th}}$ order with $f_0$ being the equilibrium distribution. We can use the deviation from equilibrium given by $\delta f=f-f_0$ to derive the evolution equation for $\pi^{\mu\nu}$ as,
\begin{align}
    \pi^{\mu\nu}= \Delta^{\mu\nu}_{\alpha\beta}\int \mathrm{dP} p^\alpha p^\beta (\delta f+\delta \bar{f}),
\end{align}
where $\delta f$ and $\delta \bar{f}$ are for particles and anti-particles respectively. In the present study, we used the newly developed Extended Relaxation Time Approximation (ERTA) framework~\cite{4s} to obtain the non-equilibrium distribution. The Boltzmann equation in the presence of a magnetic field can be written as,
\begin{equation}
    p^\mu \partial_\mu f - qB^{\sigma \nu} p_\nu \frac{\partial f}{\partial p^\sigma} = -\frac{(u \cdot p)}{\tau_R(x,p)} (f-f^*_0).
\end{equation}
where $f_0=e^{-\beta(u\cdot p)+\alpha}$ and $f^*_0=e^{-\beta^*(u^*\cdot p)+\alpha^*}$ with $\beta^*=\frac{1}{T^*}$ and $\alpha^*=\frac{\mu^*}{T^*}$. 
Defining  $T^*=T+\delta T$, $\mu^*=\mu+\delta \mu$, and $u_\mu^*=u_\mu+\delta u_\mu$, these necessary counter-terms can be deduced by requiring the matching conditions $n=n_0$ and $\epsilon=\epsilon_0$ to be valid in the Landau frame where the fluid velocity is given by $u_\nu T^{\mu\nu}=\epsilon u^\mu$. Under the assumption of $E^\mu = 0$ (non-resistive case), the electromagnetic field tensor becomes,
\begin{equation}\label{3.4}
    F^{\mu\nu} \rightarrow B^{\mu\nu} = \epsilon^{\mu\nu\alpha\beta}u_\alpha B_\beta = -B b^{\mu\nu},
\end{equation}
where $B^\mu B_\mu = -B^2 $ and $b^\mu=\frac{B^\mu}{B}$ with $b^\mu u_\mu=0$ and $b^\mu b_\mu=-1$. With the relaxation time parameterized as:
\begin{align}\label{m2.38}
    \tau_R(x,p)=\tau_0(x) \Big(\frac{u\cdot p}{T}\Big)^\ell, && \text{where,}\,\,\, \tau_0(x)=\bar{\kappa}/T,
\end{align}
we obtain the evolution equation for $\pi^{\mu\nu}$ for a non-resistive fluid in the presence of magnetic field~\cite{5s} as,
\begin{align}\label{mag}
    \dot{\pi}^{\langle\mu\nu\rangle} &+ \frac{\pi^{\mu\nu}}{\tau_\pi} =  2\beta_\pi \sigma^{\mu\nu} - \frac{4}{3} \pi^{\mu\nu}\theta +2\pi_\gamma^{\langle\mu}\omega^{\nu\rangle\gamma} - \tau_{\pi\pi}\pi_\gamma^{\langle\mu}\sigma^{\nu\rangle\gamma}\nonumber \\
    &-\tau_{\pi n} n^{\langle \mu}\dot{u}^{\nu\rangle}+\lambda_{\pi n} n^{\langle \mu}\nabla^{\nu\rangle}\alpha+ l_{\pi n} \nabla^{\langle\mu}n^{\nu\rangle}+\delta_{\pi B}\Delta^{\mu\nu}_{\eta \beta}qBb^{\gamma\eta}g^{\beta\rho}\pi_{\gamma\rho}\nonumber\\
    &-qB\tau_{\pi n B}\dot{u}^{\langle \mu}b^{\nu \rangle \sigma}n_\sigma-qB\lambda_{\pi n B} n_\sigma b^{\sigma \langle \mu}\nabla^{\nu \rangle}\alpha - q\tau_0\delta_{\pi n B} \nabla^{\langle \mu}\left(B^{\nu\rangle\sigma}n_\sigma\right).
\end{align}
We refer to Ref.[5] for the detailed derivation and definition of all the coefficients and associated tensorial structures. These modified transport coefficients within the ERTA setup depend on the momentum dependence parameter $\ell$. Magnetic field effects are entering through the last four terms of the Eq.(\ref{mag}). The first order limit of the above equation leads to the Navier-Stoke's limit but the presence of magnetic fields introduces anisotropy in the system which leads to the first order shear transport coefficient splitting into five components. As depicted in figure \ref{fig_sunny}, the ERTA estimation of magnetic field dependent shear coefficients significantly modifies the conventional RTA approach. 

Further, we compare the ERTA results with the exact solutions $\lambda \phi^4$ theory in the case of vanishing magnetic field. The comparison of the our results with those found by solving Boltzman equation exactly for the case of a self interacting $\lambda \phi^4$ theory \cite{6s,7s} which corresponds to $\ell=1$ is shown in table 1. We see that there is a significant agreement between the values of the transport coefficients from our formula and from the ones derived exactly.

\begin{table}[h]
\tbl{Comparison of ERTA results for various transport coefficients with exact calculations from $\lambda \phi^4$ theory}
{\begin{tabular}{@{}cccc@{}} \toprule
Coefficients & RTA results & ERTA results &
$\lambda \phi^4$ results \\
& ($\ell =0$) & ($\ell=0$) & (exact) \\ \colrule
$\tau_\pi$ \hphantom{00} & \hphantom{0} $\tau_c$ & \hphantom{0} $\frac{24 d_g}{g n_0 \beta^2}$ & $\frac{72}{g n_0 \beta^2}$ \\
$\eta$ \hphantom{00} & \hphantom{0} $\frac{4P\tau_c}{5}$ & \hphantom{0} $\frac{16 d_g }{g \beta^3}$ & $\frac{48}{g \beta^3}$ \\
$\kappa$ \hphantom{00} & \hphantom{0}$\frac{n_0 \tau_c}{12}$ & \hphantom{0} $\frac{d_g}{g\beta^2}$ \hphantom{0} & $\frac{3}{g \beta^2}$ \hphantom{0} \\
$\delta_{\pi\pi}$ \hphantom{00} & \hphantom{0}$\frac{4}{3}$ & \hphantom{0} $\frac{4}{3}$ \hphantom{0} & $\frac{4}{3}$ \hphantom{0}\\
$\tau_{\pi\pi}$ \hphantom{00} & \hphantom{0}$\frac{10}{7}$ & \hphantom{0} $2$ \hphantom{0} & $2$ \hphantom{0} \\
$\l_{\pi n}$ \hphantom{00} & \hphantom{0} 0 & \hphantom{0} $-\frac{4}{3\beta}$ \hphantom{0} & $-\frac{4}{3\beta}$ \hphantom{0}\\
$\tau_{\pi n}$ \hphantom{00} & \hphantom{0} 0 & \hphantom{0} $-\frac{16}{3\beta}$ \hphantom{0} & $-\frac{16}{3\beta}$ \hphantom{0}\\
$\lambda_{\pi n}$ \hphantom{00} & \hphantom{0}$0$ & \hphantom{0} $\frac{2}{3\beta}$ \hphantom{0} & $\frac{5}{6 \beta}$ \hphantom{0}\\
\botrule
\end{tabular}}
\begin{tabnote}
We see that our formulas for the transport coefficients reduces correctly to the ones for RTA ($\ell=0$) and they're also close to the ones calculated from the exact theory \cite{7s}.
\end{tabnote}
\end{table}

We verified the consistency of the ERTA results with other parallel approaches and with the exact Boltzmann
predictions

\newcommand{\pvec}{{\boldsymbol p}}
\newcommand{\kms}{km\,s$^{-1}$}
\newcommand{\msun}{$M_\odot}

\section{Longitudinal Spin Polarization in a Thermal Model with Dissipation}

\author{Soham Banerjee}

\bigskip

\begin{abstract}
In this work, we study the longitudinal spin polarization of $\Lambda$ hyperons produced in relativistic heavy-ion collisions. We use a relativistic kinetic theory framework that incorporates spin degrees of freedom treated classically, combined with the freeze-out parametrization from previous investigations. This approach allows us to include dissipative corrections (arising from thermal shear and gradients of thermal vorticity) in the Pauli-Lubanski vector, which determines the spin polarization and can be directly compared with experimental data. As in similar earlier studies, achieving a successful description of the data requires additional assumptions—in our case, the use of projected thermal vorticity and an appropriately adjusted spin relaxation time ($\tau_s$). Our analysis indicates that $	\tau_s \sim 5$ fm/$c$, which is comparable with other estimates.
\end{abstract}




\subsection{Introduction:}

One of the recent motivations for studying non-central heavy-ion collisions is to study the spin polarization effects of various produced particles, for example, $\Lambda$ hyperons~\cite{STAR:2017ckg, STAR:2018gyt, STAR:2019erd} and vector mesons~\cite{ALICE:2019aid}. In particular, there is a lot of interest in the measurement of the longitudinal spin polarization of $\Lambda$'s produced in Au-Au collisions at the beam energy of $\sqrt{s_{\rm NN}} = 200$~GeV \cite{STAR:2019erd}, as well as in developing related theoretical models ~\cite {amaresh, Florkowski:2018ahw,Hattori:2019lfp, Bhadury:2020cop, Wagner:2022amr, Weickgenannt:2023btk}. In this case, the data indicates a quadrupole structure of the longitudinal polarization in the transverse momentum plane, which disagrees in sign with most theoretical predictions (the so-called sign problem)~\cite{Becattini:2017gcx}.

In the paper \cite{soham}, we continue some of the earlier studies of the longitudinal polarization \cite{Florkowski:2019voj, Florkowski:2021xvy} that use thermal model to parametrize freeze-out conditions in Au-Au collisions. A novel feature of the current analysis is that we incorporate dissipative corrections to the Pauli-Lubanski (PL)vector. They are determined within the framework of kinetic theory, which treats spin degrees of freedom classically~\cite{Florkowski:2018fap}.

In polarization studies, the spin polarization tensor $\omega_{\mu\nu}$ is commonly taken as the thermal vorticity \begin{equation} \varpi_{\mu\nu} = -\frac{1}{2}(\partial_\mu \beta_\nu - \partial_\nu \beta_\mu), \end{equation} where $\beta^\mu = u^\mu / T$ is the ratio of flow velocity to temperature. This yields the Pauli-Lubanski vector involving the spin equilibrium distribution function. The kinetic theory approach allows corrections to equilibrium distributions, obtained via the relaxation time approximation (RTA)\cite{Bhadury:2020puc, Bhadury:2020cop, Bhadury:2022ulr}, naturally leading to contributions from thermal shear \begin{equation} \xi_{\mu\nu} = \frac{1}{2}(\partial_\mu \beta_\nu + \partial_\nu \beta_\mu) \end{equation} and gradients of thermal vorticity.

We use projected thermal vorticity, setting $\varpi_{0i} = 0$, consistent with a non-relativistic treatment gives the correct sign of longitudinal spin polarization's quadrupole structure. We show ~\cite{soham} that including dissipative corrections and fitting the spin relaxation time ($\tau_s \sim 5$ fm/$c$) resolves these discrepancies.

\subsection{Spin observables at freeze-out}
The mean spin polarization of $\Lambda$ hyperons is found from the Pauli-Lubanski (PL) vector:
\begin{equation}
    E_p \frac{d \Delta \Pi_\alpha (x,p)}{d^3p} = -\frac{1}{2} \epsilon_{\alpha \mu \nu \beta} \Delta \Sigma_\lambda E_p \frac{dS^{\lambda,\mu\nu}}{d^3p} \frac{p^\beta}{m},
    \label{eq:PL1}
\end{equation}
where $p^\mu = (p^0, \pvec)$ is the $\Lambda$ four-momentum with $p^0 = E_p = \sqrt{m^2 + \pvec^2}$, $m = 1.116$~GeV is the $\Lambda$ mass, $\Delta \Sigma$ is the infinitesimal freeze-out element, and $S^{\lambda,\mu\nu}$ is the spin tensor.

Integrating Eq.~(\ref{eq:PL1}) gives the momentum density of the PL vector as shown in ~\cite{soham}:
\begin{align}
    E_p \frac{d \Pi_\tau (p)}{d^3p} &= - \frac{1}{2} \frac{\cosh \xi}{(2\pi)^3 m}  \int \Delta \Sigma \cdot p\, e^{-\beta \cdot p}\, \epsilon_{\tau \mu \nu \beta } p^\beta  \times \left[ \left( 1 + \chi \right) \omega^{\mu \nu }  - \frac{\tau_s}{u \cdot p} \; p \cdot \partial\, \omega^{\mu \nu } \right]
    \label{PL-KT}
\end{align}
where $\xi = \mu_B/T$, $\omega^{\mu\nu}$ is the spin polarization tensor, $\tau_s$ is the spin relaxation time, and $\chi = [\tau_s / (u \cdot p)] \, p^\rho p^\sigma \xi_{\sigma \rho}$. In this framework, thermal shear contributes only when coupled with $\omega^{\mu\nu}$, as spin integrals with odd $s^{\mu\nu}$ vanish.

For the spin polarization per particle, we introduce the number density:
\begin{equation}
    E_p \frac{d N(p)}{d^3p} = \frac{4 \cosh\xi}{(2\pi)^3} \int \Delta \Sigma \cdot p\, e^{-\beta \cdot p}.
\end{equation}
The quantity comparable to experimental data is the ratio:
\begin{equation}
    \langle P(\phi_p)\rangle = \frac{\int p_T dp_T E_p \frac{d \Pi^z (p)}{d^3p}}{\int d\phi_p p_T dp_T E_p \frac{d N(p)}{d^3p}}.
\end{equation}
In our model, thermal vorticity serves as a proxy for the spin polarization tensor ($\omega^{\mu\nu} \to \varpi^{\mu\nu}$).

\subsection{Results and Discussion}
We present numerical results for the $p_T$-integrated longitudinal component of the mean Pauli-Lubanski four-vector, characterizing the spin polarization of $\Lambda$ and $\overline{\Lambda}$ hyperons. The model parameters are listed in ~\cite{Florkowski:2019voj}, fitted to PHENIX data at $\sqrt{s_{\rm NN}} = 130$ GeV for three centrality classes: $0$--$15\%$, $15$--$30\%$, and $30$--$60\%$, with freeze-out temperature $T_{\rm f} = 0.165$ GeV.

Focusing on longitudinal polarization, we evaluate $\Pi_z(p_x, p_y)$ without a boost, since it is invariant under transverse boosts. We consider the central rapidity region ($y_p = 0$) and integrate transverse momentum ($p_T$) over 0--3 GeV.

\begin{figure}[t]
    \centering
    \begin{minipage}[b]{0.49\linewidth}
        \centering
        \includegraphics[width=\linewidth]{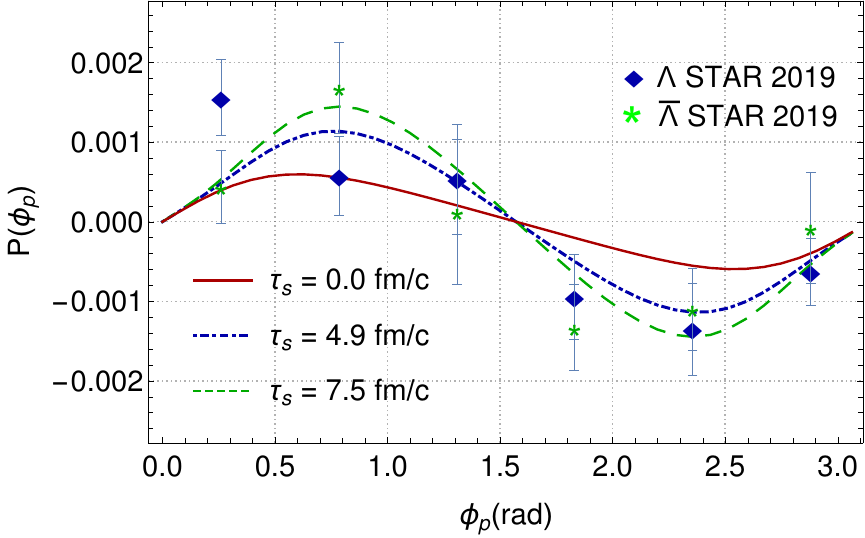}
        \caption{(Color online) Longitudinal polarization as a function of azimuthal angle $\phi_p$ for 30--60\% Au-Au collisions at $\sqrt{s_{\rm NN}} = 200$ GeV.}
        \label{fig_mom2}
    \end{minipage}%
    \hfill
    \begin{minipage}[b]{0.49\linewidth}
        \centering
        \includegraphics[width=\linewidth]{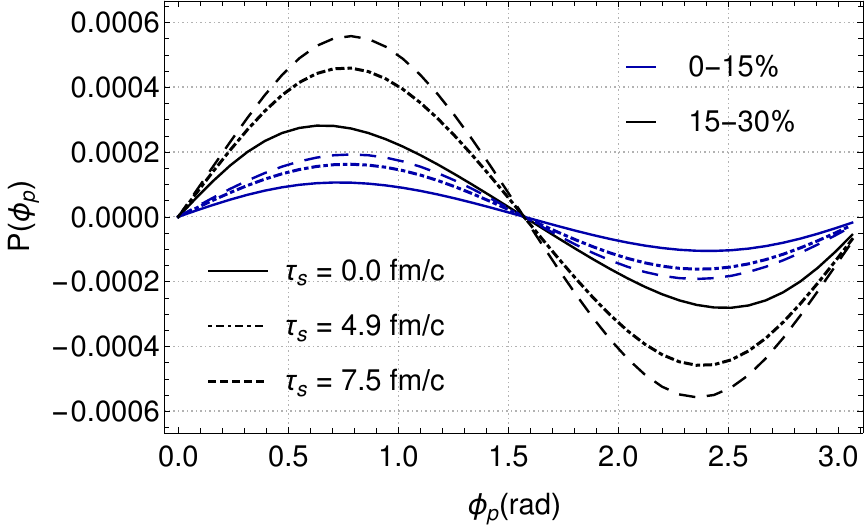}
        \caption{(Color online) Predictions for Au-Au collisions at $\sqrt{s_{\rm NN}} = 200$~GeV for $0$--$15\%$ and $15$--$30\%$ centralities.}
        \label{fig_mom1}
    \end{minipage}
    \caption{Comparison of theoretical predictions with experimental data for Au-Au collisions at $\sqrt{s_{\rm NN}} = 200$ GeV.}
    \label{fig:combined}
\end{figure}

The effect of dissipative corrections depends on the spin relaxation time $\tau_s$, which increases longitudinal polarization. A chi-squared test suggests that $\tau_s = 7.5$ fm/$c$ fits $\overline{\Lambda}$ data best, while $\tau_s = 4.9$ fm/$c$ fits $\Lambda$ data.

In Fig.~\ref{fig_mom2}, we show the longitudinal polarization as a function of azimuthal angle $\phi_p$ for 30--60\% centrality at $\sqrt{s_{\rm NN}} = 200$ GeV, compared with experimental data~\cite{STAR:2019erd}. Conversion from $\langle\cos\theta^*_p\rangle$ to helicity (PH) uses $\alpha_H = 0.732$~\cite{ParticleDataGroup:2020ssz}. The solid line shows the result without dissipative corrections, while dashed and dashed-dotted lines represent $\tau_s=7.5$ and $\tau_s=4.9$ fm/$c$, respectively. Fig.~\ref{fig_mom1} shows predictions for other centrality classes for the same system.

In this study, we investigated the longitudinal spin polarization of $\Lambda$ hyperons in relativistic heavy-ion collisions using a kinetic theory framework with classical spin treatment and freeze-out parametrization. This allowed us to include dissipative corrections from thermal shear and gradients of thermal vorticity in the Pauli-Lubanski vector, enabling direct comparison with experimental data.

We found that additional assumptions, such as using projected thermal vorticity, were needed to match the data. By fitting the spin relaxation time, we estimated it to be around $5$ fm/$c$. Our results indicate that dissipative corrections are crucial to explaining the observed longitudinal polarization.

\section{Mesonic screening masses within Gribov-quantized QCD framework}
\author{Sumit}

\bigskip

\begin{abstract}
The screening masses of mesons serve as a gauge invariant and definite order parameter of chiral symmetry restoration. 
We have studied and calculated the spatial correlation lengths of various mesonic observables using the non-perturbative Gribov resummation approach, both for quenched QCD and $(2+1)$ flavor QCD. This study draws on analogies to NRQCD effective theory, which is commonly used to analyze heavy quarkonia at zero temperature.
\end{abstract}

\subsection{Introduction}\label{Int}	
The correlation functions in temporal directions are utilized to estimate the ``real-time'' features of the plasma, such as particle production rate through spectral functions~\cite{McLerran:1984ay} while correlation functions in spatial directions give info about the length scale of the thermal fluctuation correlations. Infrared (IR) issues presence in finite temperature QCD leads to dimensionally reduced effective theory, which improves the perturbation theory and enables one to study the non-perturbative methods via a three-dimensional theory~\cite{Linde:1980tss, Appelquist:1981vg}. 
The class of observables that are more sensitive and closely related to IR physics is the correlation lengths of the mesonic operators, which are gauge-invariant quark bilinears constructed from the light quark flavors~\cite{Detar:1987kae}. The quark bilinears' two-point spatial correlation functions can determine the correlation lengths that are dominated by the screening masses at large distances, defined as the inverse of the correlation lengths. \\
The next-to-leading (NLO) correction for mesonic screening masses has been calculated using an effective theory in ref.~\cite{Laine:2003bd} and through the recent lattice QCD results presented in ref.~\cite{Bazavov:2019www} it has been found that the perturbative result of meson screening masses is comparable with the lattice results only in the high-temperature limit. Thus, no such analytic calculation in the literature can explain the lattice data for  $T \leq 2$ GeV in the low-temperature regime. We have tried to overcome this gap by using a non-perturbative resummation scheme offered by Gribov quantization, which is one of the ways to handle the infrared region of QCD~\cite{Gribov:1977wm}.   
\subsection{Setup}\label{section_2}	
The Euclidean quark lagrangian at finite temperature QCD is given by
\begin{equation}\label{quark_lag.}
	\mathcal{L}_{\mathrm{E}}^{\mathrm{Q}}=\bar{\psi}\left(\gamma_\mu \mathcal{D}_\mu+M\right) \psi,
\end{equation}
The operators among which the correlation is considered are of the following form   
\begin{equation}\label{opera_form}
	\mathfrak{O}^{a} = \bar{\psi} f^{a}\Gamma\psi,
\end{equation}
where $\Gamma$ determines different channels and $f^{a}$ provides flavor basis. 
We will consider the correlators in position space, having the following structure 
\begin{equation}\label{Corr_in_position}
	\mathcal{P}_{\mathbf{x}}\left[\mathfrak{O}^a, \mathfrak{O}^b\right] \equiv \int_0^{1 / T} \mathrm{~d} \tau\left\langle \mathfrak{O}^a(\tau, \mathbf{x}) \mathfrak{O}^b(0,0)\right\rangle
\end{equation}
Using rotational invariance and doing the average over the $x_{1} x_{2} $ surface, leads 
\begin{eqnarray}\label{corre_in_z}
	\mathcal{P}_z\left[\mathfrak{O}^a, \mathfrak{O}^b\right]&=&\int \mathrm{d}^2 \mathbf{x}_{\perp} \mathcal{P}_{\left(\mathbf{x}_{\perp}, z\right)}\left[\mathfrak{O}^a, \mathfrak{O}^b\right]
\end{eqnarray}
\begin{figure}
	\centering
	{\includegraphics[scale=.45]{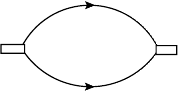}}
	\quad \quad \quad
	{\includegraphics[scale=.45]{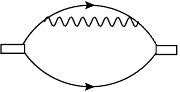}}
	\quad \quad \quad
	{\includegraphics[scale=.45]{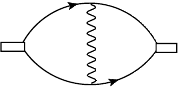}}
	\caption{The graphs which contribute to meson correlation function: (left) free theory correlator (middle) quark self-energy graph (right) interaction of quark and antiquark.}%
	\label{fig:1}%
\end{figure}
In the high-temperature limit, the free theory correlator shown in figure~(\ref{fig1(a)}) has a corresponding function $A_{3\mathrm{d}}\left(2 p_n\right)$ which is given by
\begin{eqnarray}\label{A3d_func}
	A_{3\mathrm{d}}\left(2 p_n\right) 
	&=&\frac{i}{8 \pi q} \ln \frac{2 p_n-i q}{2 p_n+i q}.
\end{eqnarray}
The singularity of the above function appears at $2p_{n}$. Thus, the correlator dominates at large distances for the zeroth Matsubara frequencies defined as $\pm p_{0} = \pm \pi T$.
\subsection{Next to leading order for flavor non-singlet correlator}\label{Section_3}
Beyond $\mathcal{O}(T)$, the relevant diagrams are
shown in figure~(\ref{fig1(b)}) and (\ref{fig1(c)}). In principle, many other higher-order graphs need to be taken into account, and for that, some resummation scheme is necessary. The effective theory approach provides a convenient way to perform such resummations~\cite{Laine:2003bd}. \\
Only the lowest fermionic modes $p_{0} \equiv \pi T$ need to be considered, as mentioned above. Writing the quark lagrangian for the field $\psi(x)$ having Matsubara mode $\omega_{n}$ along with the decomposition of Dirac spinor in terms of $\chi$ and $\phi$ and then doing an expansion in powers of $ 1/p_{0}$, the effective action for the fermionic mode becomes
\begin{eqnarray}\label{tree_level_action}
	\mathcal{S}_\psi^{\text {eff}} \!\!&=&\!\! \int d^3 x \{ i \chi^{\dagger}\left[p_{0}-g_{\mathrm{E}} A_0+\mathcal{D}_3-\frac{1}{2 p_{0}}\left(\mathcal{D}_k^2+\frac{g_{\mathrm{E}}}{4 i}\nonumber\right.\right. \left.\left. \left[\sigma_k, \sigma_l\right] G_{k l}\right)\right]\chi + i \phi^{\dagger}\left[p_{0}\nonumber \right. \\
	&-& \left. g_{\mathrm{E}} A_0-\mathcal{D}_3 -\frac{1}{2 p_{0}}\right. \left.\left.\left(\mathcal{D}_k^2+\frac{g_{\mathrm{E}}}{4 i}\left[\sigma_k, \sigma_l\right] G_{k l}\right)\right] \phi\right\}+ \mathcal{O}\bigg(\frac{1}{p_{0}^{2}}\bigg)
\end{eqnarray}
Power counting arguments show that transverse gluons can be ignored till $\mathcal{O}(g^{2}T)$. Thus, following effective langrangian is sufficient to use 
\begin{eqnarray}\label{effective_lag.}
	\mathcal{L}_{\text {eff }}^{\mathrm{\psi}}&=&i \chi^{\dagger}\left(M-g_{\mathrm{E}} A_0+\mathcal{D}_3-\frac{\nabla_{\perp}^2}{2 p_0}\right) \chi + i \phi^{\dagger}\left(M-g_{\mathrm{E}} A_0-\mathcal{D}_3-\frac{\nabla_{\perp}^2}{2 p_0}\right) \phi ,
\end{eqnarray}
where $\mathcal{D}_{3}= \partial_{3} - ig_{E} A_{3}$. To be consistent at $\mathcal{O}(g^{2}T)$, we should replace, energy i.e., $p_{0}$ by a matching coefficient $M = p_{0} + \mathcal{O}(g^{2}T)$. This matching will be done through the finite temperature Euclidean dispersion relation. On the QCD side, to find the dispersion relations, we utilized the Gribov gluon propagator, which is written as 
\begin{equation}\label{Gribov_prop}
	D_{\mu\nu}^{ab}(P)=\delta^{a b} \left(\delta_{\mu \nu}-(1-\xi)\frac{P_\mu P_\nu}{P^2}\right)\frac{P^2}{P^4+\gamma_{\mathrm{G}}^4}
\end{equation}
where $\xi$ fixes the particular gauge. QCD side dispersion relation comes out to be~\cite{Sumit:2023hjj}
\begin{equation}
		p_{3} \approx i\bigg[p_{0}-g^{2}C_{F}(I_{1}+I_{2})\bigg]
	\end{equation}	
where the integral  $I_1$ and $I_{2}$ becomes
\begin{eqnarray}\label{int_I1}
		I_1 = - \hspace{1mm} \frac{1}{p_{0}} \int_{0}^{\infty} \frac{q^{2}dq}{(2\pi)^{2}}\bigg[\frac{n^+}{E_+} +	 \frac{n^-}{E_-} \bigg], \quad \quad I_2 
		& =&\frac{1}{p_{0}}\bigg[ \frac{-T^{2}}{24} + X \bigg],
	\end{eqnarray} 
The variable $X$ is defined in ref.~\cite{Sumit:2023hjj}
Here,  $n^\pm$ represents the B.E distribution function with $ E_\pm=\sqrt{q^2\pm i\gamma_{G}^2}$.  
The pole location is $p_{3} = i M$ on the effective theory side. Thus, we get $M = p_{0}-g^{2}C_{F}(I_{1}+I_{2})$. 
Now to deal with the dynamics at $\mathcal{O}(g^{2}T)$, the modified correlator is defined as
\begin{equation}
\mathcal{P}\left(\mathbf{r}, z\right) \equiv \int_{\mathbf{R}}\left\langle\phi^{\dagger}\left(\mathbf{R}+\frac{\mathbf{r}}{2}, z\right)  \chi\left(\mathbf{R}-\frac{\mathbf{r}}{2}, z\right) \chi^{\dagger}(0) \phi(0)\right\rangle,
\end{equation}
The above correlator $\mathcal{P}\left(\mathbf{r}, z\right) $ satisfy the following partial differential equation 
\begin{equation}
	\left[\partial_z+2 M-\frac{1}{p_0} \nabla_{\boldsymbol{r}}^2+\Phi(r)\right] \mathcal{P}(r, z) = 2 N_{c} \delta(z) \delta^{(2)}(\boldsymbol{r}) ,
\end{equation}
where we obtain the one-loop static potential as~\cite{Sumit:2023hjj}
\begin{equation}
	\Phi(r)=g_{\mathrm{E}}^2 \frac{C_F}{2 \pi}\left[\ln \frac{{\gamma_{G}} r}{2}+\gamma_E-K_0\left(\gamma_{G} r\right)\right] .
\end{equation}
Here, $K_{0}$ is the modified Bessel function. The screening mass $ m  =\zeta^{-1}$, comes from
\begin{equation}\label{sch_eq.}
	\left[2 M-\frac{\nabla_{\boldsymbol{r}}^2}{p_0}+\Phi(\mathbf{r})\right] \Psi_0 = m \Psi_0 ,
\end{equation}
where $\Psi_{0}$ represents the ground state wave function. After a trivial re-scaling of the variable, the screening mass comes out to be 
\begin{equation}\label{screening_mass_final}
m -2 M \equiv g_{\mathrm{E}}^2 \frac{C_F}{2 \pi} E_0
\end{equation}
Contrary to the perturbative case~\cite{Laine:2003bd}, we find temperature dependent $E_{0}$ that saturates to a constant value at high temperature.
\begin{figure}[ht]
		\centering
		\includegraphics[scale=0.25]{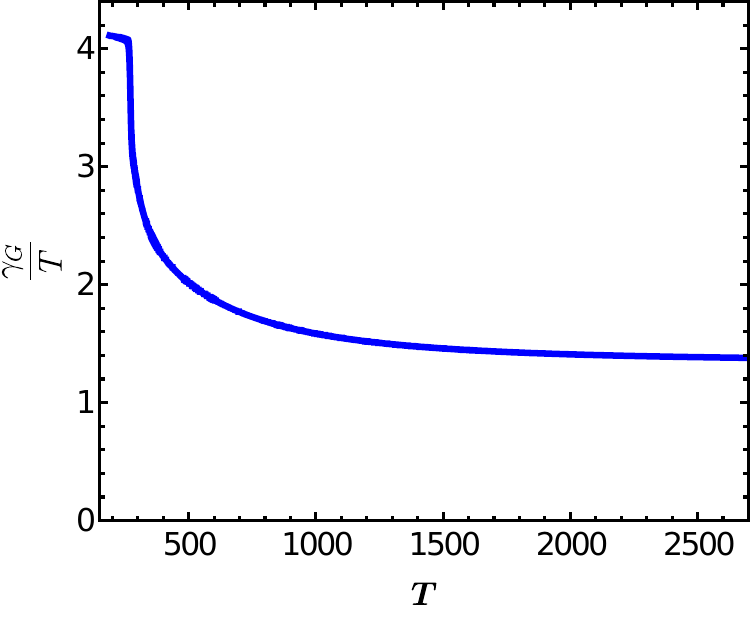}
		\includegraphics[scale=0.25]{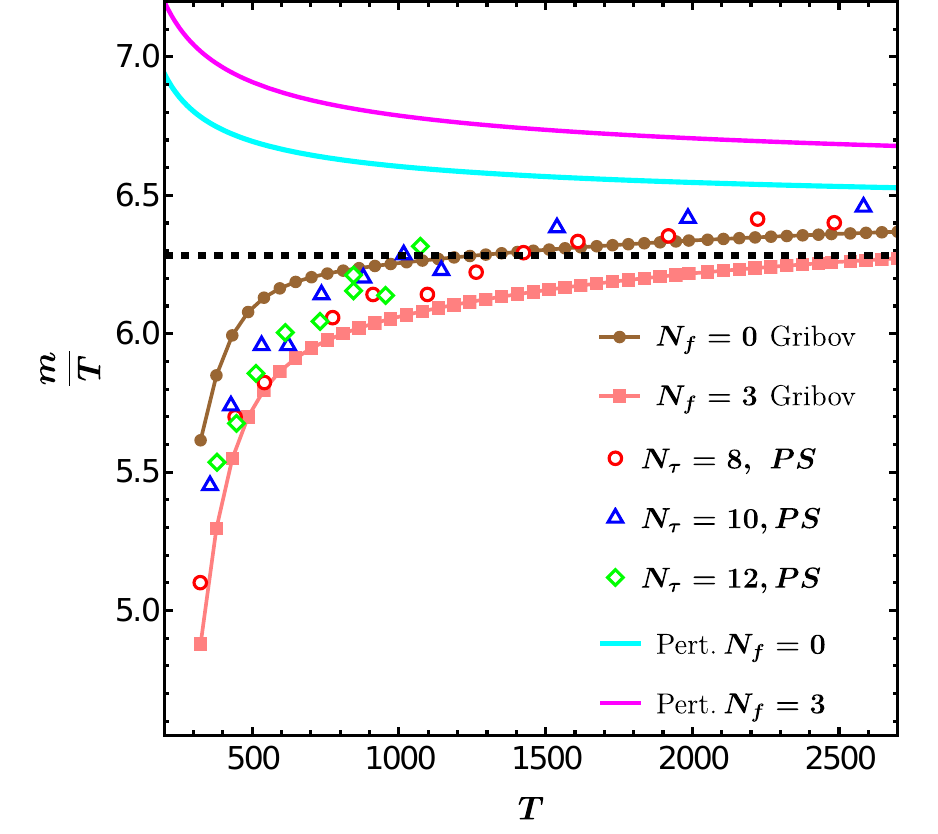}
		\caption{\textit{Left Panel}: Temperature variation of scaled Gribov mass parameter obtained using lattice (thermodynamics) data~\cite{Jaiswal:2020qmj}. \textit{Right Panel}: The temperature dependence of the scaled screening mass. The dashed line represents the free theory result from $(m =2\pi T)$. We compare the Gribov results with the perturbative and lattice results for various $N_{\tau}$. Here, $PS$ represents the pseudo scalar channel.}
\label{fig_2}	
	\end{figure}
\subsection{Results and Discussion}\label{}
The final analytic result for the screening mass reads as 
	\begin{eqnarray}\label{screening_mass_Final}
		m& =& 2 M + g^{2} T \frac{C_{F}}{2\pi} E_{0} = 2\pi T + g^{2}T \frac{C_{F}}{2\pi} \bigg(E_{0} - \frac{4\pi}{T}\big(I_{1}+I_{2}\big)\bigg)
	\end{eqnarray}
We have used the lattice-fitted running coupling from ref.~\cite{Fukushima:2013xsa}, which is given by
	\begin{equation}
		\alpha_s\left(T / T_c\right) \equiv \frac{g^2\left(T / T_c\right)}{4 \pi}=\frac{6 \pi}{11 N_{\mathrm{c}} \ln \left[a\left(T / T_c\right)\right]},\label{alpha}
	\end{equation}
where $a = 1.43$ for the infrared (IR) case and $a=2.97$ for ultarviolet (UV) case respectively. 
	
By utilizing eq.~\eqref{screening_mass_Final}, we plot the scaled screening mass $m/T$ with temperature in figure~\ref{fig_2} for quenched QCD $(N_{f}=0)$ case and for $N_{f} = 3$ case. We found a good agreement with the lattice data reported in ref.~\cite{Bazavov:2019www} and a good improvement over the perturbative results obtained in ref.~\cite{Laine:2003bd} in the low-temperature domain. The main outcome of figure~\ref{fig_2} shows that screening mass decreases significantly from the free theory results in the low-temperature region. At the same time, in the high-temperature realm, it approaches the screening mass result obtained in ref.~\cite{Laine:2003bd}.

\section{Study of Higher  Moments of Net-Strangeness Multiplicity Distribution using the AMPT Model and their Expectation for Net-Kaon using the Sub-Ensemble Acceptance Method}

\author{Rohit Kumar, Amal Sarkar}

\bigskip

\begin{abstract}
Non-monotonic behavior of higher-order moments of conserved charges serve as the signature of the QCD phase transition\cite{stephanhov}. However, due to experimental limitations, it is not possible to capture all the particles produced in heavy-ion collisions. So non-conserved charges are used as a proxy for the conserved numbers. Here higher-order moments of net strangeness have been calculated using the AMPT model and their expectation for net-kaon have been estimated using the sub-ensemble acceptance model\cite{vovchenko2020cumulants}. In this study have attempted to estimate how efficiently non-conserved charges can be used as a proxy for conserved charges.
\end{abstract}




\subsection{Introduction}

Quantum Chromodynamics (QCD) is a robust theory that describes strong interactions using basic principles. Phase transitions are fundamental and have a crucial role in understanding the behavior of matter. 
The QCD phase diagram describes different matter phases on the plane of temperature T and chemical potential $\mu_{B}$. Based on understanding so far there exists at least two QCD matter phases: a hadronic phase with confined quarks and gluons, and a QGP phase where they are unconfined and free to move independently. Lattice QCD calculations suggest a crossover region in the QGP-hadronic phase at $\mu_{B} \approx  $ 0. However, phase transition of the first order exists at larger $\mu_{B}$, as suggested by the Hydrodynamics. Models based on QCD, suggest that the first-order phase transition should end at a critical point. Theoretical studies suggest that the higher moments of the conserved quantities such as mean ($m$), standard deviation ($\sigma$), skewness ($S$) and kurtosis ($\kappa$) are affected in the vicinity of critical point. These moments (cumulants) are directly related to the thermodynamic susceptibilities which are related to correlation length. The variance($\sigma^{2}$) is related to $\xi^{2}$ skewness varies as $\xi^{4.5}$ and kurtosis as $\xi^{7}$ where $\xi$ describes the correlation length\cite{stephanhov}. In the vicinity of critical point the correlation length diverges, expected resulting long-range correlations and fluctuations in the experimentally measured moments. In this study the AMPT default version\cite{lin2005multiphase} has been used to compute net-strangeness moments at 7.7, 19.6, and 27 GeV in Au+Au collisions. Studies done very recently using the Sub-Ensemble Acceptance method on the HRG model predict a dependence of higher-order moments on experimental acceptance. The behaviour of net-strangeness has been analyzed for various volume fractions. We have analyzed cumulants of net-kaon and net-strange multiplicity and their correlation using the Sub-Ensembles Acceptance Method.

\subsection{Sub-Ensemble Acceptance method}

The Sub-Ensemble Acceptance Method enables a direct comparison between fluctuations in conserved quantities with smaller experimentally accessible subsets. In small acceptance windows, dynamical correlations are reduced\cite{vovchenko2020cumulants}. For a better understanding of large acceptance, the acceptance factor for strange particles is interpreted as the ratio of the average number of captured strange particles $\langle N^{acc}_{S}\rangle$ to the number of strange particles in full phase space $\langle N^{4\pi}_{S}\rangle$. Experiments usually analyze on net-kaon higher-order moments which are used as a proxy for the net-strangeness. The acceptance fraction $\alpha$ of the net kaon distribution for strangeness number conservation in a sub-volume can be defined as $\alpha = \langle N^{acc}_{k}\rangle / \langle N^{4\pi}_{S}\rangle$, where the mean number of kaon for a particular acceptance is given by $\langle N^{acc}_{k}\rangle$. The Ratio of $4^{th}$ order to $2^{nd}$ order moment $\kappa\sigma^{2} (\frac{C_{4}}{C_{2}})$ for distribution of charge $S_{1}$ within the subsystem of the total net Strangeness number, $S$, is defined by,
\vspace{-0.1cm}
    \begin{equation}\label{27}
\frac{C_{4}[S_{1}]}{C_{2}[S_{1}]}=(1-3\alpha\beta)\frac{\chi_{4}^{S}}{\chi_{2}^{S}}-3\alpha\beta\\
(\frac{\chi_{3}^{S}}{\chi_{2}^{S}})^{2}
\end{equation}
where $\beta=1-\alpha$.\\

\begin{figure}[hb!]
        \centering
        \includegraphics[width = 0.45\textwidth,height = 0.28\textwidth]{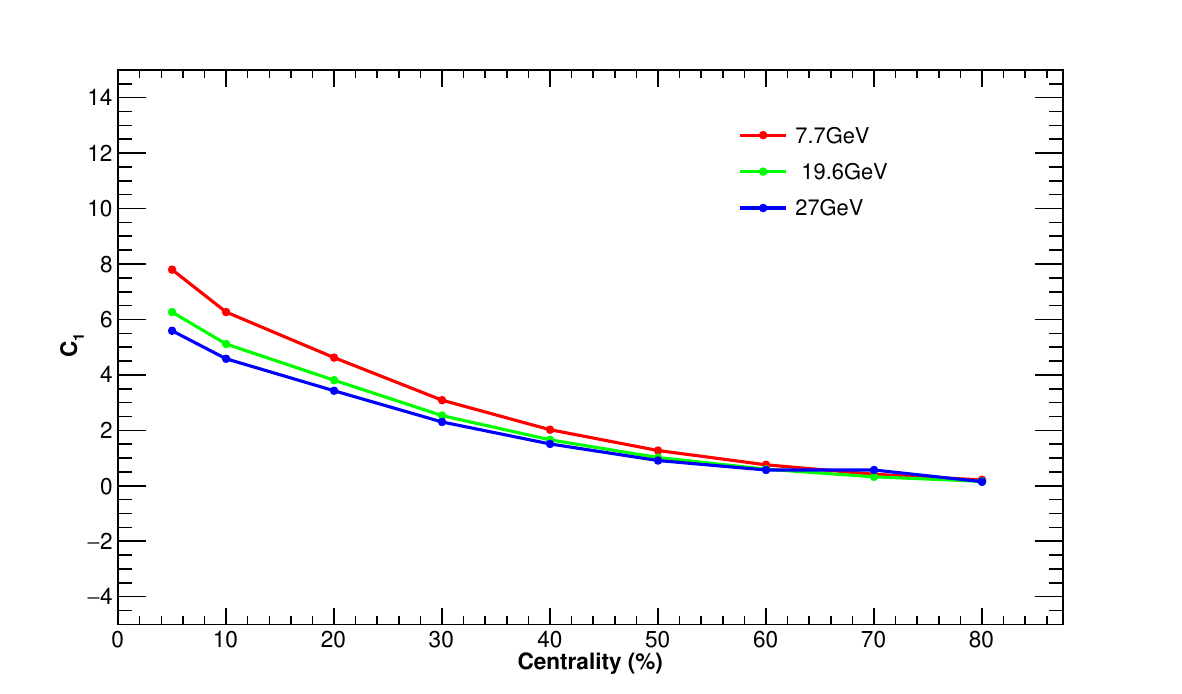}
        \includegraphics[width = 0.45\textwidth,height = 0.28\textwidth]{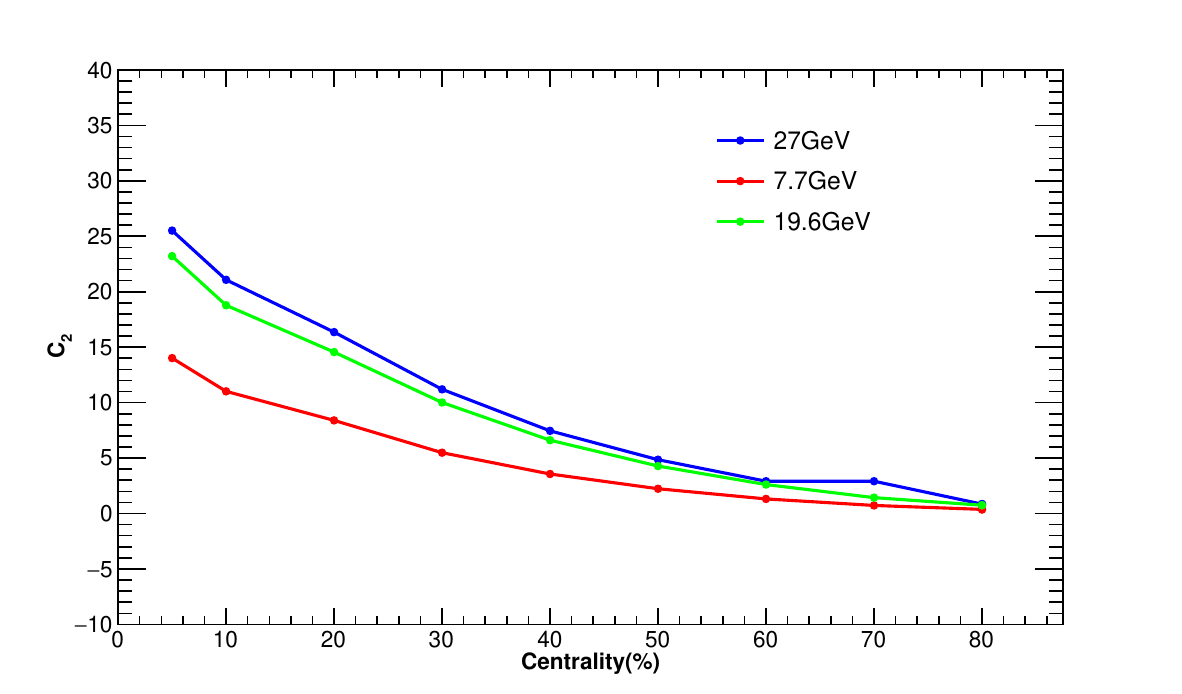}
        \includegraphics[width = 0.45\textwidth,height = 0.28\textwidth]{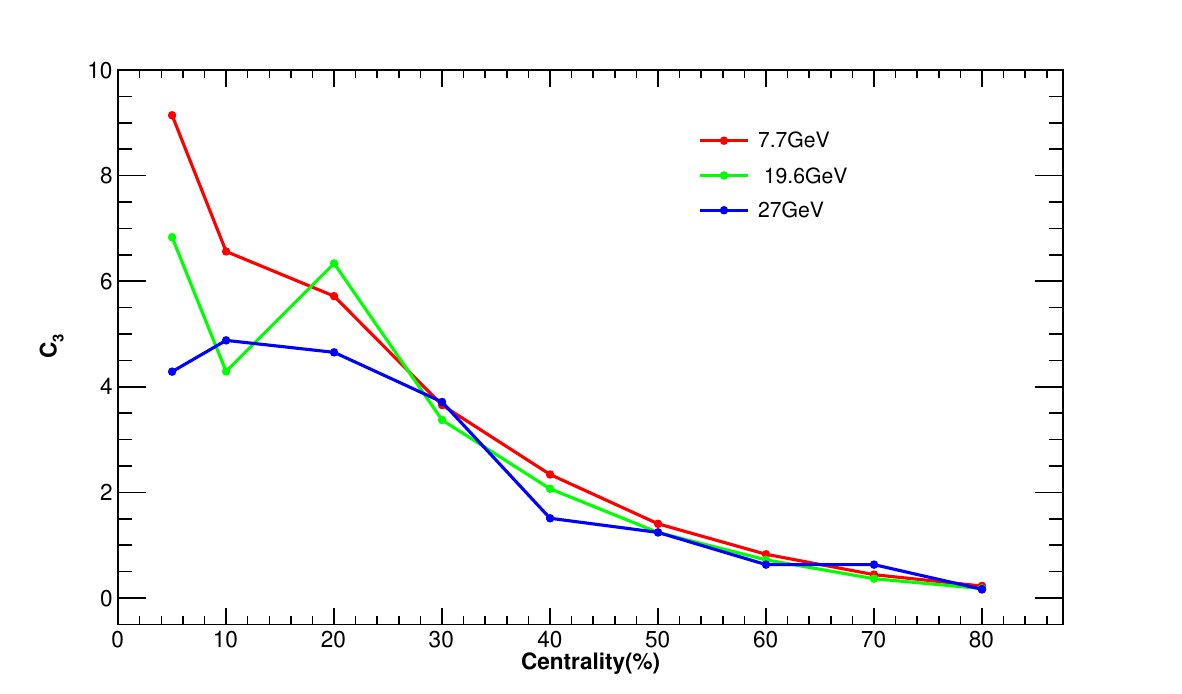}
        \includegraphics[width = 0.45\textwidth,height = 0.28\textwidth]{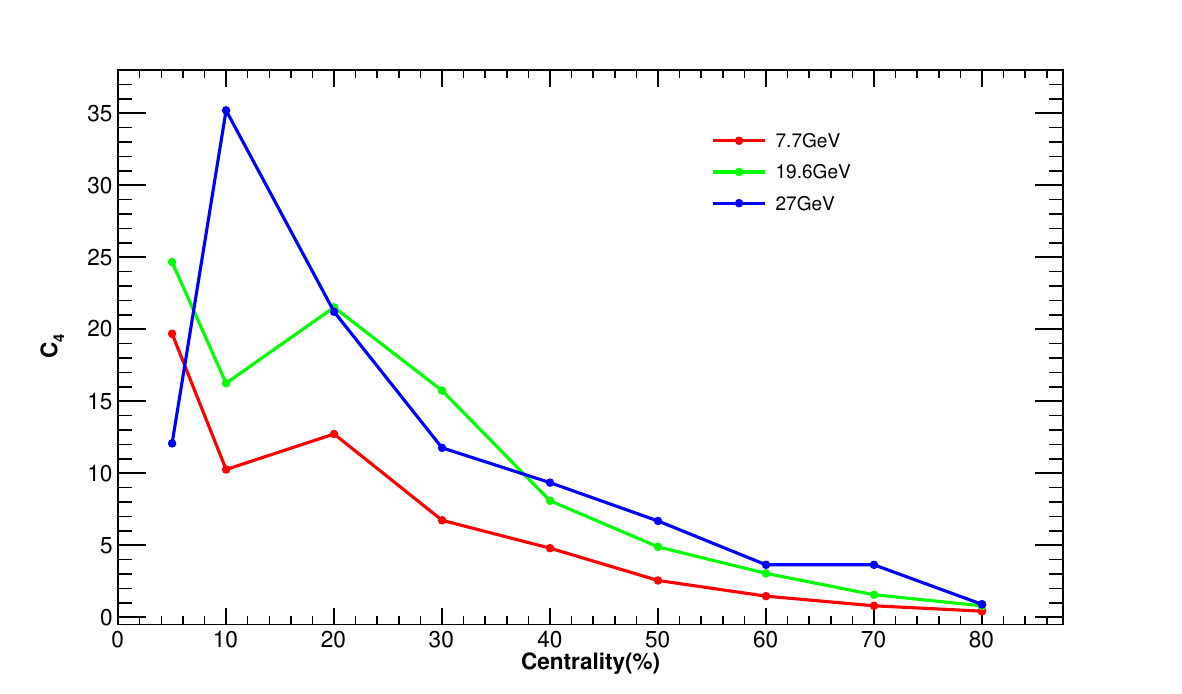}
        \vspace{-0.4cm}
        \caption{Moments for net-kaon distribution from $C_{1}$ to $C_{4}$ for nine  centralities and  different energies}
        \label{panel}
    \end{figure}

\subsection{Analysis and Results}
In this study 7, 0.9, 1.0 million events has been used for 7.7, 19.6 and 27 GeV respectively. For particle selection, the kinematic cut has been taken as $0.2 < p_{T} < 1.6$ and $|\eta|<0.5$ for comparing them to the experimental results. Cumulants up to fourth order and their volume independent ratios ($\frac{C_3}{C_2} (S\sigma), \frac{C_3}{C_2} (\kappa\sigma^{2})$) have been calculated along with the errors estimated using the delta theorem method\cite{error}.

    \begin{figure}[h]
        \centering
        \includegraphics[width = 0.45\textwidth,height = 0.28\textwidth]{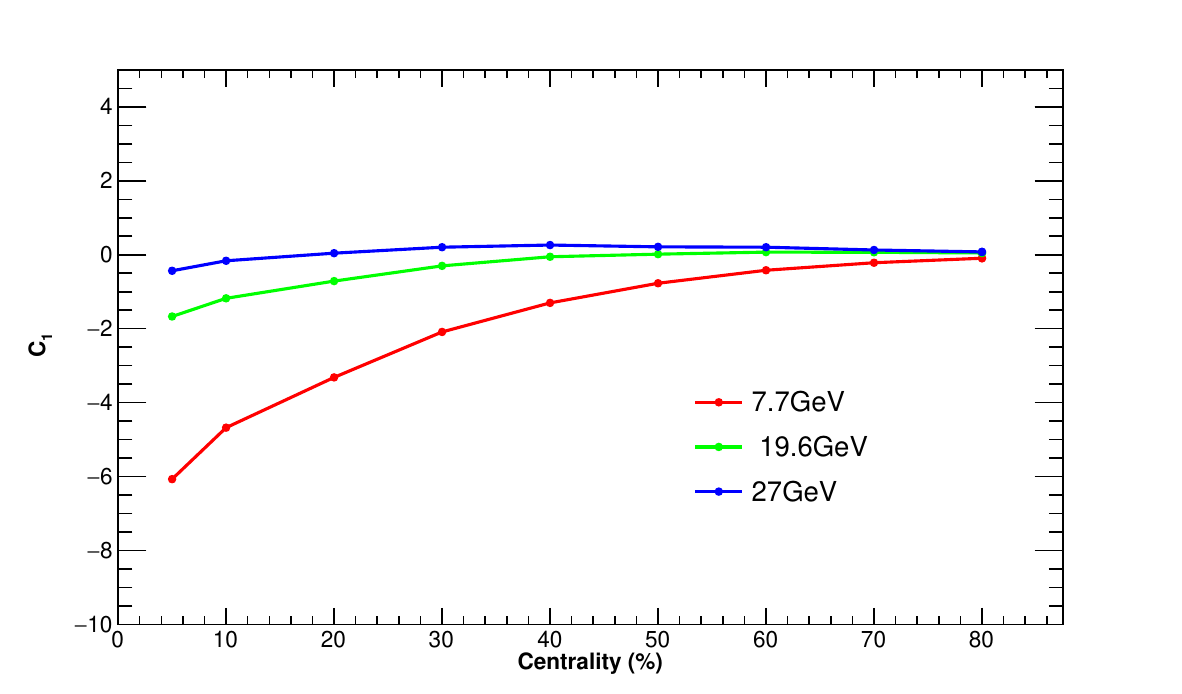}
        \includegraphics[width = 0.45\textwidth,height = 0.28\textwidth]{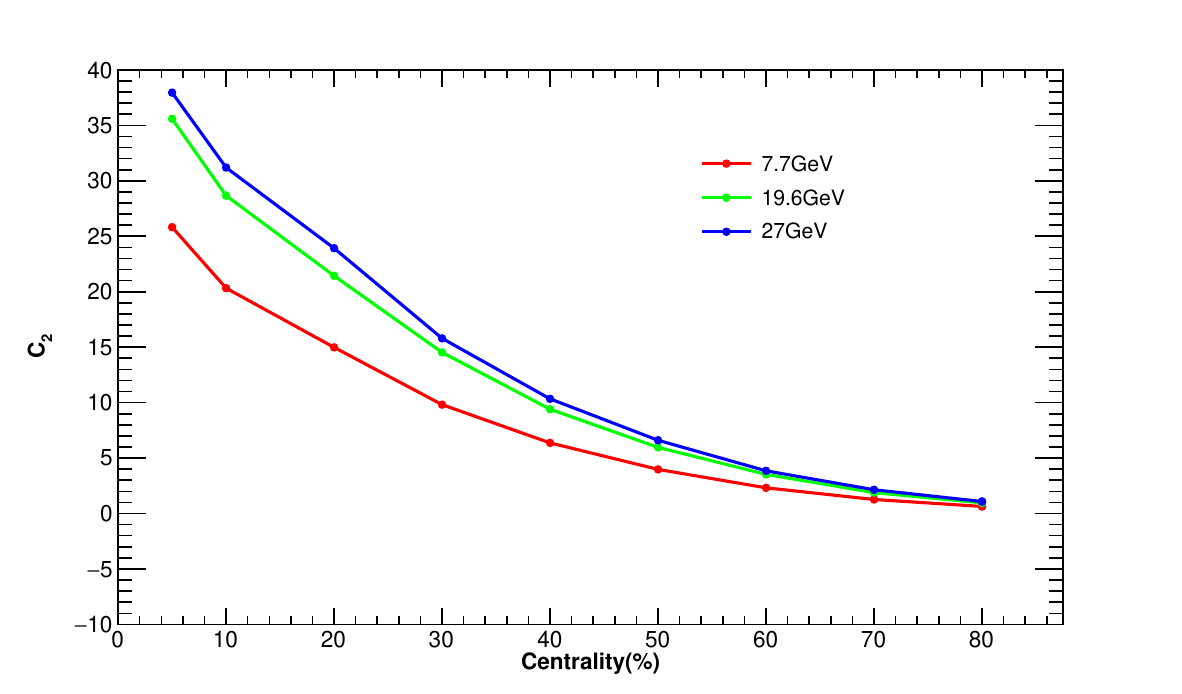}
        \includegraphics[width = 0.45\textwidth,height = 0.28\textwidth]{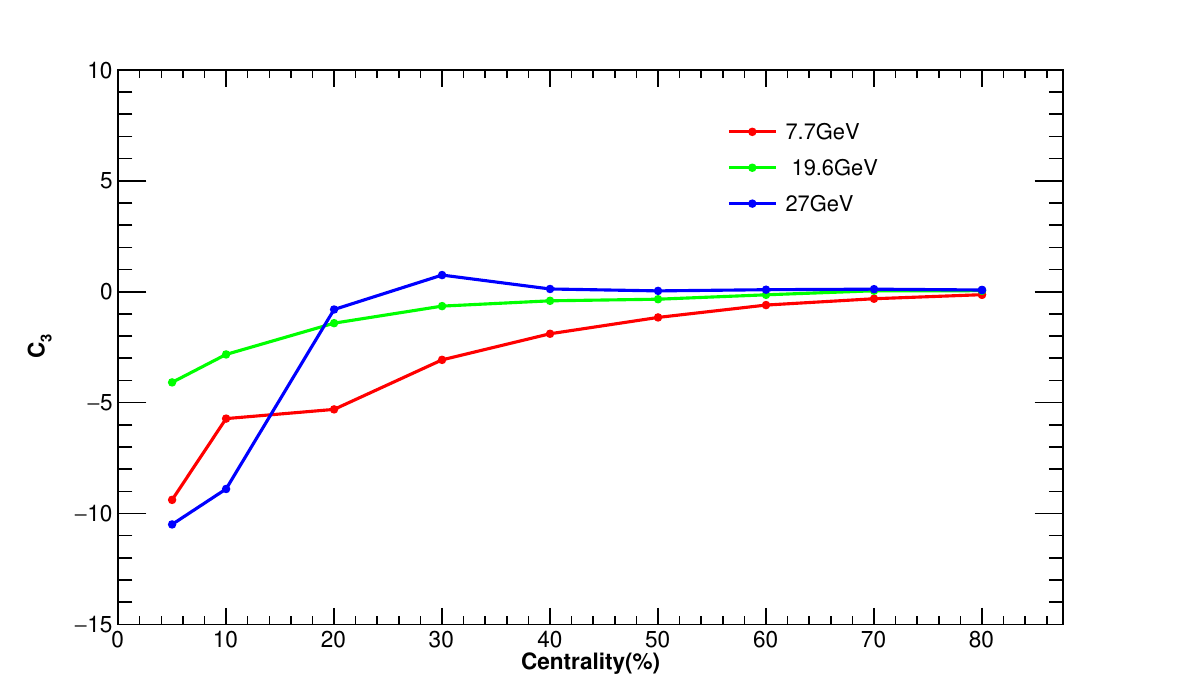}
        \includegraphics[width = 0.45\textwidth,height = 0.28\textwidth]{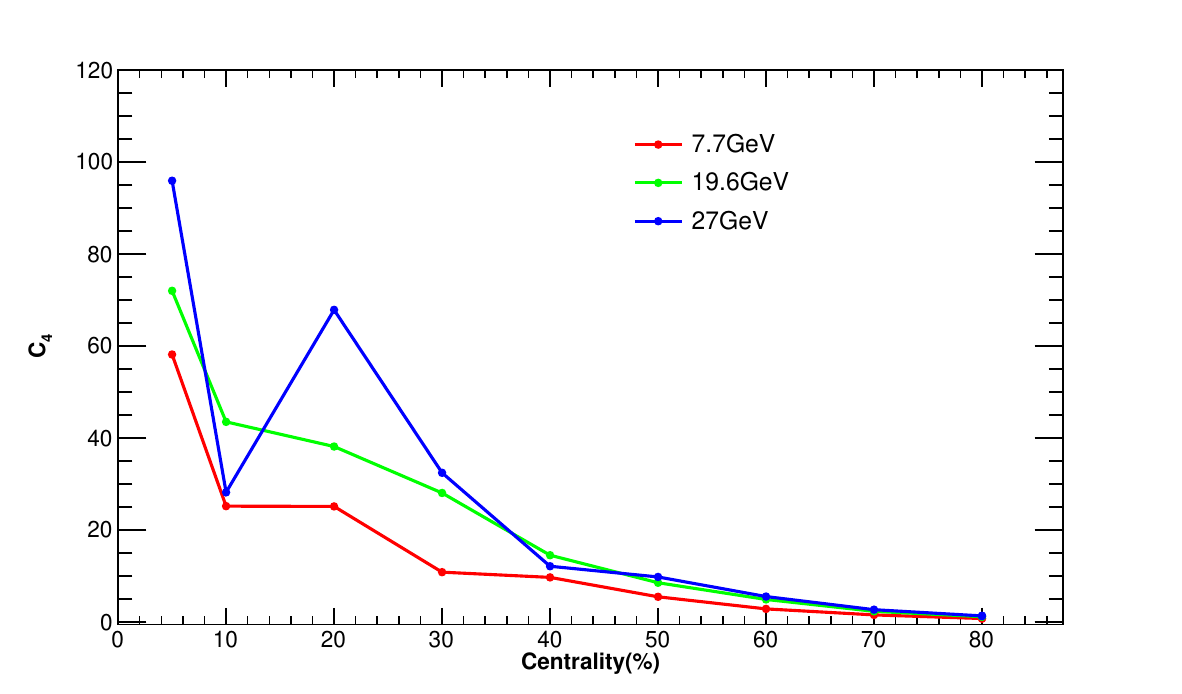}
        \vspace{-0.4cm}
        \caption{Moments for net-strangeness distribution from $C_{1}$ to $C_{4}$ for nine  centralities  three  energies}
        \label{panel1}
    \end{figure}

The higher-order moments of net-kaon and net-strangeness multiplicity distribution have been shown in Fig. \ref{panel} and Fig. \ref{panel1} respectively. 

 \begin{figure}[hb!]
        \centering
        \includegraphics[width = 0.45\textwidth,height = 0.28\textwidth]{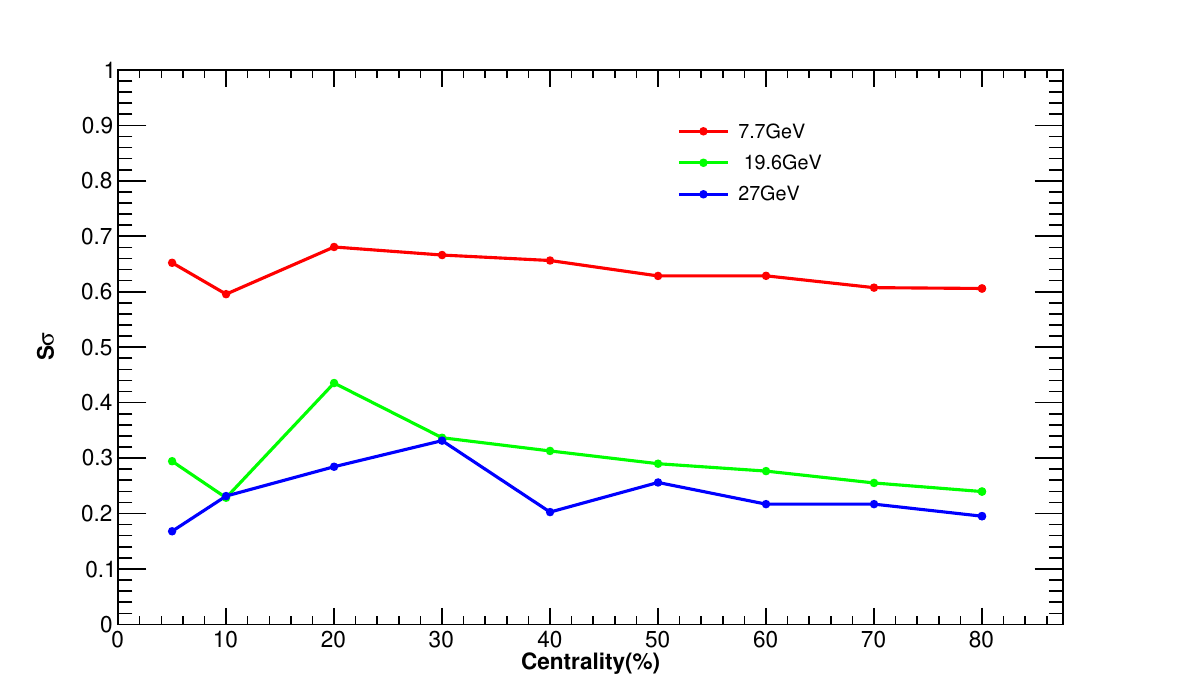}
        \includegraphics[width = 0.45\textwidth,height = 0.28\textwidth]{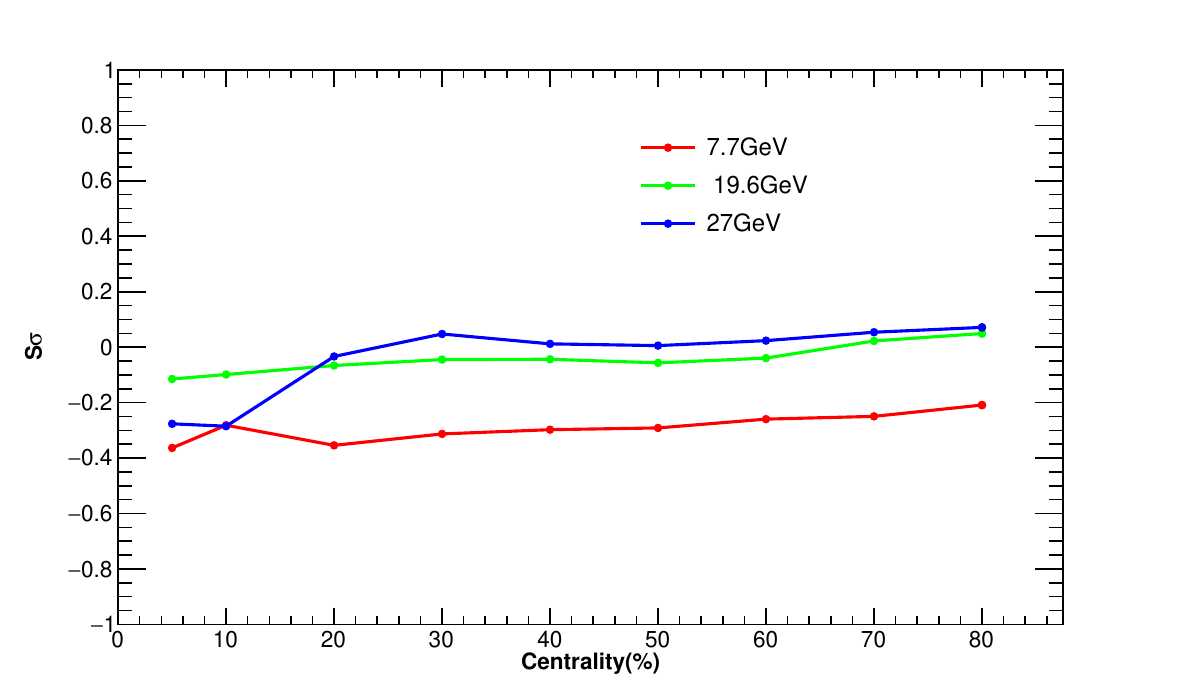}
        \includegraphics[width = 0.45\textwidth,height = 0.28\textwidth]{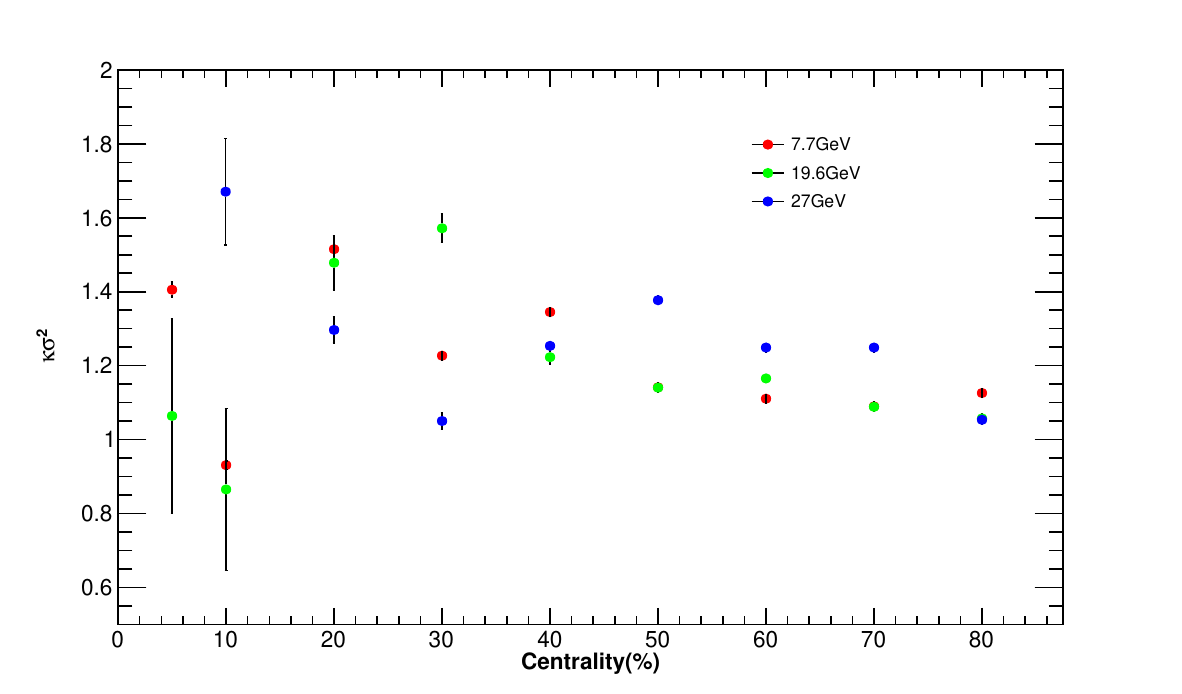}
        \includegraphics[width = 0.45\textwidth,height = 0.28\textwidth]{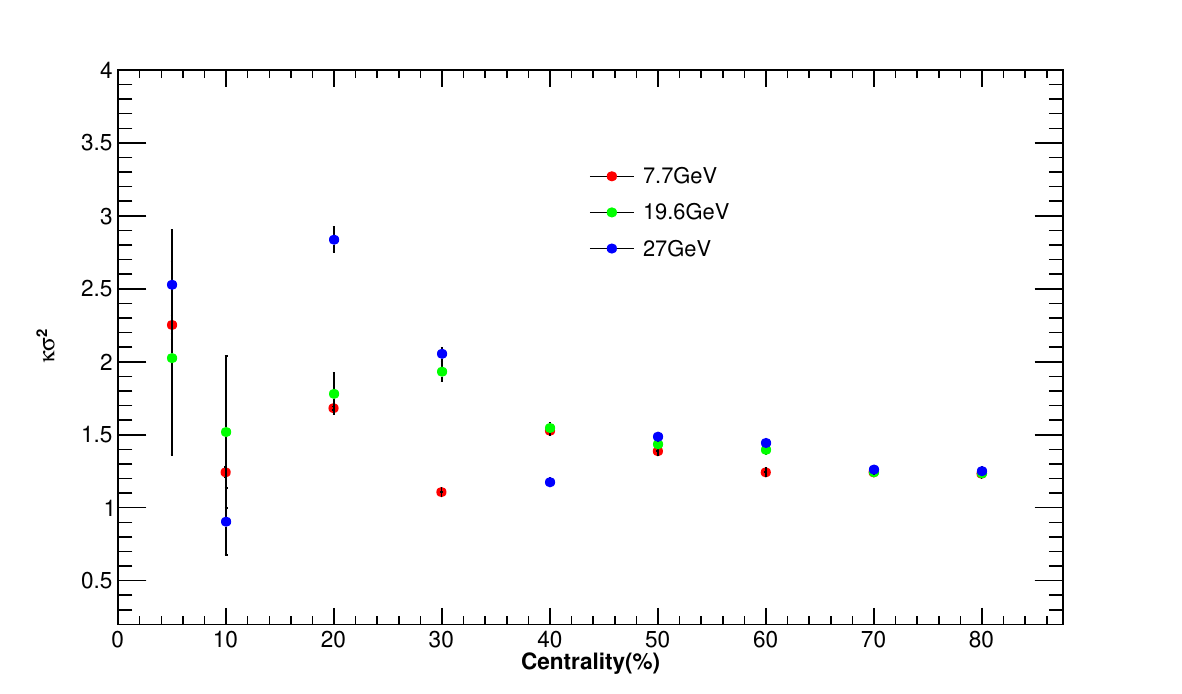}
        \vspace{-0.4cm}
        \caption{Volume independent ratio of higher moments first two figures show $S\sigma$ and second two figure show $\kappa\sigma^{2}$ for net-kaon and net-strangeness respectively}
        \label{panel3}
    \end{figure}
It can be observed that higher moments vary with centrality. $C_{1}$ and $C_{2}$ are higher in central collisions and get lower in peripheral collisions. Similarly $C_{3}$ and $C_{4}$ exhibit nearly the same behavior but some fluctuations are solely statistical.
In Fig. \ref{panel3} the plots represent the volume-independent ratio of cumulants. Due to the Skellam distribution of net-kaon multiplicity, the ratio of two even cumulants is expected to be unity. However, recent research has unveiled that fluctuations in the ratio of cumulants can also be attributed to statistical effects. Since AMPT does not contain any information about the critical behavior of strong interaction the fluctuations in the cumulants and their volume-independent ratios are due to solely statistical. Indeed, it’s worth noting that the fluctuations observed for the energy of 7.7 GeV are comparatively lower than those for the other energies. For instance, $\kappa\sigma^{2}$, which characterizes the variance of particle multiplicity fluctuations, typically requires at least 10 million events to yield statistically reliable results. The volume-independent ratio of cumulants shows less fluctuations for energy 7.7GeV which is obvious as for this energy we have much higher statistics. 

    \begin{figure}[h]
        \centering        \includegraphics[width=0.48\linewidth]{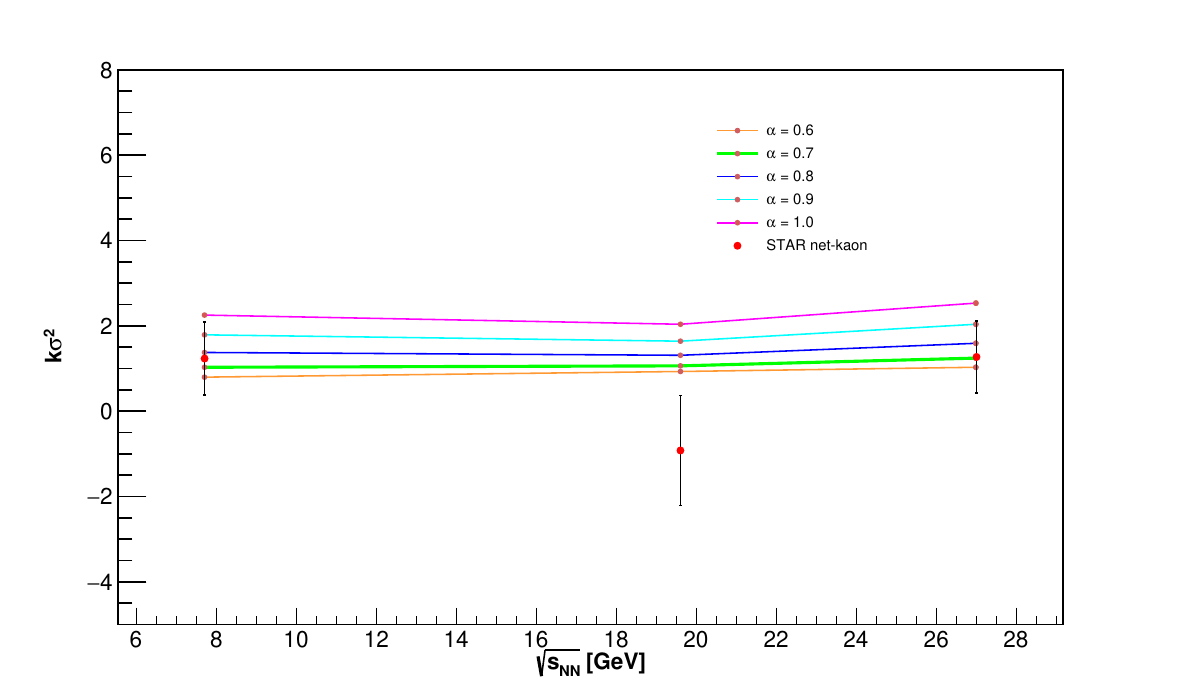}
        
        \caption{ $\alpha$ dependence of $\kappa\sigma^{2}$ for different energies}
        \label{alpha_rohit}
    \end{figure}
The fluctuations in the central collision are higher because of QGP formed\cite{braun2004particleproduction} in central collisions is a highly dynamic and
evolving system with complex interactions among its constituents. Fig. \ref{alpha_rohit} shows $\alpha$ dependence of $\kappa\sigma^{2}$ and  it is seen that the value of $\kappa\sigma^{2}$ decrease with decrease in $\alpha$. It is notable that for nearly $\alpha = $ 0.8 values of net-kaon and net-strangeness are equal.

\subsection{Summary and conclusions}

Higher order moments net-kaon and net-strangeness have been reported using the AMPT model at 7.7, 19.6 and 27 GeV. Higher-order moments strongly depend on the statistics and therefore, for $\sqrt{S_{NN}}$ = 7.7GeV show less fluctuation than 19.6GeV and 27GeV. The net-kaon distribution measured in experiments depends on the parameter $\alpha$, which defines the fluctuation measurement subsystem. With sufficiently large acceptance to reach the thermodynamic limit, the subensemble acceptance method applies to any equation of state.

\newcommand{\MB}[1]{\left|#1\right|}
\newcommand{\mb}[1]{|#1|}
\newcommand{\FB}[1]{\left(#1\right)}
\newcommand{\fb}[1]{(#1)}
\newcommand{\SB}[1]{\left\{#1\right\}}
\newcommand{\TB}[1]{\left[#1\right]}
\newcommand{\AB}[1]{\left<#1\right>}

\section{Speed of sound in magnetized nuclear matter}
\author{Rajkumar~Mondal, Nilanjan Chaudhuri, Pradip Roy, and Sourav Sarkar}

\bigskip

\begin{abstract}
We study the thermodynamic properties of magnetized nuclear matter at finite temperature and baryon chemical potential using the nonlinear Walecka model. The presence of magnetic field shifts the location of the spinodal lines and the critical endpoint (CEP) in the $T-\mu_B$ plane. Due to the directional nature of the background magnetic field, thermodynamic quantities, like the squared speed of sound, show anisotropic behavior, splitting into components along and perpendicular to the field direction. 
\end{abstract}




\subsection{Introduction}
The primary objective of relativistic heavy ion collision (HIC) experiments is to study strongly interacting matter. It is conjectured that in non-central HICs, a very strong and transient ($\sim$ few fm/c) magnetic field of the order $\sim 10^{15-18}$ Gauss or larger might be generated due to the receding charged spectators~\cite{Tuchin:2013ie}. However the finite conductivity of the medium ($\sim$few MeV) can delay the decay process allowing a nonzero magnetic field to persist even during the hadronic phase, following a phase transition or crossover from QGP (quark gluon plasma)~\cite{Kalikotay:2020snc}. The study of strongly interacting magnetized matter is also relevant for astrophysical scenarios such as neutron star research.
The speed of sound is a key quantity in all thermodynamic systems and is directly related to the thermodynamic properties of system including its equation of state (EOS). The sensitivity of the speed of sound on temperature, density, chemical potential, etc. provides crucial insights: it exhibits a local minimum at a crossover transition, while it reaches zero at the critical point and along the corresponding spinodal lines. The
variation of speed of sound with density has a substantial impact on the mass-radius relationship, cooling rate, the maximum possible mass of neutron star~\cite{Ozel:2016oaf} and tidal deformability. It has crucial impact on the frequencies of gravitational waves generated by the g-mode oscillation of a neutron star~\cite{Jaikumar:2021jbw}.

In this work, we explore the speed of sound and the liquid-gas phase transition in magnetized nuclear matter. We have choosen the nonlinear Walecka model to describe the system because of its effectiveness in describing the saturation properties and equation of state of nuclear matter,
incorporating relativistic effects. The temperature and baryon chemical potential ranges studied are particularly relevant compact astrophysical objects like neutron stars or magnetars.
\subsection{Walecka Model}
The Lagrangian density of Walecka Model for nuclear matter in presence of a background magetic field is ~\cite{Mondal:2023baz}
{\small\begin{eqnarray}
	\mathcal{L}&=&-\frac{1}{4}F_{\mu\nu}F^{\mu\nu}+\bar{\psi}\FB{i\gamma^\mu D_\mu-m_N+\gamma^0\mu_B}\psi + g_\sigma\bar{\psi}\sigma\psi-g_\omega\bar{\psi}\gamma^\mu\omega_\mu\psi\nn\\&&+\frac{1}{2}(\partial_\mu\sigma\partial^\mu\sigma-m_\sigma^2\sigma^2)-\frac{b}{3}m_N(g_\sigma\sigma)^3-\frac{c}{4}(g_\sigma\sigma)^4-\frac{1}{4}\omega_{\mu\nu}\omega^{\mu\nu}+\frac{1}{2}m_\omega^2\omega_\mu\omega^\mu~~~
\end{eqnarray}}
where $F_{\mu\nu}$ is the field tensor corresponding to the external magnetic field, $\psi$ represents the nucleon isospin doublet, $\sigma$ and $\omega$ mesons are the mediators of interactions between nucleons, $D_\mu=\partial_\mu+ieA_\mu$ with $A_\mu=(0, yB, 0, 0)$ and $\omega_{\mu\nu}=\partial_\mu\omega_\nu-\partial_\nu\omega_\mu$. 
In mean field approximation (MFA), the mass $(M_N)$ and the chemical potential $(\mu_B)$ of nucleon are modified as $M=m_N-g_\sigma\bar{\sigma}$ and $\mu^\star=\mu_B-g_\omega{\bar{\omega}}_0$ with $\bar{\sigma}=\AB{\bar\psi\psi}$ and ${\bar{\omega}}_0=\AB{\bar\psi\gamma^0\psi}$. The model parameters $g_\sigma,~g_\omega,~b,~c$ are fitted to reproduce the properties of nuclear matter at saturation in absence of magnetic field~\cite{Haber:2014ula}. Employing MFA on the Lagrangian, one can obtain the free energy of the system, given by
{\small\begin{eqnarray}
	\Omega=\frac{B^2}{2}+\frac{1}{2}m_\sigma^2\bar{\sigma}^2+\frac{b}{3}m_N(g_\sigma\bar{\sigma})^3+\frac{c}{4}(g_\sigma\bar{\sigma})^4-\frac{1}{2}m_\omega^2{\bar{\omega}}_0^2+\Omega_N~~.
\end{eqnarray}}
The nucleonic contribution $\Omega_N$ can be decomposed into $(\Omega_{\rm sea}+\Omega_{\rm TM})$ where $\Omega_{\rm sea}$ contains pure vacuum as well as the magnetic field dependent vacuum contributions and $\Omega_{\rm TM}$ is the thermo-magnetic (TM) contribution in free energy. The expressions are obtained as~\cite{Mondal:2023baz}
{\small\begin{eqnarray}
	\Omega_{\rm sea}&=&-2\int\frac{d^3k}{\FB{2\pi}^3}E-\frac{eB}{2\pi}\sum_{n=0}^{\infty}\alpha_n\int\frac{dk_z}{2\pi}E_n
	=\Omega_{\rm vac}+\Omega_{\rm vac}^B,
\\
	\Omega_{\rm TM}&=&2\beta^{-1}\int\frac{d^3k}{\FB{2\pi}^3}\sum_{a=\pm}\SB{\text{ln}(1-f^a)}+\frac{1}{\beta}\frac{eB}{2\pi}\sum_{n=0}^{\infty}\alpha_n\int\frac{dp_z}{2\pi}\sum_{a=\pm}\SB{\text{ln}(1-f_n^a)}~~.
\end{eqnarray}}
Here $f^\pm=f(E\mp\mu^\star)$, $f_n^\pm=f(E_n\mp\mu^\star)$, $\beta=\frac{1}{T}$, $f(x)=\frac{1}{e^{x/T}+1}$ is the Fermi distribution function and "+(-)"  corresponds to the fermion(anti-fermion) distribution function. Now minimizing the free energy, one can obtain dynamical mass and the effective chemical potential of nucleon. Other thermodynamic quantites can be obtained from the free energy such as pressure $p=-\Omega$, entropy density $s=-\frac{\partial\Omega}{\partial T}$, baryon number density $n_B=-\frac{\partial\Omega}{\partial\mu_B}$, magnetization $\mathcal{M}=-\frac{\partial\Omega}{\partial B}$ $etc$.
\subsection{Speed of Sound}
Due to the breaking of rotational symmetry in presence of magnetic field, the speed of sound becomes anisotropic and splits into parallel and perpendicular components with respect to the magnetic field direction~\cite{Ferrer:2022afu,Ferrer:2010wz}. The components of speed of sound, holding with $x=s/n_B$ as a constant parameter, are~\cite{Mondal:2023baz}
{\small\begin{eqnarray}
	{C_x^2}^{(\parallel)}(T,\mu_B)&=&\FB{\frac{\partial p_\parallel}{\partial \epsilon}}_x
	=\frac{\FB{\frac{\partial{p^\parallel}}{\partial T}}_{\mu_B}\FB{\frac{\partial x}{\partial\mu_B}}_T-\FB{\frac{\partial{p^\parallel}}{\partial\mu_B}}_T\FB{\frac{\partial x}{\partial T}}_{\mu_B}}{\FB{\frac{\partial\epsilon}{\partial T}}_{\mu_B}\FB{\frac{\partial x}{\partial\mu_B}}_T-\FB{\frac{\partial\epsilon}{\partial\mu_B}}_T\FB{\frac{\partial x}{\partial T}}_{\mu_B}}\label{C2xParallel}~~,
\end{eqnarray}}
{\small\begin{eqnarray}
	{C_x^2}^{(\perp)}(T,\mu_B)&=&\FB{\frac{\partial p_\perp}{\partial \epsilon}}_x
	={C_x^2}^{(\parallel)}-B\frac{\FB{\frac{\partial\mathcal{M}}{\partial T}}_{\mu_B}\FB{\frac{\partial x}{\partial\mu_B}}_T-\FB{\frac{\partial\mathcal{M}}{\partial\mu_B}}_T\FB{\frac{\partial x}{\partial T}}_{\mu_B}}{\FB{\frac{\partial\epsilon}{\partial T}}_{\mu_B}\FB{\frac{\partial x}{\partial\mu_B}}_T-\FB{\frac{\partial\epsilon}{\partial\mu_B}}_T\FB{\frac{\partial x}{\partial T}}_{\mu_B}}\label{C2xPerpendicular}.
\end{eqnarray}}
\subsection{Numerical Results}
\begin{figure}[h]
	\centerline
	{
\includegraphics[angle=-90,width=4.2cm]{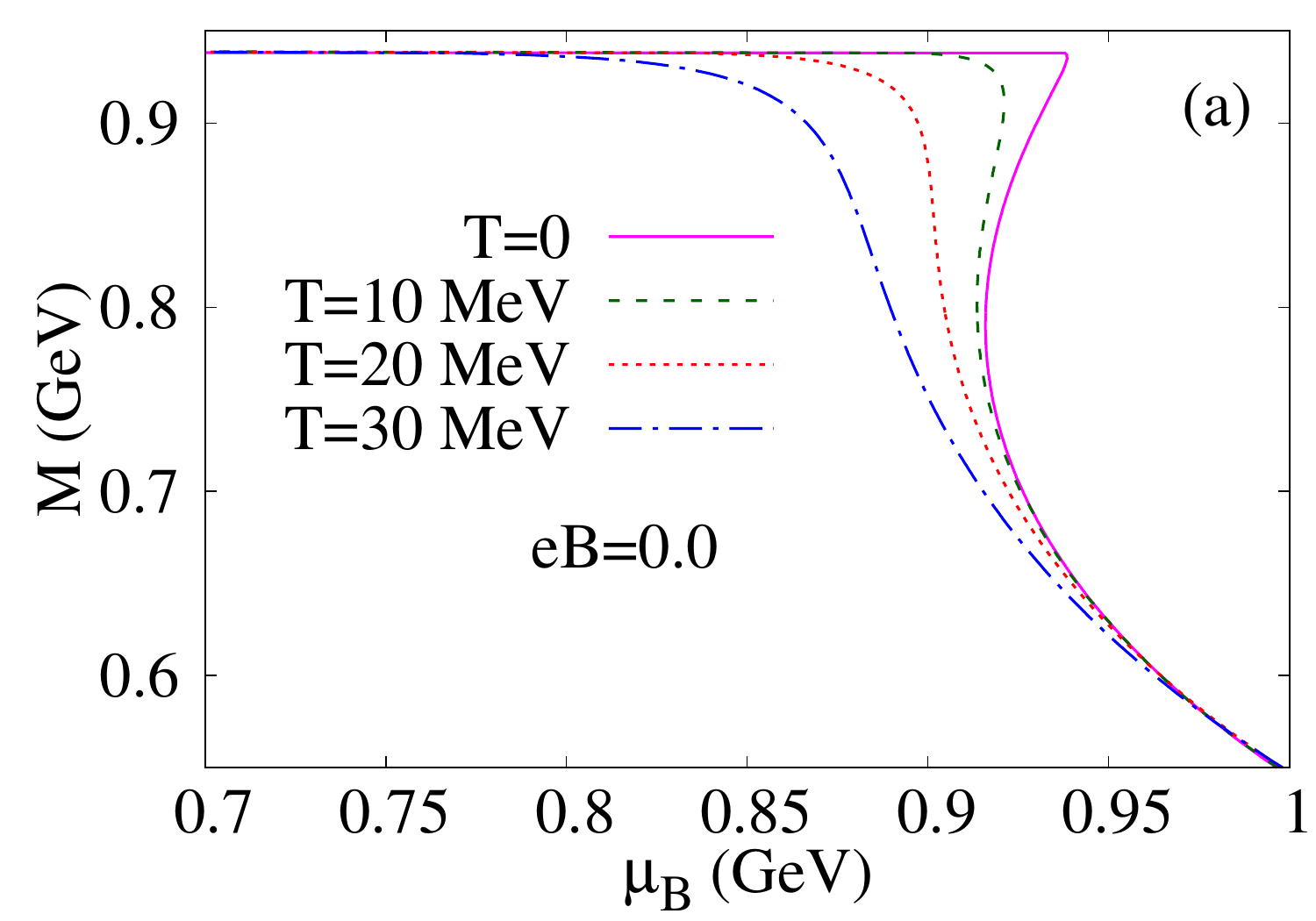}
\includegraphics[angle=-90,width=4.2cm]{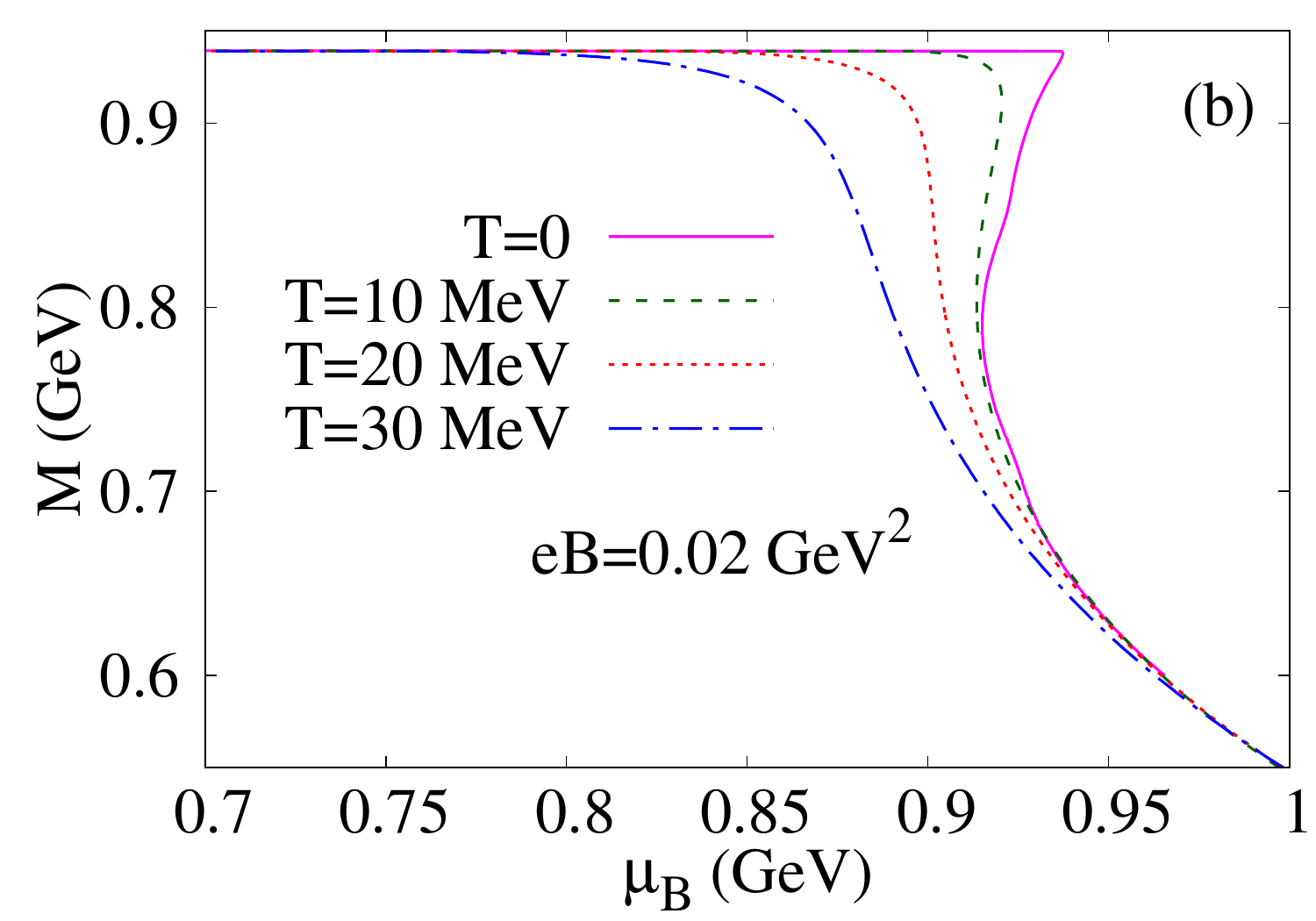}
\includegraphics[angle=-90,width=4.0cm]{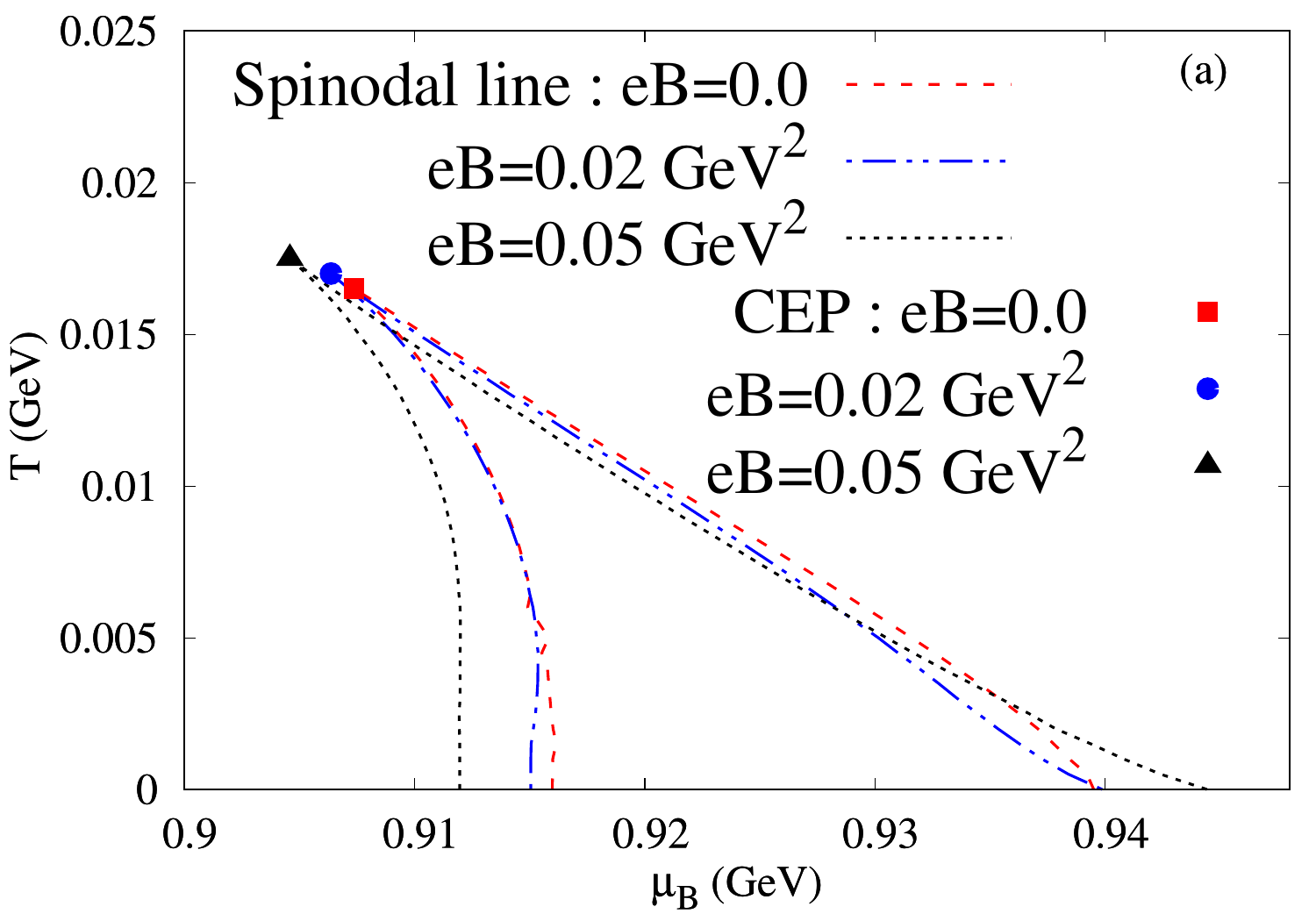}
	}
	\caption{$M$ vs $\mu_B$ for $T=0,~10,~20,~30~\rm MeV$ at (a) $eB=0$ (b) $eB=0.02~\rm GeV^{-2}$. (c) Spinodal lines and critical endpoints (CEPs) for $eB=0,~0.02,~0.05~\rm GeV^{-2}$ in $T$-$\mu_B$ plane.}
	\label{MvsMuvsTvseB}
\end{figure}

We begin with the investigations of dynamical nucleon mass as a function of 
chemical potential, temperature for $eB=0$ in Fig.~\ref{MvsMuvsTvseB}(a) and $eB=0.02~GeV^{-2}$ in Fig.~\ref{MvsMuvsTvseB}(b). At low temperatures, these figures show multiple roots for the nucleon mass within a specific range of $\mu_B$. As temperature increases, these multiple solutions disappear at a particular $T$ and $\mu_B$, known as the critical endpoint (CEP), beyond which the phase transition shifts to a crossover regime. The boundaries in $T$-$\mu_B$ plane, where multiple roots of the nucleon mass exist, are defined as the spinodal lines as shown in Fig.~\ref{MvsMuvsTvseB}(c). The region bounded by the spinodal lines is referred to as the mixed phase. It is to be noted that the spinodal lines are determined from the extrema of $\frac{\partial M}{\partial T}$. This gives rise to two distinct segments of spinodal lines which converge at the CEP. The presence of magnetic field leads to significant changes in both the spinodal lines and CEP in $T$-$\mu_B$ plane.
\begin{figure}[h]
	\centerline
	{
		\includegraphics[angle=-90,width=4.0cm]{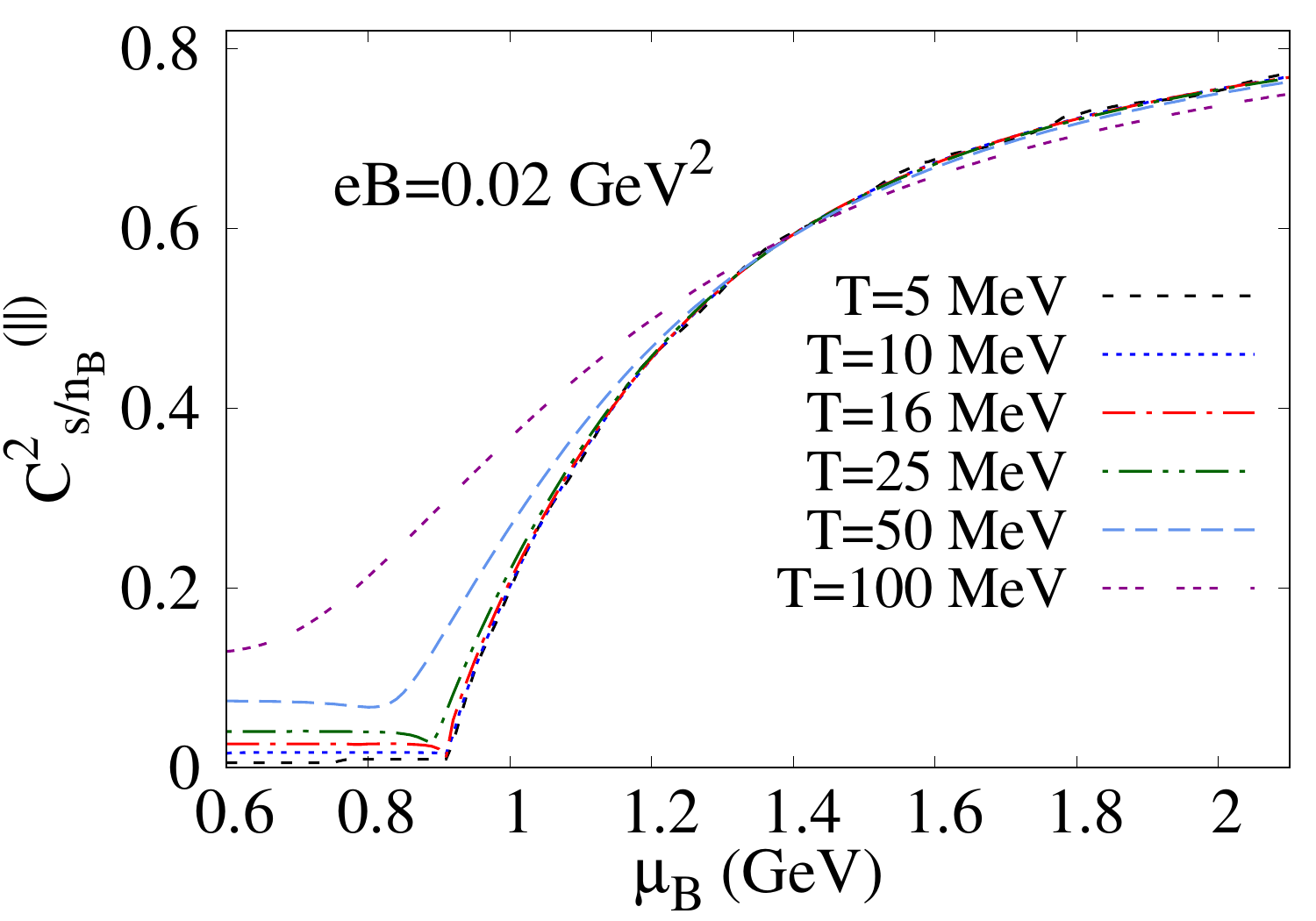}
		\includegraphics[angle=-90,width=4.0cm]{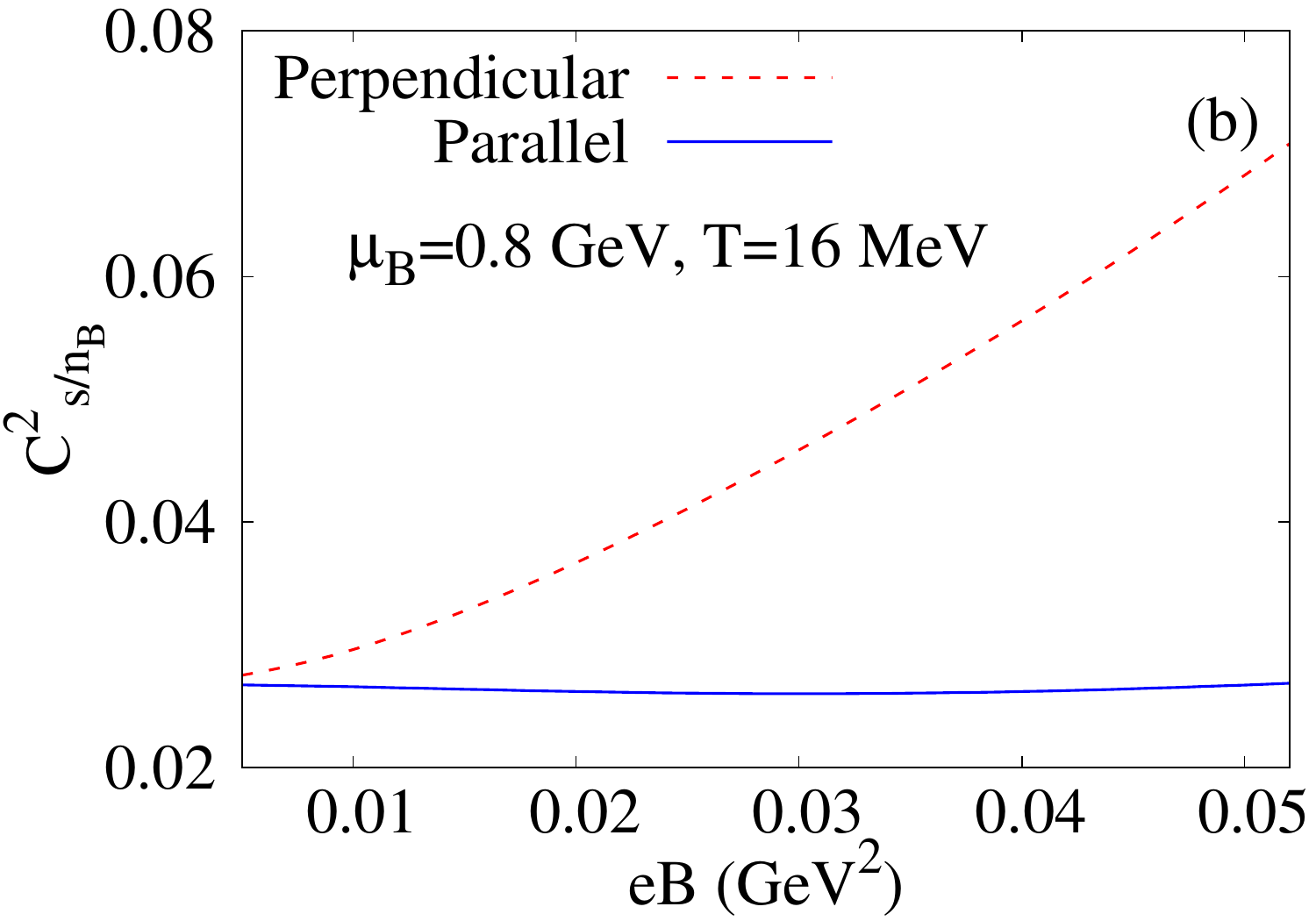}
		\includegraphics[angle=-90,width=4.0cm]{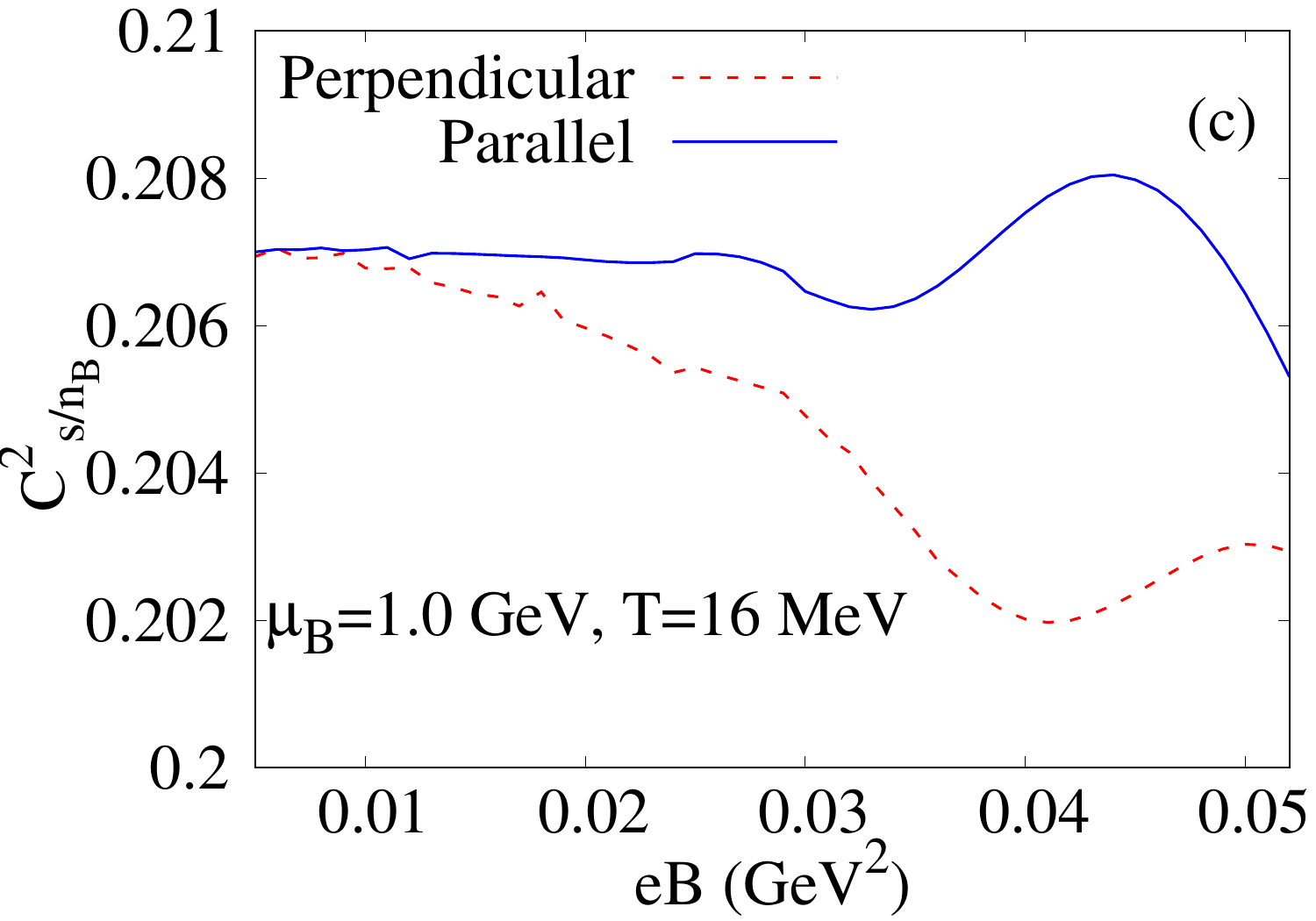}
	}
	\caption{Parallel component of squared speed of sound as a function of chemical potential in (a). Components of squared speed of sound as a function of background magnetic field for $T=16~\rm MeV$ in (b) $\mu_B=0.8~\rm GeV$, (c) $\mu_B=1.0~\rm GeV$.}
	\label{CvsMuvsT}
\end{figure}
Fig.~\ref{CvsMuvsT}(a) shows the variation of parallel component of speed of sound $(	{C_x^2}^{(\parallel)})$ with chemical potential at $eB=0.02\rm~GeV^{-2}$ for different temperatures. ${C_x^2}^{(\parallel)}$ is low at small values of $\mu_B$ and  increases at larger vlues of $\mu_B$. A sudden shift in ${C_x^2}^{(\parallel)}$ near $\mu_B=0.94$ indicates a phase transition or crossover. Figs.~\ref{CvsMuvsT}(b)-(c) show the splitting of the speed of sound as a function of the background magnetic field for two values of chemical potential: one below the phase transition and the other above it.

\subsection{Summary and Conclusion}
We have studied changes in nucleon mass, the nuclear liquid-gas phase transition,
and the squared speed of sound in magnetized nuclear matter at finite temperature and baryon chemical potential using the nonlinear Walecka model. Our results show that a magnetic field affects the locations of the critical endpoint (CEP) and spinodal lines in the \( T-\mu_B \) plane, obtained from the extrema of ${\partial M}/{\partial T}$. The magnetic field also causes anisotropy in the speed of sound, splitting into parallel and perpendicular components with respect to the magnetic field direction.
The results indicate that the speed of sound in nuclear matter can exceed $\sqrt{1/3}$ at high chemical potential, while causality is always maintained as $C_{s/n_B}^2 < 1$. This work primarily addresses symmetric nuclear matter, but the temperatures and chemical potentials considered are more relevant to neutron-star matter. The study of asymmetric nuclear matter is essential for understanding neutron-rich environments such as neutron stars. Future studies could incorporate these aspects to broaden our findings.

\def\be{\begin{eqnarray}}
	\def\ee{\end{eqnarray}}
\newcommand{\Tr}{\rm Tr}
\def\lsim{\raise0.3ex\hbox{$<$\kern-0.75em\raise-1.1ex\hbox{$\sim$}}}
\def\gsim{\raise0.3ex\hbox{$>$\kern-0.75em\raise-1.1ex\hbox{$\sim$}}}
\def\({\left(}
\def\){\right)}
\def\hmu{\hat\mu}
\def\bea {\begin{eqnarray}}
	\def\eea {\end{eqnarray}}
\def\sumintb{\sum\!\!\!\!\!\!\!\!\!\int\limits}
\def\sumintf{\sum\!\!\!\!\!\!\!\!\!\!\int\limits}
\def\sumintbb{\sum\!\!\!\!\!\!\!\!\!\int\limits}
\def\sumintbf{\sum\!\!\!\!\!\!\!\!\!\!\!\int\limits}
\def\sumintff{\sum\!\!\!\!\!\!\!\!\!\!\!\int\limits}
\def\sumintbbb{\sum\!\!\!\!\!\!\!\!\!\!\!\int\limits}
\def\sumintbbf{\sum\!\!\!\!\!\!\!\!\!\!\!\!\!\int\limits}
\def\sumintbff{\sum\!\!\!\!\!\!\!\!\!\!\!\!\!\int\limits}
\def\sumintfff{\sum\!\!\!\!\!\!\!\!\!\!\!\!\!\int\limits}
\def\Za{\frac{\zeta'(-1)}{\zeta(-1)}}
\def\Zc{\frac{\zeta'(-3)}{\zeta(-3)}}
\def \Slash{\slash \!\!\!\!}
\def \del{\partial}
\def\mn {\mu\nu}
\def\sp{\shortparallel}
\def\om {\omega}
\def\ti {\tilde}
\def\mn {\mu\nu}
\def\mr {\mu\rho}
\def\rn {\rho\nu}
\def\th {\theta}
\def\ov {\overline}
\def\sg {\sigma}

\def\bs{\!\!\!\!\!\!\!\!\!\!\!\!\!\!}
\def\bsa{\!\!\!\!\!\!\!\!\!}
\def\nn{\nonumber\\}

\section{NNLO Thermodynamics of Hot and Dense Deconfined QCD Matter}

\author{Munshi G Mustafa}

\bigskip

\begin{abstract}
In this proceedings I would like to discuss the equation of state (EoS) and other thermodynamic quantities of hot and dense QCD matter created in high energy heavy-ion collisions (known as Quark-Gluon Plasma (QGP)) within three-loop (Next-to-Next-Leading Order(NNLO))  Hard Thermal Loop Perturbation Theory (HTLpt) and compare the results with recent lattice QCD data.
\end{abstract}


\ccode{PACS numbers:}


\subsection{Introduction}
The determination of the equation of state (EOS) of QCD matter is extremely important in QGP phenomenology. Various effective models exists to describe the EOS of strongly interacting matter; however, one would prefer to utilize systematic first-principles QCD methods. Currently, the most reliable approach for determining the EOS is lattice QCD. At present, lattice calculations can be carried out at arbitrary temperature, however, they are limited to relatively small chemical potentials~\cite{Borsanyi:2012cr}.  Alternatively, perturbative QCD (pQCD) can be applied at high temperature and/or chemical potentials where the strong coupling ($g^2=4\pi \alpha_s$) is small in magnitude, and non-perturbative effects are expected to be small.  However, due to infrared singularities in the gauge sector~\cite{Linde:1978px,Linde:1980ts}, the perturbative expansion of the finite-temperature and density QCD partition function breaks down at order $g^6$ requiring 
non-perturbative input, albeit through a single numerically computable number. Up to order $g^6\ln(1/g)$, it is feasible to calculate the necessary coefficients using analytic (resummed) perturbation theory~\cite{Kajantie:2002wa}. However, one finds in practice that a strict expansion in the coupling constant converges only for temperatures many orders of magnitude higher than those relevant for heavy-ion collision experiments. The source of the poor convergence comes from contributions from soft momenta, $p \sim gT $. This suggests that one needs a reorganization of finite-temperature/density perturbation theory that treats the soft sector more carefully.

There are various ways of reorganizing the finite temperature/chemical potential perturbative series. Here we will focus on a method called hard-thermal-loop perturbation theory (HTLpt)~\cite{Braaten:1989mz,Haque:2024gva,Mustafa:2022got,Andersen:1999fw,Andersen:1999sf}. For scalar field theories one can use a simpler variant called “screened perturbation theory” (SPT)~\cite{Karsch:1997gj,Chiku:1998kd} which was inspired in part by variational perturbation theory. A gauge-invariant generalization of SPT called HTLpt was developed by Andersen, Braaten, and Strickland over two decades ago~\cite{Andersen:1999fw,Andersen:1999sf}. Since then HTLpt has been used to calculate thermodynamic functions at one loop order~\cite{Andersen:1999sf,Andersen:2002ey,Andersen:2003zk}, two loop order~\cite{Andersen:2002ey,Andersen:2003zk}, and three loop order at zero chemical potential as well as at finite chemical potential(s)~\cite{Andersen:2011sf,Haque:2013sja,Haque:2014rua}.

\subsection{NNLO (3-loop) HTL Thermodynamics}

\begin{figure}[h]
	\begin{center}
		\includegraphics[width=8cm,height=3cm]{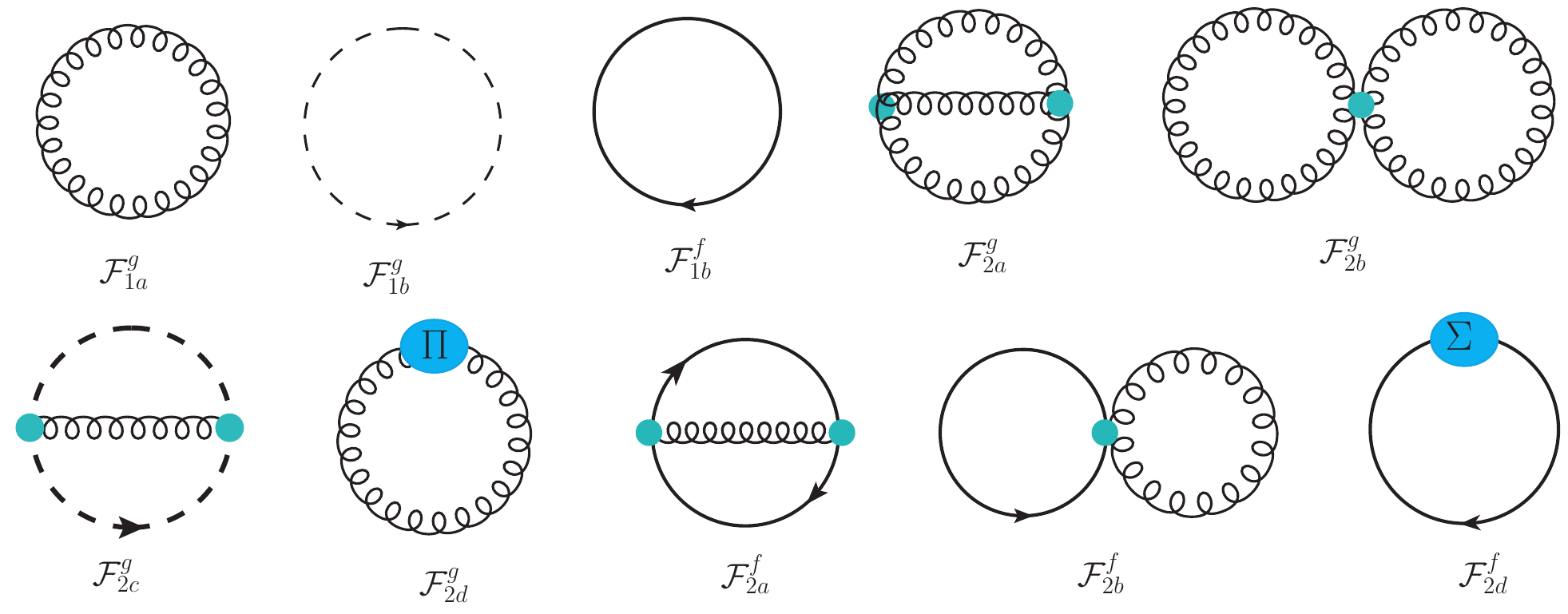}
		\caption{Diagrams containing fermionic lines relevant for NLO thermodynamics potential in HTLpt with finite chemical potential. Shaded circles indicate HTL $n$-point functions.}
		\label{diagramfig}
	\end{center}
\end{figure}

The diagrams needed for the computation of the HTLpt thermodynamic potential through NNLO are listed in 
Figs.~\ref{diagramfig} and \ref{feyn_diag3}. The shorthand notations used in Fig.~\ref{feyn_diag3}
have been explained in Fig.~\ref{shorthand}.

\begin{figure}[ht]
\begin{center}
	\caption{Three loop HTL Feynman diagrams that will contribute to the thermodynamic potential.}
	\label{feyn_diag3}
\end{center}
\end{figure}
\begin{figure}[t]
\begin{center}
	\includegraphics[width=10cm]{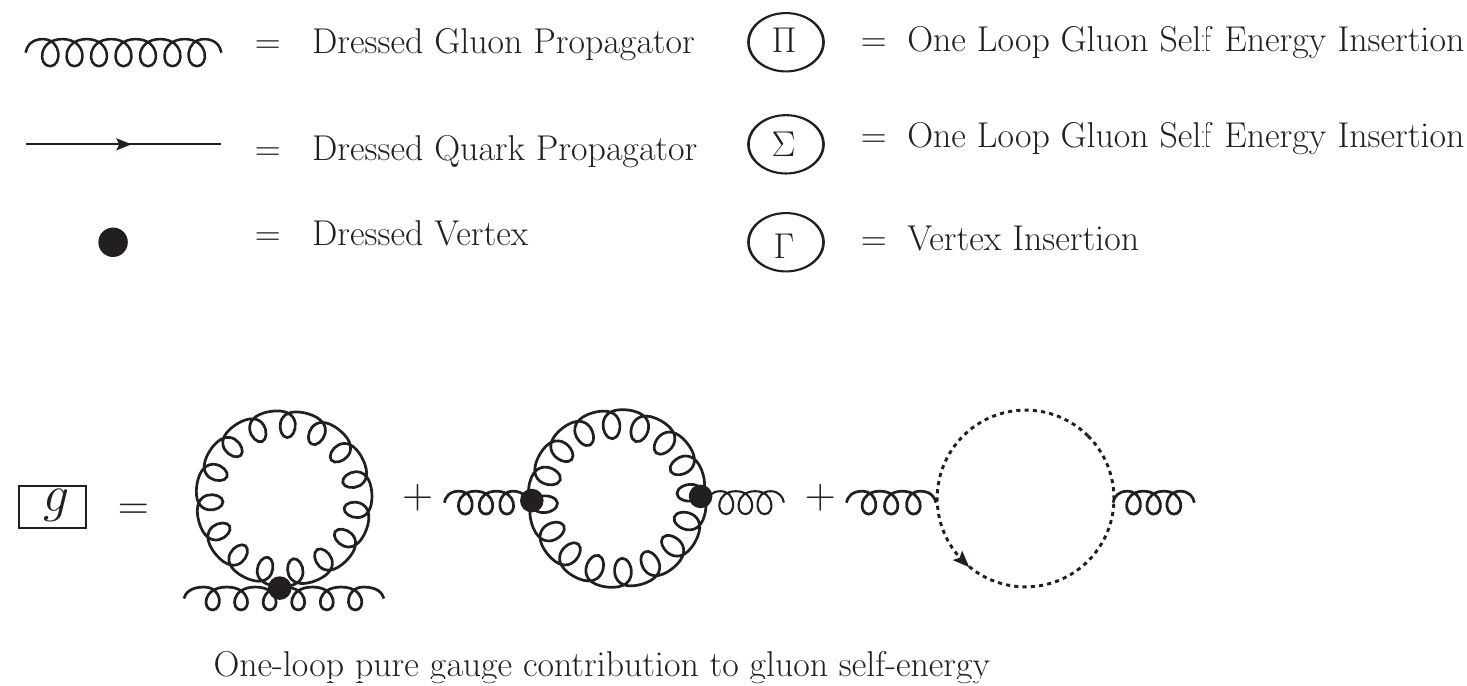}
	\caption{The shorthand notations used in Fig.~\ref{feyn_diag3}.}
	\label{shorthand}
	\end{center}
\end{figure}
\subsection{NNLO Thermodynamic Potential:}
The NNLO thermodynamic potential is obtained~\cite{Haque:2013sja,Haque:2014rua}  by adding all the Feynman diagrams in Figs.~\ref{diagramfig} and \ref{feyn_diag3} as
{\small {
\begin{eqnarray}
&&\frac{\Omega_{\rm NNLO}}{\Omega_0}
= \frac{7}{4}\frac{d_F}{d_A}\frac{1}{N_f}\sum\limits_f\(1+\frac{120}{7}\hat\mu_f^2+\frac{240}{7}\hat\mu_f^4\)
-\frac{s_F\alpha_s}{\pi}\frac{1}{N_f}\!\sum\limits_f\bigg[\frac{5}{8}\left(5+72\hat\mu_f^2+144\hat\mu_f^4\right)
-\frac{15}{2}\left(1+12\hat\mu_f^2\right)\hat m_D \nn
&-&\frac{15}{2}\bigg(2\ln{\frac{\hat\Lambda}{2}-1
	-\aleph(z_f)}\Big)\hat m_D^3
+90\hat m_q^2 \hat m_D\bigg] + \frac{s_{2F}}{N_f}\left(\frac{\alpha_s}{\pi}\right)^2\sum\limits_f\bigg[\frac{15}{64}\bigg\{35-32\(1-12\hat\mu_f^2\)\frac{\zeta'(-1)} {\zeta(-1)}+472 \hat\mu_f^2\nn
&+&1328  \hat\mu_f^4
+ 64\Big(-36i\hat\mu_f\aleph(2,z_f)+6(1+8\hat\mu_f^2)\aleph(1,z_f)+3i\hat\mu_f(1+4\hat\mu_f^2)\aleph(0,z_f)\Big)\bigg\}- \frac{45}{2}\hat m_D\left(1+12\hat\mu_f^2\right)\bigg] \nn
&+& \left(\frac{s_F\alpha_s}{\pi}\right)^2
\frac{1}{N_f}\sum\limits_{f}\frac{5}{16}\Bigg[96\left(1+12\hat\mu_f^2\right)\frac{\hat m_q^2}{\hat m_D}
+\frac{4}{3}\(1+12\hat\mu_f^2\)\(5+12\hat\mu_f^2\)
\ln\frac{\hat{\Lambda}}{2}
+\frac{1}{3}+4\gamma_E+8(7+12\gamma_E)\hat\mu_f^2\nn
&+&112\mu_f^4
- \frac{64}{15}\frac{\zeta^{\prime}(-3)}{\zeta(-3)}-
\frac{32}{3}(1+12\hat\mu_f^2)\frac{\zeta^{\prime}(-1)}{\zeta(-1)}
- 96\Big\{8\aleph(3,z_f)+12i\hat\mu_f\aleph(2,z_f)-2(1+2\hat\mu_f^2)\aleph(1,z_f)\nn
&-& i\hat\mu_f\aleph(0,z_f)\Big\}\Bigg]  + \left(\frac{s_F\alpha_s}{\pi}\right)^2
\frac{1}{N_f^2}\sum\limits_{f,g}\Bigg[\frac{5}{4\hat m_D}\left(1+12\hat\mu_f^2\right)\left(1+12\hat\mu_g^2\right)
+90\Bigg\{ 2\left(1 +\gamma_E\right)\hat\mu_f^2\hat\mu_g^2
-\Big\{\aleph(3,z_f+z_g)\nn
&+&\aleph(3,z_f+z_g^*)
+s_F 4i\hat\mu_f\left[\aleph(2,z_f+z_g)+\aleph(2,z_f+z_g^*)\right]
- 4\hat\mu_g^2\aleph(1,z_f)-(\hat\mu_f+\hat\mu_g)^2\aleph(1,z_f+z_g)- (\hat\mu_f-\hat\mu_g)^2\nn
&\times& \aleph(1,z_f+z_g^*) - 4i\hat\mu_f\hat\mu_g^2\aleph(0,z_f)\Big\}\Bigg\}-\frac{15}{2}\(1+12\hat\mu_f^2\)\(2\L-1-\aleph(z_g)\)  \hat 
m_D\Bigg]
+ \left(\frac{c_A\alpha_s}{3\pi}\right)\left(\frac{s_F\alpha_s}{\pi N_f}\right)\nn
&\times& \sum\limits_f\Bigg[
-\frac{235}{16}\Bigg\{\bigg(1+\frac{792}{47}\hat\mu_f^2+\frac{1584}{47}\hat\mu_f^4\bigg)\ln\frac{\hat\Lambda}{2}
-\frac{144}{47}\(1+12\hat\mu_f^2\)\ln\hat m_D+\frac{319}{940}\left(1+\frac{2040}{319}\hat\mu_f^2+\frac{38640}{319}\hat\mu_f^4\right)   
\nn
&-&\frac{24 \gamma_E }{47}\(1+12\hat\mu_f^2\)-\frac{44}{47}\(1+\frac{156}{11}\hat\mu_f^2\)\frac{\zeta'(-1)}{\zeta(-1)}
-\frac{268}{235}\frac{\zeta'(-3)}{\zeta(-3)}
-\frac{72}{47}\Big[4i\hat\mu_f\aleph(0,z_f)+\left(5-92\hat\mu_f^2\right)\aleph(1,z_f)\nn
&+& 144i\hat\mu_f\aleph(2,z_f)
+52\aleph(3,z_f)\Big]\Bigg\}
+ \frac{15}{2\hat m_D}\(1+12\hat\mu_f^2\)+90\frac{\hat m_q^2}{\hat m_D}
+\frac{315}{4}\Bigg\{\(1+\frac{132}{7}\hat\mu_f^2\)\ln \frac{\hat \Lambda}{2}
\nonumber\\
&+&\frac{11}{7}\(1+12\hat\mu_f^2\)\gamma_E+\frac{9}{14}\(1+\frac{132}{9}\hat\mu_f^2\)
+\frac{2}{7}\aleph(z_f)\Bigg\}\hat m_D 
\Bigg]
+ \frac{\Omega_{\rm NNLO}^{\rm YM}}{\Omega_0} \, ,
\label{finalomega}
\end{eqnarray}
}}
where $d_F=N_f N_c$ and $d_A=N_c^2-1$ , $c_A = N_c=3$ with $N_c$ is the number of
colours and  $s_F=N_f/2$. The sums over $f$ and $g$ include all quark flavours, $z_f = 1/2 - i \hat{\mu}_f$, $\Omega_0$ is the ideal gas value, and $\Omega_{\rm NNLO}^{\rm YM}$ is the pure-glue contribution is obtained in Ref~\cite{Andersen:2011sf}. We the express thermodynamic potential in terms of dimensionless variables: scaled Debye mass $\hat{m}_D = m_D/(2\pi T)$, scaled quark mass $\hat{m}_q = m_q/(2\pi T)$, $\hat{\mu} = \mu/(2\pi T)$,  $\hat{\Lambda} = \Lambda/(2\pi T)$ and $\hat{\Lambda}_g = \Lambda_g/(2\pi T)$ where $\Lambda$ and  $\Lambda_g$ are renormalization scales  for quark and gluon respectively as discussed in  Refs~\cite{Andersen:2011sf,Haque:2013sja,Haque:2014rua}.
\subsubsection{Pressure:}
The QGP pressure can be obtained directly from the thermodynamic potential in Eq.~\ref{finalomega} as
%
${\cal P}(T,\Lambda,\mu)=-\Omega_{\rm NNLO}(T,\Lambda,\mu) \,$ ,
%
where $\Lambda$ above is understood to include both scales $\Lambda_g$ and $\Lambda$. We note that in the ideal gas limit, the pressure becomes
%
${\cal P}_{\rm ideal}(T,\mu)=\frac{d_A\pi^2T^4}{45}\left[1+\frac{7}{4}\frac{d_F}{d_A}\left(1+\frac{120}{7}\hat\mu^2
+\frac{240}{7}\hat\mu^4\right)\right]$ .
%
\subsubsection{Energy density:}
The energy density can be derived as follows:
%
${\cal E}=T\frac{\partial{\cal P}}{\partial T}+\mu\frac{\partial{\cal P}}{\partial \mu}-{\cal P} \,$ .
%
We note that in the ideal gas limit, the entropy density becomes
%
${\cal E}_{\rm ideal}(T,\mu)=\frac{d_A\pi^2T^4}{15}\left[1+\frac{7}{4}\frac{d_F}{d_A}\left(1+\frac{120}{7}\hat\mu^2
+\frac{240}{7}\hat\mu^4\right)\right] $.
\subsubsection{Trace anomaly:}
The trace
anomaly is defined as ${\cal I}=({\cal E}-3{\cal P})$. In the ideal gas limit, the trace anomaly approaches to 
zero as the energy density equals three times the pressure. Nonetheless, upon incorporating interactions, the trace anomaly, also known as the interaction measure, deviates from zero.
\subsubsection{Speed of sound:}
Another quantity which is phenomenologically interesting is the speed of sound.  The speed of sound is defined as
$
c_s^2=\frac{\del{\cal P}}{\del{\cal E}} \, .
$
%
\subsubsection{Quark number susceptibilities:}
Having the complete thermodynamic potential as a function of temperature and chemical potential(s) allows us 
to calculate the quark number susceptibility (QNS). Generally, one can introduce an individual chemical potential 
for each quark flavour, forming $N_f$-dimensional vector $\bm{\mu}\equiv(\mu_1,\mu_2,...,\mu_{N_f})$.  
By taking derivatives of the pressure with respect to chemical potentials in this set, we obtain the quark 
number susceptibilities\,
%
$
\chi_{ijk\,\cdots}\left(T\right) \equiv \left. \frac{\partial^{i+j+k+ \, \cdots}\; {\cal P}\left(T,\bm{\mu}\right)}
{\partial\mu_u^i\, \partial\mu_d^j \, \partial\mu_s^k\, \cdots} \right|_{\bm{\mu}=0} \,$ .
%
Below we will use a shorthand notation for the susceptibilities by specifying derivatives by a string of quark flavours 
in superscript form, e.g. $\chi^{uu}_2 = \chi_{200}$, $\chi^{ds}_2 = \chi_{011}$, $\chi^{uudd}_4 = \chi_{220}$, etc.

\subsubsection{Result}
\begin{figure}[h]
\begin{center}
	\includegraphics[width=14cm,height=13cm]{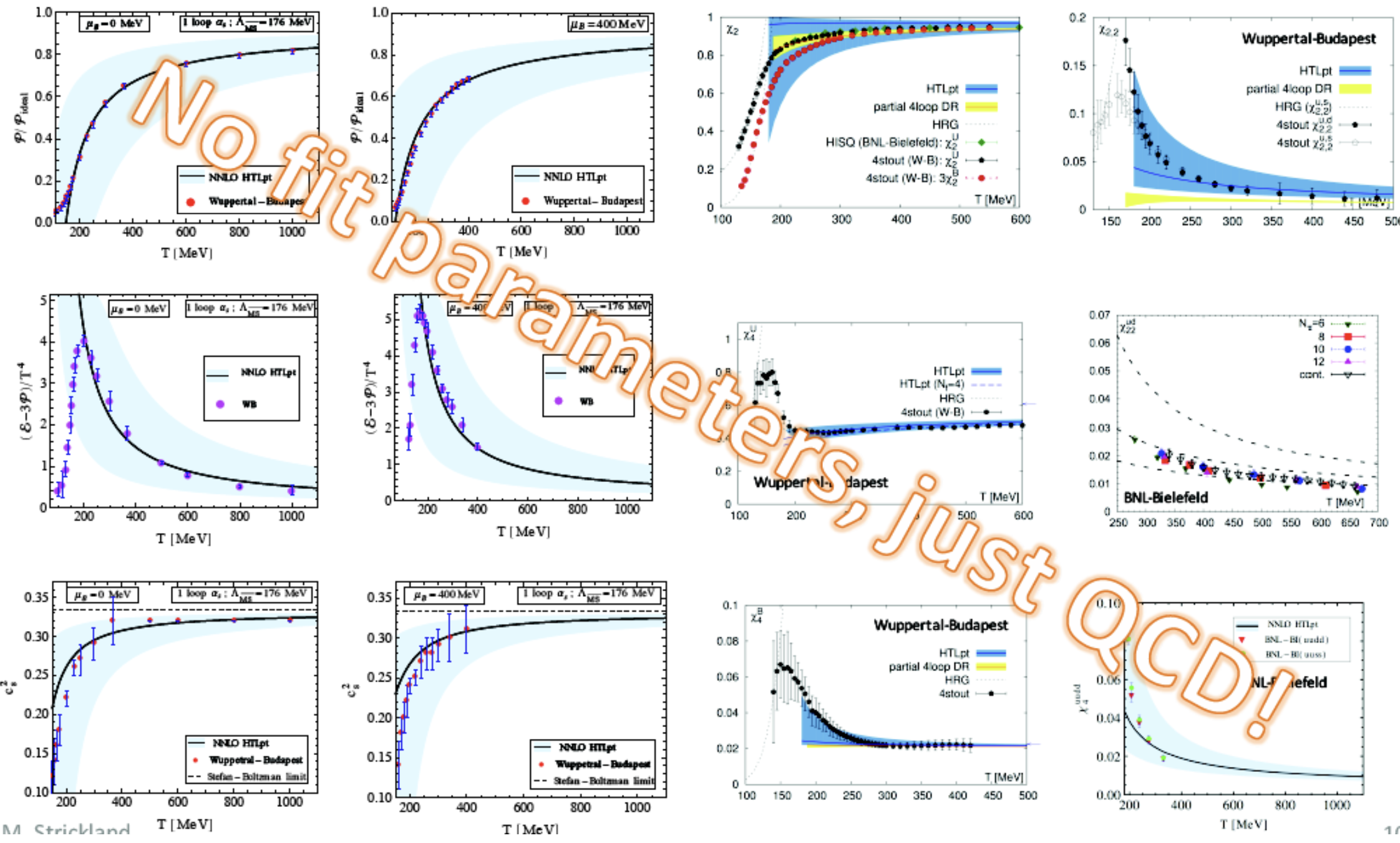}
	\caption{The NNLO HTLpt resummed result at finite T and chemical potential(s) is a renormalized and completely analytic expression that reproduces a host of lattice data for T $>$ 200 MeV. [Picture Courtesy: M. Strickland].}
	\label{all_nnlo}
	\end{center}
\end{figure}
In this subsection, we present our final results for the NNLO HTLpt pressure, energy density, entropy density, trace anomaly, speed of sound and various order quark number susceptibilities. We illustrate our NNLO result utilising the one loop  running coupling.  Regarding the renormalisation scale, we use separate scales, $\Lambda_g$ and $\Lambda$, for purely-gluonic and fermionic graphs, respectively.  We set the central values of these renormalization scales to be $\Lambda_g = 2\pi T$ and $\Lambda=2\pi \sqrt{T^2+\mu^2/\pi^2}$.  In all plots, the thick lines denote the result obtained using these central values and the light-blue band signifies the variation in the result under when both scales are varied by a factor of two, e.g. $\pi T \leq \Lambda _g \leq 4 \pi T$. For all numerical results below, we employ $c_A = N_c=3$ and $N_f=3$. We employ the Braaten-Nieto~\cite{Braaten:1995cm} mass prescription for the Debye mass $m_D$.  Following
Ref.~\cite{Andersen:2011sf} we take quark mass, $m_q=0$, which is the three loop variational solution.

We recall that the NLO result from two-loop HTLpt result deviates from the pQCD result through order $\alpha_s^3\ln\alpha_s$ at low temperatures, indicating a modest improvement over pQCD in terms of convergence and sensitivity to the renormalisation scale. Both LO and NLO HTLpt results~\cite{Haque:2012my}: exhibit less sensitivity to the choice of renormalisation scale compared to weak coupling results at successive orders of approximation. It's worth noting that although the 2-loop calculation improves upon the LO results by rectifying overcounting, it does so by pushing the problem to higher orders in $g$. This asymmetry in loop and coupling expansion in HTLpt results in contributions from higher orders in coupling at a given loop order, leading to overcounting at ${\cal O}(g^4)$ and ${\cal O}(g^5)$). A NNLO HTLpt (3-loop) calculation addresses this issue for various thermodynamic quantities through ${\cal O}(g^5)$, ensuring that the HTLpt results, when expanded in a strict power series in $g$, reproduce the perturbative result order-by-order through ${\cal O}(g^5)$. Extending the NLO expansion to NNLO provides new estimates for a comprehensive set of thermodynamic quantities and various order QNS with remarkable accuracy, agreeing with Lattice QCD\cite{Bazavov:2009zn,Borsanyi:2010cj,Borsanyi:2012cr} data down to a temperature of 200 MeV, and exhibiting complete analyticity with no fit parameters and gauge independence.

\section{Selected Highlights from Heavy-ion Experiments}

\author{Lokesh Kumar \\}

\begin{abstract}
	High energy heavy-ion collision experiments such as STAR and
        ALICE aim to create a strongly interacting matter called
        Quark-Gluon Plasma and study its properties. Understanding the dynamics of
        these collisions starting from the time when they happen is another
        important aspect of these experiments. Recent focus has also
        shifted to map the QCD phase diagram, the phase diagram of
        strongly interacting matter, and to search for QCD critical
        point and phase transition in the phase diagram. 
        We report on recent selected results from STAR and ALICE
        experiments covering different $\mu_B$ regions. The results include temperature measurement of
        QGP at RHIC, current status of the critical
        point search, (strange) particle production, and elliptic
        flow. In addition, an interesting observation regarding
        the observation of elliptic flow in lighter systems at $\mu_B\sim0$ is presented. The physics implications of these results are
        discussed. 
\end{abstract}



\subsection{Introduction}

\begin{figure}[h]
                \begin{center}
  \includegraphics[width = 8cm]{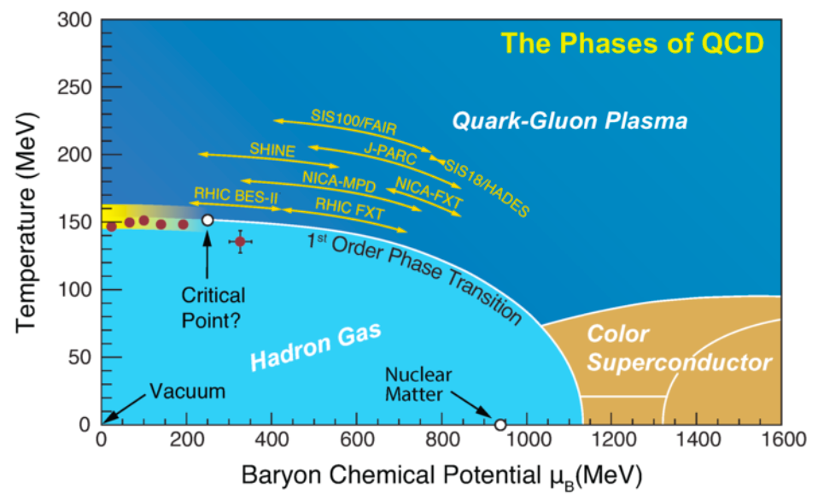}
                \vspace{-0.3cm}
		\caption{Conjectured QCD phase
                  diagram~\cite{qcd_ph}. It consists of various phase
                  structures such as critical point and first-order
                  phase boundary. Experimentally extracted freeze-out
                  points are shown by the red solid circles~\cite{STAR:2017sal}.
Also shown are the different regions covered by various experiments. 
                }\label{Fig:0}
                \end{center}
                 \vspace{-0.7cm}                
  \end{figure}

The Solenoidal Tracker At RHIC (STAR)~\cite{STAR:2005gfr} at Relativistic Heavy-Ion
Collider (RHIC), Brookhaven National Laboratory (BNL), USA and A Large Ion Collider
Experiment (ALICE)~\cite{ALICE:2008ngc} at Large Hadron Collider
(LHC), CERN, Switzerland, explore different regions of the Quantum
Chromodynamics (QCD) phase diagram. 
The phase
        diagram is usually plotted between temperature and baryon
        chemical potential ($\mu_B$) as shown in
        Fig.~\ref{Fig:0}. Both ALICE and STAR experiments cover
        different ($\mu_B$) regions of the phase diagram and hence are
        complementary to each other.
The STAR experiment covers the energy range in center-of-mass energy $\sqrt{s_{NN}}=$ 3.0
-- 200 GeV which corresponds to the $\mu_B$ region $\sim 25-720$
MeV~\cite{Cleymans:2005xv}. This region may be further divided into
(i) $\sqrt{s_{NN}} < $ 7.7 GeV ($\mu_B > 420$ MeV), which is
suitable to study particle production, search for exotic particles,
critical point search, first-order phase transition, and turn-off of
QGP signals. The other region (ii) $\sqrt{s_{NN}} = $ 7.7--200 GeV
($\mu_B = 25-420$ MeV), in addition to the (i) is suitable for studying QGP properties and dependence of the properties
on different systems. 
The ALICE experiment covers the energy range $\sqrt{s_{NN}}=$
2.76--5.5 TeV which corresponds to the $\mu_B\sim 0$
MeV. In addition to (ii) of STAR, ALICE experiment
provides a unique feature of high multiplicity events in pp collisions
that mimic the results similar to heavy-ion collisions. 

\subsection{Results and discussions}\label{Sec:alice}
We first start with presenting the temperature achieved at RHIC for
$\mu_B =$ 85--156 MeV. The
dielectrons (electron-positron pairs) are considered an excellent
probe for temperature of hot and dense QCD matter formed in heavy-ion
collisions. The system formed in heavy-ion collisions expands and
cools down rapidly. Throughout its expansion, it radiates both photons
and leptons pairs. The different ranges of the dielectron energy and
mass spectra are dominated by the radiation from different stages of
expansion.  The invariant mass, $M_{ee}$, of dielectron is a frame
independent variable and also immune to blue-shift effects and may
provide true information of the temperature of QGP at different stages
of the evolution. Dielectrons from the Low Mass Region (LMR), 0.4--1.02
GeV/$c^2$, mainly come from the hadronic phase. Those from the
Intermediate Mass Region (IMR), 1.02 -- 3.1 GeV/$c^2$, are emitted
from the QGP phase.  
\begin{figure}[h]
                \begin{center}
  \includegraphics[width = 7.3cm]{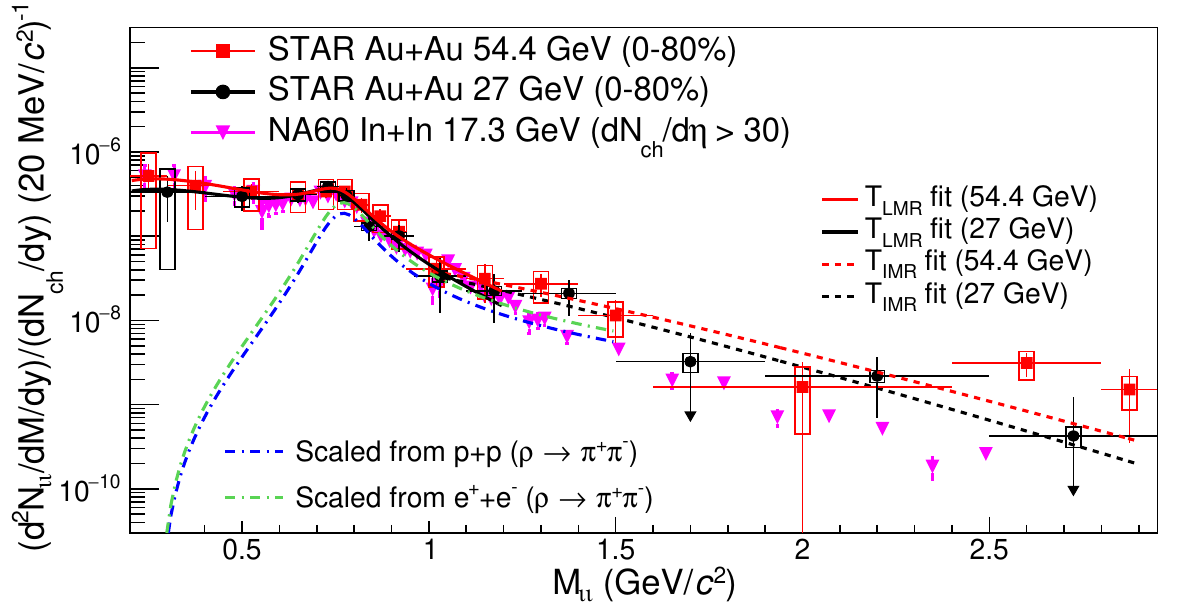}
 \includegraphics[width = 5.2cm]{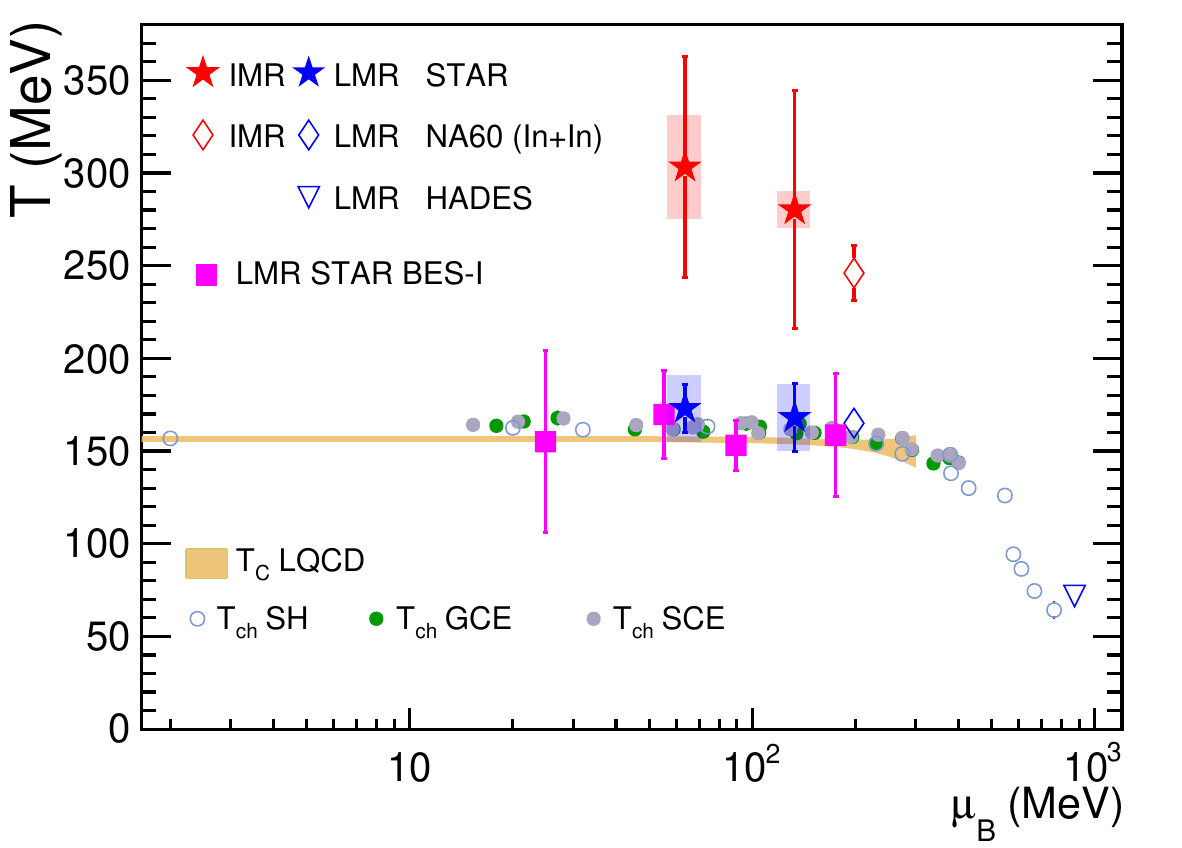}
                \vspace{-0.7cm}
		\caption{Left: Physical background (hadronic cocktail) subtracted thermal dielectron mass spectra from 54.4
GeV and 27 GeV compared with that of NA60
thermal dimuon data~\cite{STAR:2024bpc}. Right: Temperature versus $\mu_B$ showing
extracted temperatures from dileptons compared to critical point and
freeze-out temperatures~\cite{STAR:2024bpc}.}\label{Fig:1}
                \end{center}
                 \vspace{-0.7cm}                
  \end{figure}
Figure~\ref{Fig:1} (left) shows the thermal dielectron mass spectra from 54.4
GeV and 27 GeV compared with NA60
thermal dimuon data.
Dashed lines show the fitting curves for the corresponding
temperature extractions, $ (a BW+b M^{3/2}) e^{-M/T}$ for LMR
, where $a$ and $b$ are parameters, and $M^{3/2} e^{-M/T}$ for IMR.
The extracted $T_{\rm{LMR}}$ is $167 \pm 21$ (stat.) $\pm 18$ (syst.)
MeV and
$172 \pm 13$ (stat.) $\pm 18 (syst.)$ MeV for the Au+Au collisions at
$\sqrt{s_{NN}} =$ 27 and 54.4 GeV, respectively. For NA60 data the
extracted temperature is 165 $\pm$ 4 MeV. It is observed that the
$T_{\rm{LMR}}$ for different energies and systems are consistent with
each other and close to the critical temperature ($\sim 156$ MeV)
and the freeze-out temperature as shown in Fig.~\ref{Fig:1}
(right). The extracted 
$T_{\rm{IMR}}$ is $280 \pm 64$ (stat.) $\pm 10$ (syst.)
MeV and
$303 \pm 59$ (stat.) $\pm 28 (syst.)$ MeV for the Au+Au collisions at
$\sqrt{s_{NN}} =$ 27 and 54.4 GeV, respectively. $T_{\rm{IMR}}$ is
$245 \pm 17$  for NA60 In+In data. These values are higher than
the critical temperature suggesting that the dileptons emission in this mass
region is predominantly from the partonic phase. 


\begin{figure}[h]
	\begin{center}
		\includegraphics[width = 6.3cm]{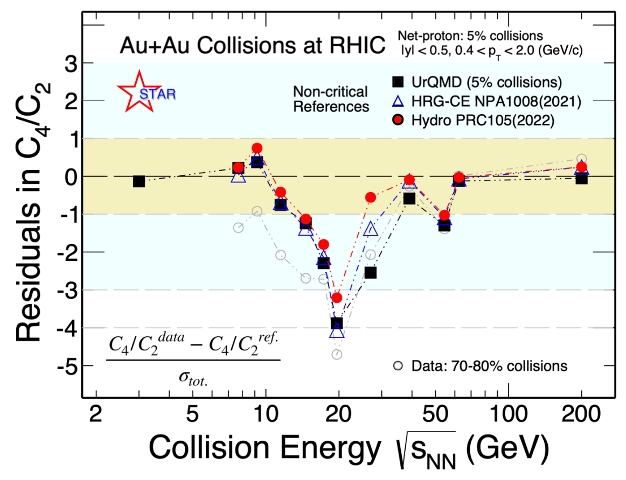}
		\includegraphics[width =6.2cm]{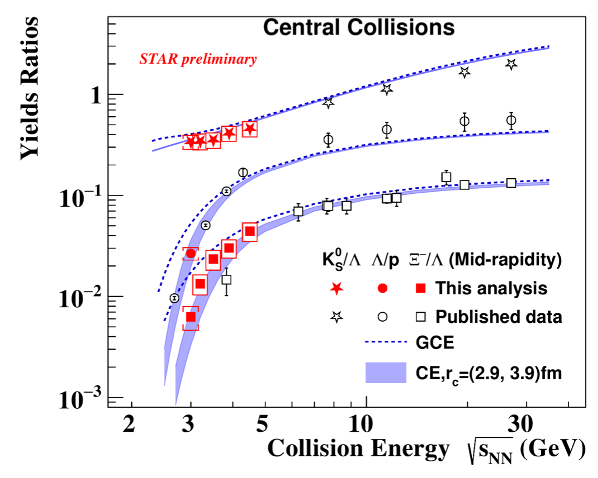}          
                \vspace{-0.7cm}
		\caption{Left: Residuals in cumulant ratio $C_4/C_2$
                  from experimental data with respect to models not
                  including the critical point plotted 
                  as a function of collision energy~\cite{cp_sqm24}.
Right: Particle yield ratios of $K^0_S/\Lambda$, $\Lambda/p$, and
$\Xi^-/\Lambda$ as a function of collision energy along with 
thermal models (CE and GCE) expectations~\cite{str_fix_sqm24}.
}\label{Fig:2}
	\end{center}
                \vspace{-0.7cm}
      \end{figure}
We now present the current status of the critical point search at RHIC
corresponding to $\mu_B=$ 25--420 MeV. The ratios of higher-order cumulants ($C_n$) of conserved quantities
such as net-baryons (protons) are proposed as probe for the critical
point search and can be compared directly with the ratios of
susceptibilities from the lattice QCD~\cite{}. It is expected that
$C_4/C_2$ when studied as a function of collision energy would show a non-monotonic
behavior near the critical point~\cite{}. 
Figure~\ref{Fig:2} (left) shows the residuals in $C_4/C_2$ from STAR experiment
with respect to models not including the critical point as a function of
collision energy. In this figure, the deviation in $C_4/C_2$ for net-protons in 0--5\%
central and 70--80\% peripheral collisions with respect to the models that do not include the
critical point such as UrQMD, hydrodynamics, and thermal models, is
plotted as a function of collision energy. The 0--5\% data shows a minimum
around $\sqrt{s_{NN}}=20$ GeV compared to reference models and
70--80\% collisions. The significance of deviation is around 3.2--4.7$\sigma$ at 20
GeV.

We now move towards the high $\mu_B$ ($>$ 420 MeV) region.
Figure~\ref{Fig:2} (right) shows the particle yield ratios of $K^0_S/\Lambda$, $\Lambda/p$, and
$\Xi^-/\Lambda$ as a function of collision energy. Results from fixed
target collisions at $\sqrt{s_{NN}}=$ 3 -- 4.5 GeV are also shown
($\mu_B =$ 590--720 MeV). These results are compared with the thermal
models, such as grand canonical (GCE) and the canonical (CE)
ensembles. The GCE represents the scenario in which quantum numbers
(baryon number, charge number, strangeness number)
are conserved on an average, while CE corresponds to the exact conservation of the
quantum numbers. From figure, it is seen that at relatively higher energies, both GCE and CE give
consistent results, however, when we move towards the lower energies
($\sqrt{s_{NN}} < 5$ GeV) the GCE fails to explain the data
while CE explains all the particle ratios with a strangeness
canonical radius $r_c=2.9-3.9$ fm. This suggests that the medium
properties of the system change at high $\mu_B$.

\begin{figure}[h]
	\begin{center}
		\includegraphics[width =12.5cm]{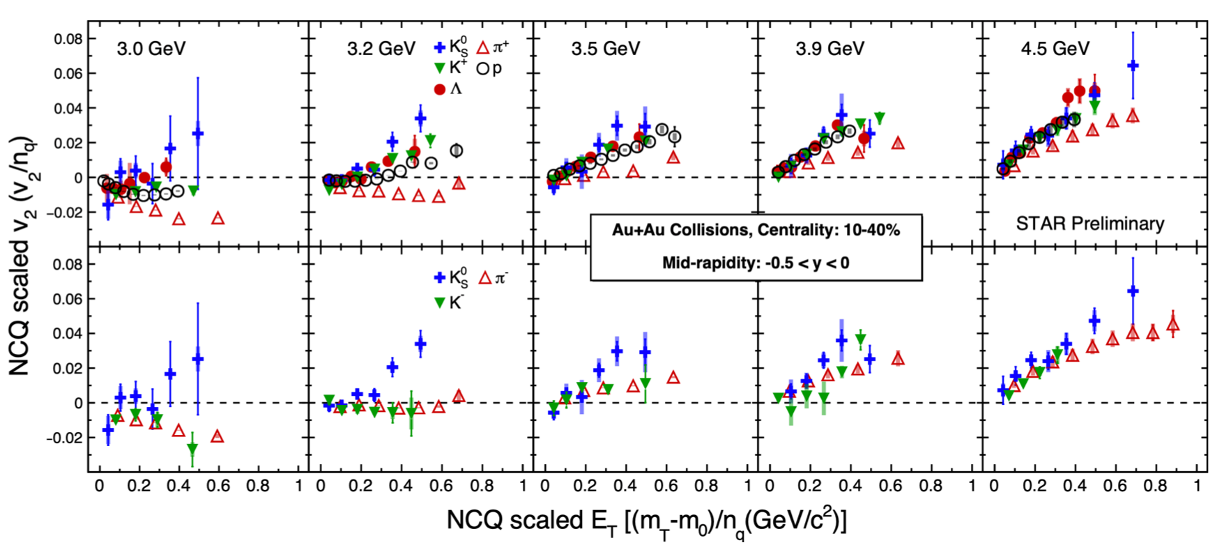}          
                \vspace{-0.3cm}
		\caption{ NCQ scaled $v_2$ versus NCQ scaled
                  transverse energy for various mesons and baryons in
                  Au+Au collisions at $\sqrt{s_{NN}}=3.0-4.5$ GeV~\cite{v2_fix_sqm24}.
		}\label{Fig:3}
	\end{center}
                \vspace{-0.7cm}
      \end{figure}
Continuing further discussion in the high $\mu_B$ region, Fig.~\ref{Fig:3}
shows the Number of Constituent Quarks (NCQ) scaled elliptic flow
($v_2$) as a function of NCQ scaled transverse energy ($E_T$),
i.e. ($m_T-m_0$), for various mesons and baryons in mid-central Au+Au collisions at $\sqrt{s_{NN}}=3.0-4.5$ GeV.
At top RHIC energy, this measurement shows that all baryons and mesons
follow a single curve or behavior at the intermediate transverse
momentum ($p_T$). It is
termed as NCQ-scaling and 
suggests that the observed flow is due to the partons and hence
considered as one of the signatures of the QGP formation at top RHIC
energy. Breaking of NCQ scaling would imply absence of QGP formation
or hadronic nature of the system formed in the collisions. From
figure, we observe that at $\sqrt{s_{NN}}=3.0$ GeV ($\mu_B$=720 MeV), the NCQ-scaling is
clearly broken suggesting that hadronic interactions dominate at this
energy. Going from 3 to 4.5 GeV in energy, it seems that the NCQ scaling is
getting restored, however not fully, suggesting that partonic interactions are getting
dominated. 

\begin{figure}[h]
	\begin{center}
		\includegraphics[width = 5.5cm]{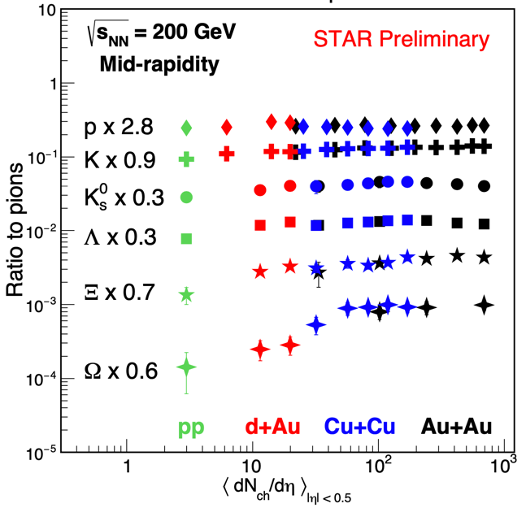}
		\includegraphics[width =7.0cm]{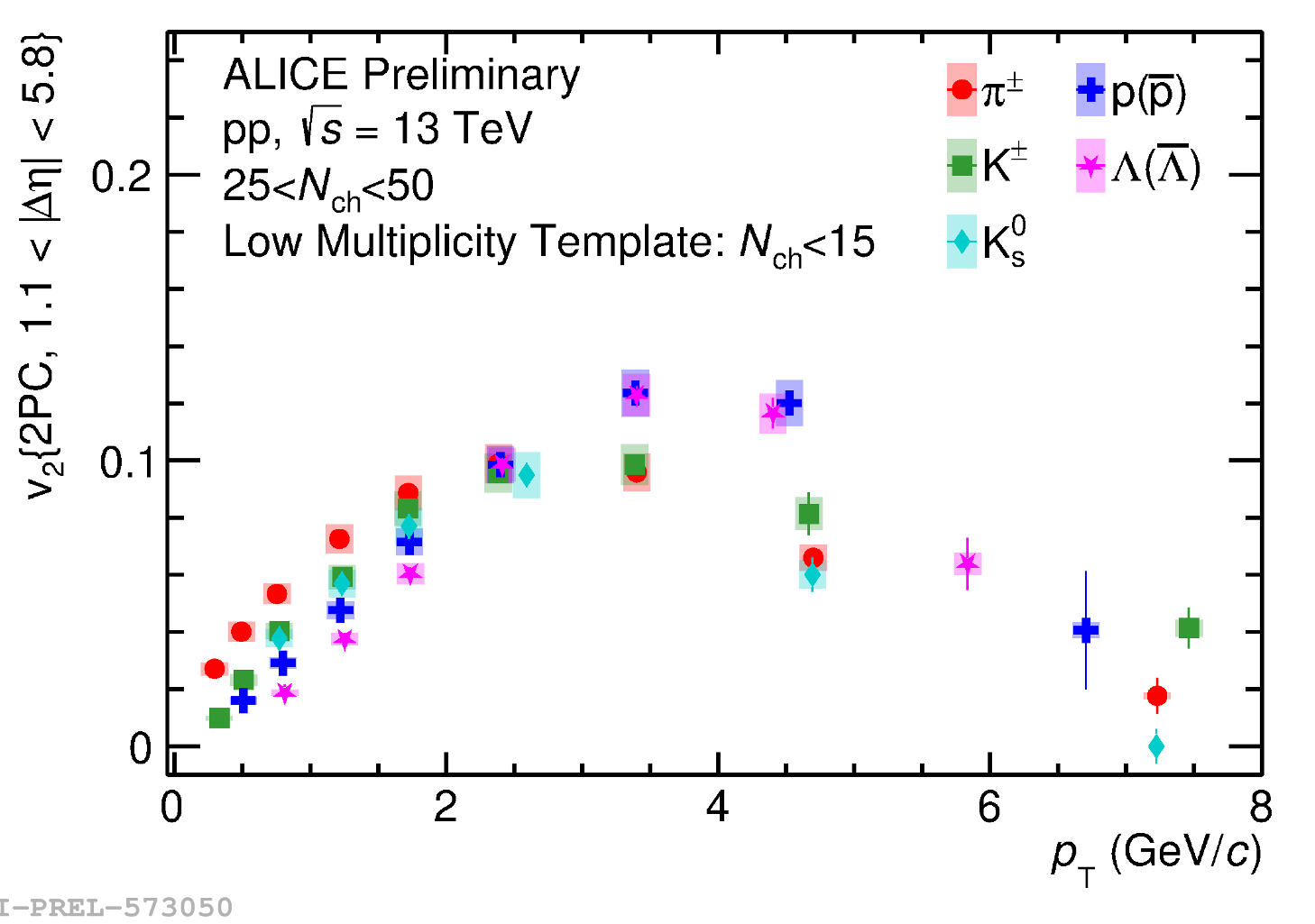}          
                \vspace{-0.6cm}
		\caption{Left: Ratios of various particles' yields to
                  pions as a function of charged particle
                  multiplicity~\cite{str_dau_sqm24}. 
Right: $v_2$ as a function of $p_T$ for high multiplicity pp
collisions at $\sqrt{s}=13$ TeV for various mesons and baryons~\cite{v2_pp_sqm24}.
}\label{Fig:4}
	\end{center}
                \vspace{-0.7cm}
      \end{figure}
Now we move to low $\mu_B$ side, starting with $\mu_B = 25$ MeV
($\sqrt{s_{NN}}=200$ GeV).
Figure~\ref{Fig:4} (left) shows the ratios of various particles' yields to
                  pions as a function of charged particle
                  multiplicity $\langle dN_{ch}/d\eta
                  \rangle_{|\eta|<0.5} $ for pp, $d+$Au, Cu+Cu, and Au+Au
                  collisions at  $\sqrt{s_{NN}}=200$ GeV. We observe
                  that the particle ratios follow a smooth trend from
                  pp to $d+$Au, to heavy-ion collisions. This suggests
                  that the particle production is independent of the
                  system size and is only dependent on the charged
                  particle multiplicity. Ratios involving strange
                  hadrons show an increase from pp to heavy-ion
                  collisions suggesting the strangeness
                  enhancement. Ratios involving strange particles with
                  more strangeness content (e.g. $\Omega$) show more
                  increase from pp to heavy-ions. 

                  Lastly, we discuss the results related to $\mu_B
                  \sim 0$  region. We stated above that
                  NCQ-scaling is considered as one of the signatures
                  of the partonic phase. This also implies that if we
                  plot $v_2$ as a function of $p_T$ without
                  scaling with NCQ, we would observe baryon-meson
                  splitting at intermediate $p_T$. The pp collisions are
                  assumed to be a reference where we do not expect the
                  QGP formation. Figure~\ref{Fig:4} (right) shows the
                  $v_2$ as a function of $p_T$ for high multiplicity
                  pp collisions ($25<N_{ch}<50$) at $\sqrt{s} =$ 13
                  TeV.  Results are shown for various mesons ($\pi, K,
                  K^0_{S})$ and baryons (p, $\Lambda$). The results
                  show a splitting between all mesons and baryons at
                  intermediate $p_T$. This observation is consistent
                  with the expectation of the partonic phase.
                  
\subsection{Summary}
We have reported the selected results from heavy-ion collision
experiments. The results are mainly from the STAR and ALICE
experiments. We have discussed the results based on different $\mu_B$
regions. We presented the extracted temperature for the LMR and IMR
regions from the dielectron invariant mass spectra in Au+Au collisions
at $\sqrt{s_{NN}} =$ 27 and 54.4 GeV ($\mu_B =$ 85--156 MeV). We observe that
the extracted $T_{\rm{LMR}}$ is consistent with the critical
temperature (156 MeV) and the freeze-out temperature. The extracted
$T_{\rm{IMR}}$ is higher than the critical temperature suggesting that
the dileptons emission in this mass region is mainly from the partonic
phase. Then we discussed the current status of the critical point
search ($\mu_B =$ 25--420 MeV). We observed that the net-proton $C_4/C_2$
for 0--5\%  central collisions shows a minimum
around $\sqrt{s_{NN}}=20$ GeV compared to reference models and
70--80\% collisions. The significance of deviation is around 3.2--4.7$\sigma$ at 20
GeV. After that, we discussed the results corresponding to
high-$\mu_B$ ($>$ 420 MeV). We presented particle yield ratios of $K^0_S/\Lambda$, $\Lambda/p$, and
$\Xi^-/\Lambda$ as a function of collision energy and compared with
the GCE and CE models.
We observed that at lower energies
($\sqrt{s_{NN}} < 5$ GeV)  the GCE fails to explain the data
while CE explains all the particle ratios with a strangeness
canonical radius $r_c=2.9-3.9$ fm. This suggests that the medium
properties of the system change at high $\mu_B$. In the still
high-$\mu_B$ region, the 
NCQ-scaling is broken at  $\sqrt{s_{NN}}=3.0$ GeV ($\mu_B$=720 MeV)
suggesting that hadronic interactions dominate at this energy. Going
from 3 to 4.5 GeV in energy, the NCQ scaling starts to get restored suggesting
that partonic interactions are getting dominated. Towards the
low-$\mu_B \sim25$ MeV, we observe that ratios of yields of various
particles to pion yields as a function of charged particle
multiplicity show a smooth evolution which is independent of system
size.
Ratios involving strange hadrons show an increase from pp to heavy-ion
collisions suggesting the strangeness enhancement and particles with
more strangeness content (e.g. $\Omega$) show more increase from pp to
heavy-ions. For $\mu_B \sim0$,  $v_2$ as a function of $p_T$ for high
multiplicity pp collisions (25 $< N_{ch} <$ 50) at $\sqrt{s} =$ 13 TeV shows a splitting between mesons and baryons at intermediate $p_T$. This observation is consistent with the expectation of the partonic phase.


%
\newpage
\section*{Acknowledgments}
This writeup is a compilation of the contributions presented at the "Hot QCD Matter 2024 conference" held from July 1-32, 2024, in Mandi, India. 
Amaresh Jaiswal extends his gratitude to the organizers of the Hot QCD Matter 2024 conference for hosting such a successful and well-organized event and insisting on having his contribution for the proceedings.
D.B. is supported by the Department of Science and Technology, Government of INDIA, under the SERB National Post-Doctoral Fellowship Reference No. PDF/2023/001762. P.P. is supported by the U.S. Department of Energy, Office of Science, through Contract No. DE-SC0012704.
AD would like to acknowledge the New Faculty Seed Grant (NFSG), NFSG/PIL/2024/P3825, provided by the Birla Institute of Technology and Science, Pilani, India. HM would like to express his gratitude to the organizers of HOT QCD MATTER conference at IIT Mandi for warm hospitality.
Munshi G. would like to acknowledge Najmul Haque, Aritra Bandyopadhyay, Jen Andersen, Michael Strickland and Nan Su for fruitful collaboration on this work. The author would also like to acknowledge the travel support  and hospitality received from the organisers of the second Hot QCD Matter Conference held in Indian Institute of Technology, Mandi, Himachal Pradesh, India, during July 1-3, 2024.
Lokesh Kumar would like to thank the organizers of the Hot QCD Matter 2024 and IIT Mandi
for invitation to discuss these results
 and thanks to the STAR and ALICE collaborations. The support from the DST project
No. SR/MF/PS-02/2021-PU (E-37120) is acknowledged.
RC expresses gratitude to the organizing committee for their support in attending the conference. Special thanks to Sanchari Thakur and Pingal Dasgupta for providing the figures.
Sumit would like to thank Najmul Haque and Binoy Krishna Patra for their collaboration and helpful discussion.
 C.W.A. and T.Z.W acknowledge the DIA
programme. This work was partly supported by the Doctoral
Fellowship in India (DIA) programme of the Ministry of Education, Government of India. S.N. and A.D. gratefully acknowledge the Ministry of Education (MoE), Government of India. 
Soham would like to thank Samapan Bhadury, Wojciech Florkowski, Amaresh Jaiswal, and Radoslaw Ryblewski for their valuable collaboration and support. Prabhakar Palni would like to acknowledge the support from the SERB Seminar/Symposia Scheme (File no. SSY/2024/000612) and the IIT Mandi SRIC seed grant support (Ref. No. IITM/SG/2024/01-2348). Anuraag Rathore and Prabhakar Palni would like to acknowledge the School of Physical Sciences IIT Mandi and Param Himalaya computing facility. Arvind Khuntia gratefully acknowledge and thank INFN Bologna, Italy for INFN postdoctoral fellowship and support.


\end{document}